\documentstyle[12pt,a4,amstex,fleqn,float,epsfig,graphics]{report}

\begin{document} \setlength{\unitlength}{1mm}


\def\beqd            {\begin{displaymath}}
\def\eeqd            {\end{displaymath}}
\def\baa             {\begin{array}}
\def\eaa             {\end{array}}
\def\beqaa           {\begin{eqnarray}}
\def\eeqaa           {\end{eqnarray}}
\def\beqaad          {\begin{eqnarray*}}
\def\eeqaad          {\end{eqnarray*}}

\def\bce             {\begin{center}}
\def\ece             {\end{center}}
\def\btabu           {\begin{tabular}}
\def\etabu           {\end{tabular}}

\def\ind             {\indent}
\def\noi             {\noindent}
\def\nn              {\nonumber}

\newcommand{\eq}[1]  {\mbox{(\ref{eq:#1})}}
\newcommand{\fig}[1] {\mbox{Fig.~\ref{fig:#1}}}   
\newcommand{\Fig}[1] {\mbox{Figure~\ref{fig:#1}}}

\def\a               {\alpha}
\def\b               {\beta}
\def\g               {\gamma}
\def\d               {\delta}
\def\l               {\lambda}
\def\s               {\sigma}
\def\t               {\theta}
\def\x               {\chi}

\def\G               {\Gamma}
\def\D               {\Delta}
\def\L               {\Lambda}

\def\ti              {\tilde}
\def\pr              {^{\prime}}
\def\kr              {^{\dagger}}

\def\mn              {\mu\nu}
\def\gmu             {\g^\mu}
\def\gnu             {\g^\nu}
\def\gmn             {\g_{\mu\nu}}

\def\de              {\partial}
\def\demu            {\partial_{\mu}}
\def\delr            {\!\stackrel{\leftrightarrow}{\partial^\mu}\!}


\def\sq              {\ti q}
\def\sqi             {\ti q_i^{}}
\def\sqj             {\ti q_j^{}}
\def\asqi            {\ti q_i^*}
\def\sqL             {\ti q_{L}^{}}
\def\sqR             {\ti q_{R}^{}}
\def\sqe             {\ti q_{1}^{}}
\def\sqz             {\ti q_{2}^{}}
\def\sqp             {\ti q^{\prime}}
\def\qp              {q^{\prime}}

\def\st              {\ti t}
\def\sb              {\ti b}

\def\sqbar  {\Bar{\Tilde q}^{}}
\def\stbar  {\Bar{\Tilde t}}
\def\sbbar  {\Bar{\Tilde b}}

\def\su              {\ti u}
\def\sd              {\ti d}
\def\sui             {\ti u_i^{}}
\def\suj             {\ti u_j^{}}
\def\sdi             {\ti d_i^{}}
\def\sdj             {\ti d_j^{}}

\def\sl              {\ti \ell}
\def\se              {\ti e}
\def\smu             {\ti \mu}
\def\stau            {\ti \tau}
\def\snu             {\ti \nu}

\def\ch              {\ti \x^+}
\def\chpm            {\ti \x^\pm}
\def\chp             {\ti \x^+}
\def\chpc            {\ti \x^{+c}}
\def\chpb            {\bar{\ti \x}^+}
\def\chpcb           {\bar{\ti \x}^{+c}}
\def\chm             {\ti \x^-}

\def\nt              {\ti \x^0}
\def\ntb             {\bar{\ti \x}^0}

\def\Hk              {H_{\!k}^{}}

\def\sg              {\ti g}

\def\qp              {q ^{'}}
\def\bq              {\bar{q}}


\newcommand{\msq}[1]   {m_{\ti q_{#1} }}
\newcommand{\msqp}[1]  {m_{\ti q'_{#1} }}
\newcommand{\mst}[1]   {m_{\ti t_{#1} }}
\newcommand{\msb}[1]   {m_{\ti b_{#1} }}

\newcommand{\msl}[1]   {m_{\ti \ell_{#1} }}
\newcommand{\mstau}[1] {m_{\ti \tau_{#1} }}
\newcommand{\msnu}     {m_{\ti \nu }}

\newcommand{\mnt}[1]   {m_{\ti \x^{0}_{#1} }}
\newcommand{\mch}[1]   {m_{\ti \x^{+}_{#1} }}
\newcommand{\mchp}[1]  {m_{\ti \x^{+}_{#1} }}
\newcommand{\mchm}[1]  {m_{\ti \x^{-}_{#1} }}

\def\msg             {m_{\sg}}
\def\mV              {m_{\!V}}
\def\mH              {m_{\!H}}


\def\tW              {\t_W}
\def\sW              {\sin^2\t_W}
\def\tsq             {\t_{\ti q}}
\def\tst             {\t_{\st}}
\def\tsb             {\t_{\sb}}
\def\sth             {\sin\t}
\def\cth             {\cos\t}
\def\sst             {\sin\theta_{\ti t}}
\def\cst             {\cos\theta_{\ti t}}
\def\ssb             {\sin\theta_{\ti b}}
\def\csb             {\cos\theta_{\ti b}}


\def\aik             {a_{ik}}
\def\bik             {b_{ik}}
\def\lij             {\ell_{ij}}
\def\kij             {k_{ij}}

\def\hg              {^{(g)}}
\def\hsg             {^{(\sg )}}
\def\hsq             {^{(\sq )}}

\def\CL              {C_{\!qL}^{}}
\def\CR              {C_{\!qR}^{}}
\def\PL              {P_L^{}}
\def\PR              {P_R^{}}


\def\lag             {{\cal L}}
\def\Lag             {{\cal L}}
\def\lum             {{\cal L}}
\def\R               {{\cal R}}
\def\Rsq             {{\cal R}^{\sq}}
\def\Rst             {{\cal R}^{\st}}
\def\Rsb             {{\cal R}^{\sb}}
\def\M               {{\cal M}}
\def\Oas             {{\cal O}(\alpha_{s})}
\def\Vcal            {{\cal V}}
\def\Wcal            {{\cal W}}


\def\rzw             {\sqrt{2}}

\def\ee              {e^+e^-}
\def\mumu            {\mu^+\mu^-}

\def\Emiss           {E\llap/}
\def\Etmiss          {E_T^{\rm miss}}
\def\ptmiss          {p_{T}^{\rm miss}}
\def\infinity        {\infty}

\def\BR              {\rm BR}
\def\ev              {\:{\rm eV}}
\def\mev             {\:{\rm MeV}}
\def\gev             {\:{\rm GeV}}
\def\tev             {\:{\rm TeV}}
\def\cm              {\:{\rm cm}}
\def\fb              {\:{\rm f\/b}}
\def\fbi             {\:{\rm fb}^{-1}}
\def\pbi             {\:{\rm pb}^{-1}}
\def\hc              {{\rm h.c.}}
\def\cc              {{\rm c.c.}}

\def\eg              {{\em e.g.}}
\def\Eg              {{\em E.g.}}
\def\ie              {{\em i.e.}~}
\def\etal            {{\em et al.}}
\def\etc             {{\em etc.}}

\newcommand{\medf}[2] {{\footnotesize{\frac{#1}{#2}} }}
\newcommand{\smaf}[2] {{\textstyle \frac{#1}{#2} }}
\def\onesq            {{\textstyle \frac{1}{\sqrt{2}} }} 
\def\onehf            {{\textstyle \frac{1}{2} }} 
\def\oneth            {{\textstyle \frac{1}{3} }}  
\def\twoth            {{\textstyle \frac{2}{3} }}
\def\onefo            {{\textstyle \frac{1}{4} }}
\def\forth            {{\textstyle \frac{4}{3} }}


\newcommand{\ZPC}[1]  {{\em Z. Phys.}    {\bf C#1}}
\newcommand{\PLB}[1]  {{\em Phys. Lett.} {\bf B#1}}
\newcommand{\NPB}[1]  {{\em Nucl. Phys.} {\bf B#1}}
\newcommand{\PRD}[1]  {{\em Phys. Rev.}  {\bf D#1}}
\newcommand{\PRep}[1] {{\em Phys. Rep.}  {\bf #1}}
\newcommand{\PRL}[1]  {{\em Phys. Rev. Lett.} {\bf #1}}
\newcommand{\PLett}[1]  {{\em Phys. Lett.} {\bf #1}}


\newcommand{\imat}{{\rm 1\kern-.12em
\rule{0.3pt}{1.5ex}\raisebox{0.0ex}{\rule{0.1em}{0.3pt}}}}

\newcommand{\gsim}{\;\raisebox{-0.9ex}
           {$\textstyle\stackrel{\textstyle >}{\sim}$}\;}

\newcommand{\lsim}{\;\raisebox{-0.9ex}{$\textstyle\stackrel{\textstyle<}
           {\sim}$}\;}

\newcommand{\poss}[2] { \left\{ \!\! 
   {\footnotesize\begin{array}{r} #1 \\ #2 \end{array}} \!\!\right\} }

\newcommand{\mbf}[1] {\mbox{\boldmath ${\bf #1}$ }}

\thispagestyle{empty}

\begin{center}

\vspace*{1.9cm}

\LARGE
{\bf\boldmath Stop and Sbottom Phenomenology \\[3mm] in the MSSM}
  
\vspace{1.4cm}

\Large 
{\sc Sabine Kraml}\\[4mm]

\small
{\it Institut f\"ur Hochenergiephysik \\
     \"Osterreichische Akademie der Wissenschaften \\
     A--1050 Wien, Nikolsdorfer Gasse 18}

\end{center}

\vfill
\hphantom{.}\hfill hep--ph/9903257

\clearpage

\begin{titlepage} 
\begin{center}

\vspace*{-0.5cm}  

{\LARGE \sc Dissertation}   

\vspace{1.6cm}

\LARGE
{\bf\boldmath Stop and Sbottom Phenomenology \\[3mm] in the MSSM}
  
\vspace{1.5cm}

\normalsize
ausgef\"uhrt zum Zwecke der Erlangung des akademischen Grades eines\\
Doktors der technischen Wissenschaften unter der Leitung von

\vspace{1cm}

\large  
Univ. Prof. Dr. Walter Majerotto\\[3mm]

\normalsize
E 136\\
Institut f\"ur Theoretische Physik der TU Wien\\
{\footnotesize und}\\
Institut f\"ur Hochenergiephysik der \"OAW 

\vspace{1cm}

\normalsize
eingereicht an der Technischen Universit\"at Wien\\
Technisch--naturwissenschaftliche Fakult\"at 

\vspace{6mm}

von 

\vspace{6mm}

\Large 
{\sc DI Sabine Kraml}\\[3mm]

\normalsize
Matr.\,Nr. 8925578\\
A--1120 Wien, Meidlinger Hauptstr. 38/7 \\
Email: kraml{\footnotesize @@}hephy.oeaw.ac.at 

\end{center}

\vfill

\normalsize
Wien, am 13. Jan. 1999 \hfill \underline{\hspace{5cm}}

\end{titlepage}


\clearpage
\thispagestyle{empty}
\vspace*{4cm}
\begin{flushright}
\emph{Alle Tr\"aume des Lebens beginnen in der Jugend}\\
{\scriptsize Heinrich Harrer}
\end{flushright}
\clearpage

\pagenumbering{roman}

\section*{Preface}
\addcontentsline{toc}{section}{Preface}

In the past years there has been considerable enthusiasm 
for the idea that nature might be supersymmetric. 
The reasons are manyfold.  
Some people were motivated by the idea to unify bosons and fermions, 
others were attracted by the fact that local supersymmetry involves gravity. 
One of the most appealing features of supersymmetry (SUSY) is, however, 
that it provides a solution to the hierarchy problem by 
protecting the electroweak scale from large radiative corrections --- 
provided that SUSY particles exist at or below the TeV scale. 

The experimental search for supersymmetric particles, which 
is one of the primary tasks at present and future colliders, 
relies on detailed phenomenological studies. 
In this thesis, I study the phenomenology of stops and sbottoms 
(the superpartners of the top and bottom quarks) in the framework 
of the minimal supersymmetric extension of the Standard Model (MSSM). 

\noi 
This thesis is organized as follows:

\noi
The first chapter serves as an introduction to the MSSM. 
I briefly review the model and explain why stops and sbottoms provide 
an interesting playground for exploring and testing this model. 

\noi
In the second chapter I discuss stop and sbottom production 
at lepton colliders. 
I give the analytical formulae for the production cross sections 
at $e^+e^-$ and $\mu^+\mu^-$ colliders including (general) beam 
polarization. 
A numerical analysis shows the cross sections that 
can be expected at LEP2, an $e^+e^-$ Linear Collider, and a Muon 
Collider. 
In particular, I address the topics of what can be gained by using 
polarized beams and how one can determine the MSSM parameters by 
cross section measurements.
 
\noi
The third chapter is devoted to stop and sbottom decays including 
1--loop SUSY--QCD corrections. 
The calculations are done in the on--shell renormalization scheme 
using dimensional reduction. 
I present the analytical formulae for the $\Oas$ corrected decay 
widths and explain the subtleties that have to be taken into account 
in the calculations. 
Moreover, I perform a detailed numerical analysis of squark
decay widths and branching ratios at tree level and $\Oas$.  
It turns out that stops and sbottoms can have very complex 
decay patterns. 
SUSY--QCD corrections can change the individual decay widths 
and branching ratios by a few ten percent. 
Hence they are important for precision measurements.

This work grew from a series of workshops on physics  
at present and future colliders. 
Its aim is to give a comprehensive overview of the phenomenology of 
stops and sbottoms and to serve as a basis for Monte Carlo studies.

\clearpage
\section*{Acknowledgements}
\addcontentsline{toc}{section}{Acknowledgements}

Above all, I thank my `Doktorvater' Prof.~Walter~Majerotto for his 
guidance and advice, and the inspiring working atmosphere. 
He also gave me the opportunity to participate in several 
international workshops, including numerous visits to CERN.  

\noi
I thank Dr.~H.~Eberl for a close and lively collaboration. 
Likewise, I thank Prof.~A.~Bartl, Dr.~T.~Gajdosik, and Dr.~W.~Porod 
for many useful discussions. 
During the past two years I also enjoyed collaborating with 
Prof.~K.~Hidaka and Prof.~Y.~Yamada. I am grateful for this, too. 

\noi
I am indebted to the `Rechengruppe' of our institute, 
led by Dr.~G.~Walzel, for excellent computing support. 
Moreover, I thank the Vienna CMS group for their hospitality 
during my visits to CERN. 

\noi
In a work that expands over years, dissappointments  
are of course not unknown. Though, joy clearly prevailed. 
In this context I wish to thank DI~W.~Adam, Dr.~A.~Boresch, 
DI~B.~G\"odel, Dr.~R.~Veenhof, and the members of the 
SKI Dojo Wien for their company and friendship. 
   
\noi
This work was supported in part by the 
``Fonds zur F\"orderung der wissenschaftlichen Forschung'' 
of Austria, project no. P10843--PHY.

\tableofcontents 
\addcontentsline{toc}{section}{Contents}

\clearpage
\pagenumbering{arabic}

\chapter{The Minimal Supersymmetric Standard Model}

\section{Definition of the Model}

The simplest ---and most popular--- supersymmetric model is the 
straightforward supersymmetrization of the Standard Model (SM),  
where one introduces only those couplings and fields that are necessary 
for consistency. 
This is known as the {\em Minimal Supersymmetric Standard Model} (MSSM)
\cite{nilles,haberkane,barbieri}. 
The MSSM is build up as follows:

\noi
\begin{itemize}
  \item In addition to the gauge boson fields, spin--$\onehf$ ``{\sc gaugino}'' 
        fields are introduced.  
        The partners of $B_\mu$ and $W^i_{\!\mu}$ are denoted $\ti B$ and $\ti W^i$. 
        In analogy to the photon, the $Z$ and the $W^\pm$ bosons,
        one can form a photino $\ti\g$, a Z--ino $\ti Z$, 
        and $\ti W^\pm$--inos from the $\ti B$ and $\ti W^i$ fields. 
        The superpartners of the gluons are the gluinos $\sg$. 
  \item Quarks and leptons get spin--0 partners called 
        {\sc squarks} and {\sc sleptons}. 
        As there has to be a superpartner for each degree of freedom, 
        two bosonic fields are needed per SM fermion. They are named  
        ``left'' and ``right'' states $\sq_L^{}$, $\sq_R^{}$ and 
        $\ti\ell_L^{}$, $\ti\ell_R^{}$.
  \item Moreover, one needs {\sc two complex Higgs doubletts} with 
        hypercharges $\pm 1$ in order to give masses to up-- and down--type 
        quarks and leptons, and to cancel anomalies. 
        The Higgs fields are also assigned
        spin--$\onehf$ partners, the so--called {\sc higgsinos}.   
\end{itemize}
The field content of the MSSM is shown in Table~\ref{tab:mssm}. 
Supersymmetry (SUSY) \footnote{For an introduction to supersymmetric field theory in 
  general and the MSSM in particular see also \cite{drees,martin}; 
  for a textbook on supersymmetry I refer the reader to \cite{bailin-love}.} 
in its local version includes gravity; 
the resulting theory is know as \emph{supergravity}.  
The model then also includes the graviton (spin--2) 
and its fermionic partner the gravitino (spin--$\smaf{3}{2}$). 

\begin{table}[t] \center
\renewcommand{\arraystretch}{1.5}
\begin{tabular}{|l|cc|cc|}
\hline
Superfield & Particle & Spin & Superpartner & Spin \\
\hline
\hline
\quad $\:V_1$ & $B_\mu$   & 1 & $\ti B\,\,$ & $\onehf$ \\
\quad $\:V_2$ & $W^i_\mu$ & 1 & $\ti W^i$  & $\onehf$ \\
\quad $\:V_3$ & $G^a_\mu$ & 1 & $\ti g^a$ & $\onehf$ \\
\hline
\quad $\:Q$   & $(u,\,d)_L^{}$ & $\onehf$ & $(\ti u_L^{},\,\ti d_L^{})$ & 0 \\
\quad $\:U^c$ & $\bar u_R^{}$  & $\onehf$ & $\ti u_R^*$ & 0 \\
\quad $\:D^c$ & $\bar d_R^{}$  & $\onehf$ & $\ti d_R^*$ & 0 \\
\hline
\quad $\:L$   & $(\nu,\,e)_L^{}$ & $\onehf$ & $(\ti\nu_L^{},\,\ti e_L^{})$ & 0 \\
\quad $\:E^c$ & $\bar e_R^{}$    & $\onehf$ & $\ti e_R^*$ & 0 \\
\hline
\quad $\:H_1^{}$ & $(H_1^0,\, H_1^-)$ & 0 
    & $(\ti H_1^0,\, \ti H_1^-)$    & $\onehf$ \\
\quad $\:H_2^{}$ & $(H_2^+,\, H_2^0)$ & 0 
    & $(\ti H_2^+,\, \ti H_2^0)$    & $\onehf$ \\
\hline
\end{tabular}
\renewcommand{\arraystretch}{1}
\caption{Field content of the MSSM.}
\label{tab:mssm}
\end{table}

Let us now turn to the Lagrangian of the MSSM. 
Clearly, gauge interactions are determined by the gauge group, 
which is the same as in the Standard Model: 
${\rm SU}(3)_C\times {\rm SU}(2)_L\times {\rm U}(1)_Y$. 
For a concise derivation of these interactions, see 
\eg~ \cite{haberkane,drees,martin,bailin-love}. 
Masses and couplings of the matter fields 
are determined by the superpotential $\Wcal$. 
The choice of the gauge group 
constrains $\Wcal$ but does not fix it completely. 
Holding to the priciple of minimality, that means introducing 
only those terms that are necessary to build a consistent model, 
we get:
\begin{equation}
  \Wcal =  \sum_{i,j=1}^3 
  \left[ (h_E)_{ij}^{}\, H_1^{} L_i^{} E_j^c
       + (h_D)_{ij}^{}\, H_1^{} Q_i^{} D_j^c
       + (h_U)_{ij}^{}\, Q_i^{} H_2^{} U_j^c \,
  \right] + \mu H_1^{} H_2^{}
\label{eq:superpot}
\end{equation}
where $i$ and $j$ are generation indices. Contractions over 
SU(2) and SU(3) indices are understood. In particular,
\begin{equation}  
  H_1 H_2 \equiv \epsilon^{\a\b}\;H_{1\a} H_{2\b} 
  = H_1^0 H_2^0 - H_2^+ H_1^-
\end{equation}
with $\epsilon^{\a\b}$ a totally anti--symmetric tensor used to 
contract over the SU(2)$_L$ weak isospin indices $\a,\b=1,2$. 
Likewise, $H_1^{}QD^c \equiv \epsilon^{\a\b} H_{1\a}^{}Q_\b^aD_a^c$ 
where $a=1,2,3$ is a colour index, \etc.
The $3\!\times\!3$ matrices $h_D$, $h_U$, and $h_E$ are 
dimensionless Yukawa couplings giving rise to quark and lepton masses. 
Moreover, $h_D$ and $h_U$ account for the mixing between the quark 
current eigenstates as described by the CKM matrix \cite{CKM}. 
Notice that the same superpotential is obtained by requiring 
that baryon and lepton numbers be conserved (which is automatically 
fullfilled in the SM but not in the MSSM). 

\noi
The Lagrangian derived from \eq{superpot} is 
\begin{equation}
  {\cal L}_{\rm SUSY} = 
  - \bigg[\, \sum_{j,k} \,
       \frac{\de^2\Wcal}{\de\phi_j\de\phi_k}\:\psi_j\psi_k + \hc
    \bigg]
  - \sum_j\, \left|\, \frac{\de\Wcal}{\de\phi_j}\, \right|^2
\label{eq:lagsusy}
\end{equation}
where $\phi_i$ are scalar and $\psi_i$ fermion fields; 
$\Wcal$ only depends on the scalar fields. 
The first term in \eq{lagsusy} describes masses and Yukawa interactions 
of fermions, while the second term describes scalar 
mass terms and scalar interactions.

The interactions obtained in this way respect a symmetry called 
R--parity under which the ``ordinary'' fields (matter fermions, Higgs 
and gauge bosons) are even while their superpartners (sfermions, 
higgsinos and gauginos) are odd. 
As a consequence, all interactions involve an even number of SUSY 
particles (``sparticles''). 
This means that sparticles can only be produced in pairs, and 
any sparticle decay must lead to an odd number of sparticles.
Hence in the MSSM the lightest supersymmetric particle (LSP) 
is stable.
This leads to another important feature of the MSSM: 
Since the LSPs cannot decay some of them must have survived from 
the Big Bang era. Searches for so--called ``exotic isotopes'' 
have led to very stringent bounds which exclude any 
strongly or electromagnetically interacting stable particles with 
masses below a few TeV. 
The LSP of the MSSM must therefore be electrically and colour neutral 
\cite{neutral-LSP}. 
In turn, a neutral stable LSP is a good candidate for dark matter. 
For collider experiments this means that any decay chain of a sparticle 
will end in an arbitrary number of SM particles plus at least one LSP 
which escapes the detector, carring away some energy and momentum. 
In the context of the MSSM the typical SUSY signature is thus  
distinguished by missing (transverse) energy/momentum. 
Depending on different variants of the model, the LSP can be the 
lightest neutralino or the gravitino;
the sneutrino has already been ruled out \cite{hillwalkers}.  

\noi
We want to stress that R--parity is a symmetry that is 
somehow built in by hand due to our assumption of strict minimality.
In principle, the following R--parity violating terms are allowed:
\begin{equation}
  \Wcal_{\rm\not R}^{} = 
  \l LLE^c + \l' LQD^c + \l'' D^cD^cU^c + \mu' H_1 L .
\label{eq:rbreaking}
\end{equation}
Here also generation indices have been suppressed. 
Within the MSSM breaking R--parity therefore means breaking 
baryon and/or lepton number. 
This leads to a significantly different phenomenology compared 
to the R--parity conserving case. 
In particular, single sparticle production is possible, and 
the LSP is no longer stable. 
However, there are very stringent constraints on the couplings 
$\l$, $\l'$, $\l''$, and $\mu'$. 
If \eg, both baryon and lepton number were broken, this would 
lead to rapid proton decay. Together with other 
constraints \cite{Rparity} 
and the wish to embed the MSSM in a Grand Unified Theory 
this requires that the couplings in \eq{rbreaking} be at least 
very small, if not zero. 
However, in the way the MSSM is understood in this thesis, 
R--parity is conserved. 
(For the phenomenology of R--parity violating models 
see \eg~ \cite{Rparity}.) 

The scalar potential ${\cal V}$ is obtained from $F$-- and $D$--terms:
\begin{equation}
  {\cal V} = F_i^*F_i^{} + 
  \onehf\, \left[ D^aD^a + D^iD^i + (D')^2\right]
\end{equation} 
with
\begin{gather}
  F_i = \frac{\de\Wcal(\phi_i)}{\de\phi_i}, \\
\intertext{and}
  D^a = g_3^{}\,\phi^*\, T^a\, \phi\,, \quad
  D^i = \onehf\,g_2^{}\,\phi^*\, \s^i\, \phi\,, \quad  
  D'  = \onehf\,g_1^{}\,{\rm Y}_{\!\phi}\, \phi^*_i\phi_i^{}\,.
\end{gather}
$T^a=\l^a/2$ $(a=1,\ldots 8)$ with $\l^a$ the Gell-Mann--Low matrices.  
$\s^i$ $(i=1,\ldots 3)$ are the Pauli matrices and 
${\rm Y}_{\!\phi}=2(Q-I_3)$ is the hypercharge.

The Lagrangian as given in \eq{lagsusy} conserves supersymmetry. 
However, in a realistic model SUSY must be broken 
Otherwise the masses of ordinary particles and their superpartners 
would be equal, which is not the case as we know from experiment.   
As the genuine mechanism of (dynamical) SUSY breaking is not yet 
understood, we parametrize it by inserting SUSY breaking 
terms by hand into the Lagrangian. 
The terms that break SUSY {\em softly}, 
\ie do not induce quadratic divergencies, 
are \cite{softbreaking}
\begin{itemize}
  \item gaugino mass terms $-\onehf M_a \bar\l_a \l_a$, 
          where $a$ is the group index;
  \item scalar mass terms $-M_{\phi_i}^2|\phi_i|^2$; 
  \item trilinear scalar interactions $A_{ijk}\,\phi_i\phi_j\phi_k$; and
  \item bilinear terms $-B_{ij}\,\phi_i\phi_j + \hc$. 
  \footnote{In general also linear terms $-C_i\phi_i$ are allowed, 
            where $\phi_i$ is a gauge singlet field.}  
\end{itemize}
They lead to the following explicit form of the soft SUSY breaking 
Lagrangian:
\begin{equation}
\begin{split}
  -{\cal L}_{\rm soft} = 
  &\;\onehf M_1^{}\ti B \ti B + \onehf M_2^{}\ti W\ti W 
      + \onehf\,M_3^{}\,\sg\sg 
      + m_{H_1}^2 |H_1^{}|^2 + m_{H_2}^2 |H_2^{}|^2 \\
  & + M_{\ti Q}^2 |\sq_L^{}|^2 
    + M_{\ti U}^2 |\su_R^{\,c}|^2 
    + M_{\ti D}^2 |\sd_R^{\,c}|^2
    + M_{\ti L}^2 |\sl_L^{}|^2
    + M_{\ti E}^2 |\se_R^{\,c}|^2 \\
  & + \left( h_E^{} A_E^{} H_1^{} \sl_L^{}\se_R^c
           + h_D^{} A_D^{} H_1^{} \sq_L^{}\sd_R^c
           + h_U^{} A_U^{} H_2^{} \sq_L^{}\su_R^c
           + B\mu H_1 H_2 + \hc \right). \\
\end{split}
\label{eq:lagsoft}
\end{equation}
$M_1^{}$, $M_2^{}$, and $M_3^{}$ are the U(1), SU(2), and SU(3) 
gaugino masses, respectively. 
$m_{H_1}^2$, $m_{H_2}^2$, and $B\mu$ are mass terms for the Higgs fields. 
The scalar mass terms  
$M_{\ti Q}^2$, $M_{\ti U}^2$, $M_{\ti D}^2$, 
$M_{\ti L}^2$, and $M_{\ti E}^2$ are in general hermitean 
$3\!\times\!3$ matrices in generation space, while 
$h_U A_U$, $h_D A_D$, and $h_E A_E$ are general $3\!\times\!3$ matrices.  
Allowing all the parameters in \eq{lagsoft} to be complex, 
we end up with 124 masses, phases and mixing angles as 
free parameters of the model. 
Notice that $\lag_{\rm soft}$ also respects R--parity. 
Indeed, a R--parity violating term in Eq.~\eq{lagsoft} 
would lead to an unstable vacuum unless the same term also appears 
in the superpotential Eq.~\eq{superpot}.

\section{Renormalization Group Equations}\label{sect:RGEs}

The physical quantities at the electroweak scale are related  
to their values at some high energy scale by renormalization 
group (RG) equations. 
We will not perform a RG analysis here but just add some 
quantitative arguments:

\noi
One of the very appealing features of the MSSM is that it allows 
for gauge coupling unification at $M_X\sim 10^{16}\gev$ \cite{guni}. 
The MSSM is thus compatible with a Grand Unified Theory (GUT).
As in GUT models the gauginos all live in the same representation 
of the unified gauge group, gaugino masses are also unified 
at scales $Q\geq M_X$. 
The 1--loop RG equations \cite{RGE}
for the gauge couplings and the gaugino 
masses are
\begin{equation}
   \frac{d}{dt}\,g_a^{} = \frac{b_a}{16\pi^2}\,g_a^3, \qquad
   \frac{d}{dt}\,M_a^{} = \frac{b_a}{8\pi^2}\,g_a^2 M_a^{}
\end{equation}
with $t=\ln (Q/M_X)$ and 
$b_a = 33/5,\:1,\:-3$ for $a=1,\:2,\:3$, respectively 
\footnote{The coefficients $b_a$ in the MSSM are different from those 
in the SM, where $b_a = \{41/10,\:-19/6,\:-7\}$, due to the richer 
particle spectrum in the loops.}.
One therefore has 
\begin{equation}
  \frac{M_1}{g_1^2} = \frac{M_2}{g_2^2} = \frac{M_3}{g_3^2} 
\end{equation}
at any RG scale, up to small 2--loop effects.

\noi
We next consider the evolution of scalar masses. 
For simplicitly we here assume that the soft masses 
of squarks and sleptons are flavour--diagonal 
\eg, $M_{\ti Q}^2={\rm diag}(M_{\ti Q_1}^2,M_{\ti Q_2}^2,M_{\ti Q_3}^2)$. 
Moreover, we neglect Yukawa and trilinear couplings of the 
first and second generation, 
\ie $h_U={\rm diag}(0,0,h_t)$, 
   $A_U={\rm diag}(0,0,A_t)$, \etc. 
The 1--loop RG equations for the masses of squarks and sleptons  
of the first two generations are then given by 
\begin{equation}
  16\pi^2 \frac{d}{dt}\,M_\phi^2 = -\sum_{a=1}^3 
  8 g_a^2 C_a^\phi\, |M_a^{}|^2
\label{eq:rgscal}
\end{equation}
with the sum running over the gauge groups. 
The $C_a^\phi$ are Casimir operators:  
$C_1^\phi = \frac{3}{5}(\frac{{\rm Y}_{\!\phi}}{2})^2$ for each scalar 
$\phi$ with hypercharge ${\rm Y}_{\!\phi}$ 
(${\rm Y}_{\!Q}=\frac{1}{3}$, ${\rm Y}_{\!u}=-\frac{4}{3}$, 
${\rm Y}_{\!\d}=\frac{2}{3}$, \etc);
$C_2^\phi = 3/4$ (0) for 
$\phi = \ti Q,\ti L$ ($\ti U^c,\ti D^c,\ti E^c$); and  
$C_3^\phi = 4/3$ (0) for 
$\phi = \ti Q,\ti U^c,\ti D^c$ ($\ti L,\ti E^c$).
The right side of Eq.~\eq{rgscal} is strictly negative, so 
$M_\phi^2$ {\em grows} when being evolved down to the low scale.
Moreover, owing to SU(3) contributions,  
squark masses grow faster than slepton masses. \\
The mass--squared parameters of the third generation squarks and sleptons 
as well as of the Higgs fields also obey \eq{rgscal} but get 
additional contributions from Yukawa and trilinear couplings: 
\begin{gather}
  16\pi^2 \frac{d}{dt}\,M_{\ti Q_3}^2 = X_t^{} + X_b^{} 
    - \frac{32}{3}\,g_3^2\,|M_3^{}|^2 
    - 6g_2^2\,|M_2^{}|^2 - \frac{2}{15}\,g_1^2\,|M_1^{}|^2, \\
  16\pi^2 \frac{d}{dt}\,M_{\ti U_3}^2 = 2X_t^{} 
    - \frac{32}{3}\,g_3^2\,|M_3^{}|^2 
    - \frac{32}{15}\,g_1^2\,|M_1^{}|^2, \\
  16\pi^2 \frac{d}{dt}\,M_{\ti D_3}^2 = 2X_b^{} 
    - \frac{32}{3}\,g_3^2\,|M_3^{}|^2 
    - \frac{8}{15}\,g_1^2\,|M_1^{}|^2, \\
  16\pi^2 \frac{d}{dt}\,M_{\ti L_3}^2 = X_\tau^{} 
    - 6 g_2^2\,|M_2^{}|^2 - \frac{3}{5}\,g_1^2\,|M_1^{}|^2, \\
  16\pi^2 \frac{d}{dt}\,M_{\ti E_3}^2 = 2X_\tau^{} 
    - \frac{24}{5}\,g_1^2\,|M_1^{}|^2, \\
\intertext{and} 
  16\pi^2 \frac{d}{dt}\,m_{H_1}^2 = 3X_b^{} + X_\tau^{} 
    - 6g_2^2\,|M_2^{}|^2 - \frac{6}{5}\,g_1^2\,|M_1^{}|^2, \\
  16\pi^2 \frac{d}{dt}\,m_{H_2}^2 =  3X_t^{} 
   - 6g_2^2\,|M_2^{}|^2 - \frac{6}{5}\,g_1^2\,|M_1^{}|^2, 
  \label{eq:rgmh2}
\end{gather}
with
\begin{gather}
  X_t^{} = 2|h_t|^2\,(m_{H_2}^2+M_{\ti Q_3}^2+M_{\ti U_3}^2) 
           + 2|A_t^{}|^2,\\
  X_b^{} = 2|h_b|^2\,(m_{H_1}^2+M_{\ti Q_3}^2+M_{\ti D_3}^2) 
           + 2|A_b^{}|^2,\\
  X_\tau^{} = 2|h_t|^2\,(m_{H_1}^2+M_{\ti L_3}^2+M_{\ti E_3}^2) 
           + 2|A_\tau^{}|^2 .
\end{gather}
$X_t$, $X_b$, and $X_\tau$ are always positive, so they {\em decrease} 
the scalar masses as one runs downwards to the low scale.   
Therefore, the soft breaking parameters of the third generation 
are in general smaller than those of the first and second generation  
[unless one starts with very different values at the high scale].\\
The RG equations for $m_{H_1}^2$ and $m_{H_2}^2$ are still a special case. 
Compared to those for the soft squark masses they do not get the large 
contributions proportional to $|M_3^{}|^2$. 
Moreover, $X_t^{}$ and $X_b^{}$ enter with larger coefficients. 
This can cause $m_{H_1}^2$ or $m_{H_2}^2$ to become negative 
near the electroweak scale \cite{inoue}. 
Since this leads to a breakdown of the electroweak symmetry 
solely by quantum corrections this is called 
{\em radiative electroweak symmetry breaking} \cite{ibanez-ross}. 
Here we note that owing to the large top quark mass one can 
expect an especially large effect from $X_t$, favoring 
$m_{H_2}^2<m_{H_1}^2$ (unless $h_t\sim h_b$).

\noi
The gaugino masses do not only enter Eqs.~\eq{rgscal} to \eq{rgmh2} 
but also the RG equations for the $A$ parameters. 
Non--zero gaugino masses at $M_X$ are therefore 
sufficient to create all the other soft SUSY breaking terms.
On the other hand, if the gaugino masses vanished at $M_X$
they would only be generated through 2--loop and higher 
order effects and hence be very small. 
 
\noi 
Another (technical) remark seems appropriate at this point: 
For computing radiative corrections in supersymmetry, 
it is important to choose regularization and 
renormalization schemes that preserve supersymmetry 
(and gauge symmetry, of course). 
Dimensional regularization (DREG), for instance, explicitly violates 
SUSY because the continuation of the number of spacetime dimensions to 
$D=4-\epsilon$ introduces a mismatch between the number of gauge 
boson and gaugino degrees of freedom.  
A solution is to perform the momentum integrals in $D=4-\epsilon$ 
dimensions while taking the vector index $\mu$ of the gauge boson 
fields over all four dimensions \cite{dred1,dred2,dred3}. 
This is known as {\em dimensional reduction} (DRED) 
and nicely respects both gauge symmetry and SUSY, 
at least up to 2--loop order. 
We will come back to this when discussing supersymmetric 
QCD corrections in Chapter~3. 

\section{Models of SUSY Breaking}

Above we have introduced explicit SUSY breaking terms because 
we are ignorant of the fundamental machanism that breaks 
supersymmerty. 
If SUSY is broken spontaneously there exists a Goldstone 
fermion called the goldstino. 
In global supersymmetry the goldstino is massless. 
In local supersymmetry (supergravity) the goldstino  
is ``eaten'' by the gravitino $(\sg_{3/2})$ which in this way 
acquires a mass $m_{3/2}$ \cite{superHiggs}. 
This is called the \emph{super--Higgs mechanism} and is completely 
analogous to the ordinary Higgs mechanism in gauge theories. 

\noi
Present models of spontaneously--broken low--energy supersymmetry
assume that SUSY is broken in a ``hidden'' or ``secluded'' sector 
which is completely neutral with respect to the SM gauge group. 
The information of SUSY breaking is then mediated to the 
``visible'' sector, which contains the MSSM,  
by some mechanism. 
There are no renormalizable tree--level interactions between the 
hidden and visible sectors.  
Two scenarios have been studied in detail: gravity--mediated 
and gauge--mediated SUSY breaking. 

In \emph{gravity--mediated SUSY breaking}, SUSY breaking is transmitted 
to the MSSM via gravitational interactions \cite{gravmed}. 
The breakdown of the symmetry occurs at ${\cal O}(10^{10})$ GeV or higher, 
and the gravitino gets a mass of the order of the electroweak scale. 
The simplest realization of such a framework is the 
\emph{minimal supergravity model} (mSUGRA) \cite{nilles,msugra}. 
In this approach, one assumes a universal gaugino mass $M_{1/2}$, 
a universal scalar mass $M_0$, and a universal cubic coupling $A_0$ 
at $M_X$. In addition, one just needs to specify $\tan\b$ and the 
sign of $\mu_0$. RGEs are then used to derive the MSSM parameters 
at the electroweak scale. 
Since mSUGRA involves only five parameters (in addition to the 18 
SM parameters) it is highly predictive and thus used for most 
experimental searches. 
However, one should keep in mind that it also highly restrictive.

\emph{Gauge--mediated SUSY breaking} (GMSB) \cite{gaugemed} models 
involve a ``secluded'' sector where SUSY is broken and 
a ``messenger'' sector consisting of particles with 
SU(3)$\times$SU(2)$\times$U(1) quantum numbers. 
The messengers directly couple to the particles of the secluded 
sector. This generates a SUSY breaking spectrum in the messenger sector. 
Finally, SUSY breaking is mediated to the MSSM via virtual exchange of 
the messengers. A basic feature of such models is that SUSY is broken 
at much lower scales than in the gravity--mediated case, typically 
at ${\cal O}(10^4\mbox{--}10^5)$~GeV. 
Moreover, the gravitino gets a mass in the eV to keV range, 
and is therefore the LSP. This can be crucial for SUSY signatures at 
collider experiments because the next--to--lightest SUSY particle (NLSP) 
will eventually decay into its SM partner plus a gravitino. 
A long--lived $\nt_1$--NLSP that decays outside the detector 
leads to the usual SUSY signature of large missing energy plus 
leptons and/or jets. If, in contrast, the decay $\nt_1\to \g\sg_{3/2}$ 
occurs inside the detector SUSY events would in addition contain photons. 
The NLSP may, however, also be a charged particle \eg, $\stau^\pm_R$. 
This would lead either to a long--lived charged particle or to 
SUSY signatures characterized by $\tau$--leptons. \\
Since gauge interactions are flavour--blind one has universal boundary 
conditions in GMSB as in mSUGRA. The low--energy spectrum is determined 
by the mass of the messengers. 
Minimal GMSB is thus even more restrictive than mSUGRA. 
In the most general case, however, both supergravity and gauge--mediated 
effects may contribute to the breaking of supersymmetry.

A more detailed discussion of SUSY breaking is beyond the scope 
of this thesis. 
For a thorough introduction to supergravity, see \eg~\cite{bailin-love}; 
for the phenomenology of mSUGRA, see \eg~\cite{DPF95,snowmass}.  
A review of gauge--mediated SUSY breaking is given in \cite{giudice}.

\section{Electroweak Symmetry Breaking}

The scalar potential for the Higgs fields --- including all soft--breaking 
terms --- is
\begin{align}
  {\cal V}_{\rm Higgs} = 
   &\:(|\mu|^2+m_{H_1}^2) \, |H_1|^2
   +  (|\mu|^2+m_{H_2}^2) \, |H_2|^2 
   + (B\mu H_1H_2 +\hc) \notag\\[1mm]
   & + \smaf{1}{8}(g_1^2+g_2^2)\,(|H_1|^2 - |H_2|^2)^2 
     + \onehf\, g_2^2\, |H_1^+H_2^{0*}+H_2^0H_1^{-*}|^2.  
\label{eq:scalhiggspot}
\end{align}
Here the terms proportional to $|\mu|^2$ come from $F$--terms, 
the quartic interactions come from $D$--terms, and 
the terms proportional to $m_{H_1}^2$, $m_{H_1}^2$ and $B\mu$ 
come from soft SUSY breaking, see Eq.~\eq{lagsoft}.

\noi
Fist, we use SU(2)$_L$ gauge transformations to rotate away any 
VEV of one of the charged Higgs fields 
\eg, $\langle H_1^-\rangle = 0$.  
Then $\de{\cal V}_{\rm Higgs}/\de H_1^- = 0$ requires 
$\langle H_2^+\rangle = 0$ as well. 
This is good because we are now sure that electric charge is 
conserved in the Higgs sector. 
We shall thus ignore the charged components in \eq{scalhiggspot}
when minimizing the potential \footnote{To be completely honest,  
this is a simplifying view: charge might still be broken 
in the absolute minimum of the full scalar 
potential due to some non--zero sfermion VEVs.}. 

\noi
Next, we choose $B\mu$, the only term in \eq{scalhiggspot} 
that depends on complex phases, to be real and positive. 
This can be done through a redefinition of the phases of 
$H_1$ and $H_2$. 
Then $\langle H_1^0\rangle$ and $\langle H_2^0\rangle$ are also real.
This means that CP is not spontaneously broken by the Higgs scalar 
potential. The Higgs mass eigenstates are thus also eigenstates of CP.

\noi
The scalar potential has a lokal minimum other than its origin if 
\begin{equation}
  (|\mu|^2+m_{H_1}^2)\,(|\mu|^2+m_{H_2}^2) < (B\mu)^2.
\label{eq:hcond1}
\end{equation}
However, this is not enough; ${\cal V}_{\rm Higgs}$ must also 
be bounded from below. This is the case if
\begin{equation}
  (|\mu|^2+m_{H_1}^2)+(|\mu|^2+m_{H_2}^2) \ge 2B\mu.
\label{eq:hcond2}
\end{equation}
These two conditions can only be 
satisfied simultaneously if $m_{H_1}\neq m_{H_2}$ --- 
implying that in the MSSM electroweak symmetry breaking 
is not possible without first breaking SUSY! 

\noi
We are finally ready to minimize the Higgs potential by 
solving $\de{\cal V}_{\rm Higgs}/\de H_1^0=
\de{\cal V}_{\rm Higgs}/\de H_2^0=0$.
The VEVs at the minimum of the potential, 
$v_1\equiv\langle H_1^0\rangle$ and 
$v_2\equiv\langle H_2^0\rangle$, are related by 
\begin{equation}
  v^2\equiv v_1^2+v_2^2 = 2\,m_Z^2/(g_1^2+g_2^2)
  \approx (174 \gev)^2 ,
\end{equation}
so that only the ratio of the two remains a free parameter.
Defining
\begin{equation}
  \tan\b\equiv v_2/v_1
\end{equation}
the minimalization conditions are given by
\begin{gather}
  |\mu|^2+m_{H_1}^2 = -B\mu\tan\b - \onehf\, m_Z^2 \cos 2\b,\\[1mm]
  |\mu|^2+m_{H_2}^2 = -B\mu\cot\b + \onehf\, m_Z^2 \cos 2\b.
\end{gather}

The two complex scalar Higgs doublets consist of eight degrees of freedom, 
three of which are eaten by the longitudinal modes of the $Z$ and 
$W$ bosons. The remaining five physical degrees of freedom form
a neutral CP--odd, two neutral CP--even and two charged Higgs bosons
denoted by $A^0$, $h^0$, $H^0$, and $H^\pm$, respectively.
They are given by
\begin{equation}
  {G^0\choose A^0} = \rzw\, 
    \left(\!\begin{array}{rl}
      \sin\b & \cos\b \\ -\cos\b & \sin\b 
    \end{array}\!\right) \, 
  {\Im(H_2^0) \choose \Im(H_1^0)},
\end{equation}
\begin{equation}
  {G^+\choose H^+} = 
    \left(\!\begin{array}{rl}
      \sin\b & \cos\b \\ -\cos\b & \sin\b 
    \end{array}\!\right) \, 
  {H_2^+\, \choose H_1^{-*}} 
\end{equation}

\noi
with $G^0$ and $G^+$ the would--be Nambu--Goldstone bosons, and \\
\begin{equation}
  {h^0\choose H^0} = \rzw\, 
    \left(\!\begin{array}{rl}
      \cos\a & \sin\a \\ -\sin\a & \cos\a 
    \end{array}\!\right) \, 
  {\Re(H_2^0)-v_2^{} \choose \Re(H_1^0)-v_1^{}} 
\end{equation}

\noi
with the Higgs mixing angle $\a$.
[$\Re(H)$ denotes the real and $\Im(H)$ the imaginary part of $H$.]
The mass eigenvalues at tree level are
\begin{align}
  m_A^2 & = 2B\mu/\sin 2\b, \\
  m_{H^\pm}^2 & = m_A^2 + m_W^2, \\
  m_{h^0\!,H^0}^2 & = \onehf\left(m_A^2+m_Z^2 \mp 
    \sqrt{(m_A^2+m_Z^2)^2 - 4m_A^2m_Z^2\cos 2\b}\right).
    \label{eq:mhtree} 
\end{align}

\begin{figure}[h!]
\center
\begin{picture}(132,64)
\put(1,1){\mbox{\psfig{figure=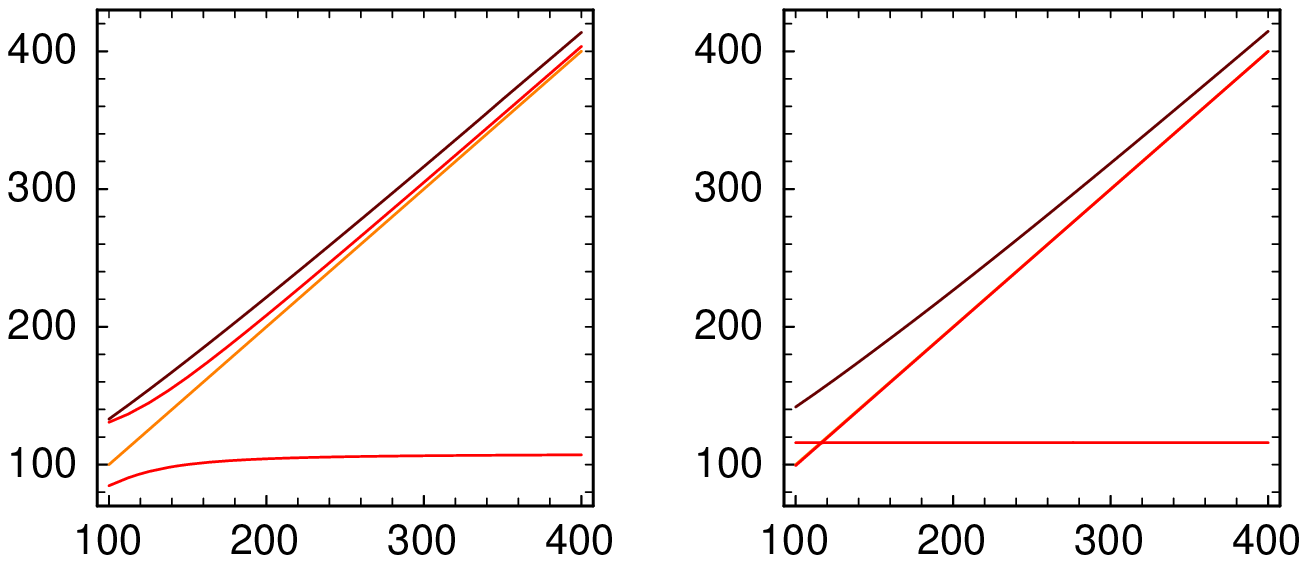,height=65mm}}}
\put(0,37){\makebox(0,0)[c]{\rotatebox{90}{$m$~[GeV]}}}
\put(39,0){\makebox(0,0)[cb]{$m_A$~[GeV]}}
\put(108,0){\makebox(0,0)[cb]{$m_A$~[GeV]}}
\put(18,55){\makebox(0,0)[bl]{$\tan\b=3$}}
\put(86,55){\makebox(0,0)[bl]{$\tan\b=30$}}
\put(38,45){\mbox{\footnotesize$H^\pm$}}
\put(36,31){\mbox{\footnotesize$A^0$}}
\put(28,20){\mbox{\footnotesize$H^0$}}
\put(27,21.3){\vector(-2,1){5}}
\put(55,18.5){\mbox{\footnotesize$h^0$}}
\put(108.5,45){\mbox{\footnotesize$H^\pm$}}
\put(97,24){\mbox{\footnotesize$H^0\!,\,A^0$}}
\put(123,19.5){\mbox{\footnotesize$h^0$}}
\end{picture}
\caption{MSSM Higgs boson masses as a function of $m_A$ for 
$\tan\b=3$ and $30$ ($M_{\ti Q}=400\gev$, $M_{\ti U}=350\gev$, 
$M_{\ti D}=450\gev$, $A_t=A_b=-300\gev$, and $\mu=400\gev$).}
\label{fig:higgsmasses}
\end{figure}

\noi
We can therefore take $m_A$ and $\tan\b$ as the free parameters 
of the MSSM Higgs sector.
From \eq{mhtree} it follows that
\begin{equation}
  m_{h^0} < {\rm min}(m_Z,m_A)\,|\cos 2\b|
\end{equation}
at tree level. However, the Higgs masses and mixing angle are 
subject to large radiative corrections. 
In this thesis we use the formulae of Ref.~\cite{radcorrh0} 
for the $(h^0\!,\,H^0)$ system and Ref.~\cite{radcorrhc} for 
the mass of $H^\pm$ 
\footnote{Notice that \cite{radcorrh0,radcorrhc} have the opposite 
          sign convention for the parameter $\mu$.}.  
Assuming that the masses of the sparticles in the loops do not 
exceed 1~TeV one obtains an upper bound for the lightest Higgs 
mass of 
\begin{equation}
  m_{h^0} \lsim 130\gev .
\end{equation}
On the other hand, $A^0\!$, $H^0\!$, and $H^\pm$ can be almost arbitrarily
heavy. If $m_A \gg m_Z$ they become nearly degenerate and decouple 
from the low--energy regime. In this case $h^0$ 
is very difficult to distinguish from a Standard Model Higgs boson.
\Fig{higgsmasses} shows the MSSM Higgs boson masses as a function 
of $m_A$ for $\tan\b=3$ and $30$. 
For the radiative corrections we have used 
$M_{\ti Q}=400\gev$, $M_{\ti U}=350\gev$, $M_{\ti D}=450\gev$, 
$A_t=A_b=-300\gev$, and $\mu=400\gev$. 

Up--type quarks get masses proportional to $v_2 = v\sin\b$; 
the masses of down--type quarks and electron--type leptons are 
proportional to $v_1 = v\cos\b$. 
At tree level the quark and lepton masses are therefore related to 
the respective Yukawa couplings by
\begin{equation}
  h_u = \frac{g m_u}{\rzw\,m_W\sin\b}, \qquad 
  h_d = \frac{g m_d}{\rzw\,m_W\cos\b}, \qquad 
  h_e = \frac{g m_e}{\rzw\,m_W\cos\b}.
\label{eq:defyuk}
\end{equation}
This is why we neglected the Yukawa couplings of the first and 
second generation, but not those of $t$, $b$, and $\tau$. 
Obviously, $h_t$ is significant due to the large top quark mass. 
Moreover, $h_t:h_b:h_\tau = m_t:m_b\tan\b:m_\tau\tan\b$ from 
\eq{defyuk}, so also the bottom and tau Yukawa couplings are
important if $\tan\b$ is large. 
In fact, certain models predict the unification of top and bottom 
(or top, bottom, and tau) Yukawa couplings for 
$\tan\b\sim m_t/m_b$.
One can, moreover, require that none of the Yukawa couplings 
become nonpertubativly large. This gives rough bounds 
of $1.2\lsim\tan\b\lsim 65$.

\section{Sparticle Mixing and Mass Eigenstates}

Once ${\rm SU}(2)_L\times {\rm U}(1)_Y$ symmetry is broken, fields with 
the same ${\rm SU}(3)_c\times {\rm U}(1)_{em}$ quantum numbers 
(and, of course, R-parity and spin) can mix with each other. 
In the Standard Model,  
$B^0$ and $W^i$ mix to $\gamma$, $Z^0$, and $W^\pm$. 
Also the Dirac masses of quarks and leptons can be understood as 
such mixing terms. 
In the MSSM, this mixing also effects squarks, sleptons, 
Higgs bosons, as well as gauginos and higgsinos. 
The lone exception is the gluino, being the only colour octet 
fermion in the model.

\noi
Indeed, masses and mixings of sparticles are of crucial importance 
both theoretically and experimentally: 
i) they determine the properties of the sparticles searched for 
and 
ii) they are directly related to the question of how SUSY is broken.  
(Notice that the main questionmarks in the MSSM come from $\lag_{\rm soft}$ 
while the couplings and all but one mass term in $\lag_{\rm SUSY}$
are linked to well known parameters of the SM!)

\subsection{Squarks} \label{sect:squarks}

In the most general case, the squark mass eigenstates are obtained 
by diagonalizing two $6\!\times\!6$ squark mass--squared matrices 
--- one for up--type and one for down--type squarks. 
However, mixing between squarks of different generations can cause
severe problems due to too large loop contributions to flavour changing 
neutral current (FCNC) processes \cite{masiero}. \\
Ignoring intergenerational mixing for the moment, 
the two general $6\!\times\!6$ squark mass--squared matrices 
decompose into a series of $2\!\times\!2$ matrices, each of 
which describes squarks of a specific flavour \cite{ellis-rudaz}:

\begin{equation}
  {\cal M}_{\sq}^2 = 
  \left( \begin{array}{cc}  \msq{L}^2 & a_q m_q \\
                            a_q m_q   & \msq{R}^2
  \end{array}\right) \;=\;
  (\R^{\sq})\kr \left( \begin{array}{cc}  \msq{1}^2 & 0 \\
                                          0         & \msq{2}^2
                \end{array}\right) \R^{\sq}
\label{eq:msqmat}
\end{equation}
with
\begin{eqnarray}
  \msq{L}^2 &=& M^2_{\ti Q} 
    + m_Z^2 \cos 2\beta\,(I_{3L}^q - e_q\sin^2\t_W) + m_q^2, \\[2mm]
  \msq{R}^2 &=& M^2_{\{\ti U,\ti D\}} 
    + e_q\,m_Z^2 \cos 2\b\,\sin^2\t_W + m_q^2, \\[1mm]
  a_q &=& A_q - \mu\, \{ \cot\b , \tan\b \}
  \label{eq:aq}
\end{eqnarray}
for $\{$up, down$\}$ type squarks, respectively. 
$e_q$ and $I_{3}^q$ are the electric charge and the third component
of the weak isospin of the squark $\sq$, and $m_q$ is the mass of the 
partner quark. 
$M_{\ti Q}$, $M_{\ti U}$, and $M_{\ti D}$ are soft SUSY breaking 
masses, and $A_q$ are trilinear couplings as in \eq{lagsoft}. 
Family indices have been neglected. 

\noi 
The off--diagonal elements of ${\cal M}_{\sq}^2$ are proportional to 
the mass of the corresponding quark. 
In the case of the 1st and 2nd generation $\sq_L^{}$ and 
$\sq_R^{}$ are therefore to a good approximation also the mass 
eigenstates. Indeed, experimental results on 
$K^0-\overline{K^0}$ and $D^0-\overline{D^0}$ mixing suggest
that the masses of up, down, charm, and strange squarks 
are highly degenerated and mixing can be neglected \cite{masiero}.
We can therefore assume
\begin{equation}
  m_{\ti u_L} = m_{\ti u_R} = m_{\ti d_L} = m_{\ti d_R} = 
  m_{\ti c_L} = m_{\ti c_R} = m_{\ti s_L} = m_{\ti s_R}.
\end{equation}
This also justifies the above assumption that intergenerational 
mixing can be neglected.

However, this does not hold for the third generation: 
Stops are expected to be highly mixed due to the large top quark mass, 
and for sbottoms mixing effects can be important if $\tan\b\,$ is large.  
In the following we discuss this left--right mixing for $\sq=\st,\sb$:
 
\noi 
According to Eq.~\eq{msqmat} ${\cal M}_{\sq}^2$ is diagonalized by a 
unitary matrix $\R^{\sq}$. Assuming that CP violating phases only occur 
in the CKM matrix, we choose $\R^{\sq}$ to be real. 
The weak eigenstates $\sq_L^{}$ and $\sq_R^{}$ are thus 
related to their mass eigenstates $\sq_1^{}$ and $\sq_2^{}$ by 
\begin{equation}
  {\sq_1^{} \choose \sq_2^{}} = {\cal R}^{\sq}\,{\sq_L^{} \choose \sq_R^{}},
  \hspace{8mm}
  {\cal R}^{\sq} = \left(\baa{rr} \cos\tsq & \sin\tsq \\ 
                                 -\sin\tsq & \cos\tsq \eaa\right) , 
\label{eq:Rsq}
\end{equation}
with $\tsq$ the squark mixing angle. 
The mass eigenvalues are given by 
\begin{equation}
  m^2_{\sq_{1\!,2}} = \onehf \left( \msq{L}^2 + \msq{R}^2
  \mp \sqrt{(\msq{L}^2 - \msq{R}^2)^2 + 4\, a^2_q m_q^2 } \,\right). 
\label{eq:sqmasseigenvalues}
\end{equation}
By convention, we choose $\sq_1^{}$ to be the lighter mass eigenstate. 
Notice, that $\msq{1}\leq\msq{L,R}\leq\msq{2}$. 
For the mixing angle $\tsq$ we require $0\leq\tsq <\pi$. We thus have  
\begin{equation} 
  \cos\tsq = 
  \frac{- a_q m_q}{\sqrt{(\msq{L}^2-\msq{1}^2)^2 + a_q^2 m_q^2}},
  \hspace{8mm} 
  \sin\tsq = 
  \frac{\msq{L}^2-\msq{1}^2}
       {\sqrt{(\msq{L}^2-\msq{1}^2)^2 + a_q^2 m_q^2}}.
\end{equation}
Moreover, $|\cos\tsq| > \onesq\,$ if $\msq{L}<\msq{R}$, and 
$|\cos\tsq| < \onesq\,$ if $\msq{R}<\msq{L}$. \\

\noi
Inverting Eqs.~\eq{msqmat} -- \eq{sqmasseigenvalues} 
one can calculate the underlying soft--breaking parameters 
from the physical squark masses and mixing angles:
\begin{align}
  M_{\ti Q}^2 & = 
    \mst{1}^2\cos^2\tst + \mst{2}^2\sin^2\tst 
    - m_Z^2\cos 2\b\, (\onehf-\twoth\sW) - m_t^2, \label{eq:MQst}\\
  M_{\ti U}^2 & = \mst{1}^2\sin^2\tst + \mst{2}^2\cos^2\tst 
                - \twoth\, m_Z^2\cos 2\b \sW - m_t^2 \label{eq:MUst}
\end{align}
in the stop sector and
\begin{align}
  M_{\ti Q}^2 & = 
    \msb{1}^2\cos^2\tsb + \msb{2}^2\sin^2\tsb 
    + m_Z^2\cos 2\b\, (\onehf-\oneth\sW) - m_b^2, \label{eq:MQsb}\\
  M_{\ti D}^2 & = \msb{1}^2\sin^2\tsb + \msb{2}^2\cos^2\tsb 
                + \oneth\, m_Z^2\cos 2\b \sW - m_b^2 \label{eq:MDsb}
\end{align}
in the sbottom sector. 
Finally, the off--diagonal element of the squark mass matrix is  
\begin{equation}
  a_q m_q = \onehf\, (\msq{1}^2-\msq{2}^2)\sin 2\tsq \,.
  \label{eq:aqinv}
\end{equation}
Notice that the parameter $M_{\ti Q}$ enters both the stop and the 
sbottom mass matrices.
Therefore, \eq{MQst} and \eq{MQsb} imply the following sum rule at 
tree level:
\begin{equation}
  m_W^2\cos 2\b = \mst{1}^2\cos^2\tst + \mst{2}^2\sin^2\tst 
  - \msb{1}^2\cos^2\tsb - \msb{2}^2\sin^2\tsb - m_t^2 + m_b^2.  
\label{eq:sqmassrelation}
\end{equation}
This shows that if $\tan\b$ and five of the six quantities 
$\mst{1}$, $\mst{2}$, $\msb{1}$, $\msb{2}$, $\tst$, $\tsb$ 
are known, the sixth can be predicted. 

In Sect.~\ref{sect:RGEs} we have learnt that the soft SUSY breaking 
parameters (and hence $\msq{L}$ and $\msq{R}$) of the 3rd generation 
are most likely different from those of the other generations. 
Left--right mixing induces an additional mass splitting, rendering 
$\sq_1^{}$ lighter and $\sq_2^{}$ heavier than $\sq_{L,R}^{}$. 
Notice that both effects are induced by large Yukawa couplings. 
The mixing angles (and Yukawa couplings) also enter the stop and 
sbottom couplings to other particles. 
Moreover, stops ---and for large $\tan\b$ also sbottoms--- give
important contributions to radiative corrections in the Higgs sector
(think \eg~ of radiative symmetry breaking!).

\noi
There are some experimental constraints on the mass splitting in the 
stop and sbottom sectors from BR$(b\to s\g)$ and $\d\rho$. 
However, these constraints are much weaker than the above 
mentioned constraints from $K^0-\overline{K^0}$ and
$D^0-\overline{D^0}$ mixing. 

\noi
All together, we now understand that the phenomenology of stops and
sbottoms can be very different from that of the other squarks.

\subsection{Sleptons} \label{sect:sleptons}

The mass matrix of the charged sleptons is completely analogous 
to that of the squarks:  
\begin{equation}
  {\cal M}_{\ti\ell}^2 = 
  \left( \begin{array}{cc}  \msl{L}^2 & a_\ell m_\ell \\
                            a_\ell m_\ell   & \msl{R}^2
  \end{array}\right) ,
  \qquad \sl = \se,\,\smu,\,\stau,
\label{eq:mslmat}
\end{equation}
with
\begin{eqnarray}
  \msl{L}^2 &=& M^2_{\ti L} - m_Z^2 \cos 2\beta\,(\onehf - \sin^2\t_W) 
                  + m_\ell^2, \\[2mm]
  \msl{R}^2 &=& M^2_{\ti E} - m_Z^2 \cos 2\b\,\sin^2\t_W 
                  + m_\ell^2, \\[1mm]
  a_\ell &=& A_\ell - \mu\,\tan\b .
  \label{eq:aell}
\end{eqnarray}
From RG evolution one expects $M^2_{\ti E} < M^2_{\ti L}$ 
and hence $\msl{R}<\msl{L}$. 
For selectrons and smuons, the ``left'' and ``right'' states 
($\se_{L,R}$ and $\smu_{L,R}$) are also the mass eigenstates. 
For staus, however, analogous arguments apply as for the sbottoms. 
If $\tan\b$ is large enough, $\stau_L^{}$ and $\stau_R^{}$ 
therefore mix to mass eigenstates 
\begin{gather}
  \stau_1^{} = \cos\t_{\stau}\stau_L^{} + \sin\t_{\stau}\stau_R^{},\\
  \stau_2^{} = \cos\t_{\stau}\stau_R^{} - \sin\t_{\stau}\stau_L^{}
\end{gather}
with $\t_{\stau}$ the mixing angle. 
All formulae given for squark mixing are also applicable in this case. 

Let us now turn to the sneutrinos. In the case of massless neutrinos, 
there is only one sneutrino, $\snu_L^{}$, with a mass 
\begin{equation}
  m_{\snu_L} = M^2_{\ti L} + \onehf\,m_Z^2 \cos 2\beta
\end{equation}
for each generation.   
If, however, neutrinos have a mass (as suggested by the Superkamiokande 
experiment \cite{superK}) there exists also $\snu_R$. 
We assume that the neutrino masses are generated by the 
see--saw mechanism   
\begin{equation}
  \big(\bar\nu_L^c,\,\bar N\big)
   \left( \begin{array}{cc} m_M & m_D \\
                            m_D & M_M     
  \end{array}\right) {\nu_L^{} \choose N} 
\end{equation} 
where $N$ is a heavy singlet field \cite{hillwalkers}.   
$m_M$ is the $\bar\nu_L^c\nu_L^{}$ Majorana mass and 
$m_D$ the $\nu_L N$ Dirac mass. 
$M_M$, the Majorana mass of $N$, is very large. 
In fact, it can be as large as the GUT scale. 
Therefore, we may expect that $\snu_R$ is also very heavy and does not 
contribute to low--energy phenomenology.

\subsection{Charginos} \label{sect:charginos}

The supersymmetric partners of the $W^\pm$ and the $H^\pm$
mix to mass eigenstates called charginos $\chp_i$ ($i=1,2$) which 
are four--component Dirac fermions. 
In order to deduce the properties of the latter we start with the 
basis 
\begin{equation}
  \psi^+ = \left( -i\l^+ \!,\, \psi^1_{H_2} \right), 
  \hspace{6mm}
  \psi^- = \left( -i\l^- \!,\, \psi^2_{H_1} \right), 
\end{equation}
where $\l^\pm = \frac{1}{\rzw}\,(\l^1 \mp i\l^2)$, $\psi^1_{H_2}$, and 
$\psi^2_{H_1}$ denote the two--component spinor fields of 
$\ti W^\pm\!,\,$ $\ti H^+\!,\,$ and $\ti H^-\!,\,$ respectively. 
The mass terms of the lagrangian of the charged gaugino--higgsino 
system can then be written as \cite{haberkane,charginos}
\begin{equation}
  \lag_{m} = -\onehf\, \left(\psi^+\!,\,\psi^- \right)\,
  \left( \begin{array}{cc} 0 & X^{\rm T} \\ X & 0 
     \end{array} \right)\,
  {\psi^+ \choose \psi^-} + \hc
\end{equation}
with \footnote{Here and in the following $M\equiv M_2^{}$.}
\begin{equation}
  X = \left( \begin{array}{cc} M & \rzw\,m_W^{}\sin\b \\
                               \rzw\,m_W^{}\cos\b & \mu
      \end{array} \right).
  \label{eq:chmassmat}      
\end{equation}
The mass matrix $X$, Eq.~\eq{chmassmat}, is diagonalized by two   
$2\!\times\!2$ unitary matrices $U$ and $V$: 
\begin{equation}
  U^* X\, V^{-1} = \M_C^{} 
  \label{eq:mchdiagA}  
\end{equation}
with $\M_C^{}$ the diagonal mass matrix. 
With the rotations 
\begin{equation}
  \x^+_i = V_{ij}\,\psi^+_j, \hspace{6mm} 
  \x^-_i = U_{ij}\,\psi^-_j, \hspace{6mm} i,j = 1,2,
\label{eq:UVdef}
\end{equation}
the mass eigenstates in Dirac notation are given by
\begin{equation}
  \chp_1 = {\x^+_1 \choose \bar\x^-_1},
  \hspace{6mm}
  \chp_2 = {\x^+_2 \choose \bar\x^-_2}.  
\end{equation}

We take $\chp_1$ to be the lighter chargino per definition. 
Moreover, assuming CP conservation, we choose a phase convention 
in which $U$ and $V$ are real. 
Following Ref.~\cite{charginos} the mass eigenvalues then are  
\begin{equation}
  M_C^{} = U\, X\, V^{-1} = \left( \begin{array}{cc} 
  \eta_1^{}\,\mchp{1} & 0 \\ 0 & \eta_2^{}\,\mchp{2} \end{array} \right)
\label{eq:mchdiag}
\end{equation}
with $\eta_i^{} = \pm 1$ the sign of the eigenvalue, and 
$\mchp{i} = |(M_C)_{ii}^{}|$ the chargino mass. 
According to Eq.~\eq{chmassmat} we have ($i=1,2$)  
\begin{equation}
  \left( M_{C}^{}\right)_{ii} = 
  \onehf\,\left[ \sqrt{(M-\mu)^2 + 2\,m_W^2\,(1+\sin 2\b)} 
     \mp \sqrt{(M+\mu)^2 + 2\,m_W^2\,(1-\sin 2\b)}\, \right].
\end{equation}
The larger mass eigenvalue is always positive, $\eta_2^{} = 1$. 
The smaller eigenvalue is positive if 
$M\!\cdot\!\mu - m_W^2\sin 2\b < 0$, and negative otherwise.

The elements $U_{ij}$ and $V_{ij}$ of the diagonalizing 
matrices can also be directly expressed by the SUSY 
parameters $M$, $\mu$, and $\tan\b$: 
\begin{eqnarray}
  & & U_{12} = U_{21} = \frac{\t_1}{\rzw}\,
    \sqrt{1 + \frac{M^2 - \mu^2 - 2\,m_W\cos 2\b}{W}} \\[2mm]
  & & U_{22} = -U_{11} = \frac{\t_2}{\rzw}\,
    \sqrt{1 - \frac{M^2 - \mu^2 - 2\,m_W\cos 2\b}{W}} \\[2mm] 
  & & V_{21} = -V_{12} = \frac{\t_3}{\rzw}\,
    \sqrt{1 + \frac{M^2 - \mu^2 + 2\,m_W\cos 2\b}{W}} \\[2mm] 
  & & V_{22} = V_{11} = \frac{\t_4}{\rzw}\,
    \sqrt{1 - \frac{M^2 - \mu^2 + 2\,m_W\cos 2\b}{W}} 
\end{eqnarray}
with
\begin{equation}
  W = \sqrt{(M^2+\mu^2+2\,m_W^2)^2 - 4\,(M\!\cdot\!\mu -m_W^2\sin 2\b)^2}
\end{equation}
and the sign factors $\t_i$, $i = 1\ldots4$, 
\begin{equation}
  \{\t_1,\,\t_2,\t_3,\t_4\} = \; \left\{ \;
  \begin{array}{ll}
    \{ 1,\,\varepsilon_{\!_B},\,\varepsilon_{\!_A},\,1 \} & 
          \ldots\;\; \tan\b > 1 \\[2mm]
    \{ \varepsilon_{\!_B},\,1,\,1,\,\varepsilon_{\!_A} \} & 
          \ldots\;\; \tan\b < 1 \\    
  \end{array}\right.
\end{equation}
where
\begin{equation}
  \varepsilon_{\!_A} = {\rm sign}(M\sin\b + \mu\,\cos\b),
  \hspace{6mm}
  \varepsilon_{\!_B} = {\rm sign}(M\cos\b + \mu\,\sin\b).
\end{equation}

In what follows we will always use the convention \eq{mchdiagA} \cite{charginos}. 
However, one can also choose $U$ and $V$ such that $\M_C^{}$ has only 
non--negative entries \cite{haberkane}. 
This is for example achieved by solving the eigenvalue 
problem for $X^{\rm T} X$,
\begin{equation}
  \M_C^2 = {\rm diag}(\mchp{1}^2,\,\mchp{2}^2) = V\,X^{\rm T} X\,V^{-1}
\end{equation}
with
\begin{equation}
  V = \left( \begin{array}{rr}
    \cos\phi_{_1} & \sin\phi_{_1} \\ -\sin\phi_{_1} & \cos\phi_{_1}
    \end{array} \right)
\end{equation}
and the matrix $U$ given by
\begin{equation}
  U = \smaf{1}{\M_C^{}}\;V\,X^{\rm T} = \left( \begin{array}{rr}
    \cos\phi_{_2} & \sin\phi_{_2} \\ -\sin\phi_{_2} & \cos\phi_{_2}
    \end{array} \right).
\end{equation}
The mass eigenvalues then are
\begin{eqnarray}
  \mchp{i}^2 &=& \onehf\,\Big[\,M^2+\mu^2+2\,m_W^2 \nn \\[1mm] 
  & & \hspace{6mm} 
  \mp\:\sqrt{(M^2-\mu^2)^2 + 4\,m_W^2 \cos^2 2\b + 
             4\,m_W^2\,(M^2 + \mu^2 + 2\,M\!\cdot\!\mu\sin 2\b)}\,\Big]\, .
\end{eqnarray}
This can be a practical approach when treating the problem 
numerically (the analytic expressions for $U$ and $V$ become rather 
complicated) and is especially useful in case of complex parameters,
see \eg~ \cite{garfield}. 
However, one has to take care that $\phi_{_1}$ and $\phi_{_2}$ are always 
in the correct sector of the unitcircle \eg, $\in [0,\pi)$.

\subsection{Neutralinos}   \label{sect:neutralinos}

The neutral gauginos and higgsinos also mix. 
Their mass eigenstates are the neutralinos $\nt_i$, $i=1\ldots 4$. 
In general, both weak and mass eigenstates are Majorana fermions; 
however, if two neutralinos are degenerated in mass they can combine 
to a Dirac spinor. 

In this thesis we choose the basis 
\begin{equation}
  \psi^0 = \left( 
  -i\l_\g,\, -i\l_Z^{},\: \psi^a_H,\: \psi^b_H\, \right)^{\!\rm T},
  \hspace{6mm} 
\end{equation}
with
\begin{equation} 
  \psi^a_H = \psi^1_{H_1}\sin\b - \psi^2_{H_2}\cos\b, \qquad
  \psi^b_H = \psi^1_{H_1}\cos\b - \psi^2_{H_2}\sin\b,
\end{equation}
and $\l_\g$, $\l_Z^{}$, $\psi^1_{H_1}$, $\psi^2_{H_2}$ the 
two--component spinors of the photino $\ti\gamma$, zino $\ti Z^{0}\!$, 
and the neutral higgsinos $\ti H^{0}_{1}$, $\ti H^{0}_{2}$, respectively. 
The mass terms of the neutral gaugino--higgsino system 
can then be written as \cite{haberkane, neutralinos}
\begin{equation}
  \lag_m = -\onehf\,(\psi^{0})^{\rm T}\: Y \:\psi^0 + \,\hc
\end{equation}
with 
\begin{equation}
  Y = \left( \begin{array}{cccc} 
  M'\,c^2_W + M\,s^2_W & (M-M')\,s_W^{} c_W^{} & 0 & 0 \\[1mm]
  (M-M')\,s_W^{} c_W^{} & M'\,s^2_W + M\,c^2_W & m_Z^{} & 0 \\[1mm]
  0 & m_Z^{} & \mu\,s_{2\b}^{} & -\mu\,c_{2\b}^{} \\[1mm]
  0 & 0 & -\mu\,c_{2\b}^{} & -\mu\,s_{2\b}^{}
  \end{array} \right)
\end{equation}
where we have used the abbreviations
$s_W^{} = \sin\tW$, $c_W^{} = \cos\tW$, 
$s_{2\b} = \sin 2\b$, $c_{2\b} = \cos 2\b$. 
$M$ and $M'$ are the SU(2) and U(1) gaugino masses, 
$M\equiv M_2^{}$, $M'\equiv M_1^{}$ (we will stick to 
this notation in the following).

The mass matrix $Y$ is diagonalized by a $4\!\times\!4$ unitary 
matrix $N$,
\begin{equation}
  N^*\,Y\,N^{-1} = \M_N^{} 
\end{equation}
with $\M_N^{}$ the diagonal mass matrix.
The mass eigenstates in two--component notation then are
\begin{equation}
  \x^0_i = N_{ij}^{}\,\psi^0_j,
  \hspace{6mm} i,j=1\ldots 4,
\end{equation}
and in four--component notation
\begin{equation}
  \nt_i = {\x^0_i \choose \bar\x^0_i}, 
  \hspace{6mm} i=1\ldots 4.
\end{equation}
As in the charged gaugino--higgsino sector one can choose $N$ such 
that $\M_N^{}$ has no negative entries by solving the squared 
eigenvalue problem, 
\begin{equation}
  \M_N^2 = {\rm diag}(\mnt{1}^2,\,\ldots,\, \mnt{4}^2) = 
  N\,Y\kr Y\,N^{-1}.
\end{equation} 
However, assuming CP conservation, we find it convenient to allow 
for negative mass eigenvalues. We can then choose a phase convention 
in which $N$ is real and orthogonal: 
\begin{equation}
  \M_N^{} =  N\:Y\,N^{-1} = {\rm diag}
  (\varepsilon_{\!_1}\mnt{1},\, \varepsilon_{\!_2}\mnt{2},\,
   \varepsilon_{\!_3}\mnt{3},\, \varepsilon_{\!_4}\mnt{4}) 
\end{equation}
with $\varepsilon_i = \pm 1$ the sign of the eigenvalue  
and $\mnt{i} = |(\M_N)_{ii}|$ the mass of the neutralino $\nt_i$ 
($0\leq\mnt{1}\leq\mnt{2}\leq\mnt{3}\leq\mnt{4}$ by convention).

In the case $M\ll|\mu|$, the two lightest neutralinos 
are dominated by their gaugino components. 
In particular, $\nt_1\sim\ti B$ and $\nt_1\sim\ti W^3$. 
Similarly, $\chpm_1$ is mostly a charged $\ti W$--ino.
For the masses one finds very roughly 
$2\mnt{1}\sim\mnt{2}\sim\mch{1}\sim M$. 
In the opposite case ($|\mu|\ll M$), $\nt_{1,2}$ and $\chpm_1$ 
are mostly higgsinos with masses close to $|\mu|$. 
Finally, for $M\sim|\mu|$ the gaugino and higgsino states 
are strongly mixed.

\section{Interaction Lagrangian}

In this section we list the relevant parts of the Lagrangian and the 
Feynman rules for squark (and related) interactions needed in this thesis. 
All interactions are given in the $(\sq_L^{},\sq_R^{})$ as well as 
in the $(\sq_1^{},\sq_2^{})$ basis.
We concentrate on the 3rd generation ($q=t,b$; $\sq=\st,\sb$).  
However, all formule also apply for the 1st and 2nd generation, 
provided one inserts the (super)CKM--Matrix elements in a proper way. 
For the 3rd generation we take 
$\rm (CKM)_{33}=(sCKM)_{33}\equiv 1$.
The corresponding expressions for (s)lepton interactions can be 
derived by the obvious replacements $q\to\ell$, $t\to\nu$, 
$b\to\ell^-$, $\sq\to\ti\ell$, $e_q\to e_\ell^{}$, etc.

\subsection{Quark -- Quark -- Gauge Boson}

\begin{equation}
  \lag_{qq\g} \;=\; -e e_q\, A_\mu\, \bar{q}\,\g^\mu\, q 
\end{equation}
\begin{align}
  \lag_{qqZ} 
    &= -\frac{g}{\cos\tW}\, Z_\mu\, \bar{q}\: \gmu\, \{
        (I_{3L}^q - e_q \sin^2\tW)\,\PL - e_q \sin^2\tW\,\PR\}\, q
        \nn \\
    &= -\frac{g}{\cos\tW}\, 
    Z_\mu\, \bar{q}\: \gmu\, (\CL\,\PL + \CR\,\PR)\, q  
\label{eq:L-qqZ}
\end{align}
\begin{equation}
  C_{\!q L,R} \,:=\, I_{3L,R}^q - e_q \sin^2\tW
\end{equation}

\begin{equation}
  \lag_{qqW} \,=\, -\frac{g}{\rzw}\, 
     (W^+_\mu\, \bar t\, \gmu\, \PL\, b + 
      W^-_\mu\, \bar b\, \gmu\, \PL\, t)  
\end{equation}

\begin{equation}
  \lag_{qqg} \,=\, 
     -g_s\, T^a_{\!rs}\,G^a_{\!\mu}\,\bar{q}_r^{}\,\gmu\, q_s^{}
\end{equation}

\bigskip

\begin{picture}(80,25)
\put(25,0){\mbox{\psfig{figure=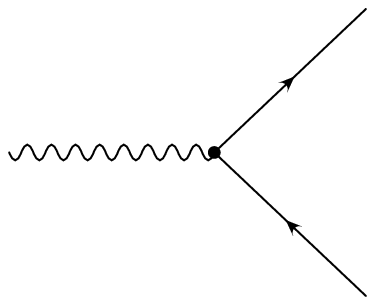,height=2.3cm}}}
\put(20,10){
  \makebox(0,0)[bl]{{\small $\g$}}}  
\put(55,22){
  \makebox(0,0)[bl]{{\small $q$}}}  
\put(55,-0.5){
  \makebox(0,0)[bl]{{\small $q$}}}  
\put(26.5,13.5){
  \makebox(0,0)[bl]{{\tiny $\mu$}}}  
\put(72,10){
  \makebox(0,0)[bl]{$-ie\,e_q\,\g^{\mu}$}}  
\end{picture} \\ 

\begin{picture}(80,25)
\put(25,0){\mbox{\psfig{figure=V_qqZ.ps,height=2.3cm}}}
\put(23,10.5){
  \makebox(0,0)[br]{{\small $Z$}}}  
\put(55,21.5){
  \makebox(0,0)[bl]{{\small $q$}}}  
\put(55,-0.5){
  \makebox(0,0)[bl]{{\small $q$}}}  
\put(26.5,13){
  \makebox(0,0)[bl]{{\tiny $\mu$}}}  
\put(72,10){
  \makebox(0,0)[bl]{$-\frac{ig}{\cos\tW}\g^{\mu}\,(\CL\PL + \CR\PR)$}}  
\end{picture} \\ 

\begin{picture}(80,25)
\put(25,0){\mbox{\psfig{figure=V_qqZ.ps,height=2.3cm}}}
\put(23,10.5){
  \makebox(0,0)[br]{{\small $W^\pm$}}}  
\put(55,21.5){
  \makebox(0,0)[bl]{{\small $q\,'$}}}  
\put(55,0){
  \makebox(0,0)[bl]{{\small $q$}}}  
\put(26.5,13){
  \makebox(0,0)[bl]{{\tiny $\mu$}}}  
\put(72,10){
  \makebox(0,0)[bl]{$-\frac{ig}{\rzw}\,\gmu\,\PL$}}  
\end{picture} \\ 

\begin{picture}(80,25)
\put(25,0){\mbox{\psfig{figure=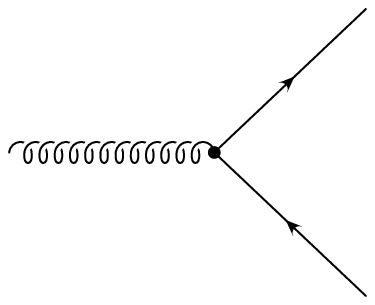,height=2.3cm}}}
\put(23,10.5){
  \makebox(0,0)[br]{{\small $g$}}}  
\put(55,21.5){
  \makebox(0,0)[bl]{{\small $q$}}}  
\put(55,0){
  \makebox(0,-0.5)[bl]{{\small $q$}}}  
\put(26.5,13.5){\makebox(0,0)[bl]{{\scriptsize $a,\mu$}}}  
\put(52,18){\makebox(0,0)[bl]{{\scriptsize $r$}}}  
\put(52,5){\makebox(0,0)[bl]{{\scriptsize $s$}}}  
\put(72,10){
  \makebox(0,0)[bl]{$-i g_s\,T^a_{rs}\,\gmu$}}  
\end{picture} \\ 

\subsection{Squark -- Squark -- Gauge Boson}

\noi (a) squark -- squark -- photon \\
\begin{eqnarray}
  \lag_{\ti{q}\ti{q}\g} 
  &=& i e e_q\, A_\mu\, 
    ( \sq_L^* \delr \sq_L^{} + \sq_R^* \delr \sq_R^{} ) \nn \\  
  &=& i e e_q\, A_\mu\, 
    ( \Rsq_{i1}\Rsq_{j1} + \Rsq_{i2}\Rsq_{j2})\, \sq_j^* \delr \sqi \nn \\  
  &=& i e e_q\,\d_{ij}\, A_\mu\: \sq_j^* \delr \sqi
\end{eqnarray}

\medskip

\noi (b) squark -- squark -- $Z^0$ \\
\begin{equation}
  \lag_{\ti{q}\ti{q} Z} 
  = \frac{ig}{\cos\tW}\, Z_\mu\,  
    (\CL\,\sq_L^*\delr\sq_L^{} + \CR\,\sq_R^*\delr\sq_R^{}) 
  = \frac{ig}{\cos\tW}\: c_{ij}^{}\, Z_\mu\,\sq_j^*\delr\sqi 
\label{eq:L-ssZ}
\end{equation}
\begin{eqnarray}
  c_{ij}^{} &:=& \CL\,\Rsq_{i1}\Rsq_{j1} + \CR\,\Rsq_{i2}\Rsq_{j2} 
  \nn\\[2mm]
  &=& \left( \begin{array}{cc}
    I^{q}_{3L}\,\cos^2\tsq - e_q \sin^2\tW 
         & -\onehf\, I^{q}_{3L}\,\sin 2\tsq \\[1mm]
    -\onehf\, I^{q}_{3L}\,\sin 2\tsq 
         & I^{q}_{3L}\,\sin^2\tsq - e_q \sin^2\tW
  \end{array} \right)_{ij}
\label{eq:cij}
\end{eqnarray}

\medskip

\noi (c) squark -- squark -- $W^\pm$ \\
\begin{eqnarray} 
  \lag_{\ti{q}\ti{q}' W} 
  &=& \frac{ig}{\rzw}\, 
  (W^+_\mu\,\st_L^*\delr\sb_L^{} + W^-_\mu\,\sb_R^*\,\delr\st_R^{}) \nn\\
  &=& \frac{ig}{\rzw}\, 
  (\R_{i1}^{\ti b}\R_{j1}^{\ti t}\,W^+_\mu\,\st_j^*\delr\sb_i^{} + 
   \R_{i1}^{\ti t}\R_{j1}^{\ti b}\,W^-_\mu\,\sb_j^*\delr\st_i^{}) 
\label{eq:csqW}
\end{eqnarray} 

\medskip

\noi (d) squark -- squark -- gluon \\
\begin{equation}
  \lag_{\ti{q}\ti{q} g}  
  = ig_s\,T^a_{\!rs}\, G_\mu^a\, 
       (\sq_{Lr}^*\delr\sq_{Ls}^{} + \sq_{Rr}^*\delr\sq_{Rs}^{}) 
  = ig_s\,T^a_{\!rs}\,\d_{ij}\,G_\mu^a\,\sq_{jr}^*\delr\sq_{is}^{}
\end{equation}

\bigskip

\noi
The corresponding Feynman rules we obtain from
\begin{equation}
  \hspace*{2cm}
  A\delr B = A\,(\de_\mu B) - (\de_\mu A)\,B \quad \to \quad 
  \sq_j^*\delr\sqi = i\,(k_i^{} + k_j^{})^\mu 
\end{equation}
where $k_i^{}$ and $k_j^{}$ are the four--momenta of $\sqi$ and $\sqj$
in direction of the charge flow.

\bigskip

\begin{picture}(80,25)
\put(25,0){\mbox{\psfig{figure=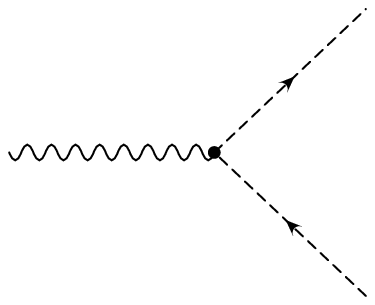,height=2.3cm}}}
\put(23,10){
  \makebox(0,0)[br]{{\small $\g$}}}  
\put(55,-0.5){
  \makebox(0,0)[bl]{{\small $\sqj$}}}  
\put(55,21.5){
  \makebox(0,0)[bl]{{\small $\sqi$}}}  
\put(45.5,18.5){
  \makebox(0,0)[br]{{\small $k$}}}  
\put(45,5){
  \makebox(0,0)[tr]{{\small $p$}}}  
\put(26.5,13.5){
  \makebox(0,0)[bl]{{\tiny $\mu$}}}  
\put(72,10){
  \makebox(0,0)[bl]{$-i e e_q\,(p + k)^\mu\,\d_{ij}$}}  
\end{picture} \\ 

\begin{picture}(80,25)
\put(25,0){\mbox{\psfig{figure=V_ssZ.ps,height=2.3cm}}}
\put(23,10.5){
  \makebox(0,0)[br]{{\small $Z$}}}  
\put(55,-0.5){
  \makebox(0,0)[bl]{{\small $\sqj$}}}  
\put(55,21.5){
  \makebox(0,0)[bl]{{\small $\sqi$}}}   
\put(26.5,13.5){
  \makebox(0,0)[bl]{{\tiny $\mu$}}}  
\put(45.5,18.5){
  \makebox(0,0)[br]{{\small $k$}}}  
\put(45,5){
  \makebox(0,0)[tr]{{\small $p$}}}  
\put(72,10){
  \makebox(0,0)[bl]{$-\frac{ig}{\cos\tW}\,c_{ij}\,(p + k)^\mu$}}  
\end{picture} \\ 

\begin{picture}(80,25)
\put(25,0){\mbox{\psfig{figure=V_ssZ.ps,height=2.3cm}}}
\put(23,10.5){
  \makebox(0,0)[br]{{\small $W^\pm$}}}  
\put(55,-0.5){
  \makebox(0,0)[bl]{{\small $\sqi$}}}  
\put(55,21.5){
  \makebox(0,0)[bl]{{\small $\sq_j'$}}} 
\put(26.5,13.5){
  \makebox(0,0)[bl]{{\tiny $\mu$}}}  
\put(45.5,18.5){
  \makebox(0,0)[br]{{\small $k$}}}  
\put(45,5){
  \makebox(0,0)[tr]{{\small $p$}}}  
\put(72,10){
  \makebox(0,0)[bl]{$-\frac{ig}{\rzw}\,
                     \Rsq_{i1}\R_{j1}^{\ti q'}\,(p + k)^\mu$}}  
\end{picture} \\ 

\begin{picture}(80,25)
\put(25,0){\mbox{\psfig{figure=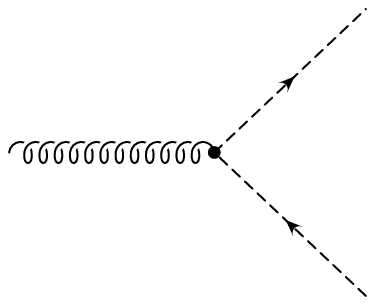,height=2.3cm}}}
\put(23,10.5){
  \makebox(0,0)[br]{{\small $g$}}}  
\put(55,-0.5){
  \makebox(0,0)[bl]{{\small $\sqj$}}}  
\put(55,21.5){
  \makebox(0,0)[bl]{{\small $\sqi$}}}  
\put(26.5,13.5){
  \makebox(0,0)[bl]{{\tiny $a,\mu$}}}  
\put(45.5,18.5){
  \makebox(0,0)[br]{{\small $k$}}}  
\put(45,5){
  \makebox(0,0)[tr]{{\small $p$}}}  
\put(52,18){\makebox(0,0)[bl]{{\scriptsize $r$}}}  
\put(52,5){\makebox(0,0)[bl]{{\scriptsize $s$}}}  
\put(72,10){
  \makebox(0,0)[bl]{$-i\,g_s\,T_{\!rs}^a\,(p + k)^\mu\,\d_{ij}$}}  
\end{picture} \\

\subsection{Quark -- Quark -- Higgs Boson}

\begin{eqnarray}
  \lag_{qqH} &=& 
  s_1^q\,h^0\,\bar{q}\,q + s_2^q\,H^0\,\bar{q}\,q 
                       + s_3^q\,A^0\,\bar{q}\,\g^5\,q  \nn\\ 
  & & \hspace{26mm} 
  +\:H^+\,\bar{t}\,(s_4^t\,\PL + s_4^b\,\PR)\,b 
  +  H^-\,\bar{b}\,(s_4^b\,\PL + s_4^t\,\PR)\,t
\end{eqnarray}
with
\begin{equation} 
  \hspace*{2cm}
  \begin{array}{ll} 
    s_1^t = -\,\frac{g\,m_t}{2\,m_W\sin\b}\,\cos\a, \hspace{16mm} &
    s_1^b = \frac{g\,m_b}{2\,m_W\cos\b}\,\sin\a, \\[2mm]
    s_2^t = -\,\frac{g\,m_t}{2\,m_W\sin\b}\,\sin\a, &
    s_2^b = -\,\frac{g\,m_b}{2\,m_W\cos\b}\,\cos\a, \\[2mm]
    s_3^t = i\,\frac{g\,m_t}{2\,m_W}\cot\b, &
    s_3^b = i\,\frac{g\,m_b}{2\,m_W}\tan\b,\\[2mm]
    s_4^t = \,\frac{g\,m_t}{\rzw\,m_W}\cot\b & 
    s_4^b = \,\frac{g\,m_b}{\rzw\,m_W}\tan\b. 
  \end{array}
\label{eq:qqHcop}
\end{equation}

\bigskip
        
\begin{picture}(80,25)
\put(25,0){\mbox{\psfig{figure=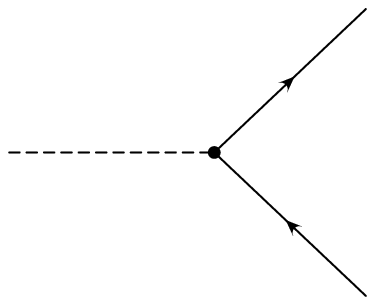,height=2.3cm}}}
\put(22,10.5){
  \makebox(0,0)[br]{{\small $h^0$}}}  
\put(54,-0.5){
  \makebox(0,0)[bl]{{\small $q$}}}  
\put(54,21.5){
  \makebox(0,0)[bl]{{\small $q$}}}  
\put(72,10){
  \makebox(0,0)[bl]{$i\,s_{1}^q$}}  
\end{picture} \\ 

\begin{picture}(80,25)
\put(25,0){\mbox{\psfig{figure=V_hqq.ps,height=2.3cm}}}
\put(22,10.5){
  \makebox(0,0)[br]{{\small $H^0$}}}  
\put(54,-0.5){
  \makebox(0,0)[bl]{{\small $q$}}}  
\put(54,21.5){
  \makebox(0,0)[bl]{{\small $q$}}}  
\put(72,10){
  \makebox(0,0)[bl]{$i\,s_{2}^q$}}  
\end{picture} \\ 

\begin{picture}(80,25)
\put(25,0){\mbox{\psfig{figure=V_hqq.ps,height=2.3cm}}}
\put(22,10.5){
  \makebox(0,0)[br]{{\small $A^0$}}}  
\put(54,-0.5){
  \makebox(0,0)[bl]{{\small $q$}}}  
\put(54,21.5){
  \makebox(0,0)[bl]{{\small $q$}}}  
\put(72,10){
  \makebox(0,0)[bl]{$i\,s_{3}^q\,\g^5$}}  
\end{picture} \\ 

\begin{picture}(80,25)
\put(25,0){\mbox{\psfig{figure=V_hqq.ps,height=2.3cm}}}
\put(22,10.5){
  \makebox(0,0)[br]{{\small $H^\pm$}}}  
\put(54,-0.5){
  \makebox(0,0)[bl]{{\small $q '$}}}  
\put(54,21.5){
  \makebox(0,0)[bl]{{\small $q$}}}  
\put(72,10){
  \makebox(0,0)[bl]{$i\,(s_4^q\,\PL + s_4^{q '}\,\PR)$}}  
\end{picture} \\

\subsection{Squark -- Squark -- Higgs Boson}
\label{sect:feyn-ssH}
%
Defining $H_{\!k}^{}=\{h^0,\,H^0,\,A^0,\,H^\pm\}$ we can write the Higgs 
interaction with squarks in the general form
\begin{equation}
  \lag_{\ti{q}\ti{q} H} = \Hk\,
    \left( \sq_L^{\b*},\, \sq_R^{\b*} \right) \, 
    \hat G_k^{\,\a} \, {\sq_L^\a \choose \sq_R^\a}
    = (G_k^{\,\a})_{ij}^{}\,\Hk\,\sq_j^{\b*}\sq_i^\a .
\end{equation}
where $\a$ and $\b$ are flavor indices. 
For $k=1,2,3$ we have of course $\a=\b$; in case of $k=4$ we have 
$\hat G_4 \equiv \hat G_4^{\,\ti u} = (\hat G_4^{\,\ti d})^{\rm T}$. 
$\hat G_k^{}$ and $G_k^{}$ are related by 
\begin{equation}
   G_k^{\,\sq} = 
   \R^{\sq}\; \hat G_{k}^{\,\sq}\; (\R^{\sq})^{\rm T},
   \qquad (k=1,2,3)
\label{eq:Gksq}
\end{equation}
\begin{equation}
   G_4^{\,\ti t} = 
   \R^{\ti t} \; \hat G_4^{} \; (\R^{\ti b})^{\rm T}, \qquad \;
   G_4^{\,\ti b} = 
   \R^{\ti b}\; (\hat G_4^{})^{\rm T}\; (\R^{\ti t})^{\rm T} 
   = (G_4^{\,\ti t})^{\rm T}.
\label{eq:G4sq}
\end{equation}

\bigskip
        
\begin{picture}(80,25)
\put(25,0){\mbox{\psfig{figure=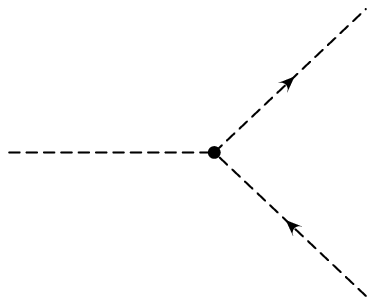,height=2.3cm}}}
\put(22,10.5){
  \makebox(0,0)[br]{\small $\Hk$}}  
\put(54,-0.5){
  \makebox(0,0)[bl]{\small $\sq_i^\a$}}  
\put(54,21.5){
  \makebox(0,0)[bl]{\small $\sq_j^\b$}}  
\put(72,10){
  \makebox(0,0)[bl]{
      $i\,\left[ \R^\a\,\hat G_k^{\,\a}\,(\R^\b)^{\rm T} \right]_{ij}    
       = i\,(G_k^{\,\a})_{ij}^{}$}}  
\end{picture} \\ 

\clearpage

\noi
The individual contributions are:

\bigskip

\noi (a) squark -- squark -- $h^0$
\begin{eqnarray}
  \lag_{\sq\sq h^0} &=& g\:\left\{\,
  \left[\,
     \smaf{m_Z}{\cos\t_W}\,(\onehf-\twoth\sW)\sin(\a\!+\!\b)
    -\smaf{m_t^2}{m_W\sin\b}\cos\a \,\right]\,h^0\st_L^*\st_L^{} 
  \right. \nn\\[2mm]
  & & \hspace{8mm} +\,\left[\,
     \twoth\smaf{m_Z}{\cos\t_W}\sW \sin(\a\!+\!\b)
    -\smaf{m_t^2}{m_W\sin\b}\cos\a \,\right]\,h^0\st_R^*\st_R^{} \nn\\[2mm] 
  & & \hspace{8mm} -\,\smaf{m_t}{2m_W\sin\b}\,
    (A_t\cos\a + \mu\sin\a)\,(h^0\st_R^*\st_L^{} + h^0\st_L^*\st_R^{}) \nn\\[2mm] 
  & & \hspace{8mm} -\,\left[\,
     \smaf{m_Z}{\cos\t_W}\,(\onehf-\oneth\sW)\sin(\a\!+\!\b)
    -\smaf{m_b^2}{m_W\cos\b}\sin\a \,\right]\,h^0\sb_L^*\sb_L^{} \nn\\[2mm]
  & & \hspace{8mm} -\,\left[\,
     \oneth\smaf{m_Z}{\cos\t_W} \sW \sin(\a\!+\!\b)
    -\smaf{m_b^2}{m_W\sin\b}\sin\a \,\right]\,h^0\sb_R^*\sb_R^{} \nn\\[2mm] 
  & & \hspace{8mm} \left.
    +\,\smaf{m_t}{2m_W\cos\b}\,
    (A_b\sin\a + \mu\cos\a)\,(h^0\sb_R^*\sb_L^{} + h^0\sb_L^*\sb_R^{}) 
    \,\right\} 
\end{eqnarray}
and thus
\begin{equation}
\hat G_1^{\,\sq} = \left(\! \begin{array}{cc}
  \frac{g\,m_Z^{}}{{\rm c}_W^{}}\,\CL\,{\rm s}_{\a+\b} 
                        -\rzw\;m_q\,h_q \poss{{\rm c}_\a}{-{\rm s}_\a} 
  & -\frac{1}{\rzw}\,h_q\,\big( A_q \poss{{\rm c}_\a}{-{\rm s}_\a} 
                                +\mu\poss{{\rm s}_\a}{-{\rm c}_\a}\big) 
  \\[4mm]
  -\frac{1}{\rzw}\,h_q\,\big( A_q \poss{{\rm c}_\a}{-{\rm s}_\a} 
                              +\mu\poss{{\rm s}_\a}{-c_\a}\big) 
  & \frac{g\,m_Z^{}}{{\rm c}_W^{}}\,\CR\,{\rm s}_{\a+\b} 
                         -\rzw\,m_q\,h_q \poss{{\rm c}_\a}{-{\rm s}_\a} 
\end{array}\! \right) 
\label{eq:GLR1}
\end{equation}

\noi 
for {\small $\Big\{\!\begin{array}{c} \mbox{\footnotesize up} \\[-1mm]
\mbox{\footnotesize down} \end{array} \!\Big\}$}
type squarks respectively. 
We use the abbreviations ${\rm c}_W^{} = \cos\tW$, 
${\rm s}_\a$ = $\sin\a$,\\[-2mm]
${\rm c}_\a$ = $\cos\a$, 
${\rm s}_{\a+\b} = \sin(\a\!+\!\b)$,  
$\CL = I_{3L}^q\!-\!e_q\sin^2\t_W^{}$, and 
$\CR = e_q\sin^2\t_W^{}$. $\alpha$ is the mixing angle in the CP even
neutral Higgs boson sector.
$h_q$ are the Yukawa couplings:   
\begin{equation}
  h_t = \frac{g\,m_t}{\sqrt{2}\:m_{W}\sin\b}, \hspace{8mm} 
  h_b = \frac{g\,m_b}{\sqrt{2}\:m_{W}\cos\b}.
\end{equation}

\bigskip

\noi (b) squark -- squark -- $H^0$
\begin{eqnarray}
  \lag_{\sq\sq H^0} &=& -g\:\left\{\,
  \left[\,
     \smaf{m_Z}{\cos\t_W}\,(\onehf-\twoth\sW)\cos(\a\!+\!\b)
    +\smaf{m_t^2}{m_W\sin\b}\sin\a \,\right]\,H^0\st_L^*\st_L^{} 
  \right. \nn\\[2mm]
  & & \hspace{14mm} +\,\left[\,
     \twoth\smaf{m_Z}{\cos\t_W}\sW \cos(\a\!+\!\b)
    +\smaf{m_t^2}{m_W\sin\b}\sin\a \,\right]\,H^0\st_R^*\st_R^{} \nn\\[2mm] 
  & & \hspace{14mm} +\,\smaf{m_t}{2m_W\sin\b}\,
    (A_t\sin\a - \mu\cos\a)\,(H^0\st_R^*\st_L^{} + H^0\st_L^*\st_R^{}) \nn\\[2mm] 
  & & \hspace{14mm} -\,\left[\,
     \smaf{m_Z}{\cos\t_W}\,(\onehf-\oneth\sW)\cos(\a\!+\!\b)
    -\smaf{m_b^2}{m_W\cos\b}\cos\a \,\right]\,H^0\sb_L^*\sb_L^{} \nn\\[2mm]
  & & \hspace{14mm} -\,\left[\,
     \oneth\smaf{m_Z}{\cos\t_W} \sW \cos(\a\!+\!\b)
    -\smaf{m_b^2}{m_W\sin\b}\cos\a \,\right]\,H^0\sb_R^*\sb_R^{} \nn\\[2mm] 
  & & \hspace{14mm} \left.
    +\,\smaf{m_b}{2m_W\cos\b}\,
    (A_b\cos\a - \mu\sin\a)\,(H^0\sb_R^*\sb_L^{} + H^0\sb_L^*\sb_R^{}) 
    \,\right\}   
\end{eqnarray}

\begin{equation}
\hat G_2^{\,\sq} = \left(\! \begin{array}{cc}
  -\frac{g\,m_Z^{}}{{\rm c}_W^{}}\,\CL\,{\rm c}_{\a+\b} 
                        -\rzw\;m_q\,h_q \poss{{\rm s}_\a}{{\rm c}_\a} 
  & -\frac{1}{\rzw}\,h_q\,\big( A_q \poss{{\rm s}_\a}{{\rm c}_\a} 
                                -\mu\poss{{\rm c}_\a}{{\rm s}_\a}\big) 
  \\[4mm]
  -\frac{1}{\rzw}\,h_q\,\big( A_q \poss{{\rm s}_\a}{{\rm c}_\a} 
                              -\mu\poss{{\rm c}_\a}{s_\a}\big) 
  & -\frac{g\,m_Z^{}}{{\rm c}_W^{}}\,\CR\,{\rm c}_{\a+\b} 
                        -\rzw\,m_q\,h_q \poss{{\rm s}_\a}{{\rm c}_\a} 
\end{array}\! \right) 
\label{eq:GLR2}
\end{equation}

\noi 
Notice that $G_2^{\,\sq}$ can be obtained from $G_1^{\,\sq}$ by the 
replacement $\a\to\a+\pi/2$, \ie ${\rm c}_\a \to {\rm s}_a$ and 
${\rm s}_\a \to -{\rm c}_a$.

\bigskip

\noi (c) squark -- squark -- $A^0$
\begin{eqnarray}
  \lag_{\ti{q}\ti{q} A^0} &=& \frac{ig}{2m_W^{}}\,\left[\,
    m_t\,(A_t\cot\b + \mu)\,(A^0\st_R^*\st_L^{} - A^0\st_L^*\st_R^{})  
  \right. \nn\\[1mm]
  & & \hspace{40mm} \left.
    +\,m_b\,(A_b\tan\b + \mu)\,(A^0\sb_R^*\sb_L^{} - A^0\sb_L^*\sb_R^{})\,
  \right]  
\end{eqnarray}
\begin{equation}
\hat G_3^{\,\sq} = i\,\smaf{g\,m_q}{2\,m_W}\, 
  \left(\! \begin{array}{cc}
     0 & - A_q\poss{\cot\b}{\tan\b} - \mu  \\[3mm]
     A_q\poss{\cot\b}{\tan\b} + \mu & 0
  \end{array}\! \right) 
\label{eq:GLR3}
\end{equation}

\noi
In this particular case we have $G_3^{\,\sq} = \hat G_3^{\,\sq}$ 
because $\hat G_3^{\,\sq}$ ist an off--diagonal and $\R^{\sq}$ is an 
unitary matrix.

\bigskip

\noi (d) squark -- squark -- $H^\pm$
\begin{eqnarray}
  \lag_{\ti{q}\ti{q} H^\pm} &=& \frac{g}{\rzw\,m_W^{}}\,\left[\,
    (m_b^2\tan\b + m_t^2\cot\b - m_W^2\sin 2\b)\,H^+\st_L^*\sb_L^{}
    +\smaf{2 m_t m_b}{\sin 2\b}\, H^+\st_R^*\sb_R^{} 
  \right. \nn\\[2mm]
  & & \hspace{4mm} \left.
    +\,m_t\,(A_t\cot\b + \mu)\,H^+\st_L^*\,\sb_R^{}
    +  m_b\,(A_b\tan\b + \mu)\,H^+\st_R^*\,\sb_L^{} \,\right] + \hc 
\end{eqnarray}

\begin{equation}
\hat G_4^{ } = \smaf{g}{\rzw\,m_W^{}}
  \left(\! \begin{array}{cc}
    m_b^2\tan\b + m_t^2\cot\b - m_W^2\sin 2\b 
                           & m_b\,(A_b\tan\b + \mu) \\[3mm]
    m_t\,(A_t\cot\b + \mu) & \frac{2 m_t m_b}{\sin 2\b}
\end{array}\! \right) 
\label{eq:GLR4}
\end{equation}

\subsection{Quark -- Squark -- Chargino} \label{sect:qsqch}
%
\begin{eqnarray}
  \lag_{q\sq\ch} 
  &=& g\,\bar t\, (-U_{1j}\,\PR + Y_t\,V_{2j}\,\PL)\, \chp_j\,\sb_L^{}
                    + g\,\bar t\,(Y_b\,U_{2j}\,\PR)\, \chp_j\,\sb_R^{} 
       \nn\\
  & & +\,g\,\bar b\, 
            (-V_{1j}\,\PR + Y_b\,U_{2j}\,\PL)\, \chpc_j\,\st_L^{}
             + g\,\bar b\, (Y_t\,V_{2j}\,\PR)\, \chpc_j\,\st_R^{} 
       \nn\\
  & & +\,g\,\chpb_j\, 
                  (-U_{1j}\,\PL + Y_t\,V_{2j}\,\PR)\, t\,\sb_L^*
                  + g\,\chpb_j\, (Y_b\,U_{2j}\,\PL)\, t\,\sb_R^* 
       \nn\\
  & & +\,g\,\chpcb_j\,(-V_{1j}\,\PL + Y_b\,U_{2j}\,\PR)\, b\,\st_L^*
                     + g\,\chpcb_j\, (Y_t\,V_{2j}\,\PL)\, b\,\st_R^* 
     \nn \\[2mm]
  &=& g\,\bar t\,
        (\ell_{ij}^{\ti b}\,\PR + k_{ij}^{\ti b}\,\PL)\,\chp_j\,\sb_i^{} + 
      g\,\bar b\,
        (\ell_{ij}^{\ti t}\,\PR + k_{ij}^{\ti t}\,\PL)\,\chpc_j\,\st_i^{} 
       \nn\\  
  & & +\, g\,\chpb_j\,
        (\ell_{ij}^{\ti b}\,\PL + k_{ij}^{\ti b}\,\PR)\,t\,\sb_i^* +
      g\,\chpcb_j\,
        (\ell_{ij}^{\ti t}\,\PL + k_{ij}^{\ti t}\,\PR)\, b\,\st_i^*
\label{eq:qsqch}
\end{eqnarray}        
The $\sqi$--$q'$--$\ch_j$ couplings $\lij^{\sq}$ and $\kij^{\sq}$ 
can be written as 
\begin{equation}
  \hspace*{21mm}
  \lij^{\sq} = \sum_n\, \Rsq_{in}\,{\cal O}_{jn}^q, \qquad
  \kij^{\sq} = \Rsq_{i1}\,{\cal O}_{j2}^{q'}
\end{equation}
with
\begin{equation}
  \hspace*{21mm}
  {\cal O}_j^t = {-V_{j1} \choose Y_t\,V_{j2}}, \qquad
  {\cal O}_j^b = {-U_{j1} \choose Y_b\,U_{j2}}.
\end{equation}
and the Yukawa factors $Y_q = h_q/g$, \ie
\begin{equation}
  \hspace*{21mm}
  Y_t = \frac{m_t}{\sqrt{2}\:m_W\sin\b}, \hspace{8mm} 
  Y_b = \frac{m_b}{\sqrt{2}\:m_W\cos\b}.
\end{equation}

\noi 
$U$ and $V$ are the $2\!{\small\times}\!2$ unitary matrices 
diagonalizing the charged gaugino--higgsino mass matrix
as defined in Sect.~\ref{sect:charginos}. 
        
\bigskip

\begin{picture}(80,25)
\put(25,0){\mbox{\psfig{figure=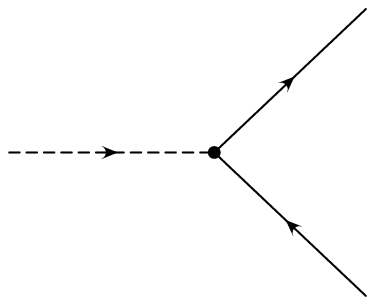,height=2.3cm}}}
\put(22,9.5){
  \makebox(0,0)[br]{{\small $\st_i$}}}  
\put(54,-0.5){
  \makebox(0,0)[bl]{{\small $\chpc_j$}}}  
\put(54,21.5){
  \makebox(0,0)[bl]{{\small $b$}}}  
\put(72,10){
  \makebox(0,0)[bl]{$ig\,(\ell_{ij}^{\ti t}\,\PR + 
                             k_{ij}^{\ti t}\,\PL)$}}  
\end{picture} \\ 

\begin{picture}(80,25)
\put(25,0){\mbox{\psfig{figure=V_sqq.ps,height=2.3cm}}}
\put(22,9.5){
  \makebox(0,0)[br]{{\small $\sb_i$}}}  
\put(54,-0.5){
  \makebox(0,0)[bl]{{\small $\chp_j$}}}  
\put(54,21.5){
  \makebox(0,0)[bl]{{\small $t$}}}  
\put(72,10){
  \makebox(0,0)[bl]{$ig\,(\ell_{ij}^{\ti b}\,\PR + k_{ij}^{\ti b}\,\PL)$}}  
\end{picture} \\ 

\begin{picture}(80,25)
\put(25,0){\mbox{\psfig{figure=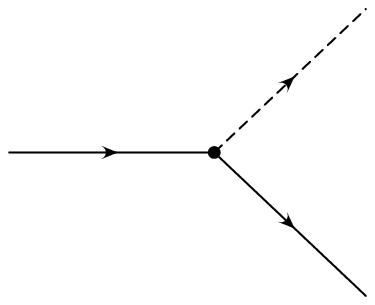,height=2.3cm}}}
\put(22,10.5){
  \makebox(0,0)[br]{{\small $b$}}}  
\put(54,-0.5){
  \makebox(0,0)[bl]{{\small $\chpc_j$}}}  
\put(54,21.5){
  \makebox(0,0)[bl]{{\small $\st_i$}}}  
\put(72,10){
  \makebox(0,0)[bl]{$ig\,(\ell_{ij}^{\ti t}\,\PL + k_{ij}^{\ti t}\,\PR)$}}  
\end{picture} \\ 

\begin{picture}(80,25)
\put(25,0){\mbox{\psfig{figure=V_qqs.ps,height=2.3cm}}}
\put(22,10.5){
  \makebox(0,0)[br]{{\small $t$}}}  
\put(54,-0.5){
  \makebox(0,0)[bl]{{\small $\chp_j$}}}  
\put(54,21){
  \makebox(0,0)[bl]{{\small $\sb_i$}}}  
\put(72,10){
  \makebox(0,0)[bl]{$ig\,(\ell_{ij}^{\ti b}\,\PL + k_{ij}^{\ti b}\,\PR)$}}  
\end{picture} \\

\subsection{Quark -- Squark -- Neutralino} \label{sect:qsqnt}
%
\begin{eqnarray}
  \lag_{q\sq\nt} 
  &=& g\,\bar q\,(f_{Lk}^q\,\PR + h_{Lk}^q\,\PL)\,\nt_k\,\sq_L^{}  
    + g\,\bar q\,(h_{Rk}^q\,\PR + f_{Rk}^q\,\PL)\,\nt_k\,\sq_R^{} 
    + \hc \nn\\[2mm]
  &=& g\,\bar{q}\,(a_{ik}^{\sq}\,\PR + b_{ik}^{\sq}\,\PL)\,\nt_k\,\sqi
    + g\,\ntb_k\,(a_{ik}^{\sq}\,\PL + b_{ik}^{\sq}\,\PR)\,q\,\sq_i^*      
\end{eqnarray}        
The $\sqi$--$q$--$\nt_k$ couplings $\aik^{\sq}$ and $\bik^{\sq}$ 
are given by
\begin{equation}
  \hspace*{20mm}
  \aik^{\sq} = \sum_n\, \Rsq_{in}\,{\cal A}_{kn}^q, \qquad
  \bik^{\sq} = \sum_n\, \Rsq_{in}\,{\cal B}_{kn}^{\,q}
\end{equation}
with
\begin{equation}
  \hspace*{20mm}
  {\cal A}_k^q = {f_{Lk}^q \choose h_{Rk}^q}, \qquad
  {\cal B}_k^q = {h_{Lk}^q \choose f_{Rk}^q},
\end{equation}
and
\begin{align}
  \hspace*{2cm}
  f_{Lk}^q &= -\rzw\,e_q \sin\tW N_{k1} 
    -\rzw\,(I_{3L}^q -e_q\sin^{2}\tW)\smaf{N_{k2}}{\cos\tW}, 
    \hspace{-1cm} \label{eq:fLk}\\
  f_{Rk}^q &= 
    -\rzw\,e_q \sin\tW (\tan\tW N_{k2} - N_{k1}), \\
  h_{Lk}^t &= \hphantom{-}Y_t \left(N_{k3}\sin\b - N_{k4}\cos\b \right)
                     =  h_{Rk}^t , \\ 
  h_{Lk}^b &= -Y_b \left(N_{k3}\cos\b + N_{k4}\sin\b \right) 
                     = h_{Rk}^b. \label{eq:hLk}
\end{align}

\noi 
$N$ is the $4\!\times\!4$ unitary matrix diagonalizing the 
neutral gaugino--higgsino mass matrix, see Sect.~\ref{sect:neutralinos}. 

\bigskip
        
\begin{picture}(80,25)
\put(25,0){\mbox{\psfig{figure=V_sqq.ps,height=2.3cm}}}
\put(22,9.5){
  \makebox(0,0)[br]{{\small $\st_i$}}}  
\put(54,-0.5){
  \makebox(0,0)[bl]{{\small $\nt_k$}}}  
\put(54,21.5){
  \makebox(0,0)[bl]{{\small $t$}}}  
\put(72,10){
  \makebox(0,0)[bl]{$ig\,(a_{ik}^{\st}\,\PR + b_{ik}^{\st}\,\PL)$}}  
\end{picture} \\ 

\begin{picture}(80,25)
\put(25,0){\mbox{\psfig{figure=V_sqq.ps,height=2.3cm}}}
\put(22,9.5){
  \makebox(0,0)[br]{{\small $\sb_i$}}}  
\put(54,-0.5){
  \makebox(0,0)[bl]{{\small $\nt_k$}}}  
\put(54,21.5){
  \makebox(0,0)[bl]{{\small $b$}}}  
\put(72,10){
  \makebox(0,0)[bl]{$ig\,(a_{ik}^{\sb}\,\PR + b_{ik}^{\sb}\,\PL)$}}  
\end{picture} \\ 

\begin{picture}(80,25)
\put(25,0){\mbox{\psfig{figure=V_qqs.ps,height=2.3cm}}}
\put(22,10.5){
  \makebox(0,0)[br]{{\small $t$}}}  
\put(54,-0.5){
  \makebox(0,0)[bl]{{\small $\nt_k$}}}  
\put(54,21.5){
  \makebox(0,0)[bl]{{\small $\st_i$}}}  
\put(72,10){
  \makebox(0,0)[bl]{$ig\,(a_{ik}^{\st}\,\PL + b_{ik}^{\st}\,\PR)$}}  
\end{picture} \\ 

\begin{picture}(80,25)
\put(25,0){\mbox{\psfig{figure=V_qqs.ps,height=2.3cm}}}
\put(22,10.5){
  \makebox(0,0)[br]{{\small $b$}}}  
\put(54,-0.5){
  \makebox(0,0)[bl]{{\small $\nt_k$}}}  
\put(54,21){
  \makebox(0,0)[bl]{{\small $\sb_i$}}}  
\put(72,10){
  \makebox(0,0)[bl]{$ig\,(a_{ik}^{\sb}\,\PL + b_{ik}^{\sb}\,\PR)$}}  
\end{picture} \\

\subsection{Interactions with Gluinos}

\noi (a) quark -- squark -- gluino \\
\begin{eqnarray}
  \lag_{q\sq\ti g} 
  &=& -\rzw\,g_s\, T^a_{\!rs} \left[
    (\bq_r^{}\,\PR\,\sg^a\,\sq_{L,s}^{} - \bq_r^{}\,\PR\,\sg^a\,\sq_{R,s}^{}) + 
    ({\bar{\sg}}^a\,\PL\,q_r^{}\,\sq_{L,s}^* 
    - {\bar{\sg}}^a\,\PL\,q_R^{}\,\sq_{R,s}^*) 
    \right] \nn\\
  &=& -\rzw\,g_s\, T^a_{\!rs} \left[
    \bq_r^{}\, (\Rsq_{i1}\PR - \Rsq_{i2}\PL) \,\sg^a\,\sq_{i,s}^{} + 
    {\bar{\sg}}^a\, (\Rsq_{i1}\PL - \Rsq_{i2}\PR) \,q_r^{}\,\sq_{i,s}^*
    \right] 
\end{eqnarray}    

\noi
Note the relative minus sign between the terms with $\sq_L^{}$ and $\sq_R^{}$: 
This is due to the facts that $\sq_R^{}$ are colour anti--triplets and  
the anti--colour generator is $-T^{a\dagger}$.

\clearpage

\noi (b) gluon -- gluino -- gluino 

\begin{equation}
  \lag_{g\ti g\ti g} = \frac{ig_s}{2}\,f_{abc}\,
  G^a_{\!\mu}\,{\bar{\sg}}^b\,\g^\mu\,\sg^c
\end{equation}

\noi
Owing to the Majorana nature of the gluino one must multiply 
by 2 to obtain the Feynman rule (or add the graph with 
$\sg \leftrightarrow \bar{\sg}$)!

\bigskip

\begin{picture}(80,25)
\put(25,0){\mbox{\psfig{figure=V_qqs.ps,height=2.3cm}}}
\put(22,10.5){
  \makebox(0,0)[br]{{\small $q$}}}  
\put(54,-0.5){
  \makebox(0,0)[bl]{{\small $\ti g$}}}  
\put(54,21.5){
  \makebox(0,0)[bl]{{\small $\sqi$}}}  
\put(27,13){\makebox(0,0)[bl]{{\scriptsize $r$}}}  
\put(51,19){\makebox(0,0)[tl]{{\scriptsize $s$}}}
\put(50,4){
  \makebox(0,0)[bl]{{\tiny $a$}}}  
\put(72,10){
  \makebox(0,0)[bl]{$-\sqrt{2}\,i\,g_s\,T^a_{\!rs}\,(\Rsq_{i1}\PL - \Rsq_{i2}\PR)$}}  
\end{picture} \\ 

\begin{picture}(80,25)
\put(25,0){\mbox{\psfig{figure=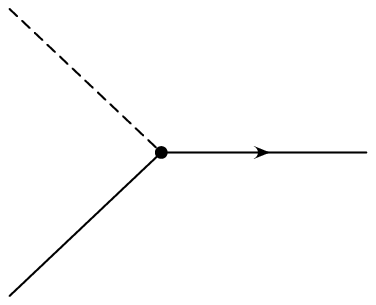,height=2.3cm}}}
\put(22,22){
  \makebox(0,0)[br]{{\small $\sqi$}}}  
\put(22,0){
  \makebox(0,0)[br]{{\small $\ti g$}}}  
\put(55.5,10.5){
  \makebox(0,0)[br]{{\small $q$}}}  
\put(24,4){
  \makebox(0,0)[bl]{{\tiny $a$}}}  
\put(25,18){\makebox(0,0)[bl]{{\scriptsize $s$}}}  
\put(50,13){\makebox(0,0)[bl]{{\scriptsize $r$}}}
\put(72,10){
  \makebox(0,0)[bl]{$-\sqrt{2}\,i\,g_s\,T^a_{\!rs}\,(\Rsq_{i1}\PR - \Rsq_{i2}\PL)$}}  
\end{picture} \\ 

\begin{picture}(80,25)
\put(25,0){\mbox{\psfig{figure=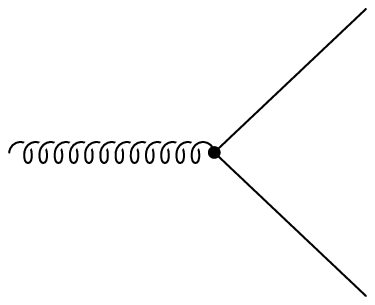,height=2.3cm}}}
\put(22,10.5){
  \makebox(0,0)[br]{{\small $g$}}}  
\put(54,-0.5){
  \makebox(0,0)[bl]{{\small $\ti g$}}}  
\put(54,21.5){
  \makebox(0,0)[bl]{{\small $\ti g$}}}  
\put(26.5,14){
  \makebox(0,0)[bl]{{\tiny $a,\mu$}}}  
\put(50.5,18){
  \makebox(0,0)[bl]{{\tiny $b$}}}  
\put(50.5,4){
  \makebox(0,0)[bl]{{\tiny $c$}}}
\put(72,10){
  \makebox(0,0)[bl]{$-g_s\,f_{abc}\,\g^\mu$}}  
\end{picture} \\

\subsection{Squark -- Squark -- Gauge Boson -- Gauge Boson}


\begin{equation}
  \lag_{\sq\sq\g\g} 
  = e^2 e_q^2\, A_\mu\,A^\mu\, 
      (\sq_L^*\,\sqL + \sq_R^*\,\sqR)  
  = e^2 e_q^2\,\d_{ij}^{}\, A_\mu\,A^\mu\, \sq_j^*\,\sqi  
\end{equation}
%
%
\begin{eqnarray}
  \lag_{\sq\sq ZZ} 
  &=& \frac{g^2}{\cos^2\tW}\, Z_\mu\,Z^\mu\,
    (C_{\!qL}^2\,\sq_L^*\,\sqL + C_{\!qR}^2\,\sq_R^*\,\sqR) \nn \\
  &=& \frac{g^2}{\cos^2\tW}\, Z_\mu\,Z^\mu\,
    (C_{\!qL}^2\Rsq_{i1}\Rsq_{j1} 
   + C_{\!qR}^2\Rsq_{i2}\Rsq_{j2})\,\sq_j^*\,\sqi \nn \\
  &=& \frac{g^2}{\cos^2\tW}\,z_{ij}\, 
         Z_\mu\,Z^\mu\,\sq_j^*\,\sqi    
\end{eqnarray} 
\begin{equation}
  \hspace*{2cm} 
  z_{ij} = \left( \begin{array}{ll}
    C_{\!qL}^2\cos^2\tsq + C_{\!qR}^2\sin^2\tsq 
    & -\onehf\,(C_{\!qL}^2-C_{\!qR}^2)\,\sin 2\tsq 
    \\[1mm]
    -\onehf\,(C_{\!qL}^2-C_{\!qR}^2)\,\sin 2\tsq 
    & C_{\!qL}^2\sin^2\tsq + C_{\!qR}^2\cos^2\tsq
  \end{array} \right)_{ij}
\end{equation}

\begin{equation}
  \lag_{\sq\sq WW} 
  = \onehf\,g^2\,W^+_\mu\, W^{-\mu}\, \sq_L^*\,\sqL 
  = \onehf\,g^2\,\Rsq_{i1}\Rsq_{j1}\, W^+_\mu\, W^{-\mu}\, \sq_j^*\,\sqi  
\end{equation}
  

\begin{equation}
  \lag_{\sq\sq\g Z} 
  = \frac{2eg}{\cos\tW}\,A_\mu\,Z^\mu\,
    (\CL\,\sq_L^*\,\sqL + \CR\,\sq_R^*\,\sqR) 
  = \frac{2eg}{\cos\tW}\,c_{ij}\,A_\mu\,Z^\mu\,\sq_j^*\,\sqi
\end{equation}            
%
%
\begin{eqnarray}
  \lag_{\sq\sq'\g W} 
  &=& \frac{eg}{\sqrt{2}}\:{\rm Y}_{\!Q}\,A_{\mu}\, 
    (W^{+\mu}\,\st_L^*\sb_L + W^{-\mu}\,\sb_L^*\st_L) \nn\\
  &=& \frac{eg}{3\sqrt{2}}\,A_{\mu}\,
    (\R_{i1}^{\ti b}\R_{j1}^{\ti t}\,W^{+\mu}\,\st_j^*\,\sb_i +
     \R_{i1}^{\ti t}\R_{j1}^{\ti b}\,W^{-\mu}\,\sb_j^*\,\st_i)
\end{eqnarray}
\begin{equation}
  \hspace*{2cm} {\rm Hypercharge~Y} = 2(e_q - I_3) \hspace{4mm}\ldots  
  \hspace{4mm} {\rm Y}_{\!Q} = \oneth  
\end{equation}
%
%
\begin{eqnarray}
  \lag_{\sq\sq' WZ} 
  &=& -\frac{g^2}{\sqrt{2}\cos\tW}\,y_Q\sin^2\tW\,Z^\mu\,
  (W^+_\mu\,\st_L^*\,\sb_L + W^-_\mu\,\sb_L^*\,\st_L) \nn \\
  &=& -\frac{g^2}{3\sqrt{2}\cos\tW}\sin^2\tW\,Z^\mu\,
    (\R_{i1}^{\ti b}\R_{j1}^{\ti t}\,W^+_\mu\,\st_j^*\,\sb_i +
     \R_{i1}^{\ti t}\R_{j1}^{\ti b}\,W^-_\mu\,\sb_j^*\,\st_i)
\end{eqnarray}
%
%
\begin{eqnarray}
  \lag_{\sq\sq gg} 
  &=& \smaf{1}{6}\,g_s^2\, G^a_{\!\mu}\,G^{a\mu}\,
          (\sq_L^*\,\sq_L^{} + \sq_R^*\,\sq_R^{})
    + \onehf\,g_s^2\,d_{abc}\, G^a_{\!\mu}\,G^{b\mu}\,T^c\,
          (\sq_L^*\,\sq_L^{} + \sq_R^*\,\sq_R^{})  \nn \\[1mm]
  &=& \onehf\,g_s^2\, (\oneth\,\d_{ab} + d_{abc}\,T^c)\, 
          G^a_{\!\mu}\,G^{b\mu}\, \sq_j^*\,\sqi  
\end{eqnarray}


\begin{equation}
  \lag_{\sq\sq\g g} 
  = 2ee_q g_s\, T^a\, G^a_{\!\mu}\,A^\mu\, (\sq_L^*\,\sq_L^{} + \sq_R^*\,\sq_R^{})
  = 2ee_q g_s\, T^a\,\d_{ij} G^a_{\!\mu}\,A^\mu\, \sq_j^*\,\sqi 
\end{equation}


\begin{equation}
  \lag_{\ti{q}\ti{q} gZ} 
  = \frac{2gg_s}{\cos\tW}\,T^a\,G^a_{\!\mu}\,Z^\mu\,
      (\CL\,\sq_L^*\,\sq_L^{} + \CR\,\sq_R^*\,\sq_R^{}) 
  = \frac{2gg_s}{\cos\tW}\,T^a\,c_{ij}\,G^a_{\!\mu}\,Z^\mu\,\sq_j^*\,\sqi     
\end{equation}
%
%
\begin{eqnarray}
  \lag_{\sq\sq' gW} 
  &=& \sqrt{2}\,g g_s\,T^a\,G^a_{\!\mu}\, 
    (W^{+\mu}\,\st_L^*\,\sb_L + W^{-\mu}\,\sb_L^*\,\st_L) \nn\\[1mm]
  &=& \sqrt{2}\,g g_s\,T^a\,G^a_{\!\mu}\,
    (\R_{i1}^{\ti b}\R_{j1}^{\ti t}\,W^{+\mu}\,\st_j^*\,\sb_i +
     \R_{i1}^{\ti t}\R_{j1}^{\ti b}\,W^{-\mu}\,\sb_j^*\,\st_i)
\end{eqnarray}

\bigskip

\begin{picture}(80,25)
\put(25,0){\mbox{\psfig{figure=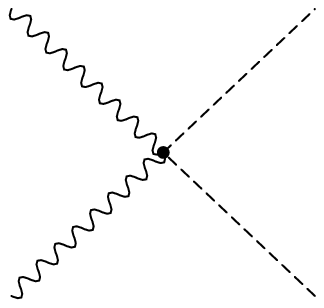,height=2.3cm}}}
\put(22,21.5){
  \makebox(0,0)[br]{{\small $\g$}}}  
\put(22,0.5){
  \makebox(0,0)[br]{{\small $\g$}}}  
\put(49,0){
  \makebox(0,0)[bl]{{\small $\sqi$}}}  
\put(49,21){
  \makebox(0,0)[bl]{{\small $\sqj$}}} 
\put(27,22){
  \makebox(0,0)[bl]{{\tiny $\nu$}}}  
\put(27,5){
  \makebox(0,0)[br]{{\tiny $\mu$}}}  
\put(72,10){
  \makebox(0,0)[bl]{$i\,e^2 e_q^2\,\d_{ij}\,g_{\mu\nu}$}}  
\end{picture} \\ 

\begin{picture}(80,25)
\put(25,0){\mbox{\psfig{figure=V_ssZZ.ps,height=2.3cm}}}
\put(22,20.5){
  \makebox(0,0)[br]{{\small $Z$}}}  
\put(22,0){
  \makebox(0,0)[br]{{\small $Z$}}}  
\put(49,0){
  \makebox(0,0)[bl]{{\small $\sqi$}}}  
\put(49,21){
  \makebox(0,0)[bl]{{\small $\sqj$}}} 
\put(27,22){
  \makebox(0,0)[bl]{{\tiny $\nu$}}}  
\put(27,5){
  \makebox(0,0)[br]{{\tiny $\mu$}}}  
\put(72,10){
  \makebox(0,0)[bl]{$\frac{ig^2}{\cos^2\tW}\,z_{ij}\,g_{\mu\nu}$}}  
\end{picture} \\ 

\begin{picture}(80,25)
\put(25,0){\mbox{\psfig{figure=V_ssZZ.ps,height=2.3cm}}}
\put(23,20.5){
  \makebox(0,0)[br]{{\small $W$}}}  
\put(23,0){
  \makebox(0,0)[br]{{\small $W$}}}  
\put(49,0){
  \makebox(0,0)[bl]{{\small $\sqi$}}}  
\put(49,21){
  \makebox(0,0)[bl]{{\small $\sqj$}}} 
\put(27,22){
  \makebox(0,0)[bl]{{\tiny $\nu$}}}  
\put(26,4){
  \makebox(0,0)[br]{{\tiny $\mu$}}}  
\put(72,10){
  \makebox(0,0)[bl]{$\frac{i g^2}{2}\,\Rsq_{i1}\Rsq_{j1}\,g_{\mu\nu}$}}  
\end{picture} \\ 

\begin{picture}(80,25)
\put(25,0){\mbox{\psfig{figure=V_ssZZ.ps,height=2.3cm}}}
\put(22,21.5){
  \makebox(0,0)[br]{{\small $\g$}}}  
\put(22,0.5){
  \makebox(0,0)[br]{{\small $Z$}}}  
\put(49,0){
  \makebox(0,0)[bl]{{\small $\sqi$}}}  
\put(49,21){
  \makebox(0,0)[bl]{{\small $\sqj$}}} 
\put(27,22){
  \makebox(0,0)[bl]{{\tiny $\nu$}}}  
\put(26,4){
  \makebox(0,0)[br]{{\tiny $\mu$}}}  
\put(72,10){
  \makebox(0,0)[bl]{$\frac{2ieg}{\cos\tW}\,c_{ij}\,g_{\mu\nu}$}}  
\end{picture} \\ 

\begin{picture}(80,25)
\put(25,0){\mbox{\psfig{figure=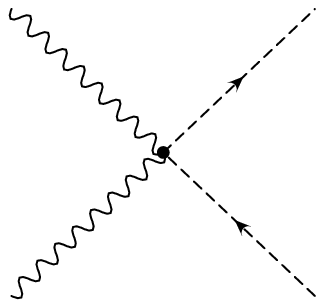,height=2.3cm}}}
\put(22,21.5){
  \makebox(0,0)[br]{{\small $\g$}}}  
\put(23,0){
  \makebox(0,0)[br]{{\small $W$}}}  
\put(49,0){
  \makebox(0,0)[bl]{{\small $\sqi$}}}  
\put(49,21){
  \makebox(0,0)[bl]{{\small $\sq_j'$}}} 
\put(27,22){
  \makebox(0,0)[bl]{{\tiny $\nu$}}}  
\put(26,4){
  \makebox(0,0)[br]{{\tiny $\mu$}}}  
\put(72,10){
  \makebox(0,0)[bl]{$\frac{ieg}{3\sqrt{2}}\,
                     \Rsq_{i1}\R_{j1}^{\sq'}\,g_{\mu\nu}$}}  
\end{picture} \\

\begin{picture}(80,25)
\put(25,0){\mbox{\psfig{figure=V_ssZW.ps,height=2.3cm}}}
\put(22,20.5){
  \makebox(0,0)[br]{{\small $Z$}}}  
\put(23,0){
  \makebox(0,0)[br]{{\small $W$}}}  
\put(49,-0.5){
  \makebox(0,0)[bl]{{\small $\sqi$}}}  
\put(49,20.5){
  \makebox(0,0)[bl]{{\small $\sq_j'$}}} 
\put(27,22){
  \makebox(0,0)[bl]{{\tiny $\nu$}}}  
\put(26,4){
  \makebox(0,0)[br]{{\tiny $\mu$}}}  
\put(72,10){
  \makebox(0,0)[bl]{$-\frac{ig^2}{3\sqrt{2}\cos\tW}\,
                     \Rsq_{i1}\R_{j1}^{\sq'}\,g_{\mu\nu}$}}  
\end{picture} \\ 

\begin{picture}(80,25)
\put(25,0){\mbox{\psfig{figure=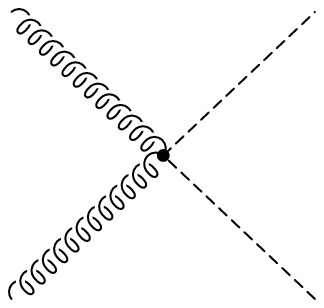,height=2.3cm}}}
\put(22,21.5){
  \makebox(0,0)[br]{{\small $g$}}}  
\put(22,0.5){
  \makebox(0,0)[br]{{\small $g$}}}  
\put(49,0){
  \makebox(0,0)[bl]{{\small $\sqi$}}}  
\put(49,20.5){
  \makebox(0,0)[bl]{{\small $\sqj$}}} 
\put(28,21){
  \makebox(0,0)[bl]{{\tiny $b,\nu$}}}  
\put(27,5){
  \makebox(0,0)[br]{{\tiny $a,\mu$}}}  
\put(72,10){
  \makebox(0,0)[bl]{$\frac{1}{2}\,i\,g_s^2\,
                     (\frac{1}{3}\d_{ab}+d_{abc}T^c)\,g_{\mu\nu}$}}  
\end{picture} \\ 

\begin{picture}(80,25)
\put(25,0){\mbox{\psfig{figure=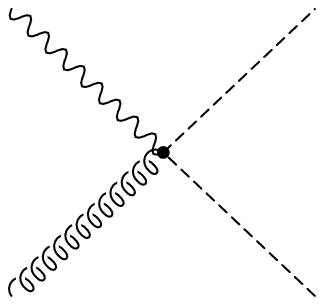,height=2.3cm}}}
\put(22,21.5){
  \makebox(0,0)[br]{{\small $\g$}}}  
\put(22,0.5){
  \makebox(0,0)[br]{{\small $g$}}}  
\put(49,0){
  \makebox(0,0)[bl]{{\small $\sqi$}}}  
\put(49,20.5){
  \makebox(0,0)[bl]{{\small $\sqj$}}} 
\put(27,22){
  \makebox(0,0)[bl]{{\tiny $\nu$}}}  
\put(27,5){
  \makebox(0,0)[br]{{\tiny $a,\mu$}}}  
\put(72,10){
  \makebox(0,0)[bl]{$2\,ie e_q g_s T^a\,\d_{ij}\,g_{\mu\nu}$}}  
\end{picture} \\ 

\begin{picture}(80,25)
\put(25,0){\mbox{\psfig{figure=V_ssGZ.ps,height=2.3cm}}}
\put(22,20.5){
  \makebox(0,0)[br]{{\small $Z$}}}  
\put(22,0.5){
  \makebox(0,0)[br]{{\small $g$}}}  
\put(49,0){
  \makebox(0,0)[bl]{{\small $\sqi$}}}  
\put(49,20.5){
  \makebox(0,0)[bl]{{\small $\sqj$}}} 
\put(27,22){
  \makebox(0,0)[bl]{{\tiny $\nu$}}}  
\put(27,5){
  \makebox(0,0)[br]{{\tiny $a,\mu$}}}  
\put(72,10){
  \makebox(0,0)[bl]{$\frac{2igg_s}{\cos\tW}\, T^a\,c_{ij}\,g_{\mu\nu}$}}  
\end{picture} \\ 

\begin{picture}(80,25)
\put(25,0){\mbox{\psfig{figure=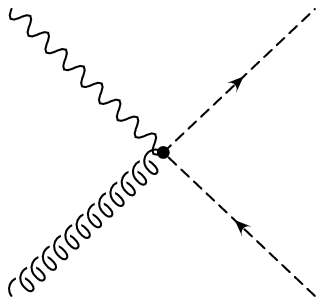,height=2.3cm}}}
\put(23,20.5){
  \makebox(0,0)[br]{{\small $W$}}}  
\put(22,0.5){
  \makebox(0,0)[br]{{\small $g$}}}  
\put(49,0){
  \makebox(0,0)[bl]{{\small $\sqi$}}}  
\put(49,20.5){
  \makebox(0,0)[bl]{{\small $\sq_j'$}}} 
\put(27,22){
  \makebox(0,0)[bl]{{\tiny $\nu$}}}  
\put(27,5){
  \makebox(0,0)[br]{{\tiny $a,\mu$}}}  
\put(72,10){
  \makebox(0,0)[bl]{$\sqrt{2}\,i\,g g_s\,T^a\,
                     \Rsq_{i1}\R_{j1}^{\sq'}\,g_{\mu\nu}$}}  
\end{picture} \\

\subsection{Four--Squark Interaction}

The four--squark interaction stems from the D--terms in the 
scalar potential:
\begin{equation}
  {\cal V} = \onehf \Big[ D^a D^a + D^i D^i + (D')^2 \Big] + \ldots 
\end{equation}
with $a=1,\ldots 8$ and $i=1,2,3$.  
For our calculations we just need the SU(3) part $\onehf\,D^a D^a$ 
which leads to: 
\begin{equation}
  \lag_{\sq\sq\sq\sq} = -\onehf\,g_s^2\,T^a_{rs} T^a_{tu} 
  \big(\sq_{L,r}^{\a *}\,\sq_{L,s}^\a - \sq_{R,r}^{\a *}\,\sq_{R,s}^\a \big)
  \big(\sq_{L,t}^{\b *}\,\sq_{L,u}^\b - \sq_{R,t}^{\b *}\,\sq_{R,u}^\b \big)
  \label{eq:Ldada}
\end{equation}  

\noi
where $\a$ and $\b$ are flavor indices.
Again a relative minus sign occurs between the $\sq_L^*\sq_L^{}$ and 
$\sq_R^*\sq_R^{}$ terms because the anti--colour generator is $-T^{a\dagger}$ 
(see interactions with gluinos).
In $(\ti q_1^{},\ti q_2^{})$ notation \eq{Ldada} reads
\begin{eqnarray}
  \lag_{\sq\sq\sq\sq}
  &=& -\onehf\,g_s^2\,T^a_{rs} T^a_{tu}\,
  (\R_{i1}^\a\R_{j1}^\a - \R_{i2}^\a\R_{j2}^\a)\,
     \sq_{j,r}^{\a *}\,\sq_{i,s}^\a \;
  (\R_{k1}^\b\R_{l1}^\b - \R_{k2}^\b\R_{l2}^\b)\,
    \sq_{l,t}^{\b *}\,\sq_{k,u}^\b  \nn \\
  &=& -\onehf\,g_s^2\,T^a_{rs} T^a_{tu}\, 
      {\cal S}_{ij}^\a\,{\cal S}_{kl}^\b\;
      \sq_{j,r}^{\a *}\,\sq_{i,s}^\a \: \sq_{l,t}^{\b *}\,\sq_{k,u}^\b
\end{eqnarray} 
with
\begin{equation}
  {\cal S}_{ij}^\a := \R_{i1}^\a\R_{j1}^\a - \R_{i2}^\a\R_{j2}^\a = 
  \left( \begin{array}{rr}  
      \cos 2\tsq  & -\sin 2\tsq  \\
      -\sin 2\tsq & -\cos 2\tsq
  \end{array} \right)_{ij}^\a
\end{equation}

\bigskip

\begin{picture}(80,25)
\put(25,0){\mbox{\psfig{figure=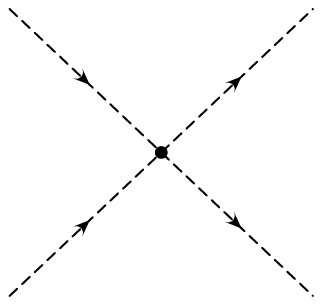,height=2.3cm}}}
\put(22,20.5){
  \makebox(0,0)[br]{{\small $\ti q_k^\b$}}}  
\put(22,0){
  \makebox(0,0)[br]{{\small $\ti q_i^\a$}}}  
\put(50,-0.5){
  \makebox(0,0)[bl]{{\small $\ti q_j^\a$}}}  
\put(50,21){
  \makebox(0,0)[bl]{{\small $\ti q_l^\b$}}} 
\put(72,10){
  \makebox(0,0)[bl]{$-i\,g_s^2\,\left[ 
     T^a_{rs}T^a_{tu}\,{\cal S}_{ij}^\a\,{\cal S}_{kl}^\b + 
     T^a_{ru}T^a_{ts}\,{\cal S}_{il}^\a\,{\cal S}_{kj}^\a  \d_{\a\b} 
     \right]$}}
\put(120,2){\mbox{\footnotesize no sum over $\a$.}} 
\end{picture} \\


\chapter{Squark Production at Lepton Colliders}

The next generation of high energy colliders 
(LHC, upgraded Tevatron, $e^+e^-$ linear colliders, 
$\mu^+\mu^-$ colliders) will explore the TeV mass range 
testing the concept of low energy supersymmetry. 
While hadron colliders are well designed for broad searches, 
it is commonly expected that for a precise determination of the 
underlying SUSY parameters a lepton collider will be necessary.

In this capter we discuss the pair production of squarks in 
$e^+e^-$ and $\mu^+\mu^-$ collisions. For squark production 
at hadron colliders, see \eg~\cite{DPF95,snowmass,hadron}.

\section{Cross Sections for {\boldmath $e^+e^-$} Colliders}

\begin{figure}[p]
\begin{picture}(170,185)
\put(11,5){\mbox{\psfig{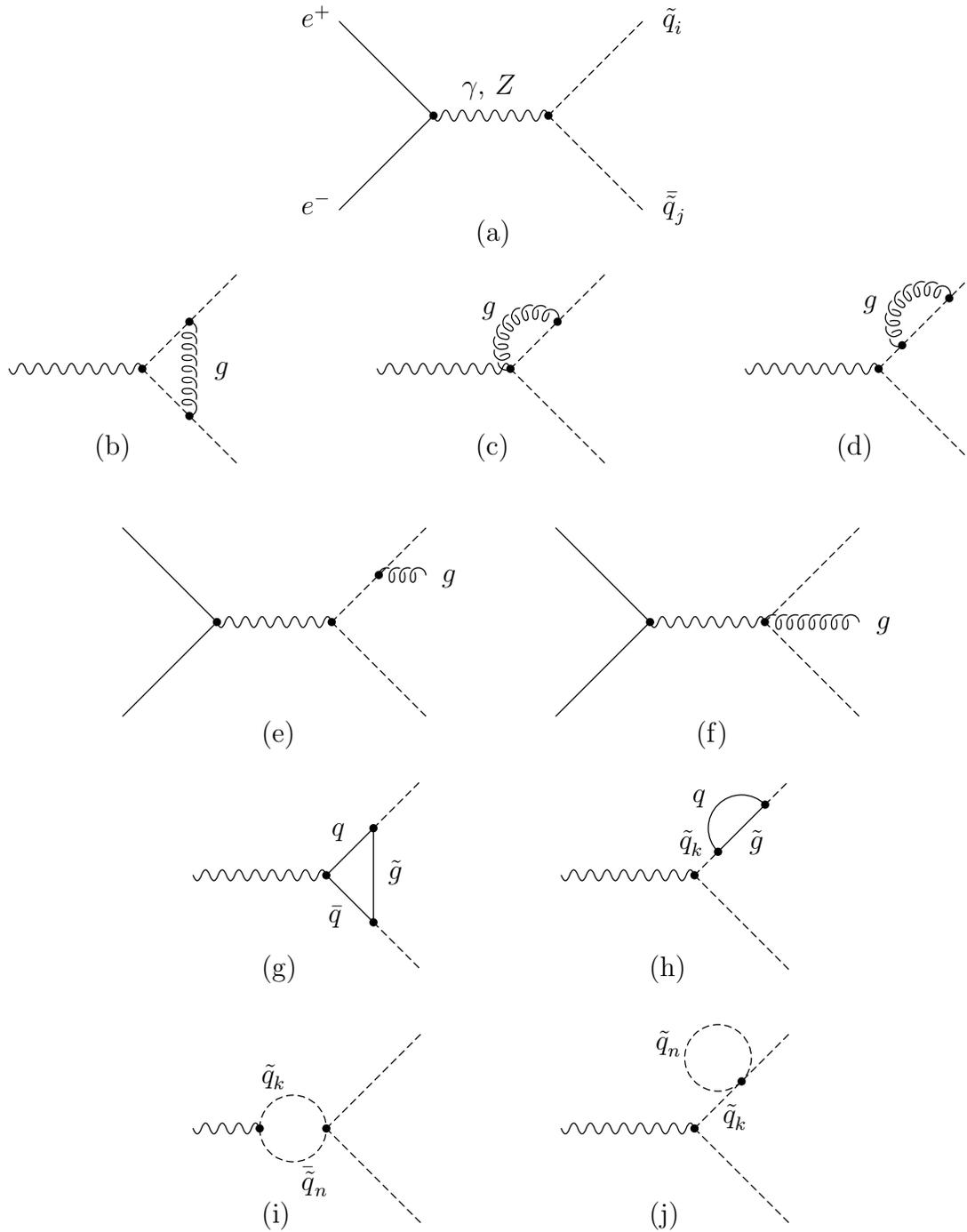}}}
\put(82,152){\mbox{(a)}}
\put(56,184){\mbox{$e^+$}}
\put(56,156){\mbox{$e^-$}}
\put(80,174){\mbox{$\g,\,Z$}}
\put(110,184){\mbox{$\sqi$}}
\put(110,156){\mbox{$\sqbar_j$}}
\put(25,120){\mbox{(b)}}
\put(43,132){\mbox{$g$}}
\put(82,120){\mbox{(c)}}
\put(83,141){\mbox{$g$}}
\put(136,120){\mbox{(d)}}
\put(140,142){\mbox{$g$}}
\put(50,77){\mbox{(e)}}
\put(77,101){\mbox{$g$}}
\put(115,77){\mbox{(f)}}
\put(142,94){\mbox{$g$}}
\put(50,42){\mbox{(g)}}
\put(60.5,63){\mbox{$q$}}
\put(60,50){\mbox{$\bar q$}}
\put(69,56.5){\mbox{$\sg$}}
\put(108,42){\mbox{(h)}}
\put(112.5,61){\mbox{$\sq_k^{}$}}
\put(114.5,68){\mbox{$q$}}
\put(123,61){\mbox{$\sg$}}
\put(50,5){\mbox{(i)}}
\put(50,26){\mbox{$\sq_k^{}$}}
\put(56,10){\mbox{$\bar{\sq}_n^{}$}}
\put(108,5){\mbox{(j)}}
\put(119,20){\mbox{$\sq_k^{}$}}
\put(109,31){\mbox{$\sq_n^{}$}}
\end{picture} 
\caption{Feynman diagrams relevant for the process 
$e^+e^-\to\sqi\bar{\sq}_j^{}$: (a) tree level, 
(b--f) gluon corrections, (g, h) gluino corrections, 
and (i, j) squark corrections. Lines that are not lettered in (b--j) 
are the same as in (a).}
\label{fig:fd-eesq}
\end{figure}

The process $e^+e^-\to \sqi\sqbar_j$ proceeds via 
$\gamma$ and $Z$ exchange, see \fig{fd-eesq}\,a.
For {\bf unpolarized beams} the cross section at tree level is given by 
\cite{hikasa-kobayashi,acd,ebm}
\begin{equation}
  \s^{tree}_U =
  \frac{\pi\a^2\kappa^3_{ij}}{s^4}\,
  \biggl[ 
    e_q^2\d_{ij} 
    + \frac{(v_e^2+a_e^2)\,c_{ij}^2 }{ 16\,{\rm c}_W^4\,{\rm s}_W^4} \, 
        D_{ZZ}^{} 
    - \frac{e_q^{}v_e^{}\,c_{ij}^{}\,\d_{ij}}{
            2\,{\rm c}_W^2\,{\rm s}_W^2 } \, D_{\g Z}^{}
  \biggr]
\label{eq:sig-eesq}
\end{equation}
where $\sqrt{s}$ is the center--of--mass energy,
$\kappa_{ij}^{} = [(s-\msq{i}^2-\msq{j}^2)^2 - 4\msq{i}^2\msq{j}^2]^{1/2}$,  
${\rm s}_W^2=\sin^2\tW$, ${\rm c}_W^2=\cos^2\tW$,
and  
\begin{equation}
  D_{ZZ}^{} = \frac{s^2}{(s-m_Z^2)^2 + \G_Z^2 m_Z^2}\,, \qquad
  D_{\g Z}^{} = \frac{s\,(s-m_Z^2)}{(s-m_Z^2)^2 + \G_Z^2 m_Z^2}\,.
\end{equation}
$v_e = 4\sin^2\tW - 1$ and $a_e = -1$ are the 
vector and axial--vector couplings of electrons to $Z$ bosons,
and $c_{ij}^{}$ is the $\sq_i\sq_j Z$ coupling as defined in \eq{cij}.
The first term of Eq.~\eq{sig-eesq} comes from pure $\g$ exchange,  
the second from pure $Z$ exchange, and the third one is due to the 
$\g$--$Z$ interference. 
Notice that the associated production of $\sq_1^{}$ and $\sq_2^{}$ 
[$i\neq j$ in \eq{sig-eesq}] proceeds only via $Z$ exchange.
The angular distribution shows the typical $\sin^2\vartheta$ shape,
where $\vartheta$ is the scattering angle of $\sqi$:
\begin{equation}
  \frac{{\rm d}\,\s^{tree}}{{\rm d}\cos\vartheta} = 
  \frac{3}{4}\,\sin^2\vartheta\;\s^{tree} .
\end{equation}

The $\g$--$Z$ interference leads to a characteristic 
dependence of the cross section on the squark mixing angle 
\cite{lep2paper,nlcpaper}. 
In case of $\sq_1^{}\sqbar_1$ production 
the cross section is maximal for $\cos\tsq=1$, \ie $\sq_1^{}=\sq_L^{}$, 
and minimal for  
\begin{equation}
  \cos^2\tsq\Big|_{min} = \frac{e_q}{I_{3L}^q}\,\sin^2\tW\,\left[\,
  1 + \left(1 - \frac{m_Z^2}{s}\right)\,
  \frac{v_e^{}}{4(v_e^2 + a_e^2)}\,\cos^2\tW\,\right] .
\label{eq:costhmin}
\end{equation}
In case of $\sq_2^{}\sqbar_2$ production 
the cross section is maximal for $\cos\tsq=0$ and minimal 
for $1-\cos^2\tsq|_{min}$.
The cross section of $\sq_1^{}\sq_2^{}$ production \footnote{Here 
  and in the following $\sq_1^{}\sq_2^{}$ means 
  $\sq_1^{}\sqbar_2$ and $\sq_2^{}\sqbar_1$ 
  \eg, $\s (\sq_1^{}\sq_2^{}) = \s (\sq_1^{}\sqbar_2) + \s (\sq_2^{}\sqbar_1)$.}
is maximal for maximal squark mixing ($\cos\tsq = \pm 1/\rzw$) and 
vanishes in case of no mixing. 
Notice, however, that the sign of $\cos\tsq$ cannot be determined from 
cross--section measurements because the latter depends only on $\cos^2\tsq$.


\noi
In case of a {\bf polarized \boldmath $e^-$ beam} the total cross section 
reads:
\begin{equation} \begin{split}
  \s\,({\cal P}_{\!-}) = \frac{\pi\a^2\kappa^3_{ij}}{s^4}\,\Big\{
  & e_q^2\,\d_{ij}
    - \frac{e_q c_{ij}^{} \d_{ij}^{}}{
             2\,{\rm s}_W^2 {\rm c}_W^2}\,
    (v_e^{} - a_e^{}\,{\cal P}_{\!-})\, D_{\gamma Z}^{} 
    \hspace*{6cm} \\
  & \hspace*{2.7cm} +\, \frac{c_{ij}^2}{16\,{\rm s}_W^4 {\rm c}_W^4}\,
    \left[ (v_e^2+a_e^2) - 
           2\,v_e^{} a_e^{}\, {\cal P}_{\!-} \right] 
    D_{ZZ}^{} \Big\}  \hspace*{-2cm}
\end{split}\end{equation}
where ${\cal P}_{\!-}$ is the degree of polarization, 
${\cal P}_{\!-} \in [-1,\,1]$.
${\cal P}_{\!-} = -0.8$, for instance, means that 80\% of the electrons are 
left--polarized while the remaining 20\% are unpolarized. 
Thus, $\s(-0.8) = 0.9\,\s_L^{} + 0.1\,\s_R^{}$ with $\s_L^{}$ and $\s_R^{}$ 
the cross sections for pure left--polarized (${\cal P}_{\!-} = -1$) and 
pure right--polarized (${\cal P}_{\!-} = 1$) $e^-$ beams, 
respectively\footnote{In the notation of \cite{nlcpaper} this 
corresponds to $\xi=-0.9$.}. 
In general terms:
\begin{equation}
   \s\,({\cal P}_{\!-}) = \onehf\,\big[ (1-{\cal P}_{\!-})\,\s_L^{}
   + (1+{\cal P}_{\!-})\,\s_R^{} \big]\,.
\end{equation}

\noi
If {\bf both the \boldmath $e^+$ and $e^-$ beams} are {\bf polarized} 
we have
\begin{equation} \begin{split}
  \s\,({\cal P}_{\!-} {\cal P}_{\!+}) = 
  \frac{\pi\a^2\kappa^3_{ij}}{s^4}\,\Big\{ 
  & e_q^2\,\d_{ij}\, (1 - {\cal P}_{\!-} {\cal P}_{\!+})
    - \frac{e_q c_{ij}^{} \d_{ij}^{}}{
             2\,{\rm s}_W^2 {\rm c}_W^2}\,
    \big[ v_e^{}\,(1 - {\cal P}_{\!-} {\cal P}_{\!+}) -
          a_e^{}\,({\cal P}_{\!-} - {\cal P}_{\!+}) \big]\, D_{\g Z}^{} 
    \hspace*{1cm} \\
  & +\, \frac{c_{ij}^2}{
             16\,{\rm s}_W^4 {\rm c}_W^4}\,
    \big[ (v_e^2+a_e^2)\,(1 - {\cal P}_{\!-} {\cal P}_{\!+})  
          - 2\,v_e^{} a_e^{}\,({\cal P}_{\!-} - {\cal P}_{\!+}) \big]\,
    D_{ZZ}^{} \Big\} \,. \hspace*{-2cm}
\end{split}\end{equation}
Here ${\cal P}_{\!-}$ denotes the polarization factor of the $e^-$ beam 
and ${\cal P}_{\!+}$ that of the $e^+$ beam; 
${\cal P}_{\!\pm} = \{-1,0,+1\}$
for $\{$left--polarized, unpolarized, right--polarized$\}$ $e^\pm$ 
beams. 


The {\bf radiative corrections} we consider for the process
$e^+e^-\to\sqi\sqbar_j$ are SUSY--QCD corrections to $\Oas$ and 
initial state radiation (ISR). 
Both have turned out to be significant.

\noi
As for the SUSY--QCD corrections, they can be split into the
standard QCD, \ie gluonic, correction $\d\s^g$, 
the correction due to gluino exchange $\d\s^{\sg}$, 
and the correction due to squark exchange $\d\s^{\sq}$:
\begin{equation}
  \s = \s^{tree} + \d\s^g + \d\s^{\sg} + \d\s^{\sq} .
\end{equation}
The corresponding Feynman diagrams are shown in \fig{fd-eesq}\,b--j.

\noi
The gluon contribution $\d\s^g$ has first been calculated in 
\cite{drees-hikasa,bhz}. It can be written as
\begin{equation}
  \d\s^g = \left[\frac{4}{3}\frac{\a_s}{\pi}\Delta_{ij}\right]\:\s^{tree}.
\end{equation}
For the explicit form of $\Delta_{ij}$, see \cite{ebm}. 
The SUSY contributions, $\d\s^{\sg}$ and $\d\s^{\sq}$, were calculated 
in \cite{acd,ebm}. 
The inclusion of the gluino and squark exchanges in the on--shell 
renormalization scheme requires, however, a proper renormalization 
of the squark mixing angle. 
This problem has been solved in Ref.~\cite{ebm} which is thus 
the first complete treatement of supersymmetric QCD corrections 
to this process.  
We will discuss the subtleties of SUSY--QCD corrections in more 
detail in Chapter~3. 
Here we just note that within the scheme of \cite{ebm} 
the counterterm for the squark mixing angle $\d\tsq$ is chosen such 
that it cancels the off--diagonal contribution of the squark 
wave--function corrections ($i\not=k$ in \fig{fd-eesq}\,h,\,j). 
The squark contribution to the correction vanishes, 
$\d\s^{\sq}=0$. 
The remaining correction due to gluino exchange is 
of the order of $\pm 1\%$ for an $e^+e^-$ collider with 
$\sqrt{s}\lsim 200$ GeV (LEP2).
However, $\d\s^{\sg}$ becomes important at higher energies as they are 
proposed for a Linear Collider: For large $\sqrt{s}$ it can be up to 
$-50\%$ of the gluon correction. Moreover, $\d\s^{\sg}$ does not scale 
with $\s^{tree}$ and thus gives an additional dependence 
on the squark mixing angle. 
See \cite{ebm,helmut-diss} for more details and a numerical analysis. 

\noi
The initial electron and positron may loose energy through photon emission.
This is known as initial state radiation (ISR). 
The actual $e^+e^-$ annihilation thus takes place at the reduced 
center--of--mass energy $\hat s = s(1-x)$ with $x$ the energy fraction 
carried away by the photon(s) \cite{peskin}:  
\begin{equation}
  \s = \int_0^1 dx 
  \left[ \b\,x^{\b-1} (1+\frac{3}{4}\b) - \b(1-\frac{x}{2}) \right]\,
  \left[ 1+\frac{2\a}{\pi} 
              \left(\frac{\pi^2}{6}-\frac{1}{4}\right) \right]\:
  \s_0(s(1-x))
\label{eq:isr}
\end{equation}
where $\s_0$ is the cross section without ISR correction and
$\b = 2\a/\pi\,(\log s/m_e^2 - 1)$. 
For the numerical evaluation of the ISR correction  
use the Monte Carlo routine PHOISR \cite{phoisr}.

\noi 
Beamstrahlung also reduces the center--of--mass energy of the 
$e^+ e^-$ collision. However, while ISR photons can in principle 
be dectected, beamstrahlung leads to an uncertainity in $\sqrt{s}$.
We thus leave this effect to Monte Carlo studies. 


\subsection{LEP2}

In 1995, after seven years of high precision measurements at the $Z$ pole, 
the LEP accelerator was upgraded to its second stage 
LEP2 \cite{lep2workshop}. One of the main motivations was the search for
a light Higgs boson and for SUSY particles. 
In particular, LEP2 covers an energy range that is hard to explore 
at LHC.

The LEP2 runs of 1995 to 1998 are listed in 
Table~\ref{tab:lep2runs} \cite{wolfgang}. 
The limits for SUSY searches presented at the 
ICHEP'98 conference in Vancouver are 
\cite{desch-vancouver,pp-vancouver}: 
\begin{itemize}
\item $m_{h^0\!,A^0} \gsim 80\gev$, $m_{H^0} \gsim 90\gev$, and 
      $m_{H^+}\gsim 59\gev$, 
\item $\mnt{1} \gsim 28\gev$, 
\item $\mch{1} \gsim 90\:(57)\gev$ for a gaugino--like $\ch_1$ 
      and $\tan\b=\rzw$ (40), \\
      $\mch{1} \gsim 62\gev$ for a higgsino--like $\ch_1$,
\item $m_{\ti e_R} \gsim 85\gev$, $m_{\ti\mu_R} \gsim 78\gev$,
      $m_{\ti\tau_R} \gsim 72\gev$, 
\item $\mst{1} \gsim 87\:(84)\gev$ for $\tst=0^\circ \:(56^\circ)$, and
      $\msb{1} \gsim 87\:(75)\gev$ for $\tsb=0^\circ \:(68^\circ)$.
\end{itemize}
With the data collected and analyzed since then,  
these limits have been improved by ${\cal O}(10\gev)$ \cite{LEPC50}.  

\vspace*{-3mm}
\renewcommand{\arraystretch}{1.3}
\begin{table}[h!]\begin{center}
\begin{tabular}{|l|c|r|}
\hline
  Date           & $\sqrt{s}$ & $\int\!{\cal L}dt$ {\footnotesize /exp.} \\
\hline \hline
  Nov. 95 \& Oct. 97 & 130~GeV & $6\pbi$\phantom{0} \\
                 & 136~GeV &  $6\pbi$\phantom{0} \\
  July--Aug. 96  & 161~GeV & $10\pbi$\phantom{0} \\
  Oct.--Nov. 96  & 172~GeV & $10\pbi$\phantom{0} \\
  Aug.--Nov. 97  & 183~GeV & $55\pbi$\phantom{0} \\
  May--Oct. 98   & \phantom{0}189~GeV\phantom{0} 
                 & \phantom{0}$160\pbi$\phantom{0} \\
\hline
\end{tabular}
\caption{LEP2 runs in 1995 -- 1998}
\label{tab:lep2runs}
\end{center}\end{table}
\renewcommand{\arraystretch}{1}

\begin{figure}[ht!]
\center
\begin{picture}(70,70)
\put(4,4){\mbox{\psfig{figure=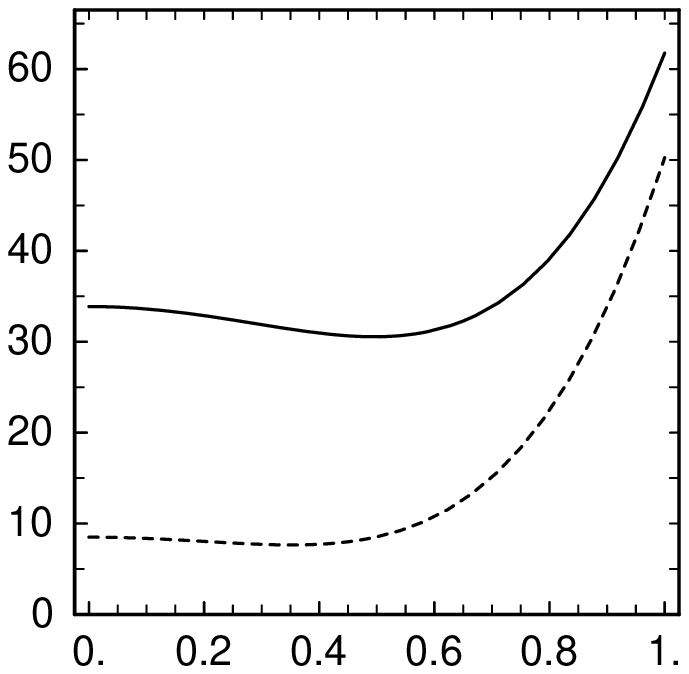,height=6.6cm}}}
\put(0,39){\makebox(0,0)[c]{\rotatebox{90}{$\sigma$~[f\/b]}}}
\put(40,0){\makebox(0,0)[cb]{$\cos\tsq$}}
\put(16,62){\mbox{\bf a}}
\put(21,42){\mbox{$\st_1\stbar_1$}}
\put(30,20){\mbox{$\sb_1\sbbar_1$}}
\end{picture}
\hspace*{10mm}
\begin{picture}(70,70)
\put(2.6,2.6){\mbox{\psfig{figure=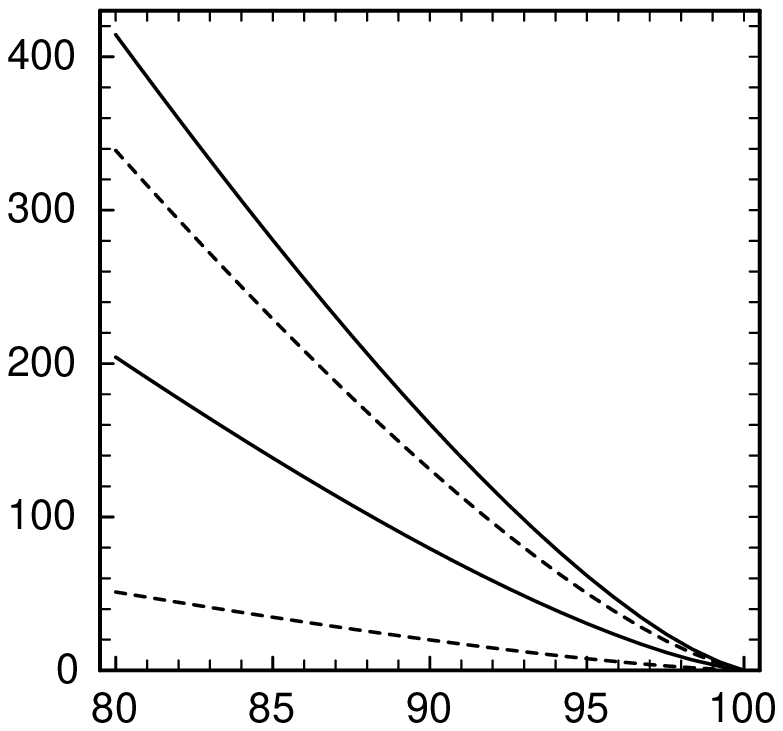,height=6.9cm}}}
\put(-1,39){\makebox(0,0)[c]{\rotatebox{90}{$\sigma$~[f\/b]}}}
\put(40,0){\makebox(0,0)[cb]{$\msq{1}$~[GeV]}}
\put(62,62){\mbox{\bf b}}
\put(45,50){\mbox{$\st_1\stbar_1$}}
\put(44,48){\vector(-2,-1){11}}
\put(46,47.5){\vector(-1,-3){8}}
\put(16,22){\mbox{$\sb_1\sbbar_1$}}
\put(21,21){\vector(1,-1){5}}
\put(23,25){\vector(1,1){10}}
\end{picture}
\caption{Cross sections for $e^+e^-\to\st_1\bar{\st}_1$ (full lines) 
and $e^+e^-\to\sb_1\bar{\sb}_1$ (dashed lines) at $\sqrt{s} = 200\gev$ 
(a) as a function of $\cos\tsq$ for $\msq{1}=95\gev$ and
(b) as a function of $\msq{1}$ for minimal and maximal 
coupling to the $Z$ boson.}
\label{fig:lep200}
\end{figure}

In 1999 and 2000 LEP2 will be operated at $\sqrt{s}\simeq 200\gev$ 
\cite{cernpresslep2}. 
Figure \ref{fig:lep200}\,a shows the 
$\st_1\stbar_1$ and $\sb_1\sbbar_1$ production cross sections 
at this energy for $\msq{1} = 95\gev$ as a function of $\cos\tsq$ 
($\sq=\st,\sb$).  
According to \eq{costhmin} $\s(\st_1\stbar_1)$ has its minimum at
$|\cst|\simeq 0.5$ and $\s(\sb_1\sbbar_1)$ at $|\csb|\simeq 0.35$.
In Fig.~\ref{fig:lep200}\,b the corresponding ranges of 
$\st_1^{}$ and $\sb_1^{}$ production cross sections are shown as a 
function of the squark masses. For the SUSY--QCD corrections we have 
taken $\mst{2}=\msb{2}=\msg=500$ GeV in both plots.

Assuming $\mst{1} < m_{\sl}$ the main decay modes of 
$\st_1$ are $\st_1\to c\,\nt_1$ and $\st_1\to b\,\chp_1$. 
The latter decay has practically 100\% branching ratio if it is 
kinematically allowed.
As $\chp_1$ further decays into $\nt_1\,\ell^+\bar\nu$ or
$\nt_1\, q\bar q^\prime$ the signature is two acoplanar $b$ jets 
accompanied by two charged leptons + large missing energy ($\Emiss$), 
or single lepton + jets + $\Emiss$, or jets + $\Emiss$. 
Here the $b$ tagging technique can be used to extract the signal. 
However, in this case the $\chp_1$ will most likely be discovered first. 
On the other hand, if $\mst{1} < m_{\chp_1} + m_b$ the flavour changing 
decay $\st_1 \to c\,\nt_1$ has practically 100\% branching ratio. 
(The decay $\st_1\to b\,f\!\bar f'\nt_1$, proceeding via a virtual chargino, 
is negligible \cite{hikasa-kobayashi}.) 
The signature is then two acoplanar jets + $\Emiss$. 
Quite generally, the invisible energy is larger for $\,\st_1\to c\,\nt_1$ 
than in case of $\,\st_1\to b\,\chp_1$.
If, however, $m_{\ch_1}+m_b > \mst{1} > m_{\sl^+(\snu)} + m_b \:(+m_\ell)$ 
the decays $\st_1\to b\,\nu\sl^+$ or $\st_1\to b\,\ell^+\snu$, 
proceeding via a virtual $\chp_1$, can compete with the decay into 
$c\,\nt_1$ \cite{hikasa-kobayashi,porod-woermann,werner-diss,werner-stop3}. 
In this case the signature is $2b + 2\ell + \Emiss$ (or 
$b + \ell^+ + jet + \Emiss$ or $jets + \Emiss$). 

\noi
The main decay modes of $\sb_1$ are $\sb_1\to b\,\nt_1$ and 
$\sb_1\to b\,\nt_2$, the second decay being possible in the parameter 
region approximately given by $M < m_{\sb_1}-m_b$ or $|\mu| < m_{\sb_1}-m_b$. 
For the $b\,\nt_1$ channel the signature is two acoplanar $b$ jets + 
missing energy $\Emiss$. If the $\sb_1$ decays into $b\,\nt_2$ the 
$b$ jets are acompanied by additional jets and/or leptons from 
$\nt_2\to\nt_1\,q\bar q$ and/or $\nt_2\to\nt_1\,\ell\bar\ell$. 
$b$ tagging will help to enhance the signal. 

\noi
If the lifetime of the squark is longer than the typical  
hadronization time of ${\cal O}(10^{-23}\,s)$, i.~e. $\Gamma \lsim 0.2$ GeV, 
it hadronizes first into a colourless $(\sq_1^{} \bar q)$ or $(\sq_1^{}qq)$ 
bound state before decaying \cite{bhz}. 
This is generally expected in case of $\,\st_1\to c\,\nt_1$ and 
$\st_1\to b\,\nu\sl^+\!,\; b\,\ell^+\snu$ since these decays involve the 
electroweak coupling twice \cite{hikasa-kobayashi}, 
but can also occur for $\st_1\to b\,\chp_1$ and $\sb_1\to b\,\nt_{1,2}$ 
depending on the nature of $\chp_1$ and $\nt_1$, and the squark mixing 
angles \cite{lep2paper}, see also Chapter~3.

\subsection{Linear Colliders with \boldmath $\sqrt{s}$ = 0.5 -- 2 TeV}

A future $e^+e^-$ Linear Collider  
will in many aspects be complementary to the CERN~LHC. 
Its physics capabilities have been studied in 
\cite{LCph1,LCph2,LCph3,LCph4}, see also \cite{DPF95,snowmass}. 
The virtues of an $e^+e^-$ Linear Collider are:
\begin{itemize}
  \item a very clean environment,
  \item flexible centre--of--mass energies, and 
  \item high polarization of the $e^-$ and possibly also of the $e^+$ beams. 
\end{itemize}
One can thus tune an $e^+e^-$ Linear Collider for many different purposes. 
For instance, one can optimize $\sqrt{s}$ for specific production processes. 
Moreover, one can use beam polarization to enhance signals and to suppress backgrounds. 
One can also make complementary measurements by using both polarization states 
of $e^-$ and $e^+$. 
This may be of great advantage for testing the TeV range in a conclusive form. \\
There are various Linear Collider projects under study in Europe, Japan, and 
the USA \cite{LCinfo}. 
A majer breakthrough was achieved recently when it was realized that an integrated 
luminosity of 
${\cal L}=300\fbi$ at $\sqrt{s}=500\gev$ and of 
${\cal L}=500\fbi$ at $\sqrt{s}=800\gev$ can be reached with the TESLA 
design \cite{nlclumi} (assuming a Snowmass year of $10^7\,s$ for running). 

\noi
In this section, we present the cross sections of stop and sbottom pair production 
at an $e^+e^-$ Linear Collider for unpolarized beams, polarized $e^-$ beams, 
and polarized $e^+$ and $e^-$ beams. 
For squark production in $e^+e^-$ annihilation with unpolarized beams and 
polarized $e^-$ beams see also \cite{nlcpaper,desy123D}.

\begin{figure}[H]
\center
\begin{picture}(70,70)
\put(4.8,5){\mbox{\psfig{figure=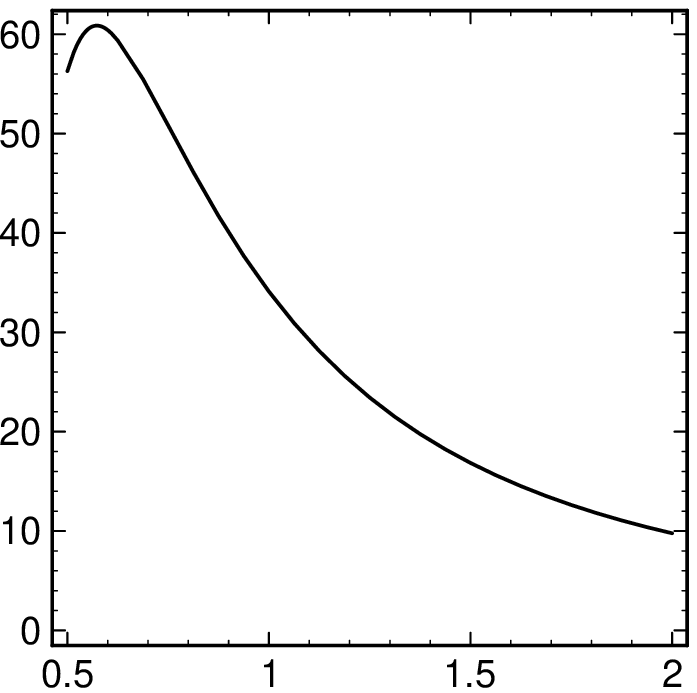,height=6.5cm}}}
\put(40,0){\makebox(0,0)[bc]{{$\sqrt{s}$~[TeV]}}}
\put(0,40){\makebox(0,0)[c]{\rotatebox{90}{$\sigma(\st_1\stbar_1)$~[f\/b]}}}
\put(62,62){\mbox{\bf a}}
\end{picture}
\hspace*{10mm}
\begin{picture}(70,70)
\put(4,5){\mbox{\psfig{figure=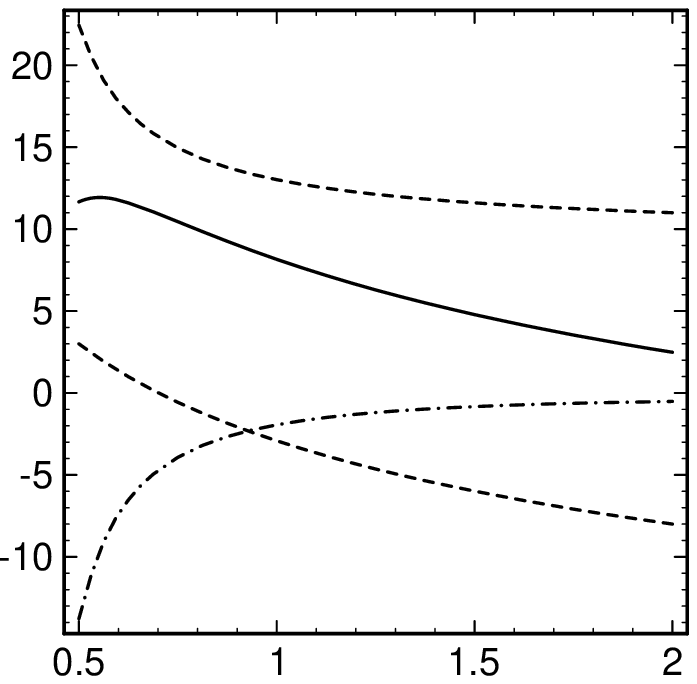,height=6.6cm}}}
\put(40,0){\makebox(0,0)[bc]{$\sqrt{s}$~[TeV]}}
\put(0,40){\makebox(0,0)[c]{\rotatebox{90}{$\Delta\sigma/\sigma^{tree}$~[\%]}}}
\put(62,62){\mbox{\bf b}}
\put(31,55){\mbox{\footnotesize gluon}}
\put(40,34){\mbox{\footnotesize gluino}}
\put(15,37){\mbox{\footnotesize ISR}}
\put(40,45){\mbox{\footnotesize total corr.}}
\end{picture}
\caption{(a) Total cross section and (b) radiative corrections relativ 
to the tree--level cross section for $\ee\to\st_1\bar{\st}_1$ as a 
function of $\sqrt{s}$ for $\mst{1}=180\gev$ and $\cst=0.7$ 
($\mst{2}=300\gev$, $\msg=300\gev$)}
\label{fig:sqrts-nlc}
\end{figure}

The $\sqrt{s}$ dependence of the $\ee\to \st_1\stbar_1$ cross section 
is shown in \fig{sqrts-nlc}\,a
for $m_{\ti t_1}=180$~GeV, $\cos\theta_{\ti t}=0.7$ and unpolarized beams. 
The effects of SUSY--QCD corrections from gluon and gluino exchange 
and of initial state radiation are demonstrated in \fig{sqrts-nlc}\,b.
Note that at high energies the gluino exchange contribution
has the opposite sign of the gluon exchange contribution, 
and the absolute values are increasing with $\sqrt{s}$.
The effect due to initial state radiation turns out to be 
of the order of 10\%. The sum of all corrections can well exceed 10\%.
The effects are similar for $\ee\to \sb_1\sbbar_1$. 

\clearpage

\begin{figure}[p]
\center
\begin{picture}(70,77)
\put(4,4){\mbox{\psfig{figure=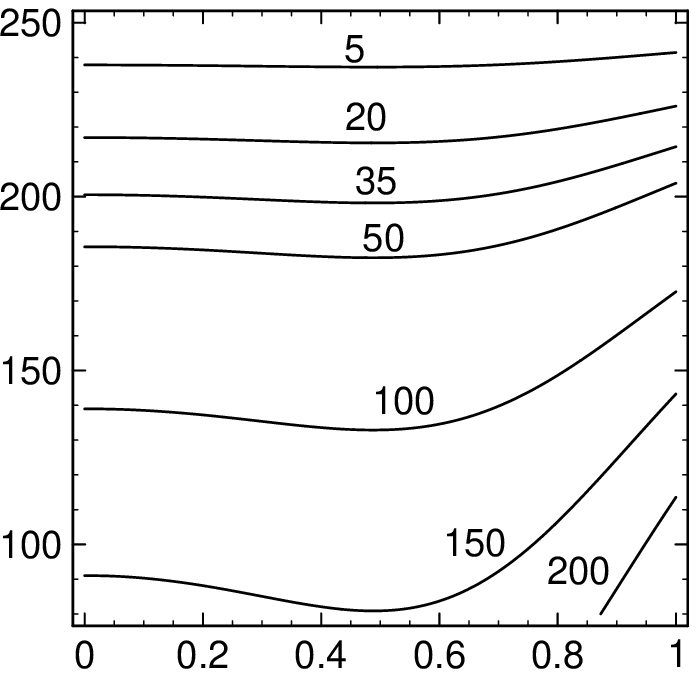,height=6.6cm}}}
\put(41,72){\makebox(0,0)[cb]{(a)~$\:\s\,(\ee\to\st_1\stbar_1)$ [fb]}}
\put(0,39){\makebox(0,0)[c]{\rotatebox{90}{$\mst{1}$~[GeV]}}}
\put(41,0){\makebox(0,0)[cb]{$\cst$}}
\end{picture}
\hspace*{10mm}
\begin{picture}(70,77)
\put(4,4){\mbox{\psfig{figure=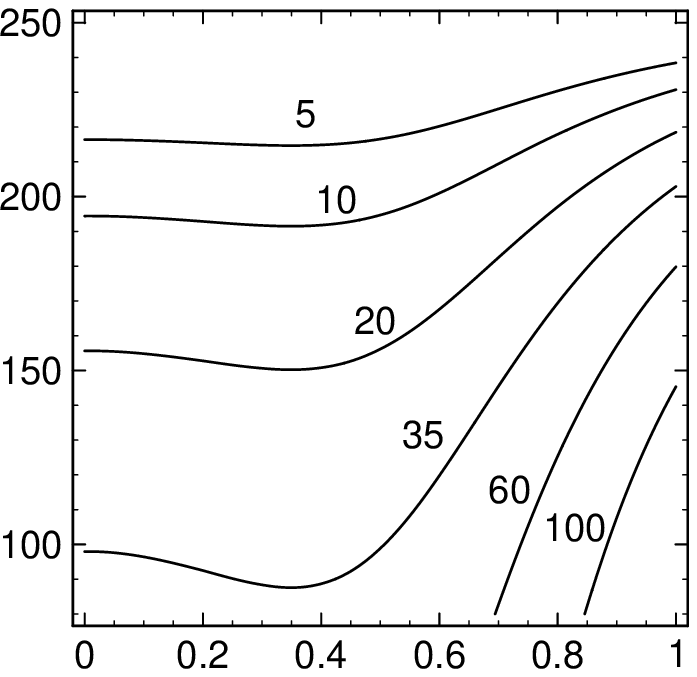,height=6.6cm}}}
\put(41,72){\makebox(0,0)[cb]{(b)~$\:\s\,(\ee\to\sb_1\sbbar_1)$ [fb]}}
\put(0,39){\makebox(0,0)[c]{\rotatebox{90}{$\msb{1}$~[GeV]}}}
\put(41,0){\makebox(0,0)[cb]{$\csb$}}
\end{picture}
\caption{Iso--cross section lines of (a) $\ee\to\st_1\bar{\st_1}$ 
and (b) $\ee\to\sb_1\bar{\sb}_1$ for $\sqrt{s}=500\gev$ and unpolarized beams.}
\label{fig:cont11-nlc}
\end{figure}

\Fig{cont11-nlc}\,a 
shows contour lines of the total $\ee\to\st_1\stbar_1$ cross section 
in the $\mst{1}-\cst$ plane for $\sqrt{s} = 500$~GeV and unpolarized beams. 
Analogously, \fig{cont11-nlc}\,b shows contour lines of the total 
$\ee\to\sb_1\sbbar_1$ cross section in the $\msb{1}-\csb$ plane.
For the calculation of the SUSY--QCD radiative corrections we have 
used $\mst{2}=\msb{2}=\msg=300$~GeV.
The $\sb_1\sbbar_1$ cross section is about two to four times smaller 
than the $\st_1\stbar_1$ cross section. 

\begin{figure}[p]
\center
\begin{picture}(70,76)
\put(4,4){\mbox{\psfig{figure=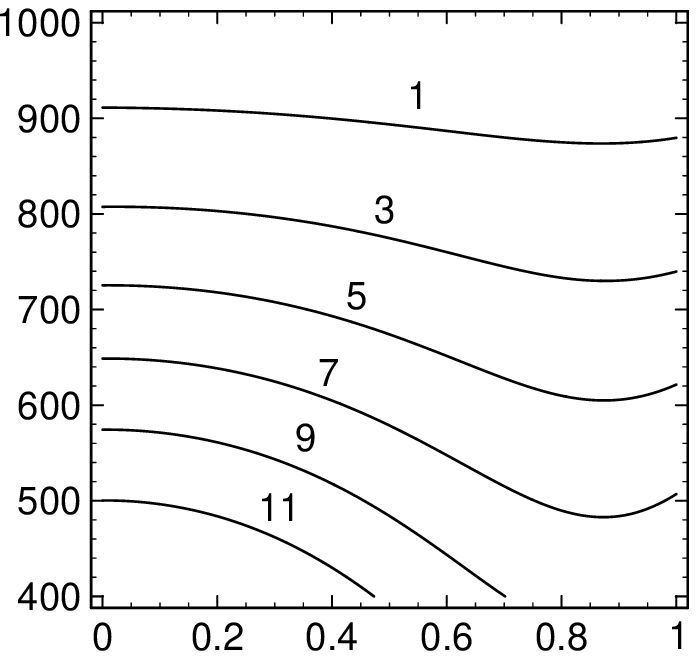,height=6.6cm}}}
\put(41, 72){\makebox(0,0)[cb]{(a)~$\:\s\,(\ee\to\st_2\stbar_2)$ [fb]}}
\put(0,39){\makebox(0,0)[c]{\rotatebox{90}{$\mst{2}$~[GeV]}}}
\put(41,0){\makebox(0,0)[cb]{$\cst$}}
\end{picture}
\hspace*{10mm}
\begin{picture}(70,76)
\put(3,3){\mbox{\psfig{figure=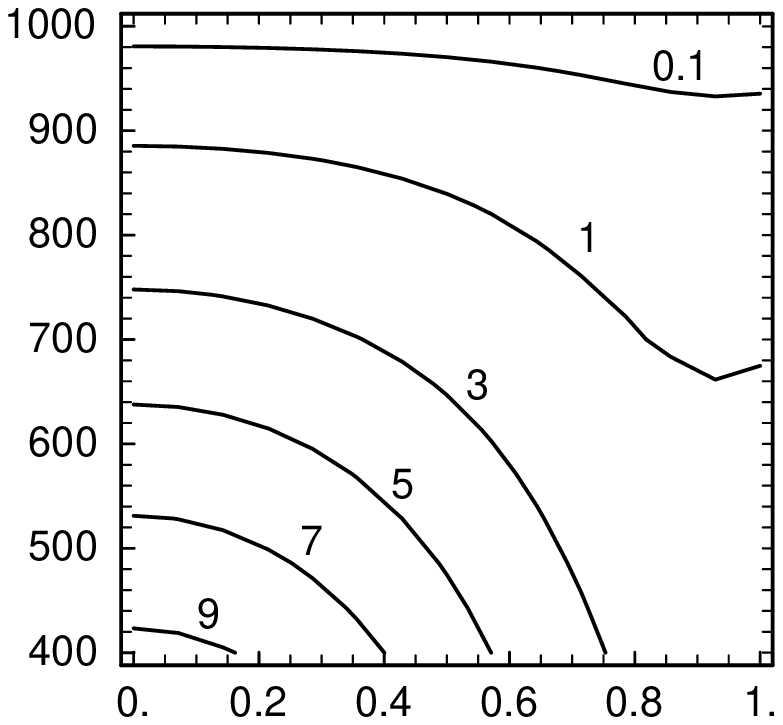,height=6.8cm}}}
\put(41, 72){\makebox(0,0)[cb]{(b)~$\:\s\,(\ee\to\sb_2\sbbar_2)$ [fb]}}
\put(0,39){\makebox(0,0)[c]{\rotatebox{90}{$\msb{2}$~[GeV]}}}
\put(41,0){\makebox(0,0)[cb]{$\csb$}}
\end{picture}
\caption{Iso--cross section lines of (a) $\ee\to\st_2\bar{\st_2}$ 
and (b) $\ee\to\sb_2\bar{\sb}_2$ for $\sqrt{s}=2$ TeV and unpolarized beams.}
\label{fig:cont22-nlc}
\end{figure}

Analgous contour lines of $\st_2\stbar_2$ and $\sb_2\sbbar_2$ 
production cross sections at $\sqrt{s}=2$~TeV are shown in 
\fig{cont22-nlc}. 
Here we have used $m_{\st_1,\sb_1}=300$~GeV and $\msg=700$~GeV for the 
radiative corrections.


\clearpage

\begin{figure}[h!]
\center
\begin{picture}(70,70)
\put(4,4){\mbox{\psfig{figure=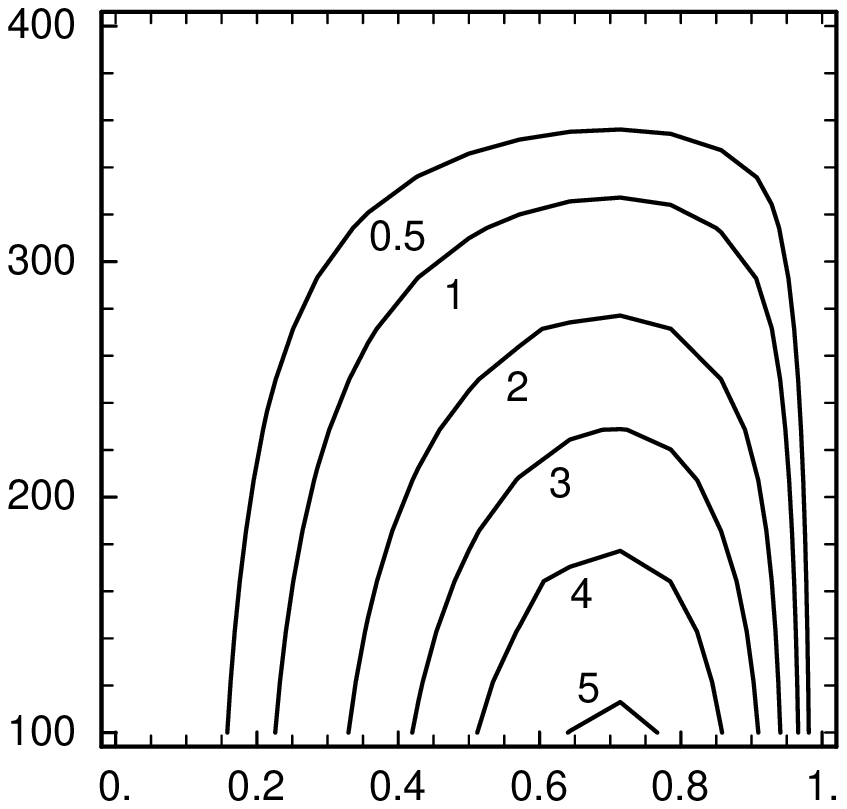,height=6.7cm}}}
\put(0,39){\makebox(0,0)[c]{\rotatebox{90}{$\mst{1}$~[GeV]}}}
\put(41,0){\makebox(0,0)[cb]{$\cst$}}
\put(17,62){\mbox{\bf a}}
\end{picture}
\hspace*{10mm}
\begin{picture}(70,70)
\put(4,4.5){\mbox{\psfig{figure=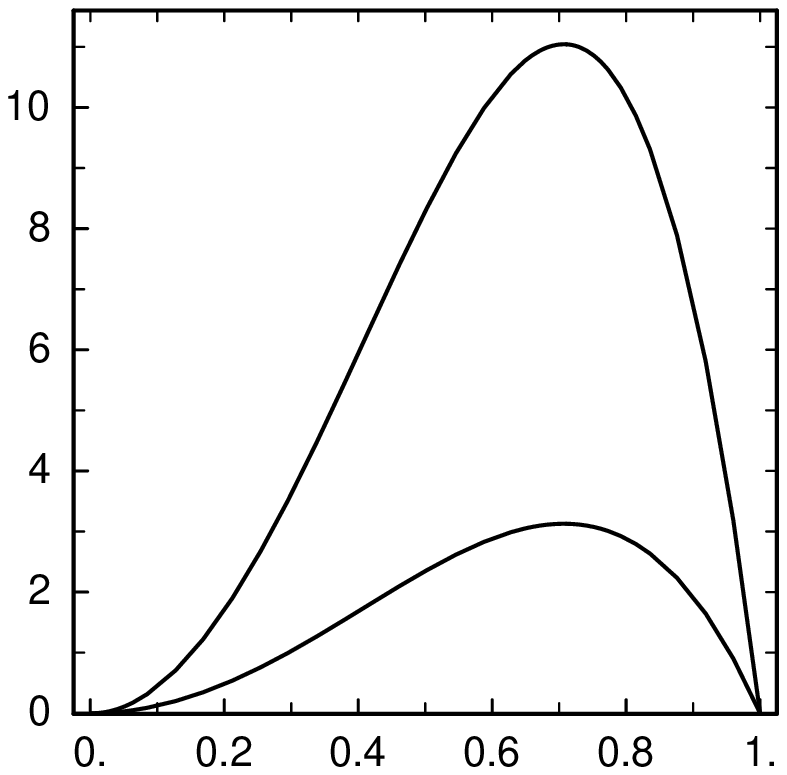,height=6.5cm}}}
\put(0,39){\makebox(0,0)[c]{\rotatebox{90}{$\s(\sb_1\sb_2)$~[fb]}}}
\put(41,0){\makebox(0,0)[cb]{$\csb$}}
\put(15,62){\mbox{\bf b}}
\put(20,50){\mbox{\footnotesize $(300,\,320)$}}
\put(40,30){\mbox{\footnotesize $(400,\,450)$}}
\end{picture}
\caption{(a) Iso--cross section lines for $\st_1\st_2$ production 
in the $\mst{1}$--$\,\cst$ plane for $\sqrt{s}=1$~TeV,
$\mst{2}=600$~GeV, and $\msg=600$~GeV.
(b) Cross sections of $\ee\to\sb_1\sb_2$ at $\sqrt{s} = 1$~TeV 
as a function of $\csb$ for $(\msb{1},\,\msb{2}) = (300,\,320)$~GeV 
and $(400,\,450)$ GeV. Both plots are for unpolarized beams.}
\label{fig:prod12m-nlc}
\end{figure}

At the first stages of a Linear Collider pair production of the 
heavier mass eigenstate $\st_2$ ($\sb_2$) may be out of reach. 
However, the associated production of $\st_1$ and $\st_2$ 
($\sb_1$ and $\sb_2$) may be possible. 
\Fig{prod12m-nlc}\,a shows the cross sections of $\st_1\st_2$ production  
as contour lines in the $\mst{1}$--$\,\cst$ plane for
$\mst{2}=600$~GeV and $\sqrt{s}=1$~TeV. 
In \fig{prod12m-nlc} $\s\,(\ee\to\sb_1\sbbar_1)$ at $\sqrt{s}=1$~TeV 
is shown as a function of $\csb$ for $(\msb{1},\,\msb{2}) = (300,\,320)$~GeV 
and $(400,\,450)$ GeV. In both plots $\msg=600$~GeV. 
As one can see, the cross sections are large enough to be detected in
a large region of the accessable parameter space. 
Here notice that replacing the stops by sbottoms or vice versa 
in \fig{prod12m-nlc} one gets essentially the same pictures.  
The only differences are due to SUSY--QCD corrections. 
Whether or not one can distinguish the two mass eigenstates of course
depends on their decay properties. 

\begin{figure}[h!]
\center
\begin{picture}(70,70)
\put(4,4){\mbox{\psfig{figure=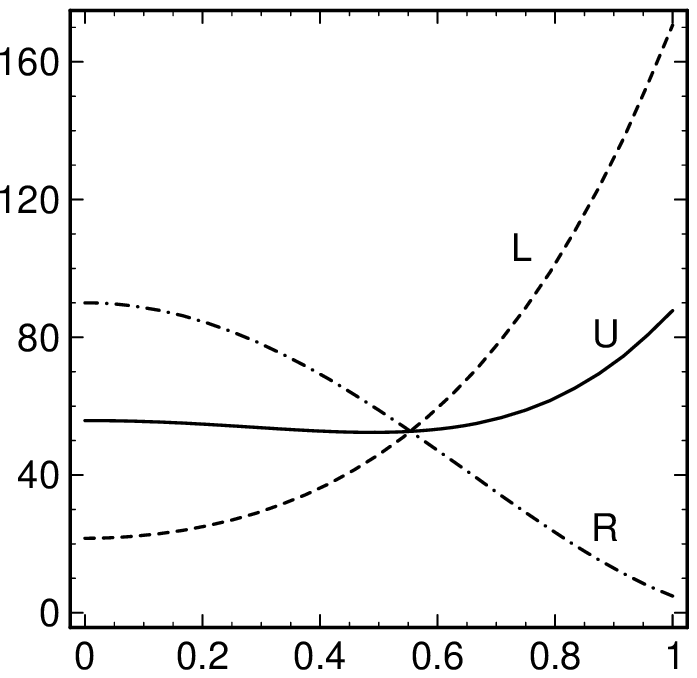,height=6.6cm}}}
\put(0,39){\makebox(0,0)[c]{\rotatebox{90}{$\sigma(\st_1\stbar_1)$~[f\/b]}}}
\put(40,0){\makebox(0,0)[cb]{$\cst$}}
\put(16,62){\mbox{\bf a}}
\end{picture}
\hspace*{10mm}
\begin{picture}(70,70)
\put(4,4){\mbox{\psfig{figure=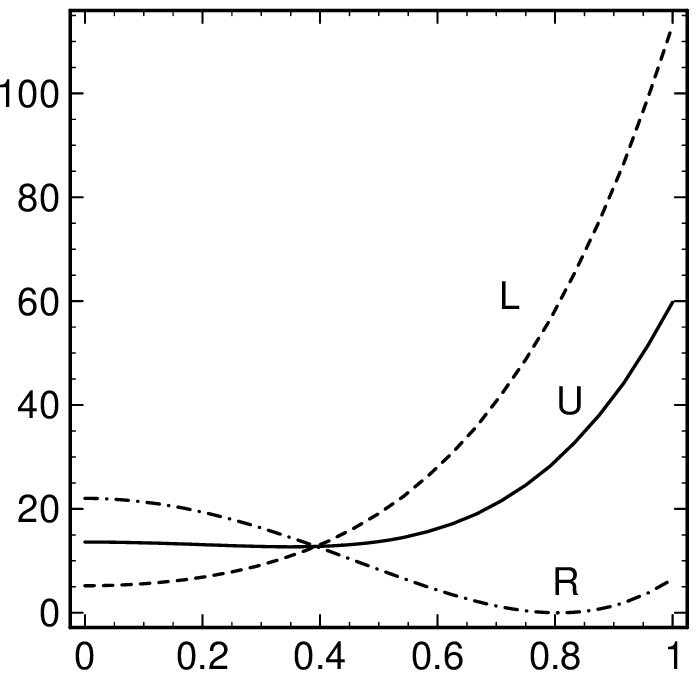,height=6.6cm}}}
\put(0,39){\makebox(0,0)[c]{\rotatebox{90}{$\sigma(\sb_1\sbbar_1)$~[f\/b]}}}
\put(40,0){\makebox(0,0)[cb]{$\csb$}}
\put(16,62){\mbox{\bf b}}
\end{picture}
\caption{$\cos\theta_{\ti t}$ dependence of the 
cross section of (a) $\ee\to\st_1\bar{\st}_1$ and 
(b) $\ee\to\sb_1\bar{\sb_1}$ 
for left-- and right--polarized as well as for 
unpolarized $e^-$ beams for $\sqrt{s} = 500$~GeV and  
$m_{\st_1,\sb_1} = 180$~GeV. }
\label{fig:pol11-nlc}
\end{figure}

\begin{figure}[h!]
\center
\begin{picture}(70,70)
\put(4,4){\mbox{\psfig{figure=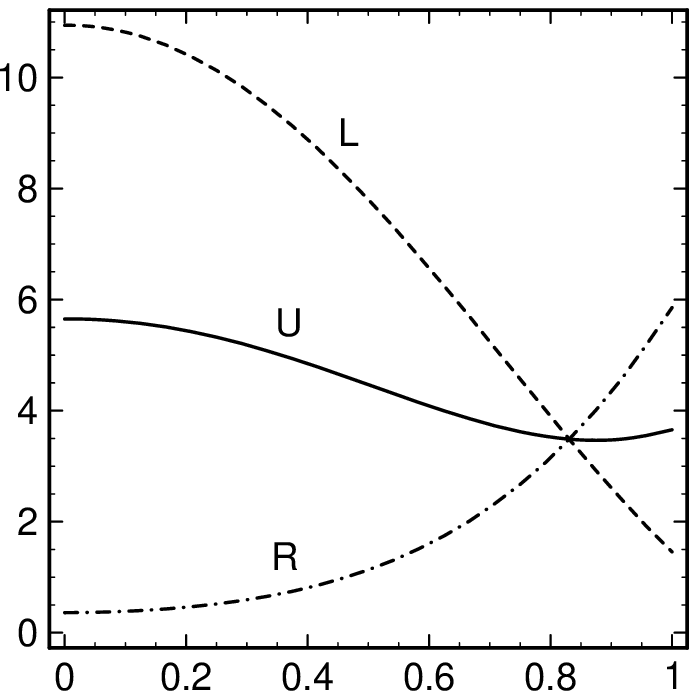,height=6.6cm}}}
\put(0,39){\makebox(0,0)[c]{\rotatebox{90}{$\sigma(\st_2\stbar_2)$~[f\/b]}}}
\put(40,0){\makebox(0,0)[cb]{$\cst$}}
\put(62,62){\mbox{\bf a}}
\end{picture}
\hspace*{10mm}
\begin{picture}(70,70)
\put(4,4){\mbox{\psfig{figure=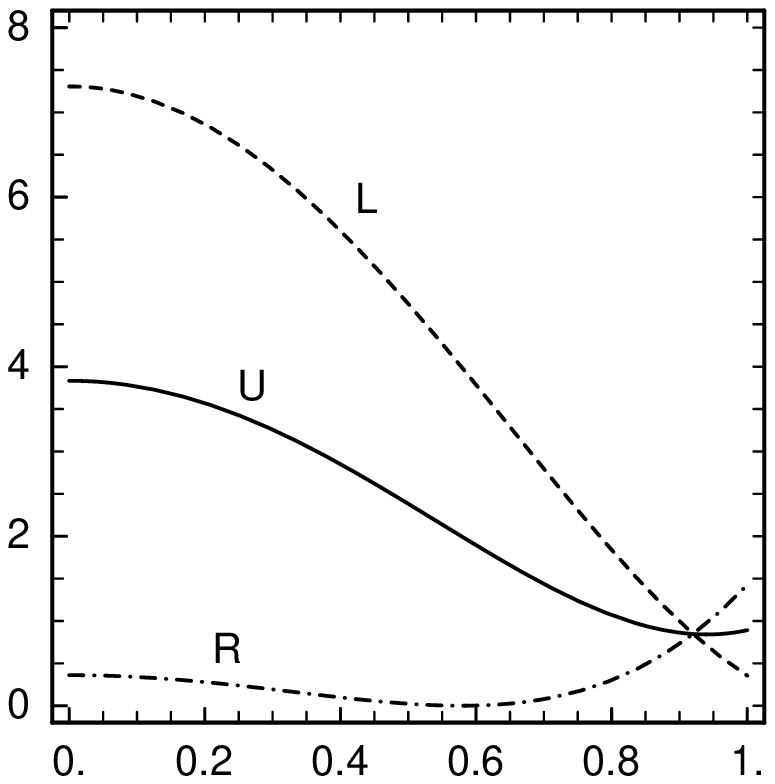,height=6.6cm}}}
\put(0,39){\makebox(0,0)[c]{\rotatebox{90}{$\sigma(\sb_2\sbbar_2)$~[f\/b]}}}
\put(40,0){\makebox(0,0)[cb]{$\csb$}}
\put(62,62){\mbox{\bf b}}
\end{picture}
\caption{$\cth_{\sq}$ dependence of the cross section of 
(a) $\ee\to\st_2\bar{\st}_2$ and (b) $\ee\to\sb_2\bar{\sb_2}$ 
for left-- and right--polarized as well as for unpolarized $e^-$ 
beams for $\sqrt{s} = 2$~TeV and $m_{\st_2,\sb_2} = 700$~GeV. }
\label{fig:pol22-nlc}
\end{figure}

Let us now turn to the effects of beam polarization. 
\Fig{pol11-nlc}\,a 
shows the $\cst$ dependence of the $\ee\to \st_1\stbar_1$ cross 
section for left-- and right--polarized as well as for unpolarized 
$e^-$ beams for $\sqrt{s}=500$~GeV and  $\mst{1} = 180$~GeV 
($\mst{2}=\msg=300$~GeV). 
For both left-- and right--polarized $e^-$ beams the cross
sections depend strongly on the mixing angle. It is important to
note that this dependence is opposite for left and right polarization. 
Therefore, experiments with polarized $e^-$ beams will allow a 
more precise determination of the mass $\mst{1}$ and the mixing 
angle $\theta_{\st}$ from cross section measurements.
This issue will be discussed in detail in the next section.  
Similar arguments hold for $\sb_1\sbbar_1$ production with polarized 
$e^-$ beams as shown in \fig{pol11-nlc}\,b. 
However, in this case the dependence on the beam 
polarization is less pronounced for $\csb\lsim 0.5$.
The influence of beam polarization on $\st_2\stbar_2$ and 
$\sb_2\sbbar_2$ production at $\sqrt{s}=2$~TeV is demonstrated 
in \fig{pol22-nlc} for $m_{\st_2,\sb_2}=700$~GeV.
Again, the $\cth_{\sq}$ dependence is much stronger for polarized
than for unpolarized beams, however, the behavior is opposite to that 
of $\ee\to \sq_1^{}\sqbar_1$. 
For the calculation of SUSY--QCD corrections we assumed 
$m_{\st_1,\sb_1}=300$~GeV and $m_{\ti g}=700$~GeV.


\clearpage

\begin{figure}[h!]
\center
\begin{picture}(70,70)
\put(4,4){\mbox{\psfig{figure=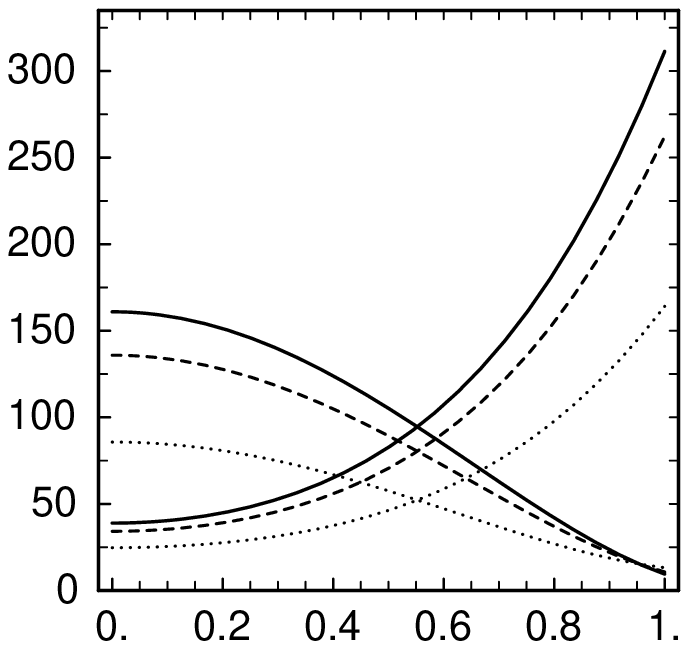,height=6.7cm}}}
\put(-1,39){\makebox(0,0)[c]{\rotatebox{90}{$\sigma(\st_1\stbar_1)$~[f\/b]}}}
\put(42,0){\makebox(0,0)[cb]{$\cst$}}
\put(18,62){\mbox{\bf a}}
\put(42,56){\mbox{\footnotesize ${\cal P}_{\!-}^{}=-0.9$}}
\put(54,54){\vector(1,-1){6}}
\put(52,54){\vector(2,-3){8}}
\put(50,54){\vector(1,-2){11}}
\put(18,47){\mbox{\footnotesize ${\cal P}_{\!-}^{}=0.9$}}
\put(22,45){\vector(0,-1){6}}
\put(24,45){\vector(0,-1){10}}
\put(26,45){\vector(0,-1){18}}
\end{picture}
\hspace*{10mm}
\begin{picture}(70,70)
\put(4,4){\mbox{\psfig{figure=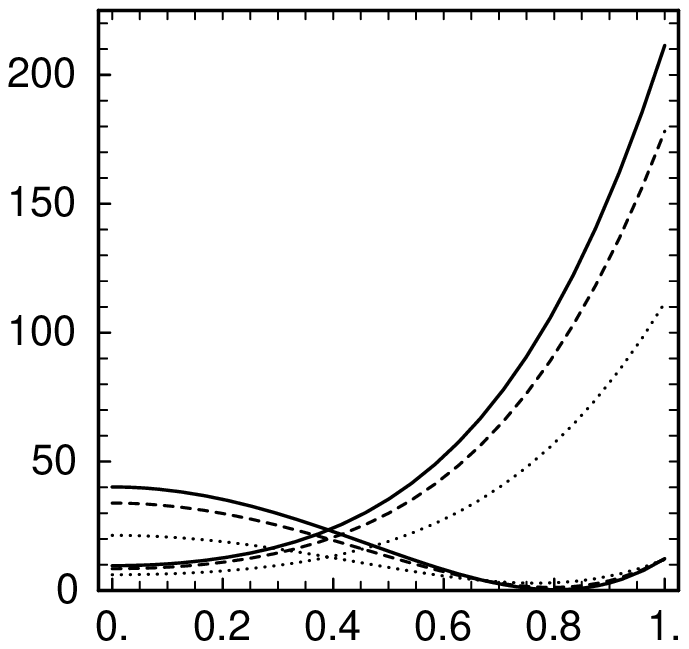,height=6.7cm}}}
\put(0,39){\makebox(0,0)[c]{\rotatebox{90}{$\sigma(\sb_1\sbbar_1)$~[f\/b]}}}
\put(42,0){\makebox(0,0)[cb]{$\csb$}}
\put(18,62){\mbox{\bf b}}
\put(42,56){\mbox{\footnotesize ${\cal P}_{\!-}^{}=-0.9$}}
\put(54,54){\vector(1,-1){7}}
\put(52,54){\vector(2,-3){9}}
\put(50,54){\vector(1,-2){11}}
\put(18,31){\mbox{\footnotesize ${\cal P}_{\!-}^{}=0.9$}}
\put(22,29){\vector(0,-1){6}}
\put(24,29){\vector(0,-1){9}}
\put(26,29){\vector(0,-1){12}}
\end{picture}
\caption{$\cos\theta_{\ti t}$ dependence of the 
cross section of (a) $\ee\to\st_1\bar{\st}_1$ and 
(b) $\ee\to\sb_1\bar{\sb}_1$ for $\sqrt{s} = 500$~GeV,   
$m_{\st_1,\sb_1} = 180$~GeV, and 90\% polarized $e^-$ beams.
The solid (dashed) lines are for 
$|{\cal P}_{\!+}|=0.9$ (0.6) and 
${\rm sign}({\cal P}_{\!+})=-{\rm sign}({\cal P}_{\!-})$. 
The dotted lines are for ${\cal P}_{\!+}=0$.}
\label{fig:bpol2-nlc}
\end{figure}

\begin{figure}[h!]
\center
\begin{picture}(70,70)
\put(4,4){\mbox{\psfig{figure=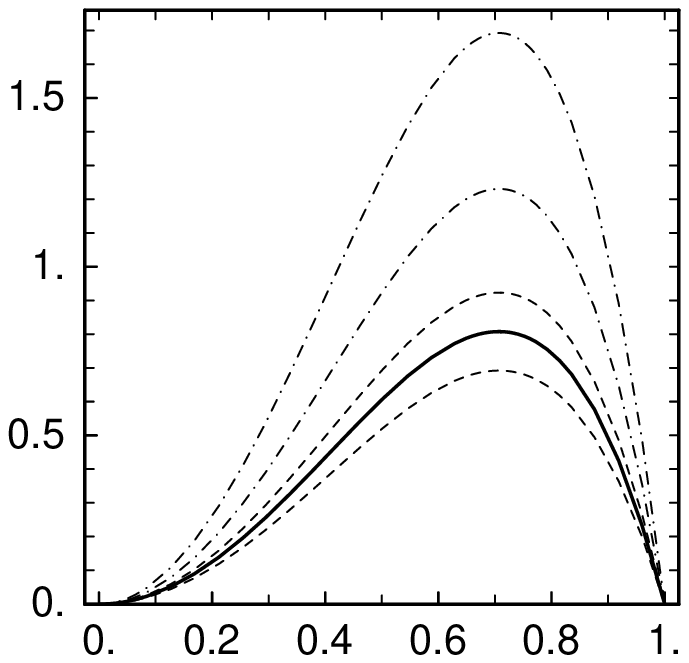,height=6.6cm}}}
\put(0,39){\makebox(0,0)[c]{\rotatebox{90}{$\sigma(\sb_1\sbbar_1)$~[f\/b]}}}
\put(41,0){\makebox(0,0)[cb]{$\csb$}}
\put(31,59){\mbox{\tiny ${\cal P}_{\!-}=-0.9$}}
\put(46,53){\mbox{\tiny ${\cal P}_{\!-}=0.9$}}
\put(45,43){\mbox{\tiny ${\cal P}_{\!-}=-0.9$}}
\put(46,29){\mbox{\tiny ${\cal P}_{\!-}=0.9$}}
\end{picture}
\caption{$\csb$ dependence of the $\ee\to\sb_1\sb_2$ cross section 
for $\sqrt{s} = 500$~GeV, $\msb{1}=230$~GeV, $\msb{2}=260$~GeV, 
and $\msg=600$~GeV. 
The solid line is for unpolarized beams, 
the dashed lines are for $|{\cal P}_{\!-}|=0.9$ and ${\cal P}_{\!+}=0$, 
and the dashdotted lines are for $|{\cal P}_{\!-}|=|{\cal P}_{\!+}|=0.9$ 
and ${\rm sign}({\cal P}_{\!+})=-{\rm sign}({\cal P}_{\!-})$.}
\label{fig:bpol3-nlc}
\end{figure}

\begin{figure}[h!]
\center
\begin{picture}(70,70)
\put(4,4){\mbox{\psfig{figure=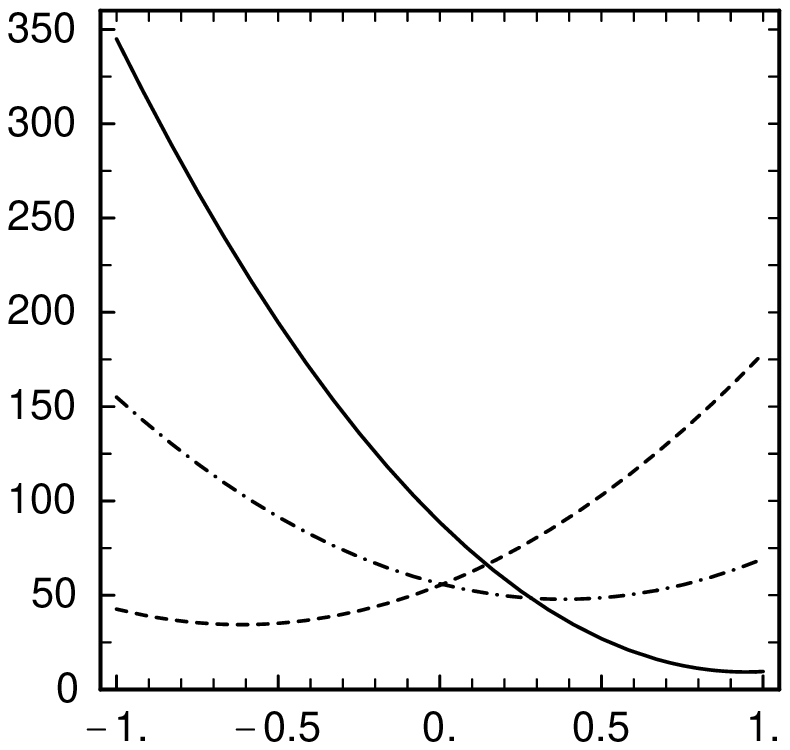,height=6.6cm}}}
\put(0,39){\makebox(0,0)[c]{\rotatebox{90}{$\sigma(\st_1\stbar_1)$~[f\/b]}}}
\put(41,0){\makebox(0,0)[cb]{${\mbf P}$}}
\put(62,62){\mbox{\bf a}}
\put(25,50){\mbox{\footnotesize $\cst=1$}}
\put(16,36){\makebox(0,0)[bl]{\rotatebox{-38}{\footnotesize $\cst=0.7$}}}
\put(49,36){\mbox{\footnotesize $\cst=0$}}
\end{picture}
\hspace*{10mm}
\begin{picture}(70,70)
\put(4,4){\mbox{\psfig{figure=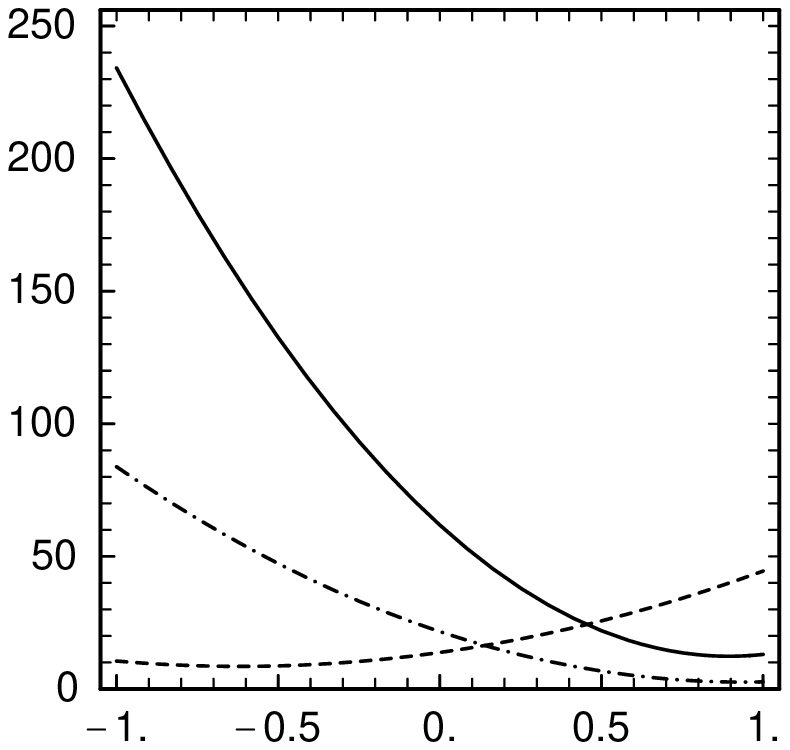,height=6.6cm}}}
\put(0,39){\makebox(0,0)[c]{\rotatebox{90}{$\sigma(\sb_1\sbbar_1)$~[f\/b]}}}
\put(41,0){\makebox(0,0)[cb]{${\mbf P}$}}
\put(62,62){\mbox{\bf b}}
\put(24,50){\mbox{\footnotesize $\csb=1$}}
\put(16,31){\makebox(0,0)[bl]{\rotatebox{-32}{\footnotesize $\csb=0.7$}}}
\put(52,22){\mbox{\footnotesize $\csb=0$}}
\end{picture}
\caption{Dependence of the 
cross section of (a) $\ee\to\st_1\bar{\st}_1$ and 
(b) $\ee\to\sb_1\bar{\sb}_1$ on the degree of polarization 
for $\sqrt{s} = 500$~GeV, $\mst{1}=\msb{1} = 180$~GeV, and 
$\mbf{P}\equiv{\cal P}_{\!-}^{}=-{\cal P}_{\!+}^{}$. }
\label{fig:bpol1-nlc}
\end{figure}

If both beams are polarized these effects are even enhanced. 
This is shown in \fig{bpol2-nlc} where we plot the cross sections of 
$\st_1$ and $\sb_1$ pair production as functions of $\cos\tsq$ 
for $\sqrt{s} = 500$~GeV, $m_{\st_1,\sb_1} = 180$~GeV, and various 
beam polarizations: 
The full lines are for 90\% polarized $e^+$ and $e^-$ beams,  
$({\cal P}_{\!-}^{},{\cal P}_{\!+}^{})$ = $(-0.9,0.9)$ and $(0.9,-0.9)$; 
the dashed lines are for 90\% polarized $e^-$ and 60\% polarized $e^+$ beams,  
$({\cal P}_{\!-}^{},{\cal P}_{\!+}^{})$ = $(-0.9,0.6)$ and $(0.9,-0.6)$; 
the dotted lines are for 90\% polarized $e^-$ and unpolarized $e^+$ beams,  
$({\cal P}_{\!-}^{},{\cal P}_{\!+}^{})$ = $(-0.9,0)$ and $(0.9,0)$. 
For the SUSY--QCD corrections we have taken $\mst{2}=\msb{2}=\msg=600$~GeV.
Analogously, \fig{bpol3-nlc} shows the cross sections of $\sb_1\sb_2$
production at $\sqrt{s} = 500$~GeV, for $\msb{1}=220$~GeV, 
$\msb{2}=240$~GeV, $\msg=600$~GeV, 90\% polarized $e^-$ beams, and 
90\% and 60\% polarized as well as unpolarized $e^+$ beams. 
The dependence on the degree of polarization can be seen 
in \fig{bpol1-nlc}: Here we plot $\s(\ee\to\st_1\stbar_1)$ and 
$\s(\ee\to\sb_1\sbbar_1)$ as function of 
$\mbf{P}\equiv{\cal P}_{\!-}^{}=-{\cal P}_{\!+}^{}$ 
for $\sqrt{s} = 500$~GeV, $m_{\st_1,\sb_1} = 180$~GeV, 
and the three cases $\sq_1^{}=\sq_L^{}$, $\sq_1^{}=\sq_R^{}$, 
and $\sq_1^{}\simeq\onehf(\sq_L^{}+\sq_R^{})$.   

\clearpage

\subsection{Determination of Soft--Breaking Parameters --- A Case Study}
\label{sect:casestudy}

In this section we estimate the experimental accuracies that can 
be obtained for stop and sbottom masses and mixing angles
at an $e^+e^-$ Linear Collider. 

\begin{figure}[h!]
\center
\begin{picture}(70,70)
\put(4,4){\mbox{\psfig{figure=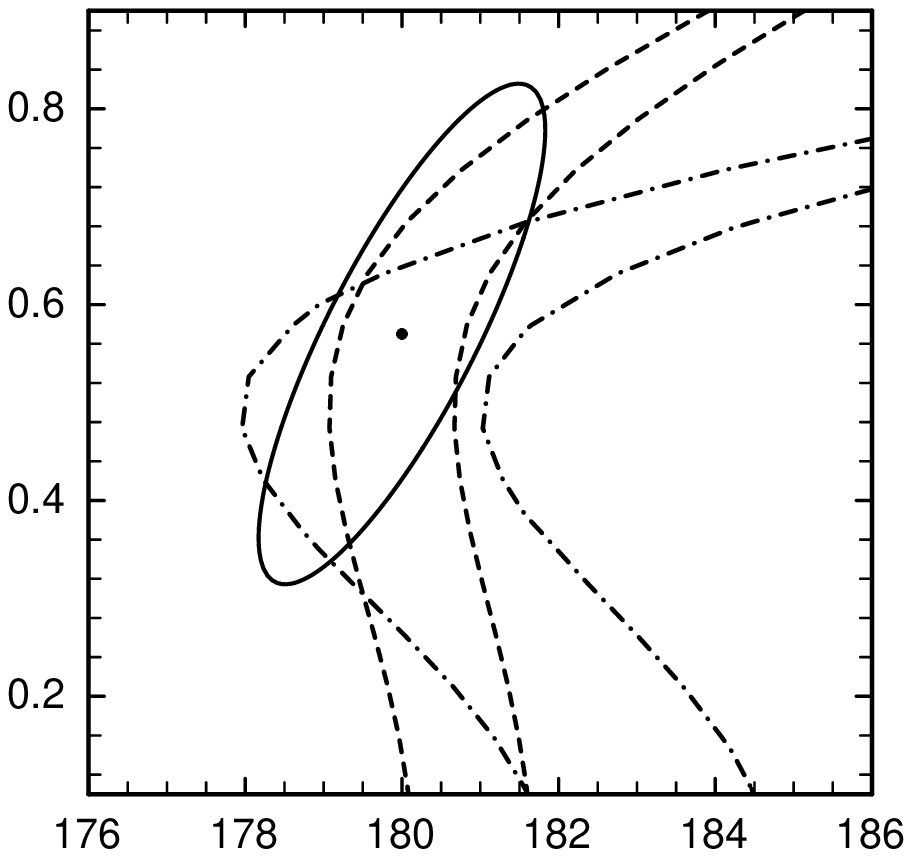,height=6.7cm}}}
\put(-1,40){\makebox(0,0)[c]{\rotatebox{90}{$\cst$}}}
\put(40,0){\makebox(0,0)[cb]{$\mst{1}$~[GeV]}}
\put(15,62){\mbox{\bf a}}
\put(13,16){\mbox{\scriptsize $\sqrt{s}=400\gev$}}
\put(46,39){\mbox{\scriptsize $\sqrt{s}=500\gev$}}
\put(32,16.5){\vector(1,0){3}}
\put(54,43){\vector(-1,1){8}}
\end{picture}
\hspace*{10mm}
\begin{picture}(70,70)
\put(4,3.5){\mbox{\psfig{figure=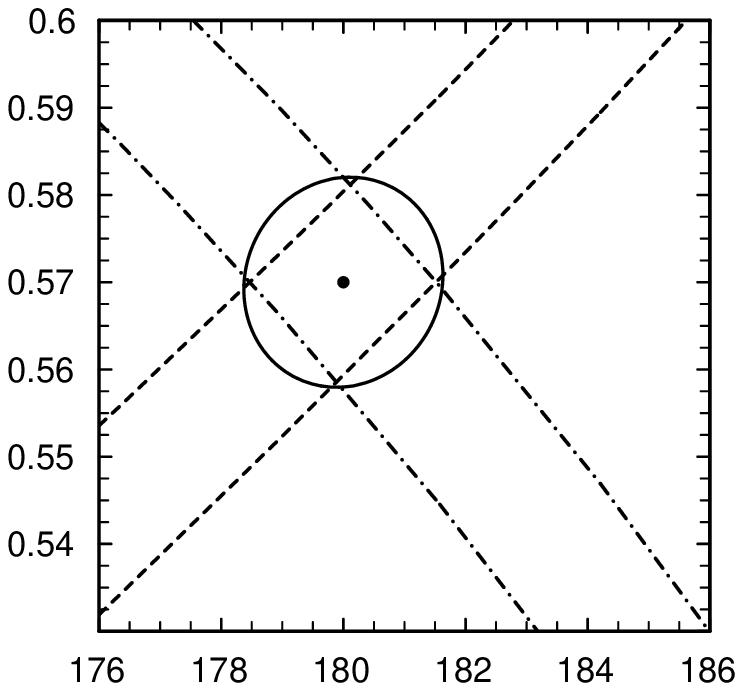,height=6.97cm}}}
\put(-1,40){\makebox(0,0)[c]{\rotatebox{90}{$\cst$}}}
\put(40,0){\makebox(0,0)[cb]{$\mst{1}~[GeV]$}}
\put(17,62){\mbox{\bf b}}
\put(22,30){\makebox(0,0)[c]{
              \rotatebox{45}{\scriptsize ${\cal P}_{\!-}=-0.9$}}}
\put(47.5,28.5){\makebox(0,0)[c]{
              \rotatebox{-50}{\scriptsize ${\cal P}_{\!-}=0.9$}}}
\end{picture}
\caption{Error bands and error ellipses for the cross section of 
$\ee\to\st_1\bar{\st}_1$ in f\/b as a function of $\mst{1}$ and $\cst$,  
(a) for unpolarized beams at $\sqrt{s}=400$ and 500~GeV and 
(b) for 90\% left-- and right--polarized beams at $\sqrt{s}=500$~GeV. 
The dot corresponds to the reference point $\mst{1}=180$~GeV and 
$\cst=0.57$.}
\label{fig:err-nlc}
\end{figure}

One possible way to determine $\mst{1}$ and $\cst$ is using 
the $\sqrt{s}$ dependence of the unpolarized $\ee\to\st_1\stbar_1$ 
total cross section. 
Let us take $\mst{1}=180$~GeV and $\cst=0.57$ as reference point,  
and $\sqrt{s}=400$~GeV and $\sqrt{s}=500$~GeV as the two reference energies.
Note that at $|\!\cst|\simeq 0.57$ the $\st_1\stbar_1$ cross 
section has its minimum. 
Motivated by the Monte Carlo study of \cite{nlcpaper,desy123E} and 
the recent big progress in the luminosity \cite{nlclumi}, 
we assume that the $\st_1\stbar_1$ production cross sections 
can be measured with accuracies of $\pm 5\%$ at $\sqrt{s}=400\gev$ 
and $\pm 2.5\%$  and at $\sqrt{s}=500\gev$. 
This leads to: 
\begin{align}
  \s_U &= 26.2 \pm 1.3 \fb \quad {\rm at}~ \sqrt{s}=400\gev, \\
  \s_U &= 59.1 \pm 1.5 \fb \quad {\rm at}~ \sqrt{s}=500\gev.
\end{align}
For the SUSY--QCD corrections we have used $\mst{2}=285\gev$ and $\msg=500\gev$.
\Fig{err-nlc}\,a  
shows the corresponding error bands and the error ellipse 
in the $\mst{1}$--$\,\cst$ plane. 
As can be seen, the mass of $\st_1$ can be determined 
with good precision with this method. 
However, the precision on the mixing angle is rather poor. 

The polarization of the $e^{-}$ beam offers the possibility of 
measuring the sfermion mixing angles with much higher accuracy. 
The cross sections of $\ee\to\st_1\stbar_1$ for 90\% left-- 
and right--polarized $e^-$ beams at the reference point $\mst{1}=180$~GeV 
and $\cst = 0.57$ for $\sqrt{s} = 500$~GeV are
\begin{align}
  \s_{-0.9}^{} &= 61.2 \pm 1.5\;{\rm f\/b}, \\  
  \s_{+0.9}^{} &= 57.1 \pm 1.4\;{\rm f\/b},
\end{align}
where we have assumed that an accuracy of $\pm 2.5\%$ 
can be achieved. 
\Fig{err-nlc}\,b  
shows the correponding error bands and the error ellipse 
in the $\mst{1}$--$\,\cst$ plane. 
The experimental accuracies obtained in this way for 
$\mst{1}$ and $\cst$ are: 
\begin{align}
  \mst{1}  &= \hphantom{.}180 \pm 1.65\gev, \\
  \cst &= 0.57 \pm 0.012.
\end{align}

We next treat the sbottom system in an analogous way. 
If $\tan\beta$ is not too large we can neglect 
left--right mixing in the sbottom sector.  
From RGEs we expect $\sb_1 = \sb_L$ and $\sb_2 = \sb_R$ in this case. 
As reference point of the sbottom system we therefore take 
$\msb{1} = 200$~GeV, $\msb{2} = 220$~GeV, and $\csb=1$. 
Based on \cite{nlcpaper} and \cite{nlclumi}, we assume that 
the cross section for $e^+e^- \to \sb_1\sbbar_1$ 
with 90\% left--polarized $e^-$ beams can be determined 
with an experimental accuracy of $\pm 2.5\%$. 
For $e^+e^- \to \sb_2\sbbar_2$ with 90\% right--polarized 
$e^-$ beams we take $\pm 8\%$ as the experimental accuracy. 
This leads to: 
\begin{align}
  \s_{-0.9}^{}\,(\ee\to\sb_1\sbbar_1) &= 82.3 \pm 2.1 \; {\rm f\/b}, \\ 
  \s_{+0.9}^{}\,(\ee\to\sb_2\sbbar_2) &= \hphantom{8}8.6 \pm 0.7\;{\rm f\/b}.
\end{align} 
Here we have assumed that sbottom and stop production can be 
distinguished. This is possible in the parameter region where the 
sbottoms mainly decay into $b\nt_{1}$. However, in the parameter 
region where the decay into $b\nt_{2}$ is possible, signatures 
similar to those of stop production could occur. 
The errors for the sbottom masses follow as:
\begin{align}
  \msb{1} &= 200 \pm 1.03\; {\rm GeV}, \\
  \msb{2} &= 220 \pm 1.93\; {\rm GeV}.
\end{align}

With these values for $\mst{1}$, $\cst$, $\msb{1}$, and 
$\msb{2}$ we can use Eq.~\eq{sqmassrelation} 
and obtain the mass of the heavier stop $\st_2$ if $\tan\b$ is 
known from other experiments. Taking, for instance, $\tan\b = 3$
leads to
\begin{equation}
  m_{\ti t_2} = 285.3 \pm 3.3\; {\rm GeV}.
\end{equation}
Confirming this value by $\st_1\st_2$ and/or $\st_2\stbar_2$ 
production at higher energies would be an independent test of the MSSM. 

Assuming that also $\mu$ is known from an other experiment 
we are now able to calculate the underlying soft SUSY breaking 
parameters $M_{\ti Q}$, $M_{\ti U}$, $M_{\ti D}$, $A_{t}$ and 
$A_{b}$ for the  squarks of the third family, 
see Eqs.~\eq{MQst} to \eq{aqinv}. 
Taking \eg, $\mu = 300\gev$ (and $\tan\beta=3$) we 
obtain the following values ($A_b=\mu\tan\b=900\gev$ since $\csb=1$):
\begin{eqnarray}
  M_{\ti Q} &=& 192.8 \pm 1.1\; {\rm GeV},   \\
  M_{\ti U} &=& 136.9 \pm 5.5\; {\rm GeV},  \\
  M_{\ti D} &=& 218.8 \pm 1.9\; {\rm GeV},  \\
  A_t &=& \!\!-31.2 \pm 8.1\; {\rm GeV} . 
\end{eqnarray}
While $M_{\ti Q}$, $M_{\ti U}$, and $M_{\ti D}$ are obtained  
with good accuracy, the error on $A_t$ is quite large. 
This, however, depends on our specific choice of $\cst$, $\tan\b$, and $\mu$.    
For $\cst=-0.57$ ($\tan\b=3$ and $\mu=300\gev$), for instance, 
we get $\mst{1} = 180\pm 1.63\gev$, $A_t = 231.2 \pm 8.1\gev$, and the  
other parameters as above.

An alternative way to determine the squark masses and mixing angles 
is the kinematical reconstruction of the squark decays. 
This has been studied in \eg,~ \cite{feng-LC}, \cite{baer-LC}, 
and \cite{daniel-LC} for the case $\st_1\to b\ch_1$. 
 
As a last remark we note that the high precision that can be 
expected at a Linear Collider indicates that radiative corrections 
should be taken into account in the Monte Carlo studies.

\clearpage
\section{Cross Sections for Muon Colliders}

The idea of a muon collider has already been brought up in the 1960's 
and 70's by Tinlot, Skrinsky, Neufer, \etal. 
However, major technical challenges have to be met in order to  
collect, accelerate, and collide muons and antimuons, which decay with a 
lifetime of $2.2\,\mu s$.
Not until 1995, at the Sausalito workshop, was it realized that with 
modern ideas and technologies a muon collider may be feasable. 
The next step further was the Muon Collider Feasibility Study 
for the Snowmass 1996 workshop \cite{mu-snowmass}.
At present, the Muon Collider Collaboration \cite{mu-coll-web} 
carries out the R\&D for a muon collider in the US.
In Europe, a muon collider is discussed within the ECFA Prospective 
Study on Muon Colliders \cite{mu-ecfa-web}, see also \cite{cernfuture}.
Specific studies are done for ${\cal O}(100)$ GeV, $0.3-0.5$ TeV, and 
multi--TeV machines with luminosities of 
$10^{31}$ to $5\times10^{34}\;cm^{-2}s^{-1}$.

\noi
The advantages of a muon collider are the following:
\begin{itemize}
  \item The effective energy of a lepton collider is much larger  
than that of a hadron collider with the same center--of--mass energy.
  \item In contrast to electrons, muons generate almost negligible 
synchotron radiation \footnote{The energy loss per revolution in a 
circular machine is proportional to $m_f^4$.}.
  \item The cross section for direct ($s$--channel) Higgs production 
in lepton--antilepton annihilation is proportional to $m_\ell^2$. 
It is thus 40000 times larger at a $\mu^+\mu^-$ 
than at an $e^+e^-$ collider.
  \item Because of the lack of bremsstrahlung (and synchotron radiation) 
energy spreads as small as 0.003\% are expected. By measuring g-2 
of the muon it should be possible to determine the absolute energy 
to an even higher accuracy. 
\end{itemize}

\noi
A muon collider can thus be circular and much smaller than $e^+e^-$ 
or hadron colliders of comparable effective energies. 
With its expected excellent energy and mass resolution 
a muon collider offers extremely precise measurements. 
Moreover, it allows for resonant Higgs production;  
in particular it may be possible to study the properties of 
relatively heavy $H^0$ and $A^0$, which can hardly be done at any 
other collider. \\
The technical challenges still to be met include 
(i) a liquid metal target for pion production, 
(ii) a new cooling technique (ionization cooling?) for the muons, and
(iii) the problem of neutrino radiation \cite{bking}.
For a more detailed introduction, see \eg~ \cite{mu-cerncourier,palmer}.  
Information about ongoing work can be found in \cite{mu-coll-web,mu-ecfa-web}.

Squark pair production in $\mu^+\mu^-$ annihilation proceeds via 
the exchange of a photon, a $Z$ boson, or a neutral Higgs boson. 
The corresponding Feynman diagrams are shown in \fig{fd-mumu}.
Provided that $m_H^{} > \msq{i}+\msq{j}$ 
($H=H^0\!,\,A^0$)\footnote{As 
the MSSM predicts $m_{h^0}\lsim 130$~GeV squark production
at the $h^0$ resonance is already excluded.},  
the process $\mu^+\mu^-\to\sqi\sqbar_j$ thus offers the 
possibility to study the squark--squark--Higgs couplings
at the  Higgs boson resonances.

For unpolarized beams the differential cross section at tree--level 
(to order $m_\mu/\sqrt{s}$) is given by
\begin{equation}
  \frac{{\rm d}\,\s}{{\rm d}\cos\vartheta} =
  \frac{3\pi\a^2\kappa_{ij}}{4s^2} \left[\,
  \frac{\kappa_{ij}^2}{s^2}\, T_{\!VV}^{}\sin^2\vartheta
  + T_{\!HH}^{}
  + \frac{\msq{i}^2-\msq{j}^2}{s}\, T_{\!VH}^a
  + \frac{\kappa_{ij}}{s}\, T_{\!VH}^b\cos\vartheta
  \,\right]
\label{eq:dsigmumu}
\end{equation}
with the kinematic function 
$\kappa_{ij}^{} = [(s-\msq{i}^2-\msq{j}^2)^2 - 4\msq{i}^2\msq{j}^2]^{1/2}$ 
and $\vartheta$ the scattering angle of $\sqi$.
$T_{\!VV}^{}$ denotes the $\g$ and $Z$ exchange contributions 
(which are the same as for $e^+e^-\to\sqi\sqbar_j$), 
$T_{\!HH}^{}$ denotes the contributions from Higgs boson exchange, 
$T_{\!VH}^a$ comes from the $Z\,$--$\,A^0$ interference, and  
$T_{\!VH}^b$ from the $(\g,Z)\,$--$\,(h^0,H^0)$ interference: 
%
%
\begin{figure}[t]
\center
\begin{picture}(135,70)
\put(5,40){\mbox{\psfig{figure=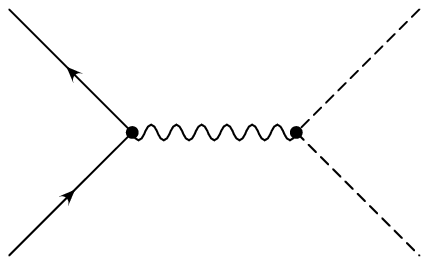,height=2.8cm}}}
\put(5,2){\mbox{\psfig{figure=FD-gZ-exchange.ps,height=2.8cm}}}
\put(75,40){\mbox{\psfig{figure=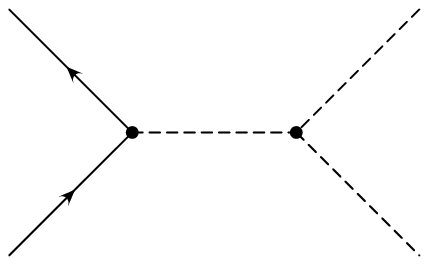,height=2.8cm}}}
\put(75,2){\mbox{\psfig{figure=FD-higgs-exchange.ps,height=2.8cm}}}
\put(129,53.5){\mbox{(a)}}
\put(-1,66.5){\mbox{$\mu^+$}}
\put(-1,40){\mbox{$\mu^-$}}
\put(53,67){\mbox{$\sqi$}}
\put(53,40){\mbox{$\sqbar_i$}}
\put(24,58){\mbox{$\g,\,Z$}}
\put(60.5,53){\mbox{$+$}}
\put(69,66.5){\mbox{$\mu^+$}}
\put(69,40){\mbox{$\mu^-$}}
\put(123,67){\mbox{$\sqi$}}
\put(123,40){\mbox{$\sqbar_i$}}
\put(93,58){\mbox{$h^0\!,\,H^0$}}
\put(129,15.5){\mbox{(b)}}
\put(-1,28.5){\mbox{$\mu^+$}}
\put(-1,2){\mbox{$\mu^-$}}
\put(53,29){\mbox{$\sq_1^{}$}}
\put(53,2){\mbox{$\sqbar_2$}}
\put(26.5,20){\mbox{$Z$}}
\put(60.5,15){\mbox{$+$}}
\put(69,28.5){\mbox{$\mu^+$}}
\put(69,2){\mbox{$\mu^-$}}
\put(123,29){\mbox{$\sq_1^{}$}}
\put(123,2){\mbox{$\sqbar_2$}}
\put(89.7,20){\mbox{$h^0\!,\,H^0\!,\,A^0$}}
\end{picture} 
\caption{Feynman diagrams for the process $\mu^+\mu^-\to\sqi\bar{\sqj}$,
            (a) $i=j$, (b) $i\neq j$.}
\label{fig:fd-mumu}
\end{figure}
%
%
\begin{equation}
  T_{\!VV}^{} =  e_q^2\,\d_{ij} 
       - \frac{e_q v_\mu^{} c_{ij}^{} \d_{ij}^{}\,s}{
               2\,{\rm s}_W^2 {\rm c}_W^2}\: \Re(d_Z^{})
       + \frac{(v_\mu^2+a_\mu^2)\,c_{ij}^2\,s^2}{
               16\,{\rm s}_W^4 {\rm c}_W^4}\: |d_Z^{}|^2 ,
\end{equation}
\begin{equation}
  T_{\!HH}^{} = \frac{h_\mu^2 s}{2 e^4} \left[ \: 
       \left| (G_1^{\sq})_{ij}^{}\sin\a\;d_h^{}
             -(G_2^{\sq})_{ij}^{}\cos\a\;d_H^{} \right|^2 
     + \left| (G_3^{\sq})_{ij}^{}\sin\b\;d_A^{} \right|^2 \right],
\end{equation}
\begin{equation}
  T_{\!VH}^a = 
  -\frac{m_\mu a_\mu h_\mu \sin\b\:c_{ij}^{}(G_3^{\sq})_{ij}^{}\,s}{
         \rzw\,e^2\, {\rm s}_W^2{\rm c}_W^2} \: \Re(d_Z^*\,d_A^{}),
\end{equation}
\begin{equation}
 \begin{split}
  T_{\!VH}^b = & \,
    \frac{2\rzw\,e_q m_\mu h_\mu \d_{ij}^{}}{e^2}\;
    \left[ (G_1^{\sq})_{ij}^{}\sin\a\;\Re(d_h^{}) - 
           (G_2^{\sq})_{ij}^{}\cos\a\;\Re(d_H^{})\right] \\
   & \hspace*{20mm} 
     -\frac{m_\mu v_\mu h_\mu\,c_{ij}^{}\,s}{
           \rzw\,e^2\, {\rm s}_W^2{\rm c}_W^2}\;
        \left[ (G_1^{\sq})_{ij}^{}\sin\a\;\Re(d_Z^* d_h^{})- 
               (G_2^{\sq})_{ij}^{}\cos\a\;\Re(d_Z^* d_H^{})\right]
 \end{split}
\end{equation}  
with $v_\mu = -1+4\sin^2\tW$ and $a_\mu=-1$ the vector and 
axial vector couplings of the muon to the $Z$ boson; 
$c_{ij}$ is the squark coupling to the $Z$, Eq.~\eq{cij}, 
$G_{\{1,2,3\}}^{\sq}$ that to $\{h^0\!,\, H^0\!,\, A^0\}$ 
as given in Section \ref{sect:feyn-ssH} 
\footnote{You may have noticed that 
$(G_3^{\sq})_{12}^{} = -(G_3^{\sq})_{21}^{}$. 
Fortunately, this difference in sign is ``repared'' by the 
term $\msq{i}^2-\msq{j}^2$ in front of $T_{\!VH}^a$ and thus 
$\s(\mu^+\mu^-\to\sq_1^{}\sqbar_2)=\s(\mu^+\mu^-\to\sq_2^{}\sqbar_1)$ 
as it should be in case of unpolarized beams.}, 
and 
\begin{equation}
  d_{X}^{} = \big[\,(s-m_{\!X}^2) + i\G_{\!\!X}^{}m_{\!X}^{}\,\big]^{-1}. 
\end{equation}

\noi
The total cross section is given by
\begin{equation}
  \s\,(\mu^+\mu^-\to\sqi\sqbar_j) = 
  \frac{3\pi\a^2\kappa_{ij}^{}}{2s^2} 
  \left[\,
    \frac{2}{3}\frac{\kappa_{ij}^2}{s^2}\: T_{\!VV}^{}
    + T_{\!HH}^{} + \frac{\msq{i}^2-\msq{j}^2}{s}\: T_{\!VH}^a
  \,\right] .
\end{equation}

\noi
Notice that the $\g$ exchange diagram does only contribute to  
$\sq_i^{}\sqbar_i$ production. On the other hand, since 
$(G_3^{\sq})_{ii}=0$, $A^0$ exchange only contributes to 
$\sq_1^{}\sq_2^{}$ production. 
Notice moreover that the $(\g,Z)\,$--$\,(h^0,H^0)$ interference in 
\eq{dsigmumu} is proportional to $\cos\vartheta$, giving rise to a 
forward--backward asymmetry. Being proportional to $m_\mu$ this asymmetry 
is, however, numerically very small, typically of the order of 
$10^{-4}$ or less. \\
Away from the Higgs boson resonances one has 
$\s(\mu^+\mu^-\to\sqi\sqbar_j)\simeq \s(e^+e^-\to\sqi\sqbar_j)$. 
However, for $\sqrt{s}\simeq m_{H^0\!,\,A^0}$ the cross section at 
a muon collider is largely enhanced compared to that of an $e^+e^-$ 
collider due to resonant $H^0$ and/or $A^0$ exchange.
While the pure gauge boson contribution shows a $\sin^2\vartheta$ 
angular dependence, the pure Higgs boson contribution does not 
depend on the scattering angle. This can be used to disentangle 
these two contributions.

However, since the muons originate from pion decays, 
they are naturally polarized.
The differential cross section for a $\mu^-$ beam with polarization 
${\cal P}_{\!-}$ and a $\mu^+$ beam with polarization ${\cal P}_{\!+}$ is:
\begin{equation}
  \frac{{\rm d}\,\s_P^{}}{{\rm d}\cos\vartheta} =
  \frac{3\pi\a^2\kappa_{ij}}{4s^2} \left[\,
  \frac{\kappa_{ij}^2}{s^2}\, {\cal T}_{\!VV}^{}\sin^2\vartheta
  + {\cal T}_{\!HH}^{} + {\cal T}_{\!VH}^{} \,\right]
\label{eq:dsigmumupol}
\end{equation}
with
\begin{equation} 
 \begin{split}
  {\cal T}_{\!VV}^{} = \; 
  & e_q^2\,\d_{ij}\, (1 - {\cal P}_{\!-} {\cal P}_{\!+})
    - \frac{e_q c_{ij}^{} \d_{ij}^{}\,s}{
             2\,{\rm s}_W^2 {\rm c}_W^2}\,
    \left[ v_\mu^{}\,(1 - {\cal P}_{\!-} {\cal P}_{\!+}) -
           a_\mu^{}\,({\cal P}_{\!-} - {\cal P}_{\!+}) \right]\,
    \Re(d_Z^{}) \hspace*{2cm} \\
  & +\, \frac{c_{ij}^2\,s^2}{
             16\,{\rm s}_W^4 {\rm c}_W^4}\,
    \left[ (v_\mu^2+a_\mu^2)\,(1 - {\cal P}_{\!-} {\cal P}_{\!+}) - 
           2\,v_\mu^{} a_\mu^{}\,({\cal P}_{\!-} - {\cal P}_{\!+}) \right]\:
    |d_Z^{}|^2 \, ,
 \end{split}
\end{equation}
\begin{equation} 
 \begin{split}
  {\cal T}_{\!HH}^{} = \:  
  & \frac{h_\mu^2 s}{2 e^4} \Big\{ \big[ \,
       \left| (G_1^{\sq})_{ij}^{}\sin\a\;d_h^{} 
             -(G_2^{\sq})_{ij}^{}\cos\a\;d_H^{} \right|^2 
     + \left| (G_3^{\sq})_{ij}^{}\sin\b\;d_A^{} \right|^2 \big] 
    \,(1 + {\cal P}_{\!-} {\cal P}_{\!+}) \\
  & + 2\sin\b\,(G_3^{\sq})_{ij} 
    \big[ (G_1^{\sq})_{ij}^{}\sin\a\;\Re(d_A^* d_h^{})- 
          (G_2^{\sq})_{ij}^{}\cos\a\;\Re(d_A^* d_H^{})\big]
    \,({\cal P}_{\!-} + {\cal P}_{\!+}) \Big\} \, ,
 \end{split} 
\end{equation}
and
\begin{equation}
 \begin{split}
  {\cal T}_{\!VH}^{} = & \:
    \frac{2\rzw\,e_q m_\mu h_\mu \d_{ij}^{}}{e^2}\:
    \left[ (G_1^{\sq})_{ij}^{}\sin\a\;\Re(d_h^{}) - 
           (G_2^{\sq})_{ij}^{}\cos\a\;\Re(d_H^{})\right] 
               \frac{\kappa_{ij}^{}}{s}\, \cos\vartheta  \\
   & -\frac{m_\mu h_\mu\,c_{ij}^{}}{
           \rzw\,e^2\, {\rm s}_W^2{\rm c}_W^2}\;
        \left[ (G_1^{\sq})_{ij}^{}\sin\a\;\Re(d_Z^* d_h^{})- 
               (G_2^{\sq})_{ij}^{}\cos\a\;\Re(d_Z^* d_H^{})\right] \\
   & \hspace{27mm} \cdot\, \big[
     \big( v_\mu - \frac{a_\mu}{2}\,({\cal P}_{\!-}-{\cal P}_{\!+}) 
       \big)\, \kappa_{ij}^{} \cos\vartheta 
     -\frac{a_\mu}{2}\,(\msq{i}^2-\msq{j}^2)\,
        ({\cal P}_{\!-}+{\cal P}_{\!+}) \big]  \\
   & -\frac{m_\mu h_\mu \sin\b\:c_{ij}^{}(G_3^{\sq})_{ij}^{}}{
            \rzw\,e^2\, {\rm s}_W^2{\rm c}_W^2} \: 
     \Big\{ (\msq{i}^2-\msq{j}^2)\, a_\mu^{}  \\
   & \hspace{18mm} - \big[ 
     (\msq{i}^2-\msq{j}^2) ({\cal P}_{\!-}-{\cal P}_{\!+})
      - \kappa_{ij}^{} ({\cal P}_{\!-}+{\cal P}_{\!+}) \cos\vartheta 
     \big]\, \frac{v_\mu^{}}{2} \,        
     \Big\} \: \Re(d_Z^*\,d_A^{}) \,. 
 \end{split}
\end{equation}  

\noi
${\cal P}_{\!\pm} = (-1,\,0,\,1)$ for (left--polarized, unpolarized, 
right--polarized) $\mu^\pm$ beams.  
The total cross section is given by:
\begin{equation}
  \s_P^{} (\mu^+\mu^-\to\sq_i^{}\sqbar_i) = 
  \frac{3\pi\a^2}{2s} \left[\,
     \frac{2}{3}\b_{ii}^3\, {\cal T}_{\!VV}^{} + \,
     \b_{ii}^{}\,{\cal T}_{\!HH}^{} \,\right] \, ,
\end{equation}
\begin{equation}
  \s_P^{} (\mu^+\mu^-\to\sq_i^{}\sqbar_j) = 
  \frac{3\pi\a^2}{2s} \left[\,
     \frac{2}{3}\b_{ij}^3\, {\cal T}_{\!VV}^{} + \,
     \b_{ij}^{}\,{\cal T}_{\!HH}^{} + 
     \b_{ij}^{}\,\hat{\cal T}_{\!ZH}^{} \,\right] 
  \qquad (i\not= j)
\end{equation}
where $\b_{ij}\equiv\kappa_{ij}/s$\, and 
\begin{equation}
 \begin{split}
  \hat{\cal T}_{\!ZH}^{} = 
  & \frac{m_\mu h_\mu\,c_{ij}^{}}{\rzw\,e^2\,{\rm s}_W^2{\rm c}_W^2}\:
    (\msq{i}^2-\msq{j}^2)\: \Big\{
    \sin\b\,(G_3^{\sq})_{ij}^{} \,
    \big[ \frac{v_\mu^{}}{2}({\cal P}_{\!-}-{\cal P}_{\!+}) 
          - a_\mu^{} \big] \;\Re(d_Z^* d_A^{}) \\
  & \hspace{16mm}+\frac{a_\mu}{2} \,
    \big[ (G_1^{\sq})_{ij}^{}\sin\a\;\Re(d_Z^* d_h^{})
         -(G_2^{\sq})_{ij}^{}\cos\a\;\Re(d_Z^* d_H^{})
    \big] ({\cal P}_{\!-}+{\cal P}_{\!+}) 
   \Big\} \,. 
 \end{split}
\end{equation}  

\noi
Note the additional $A^0$--$(h^0\!,\,H^0)$ interference 
in case of $\sq_1^{}\sq_2^{}$ production.
This interference term changes sign when one goes to the CP 
conjugate state. 
Analogous terms occur for the $Z$--$(h^0\!,\,H^0,\,A^0)$ interferences. 
Hence, $\s(\mu^+\mu^-\to\sq_1^{}\sqbar_2)\not= 
 \s(\mu^+\mu^-\to\sq_2^{}\sqbar_1)$\,  
if \,${\cal P}_{\!-}^{}\not= -{\cal P}_{\!+}^{}$\,! \\
Note moreover, that $\mu^+_L\mu^-_L$ and $\mu^+_R\mu^-_R$ combinations
only couple to Higgs bosons while $\mu^+_R\mu^-_L$ and
$\mu^+_L\mu^-_R$ combinations only couple to gauge bosons. 
Therefore, for Higgs-- and related physics it is highly preferable to
collide $\mu^+$ and $\mu^-$ of same helizities.
It is expected that a polarization of 28\% for both beams 
--- and maybe more with loss of luminosity --- 
can be achieved  (see \eg~\cite{mu-ecfa-report}).

\begin{figure}[h!]
\center
\begin{picture}(150,70)
\put(4,1){\mbox{\psfig{figure=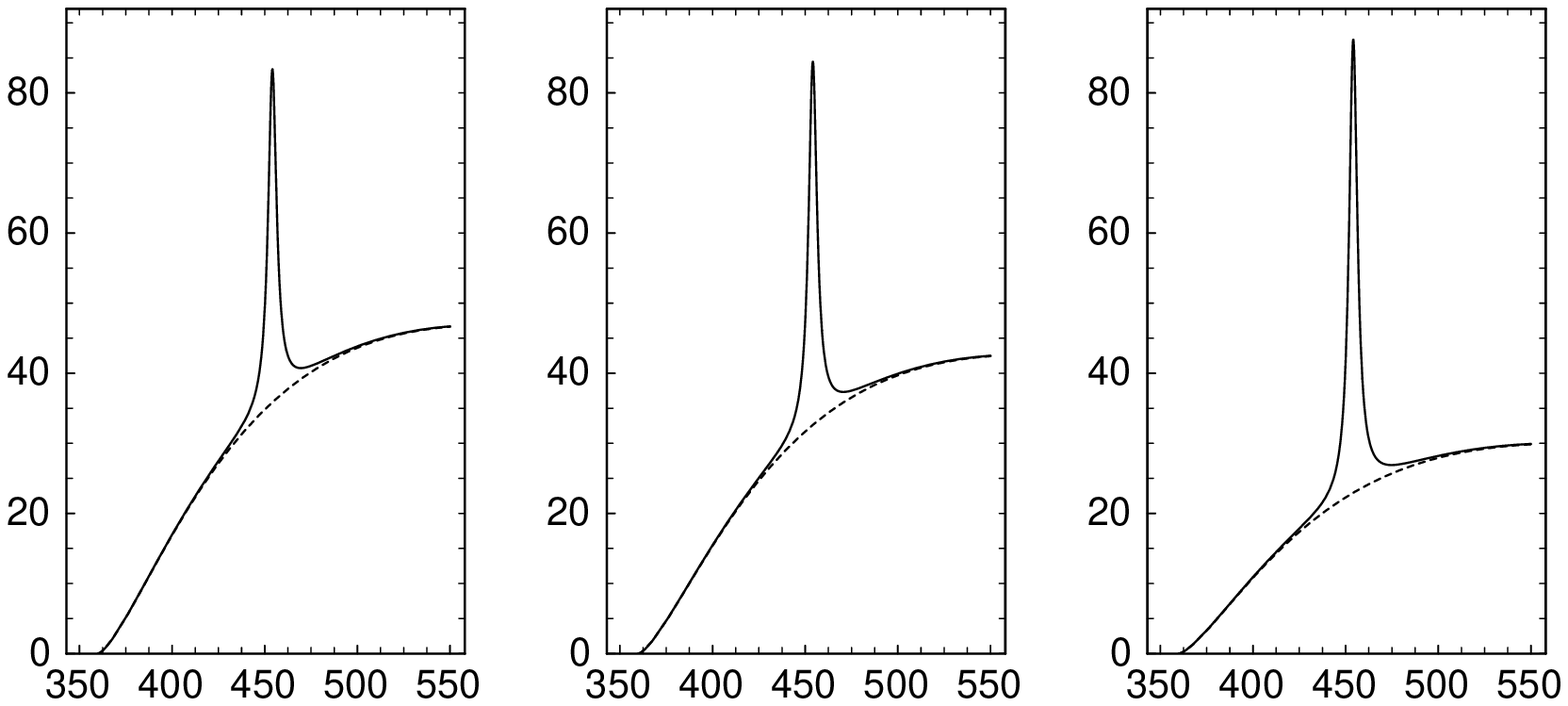,height=7.5cm}}}
\put(1,41){\makebox(0,0)[c]{\rotatebox{90}{$\sigma(\st_1\stbar_1)$~[f\/b]}}}
\put(32.5,0){\makebox(0,0)[cb]{$\sqrt{s}$~[GeV]}}
\put(81,0){\makebox(0,0)[cb]{$\sqrt{s}$~[GeV]}}
\put(129.5,0){\makebox(0,0)[cb]{$\sqrt{s}$~[GeV]}}
\put(35,16){\mbox{\footnotesize$\mbf{P=0}$}}
\put(79,16){\mbox{\footnotesize$\mbf{P=-0.3}$}}
\put(128,16){\mbox{\footnotesize$\mbf{P=-0.6}$}}
\end{picture}
\caption{Cross sections of $\mu^+\mu^-\to\st_1\bar{\st}_1$ 
for unpolarized, 30\% and 60\% left--polarized $\mu^\pm$ beams as a function 
of $\sqrt{s}$, for $\mst{1}=180$ GeV, $\mst{2}=260$ GeV, $\cst=-0.55$, 
$\msb{1}=175$ GeV, $\msb{2}=195$ GeV, $\csb=0.9$, 
$M_{\ti L}=170$ GeV, $M_{\ti E}=150$ GeV, $A_\tau=300$ GeV,
$M=140$ GeV, $\mu=300$ GeV, $\tan\b=3$, and $m_A=450$ GeV;
$\mbf{P}\equiv{\cal P}_{\!-}^{}={\cal P}_{\!+}^{}$. }
\label{fig:mustfig1}
\end{figure}
\begin{figure}[h!]
\center
\begin{picture}(150,70)
\put(4,1){\mbox{\psfig{figure=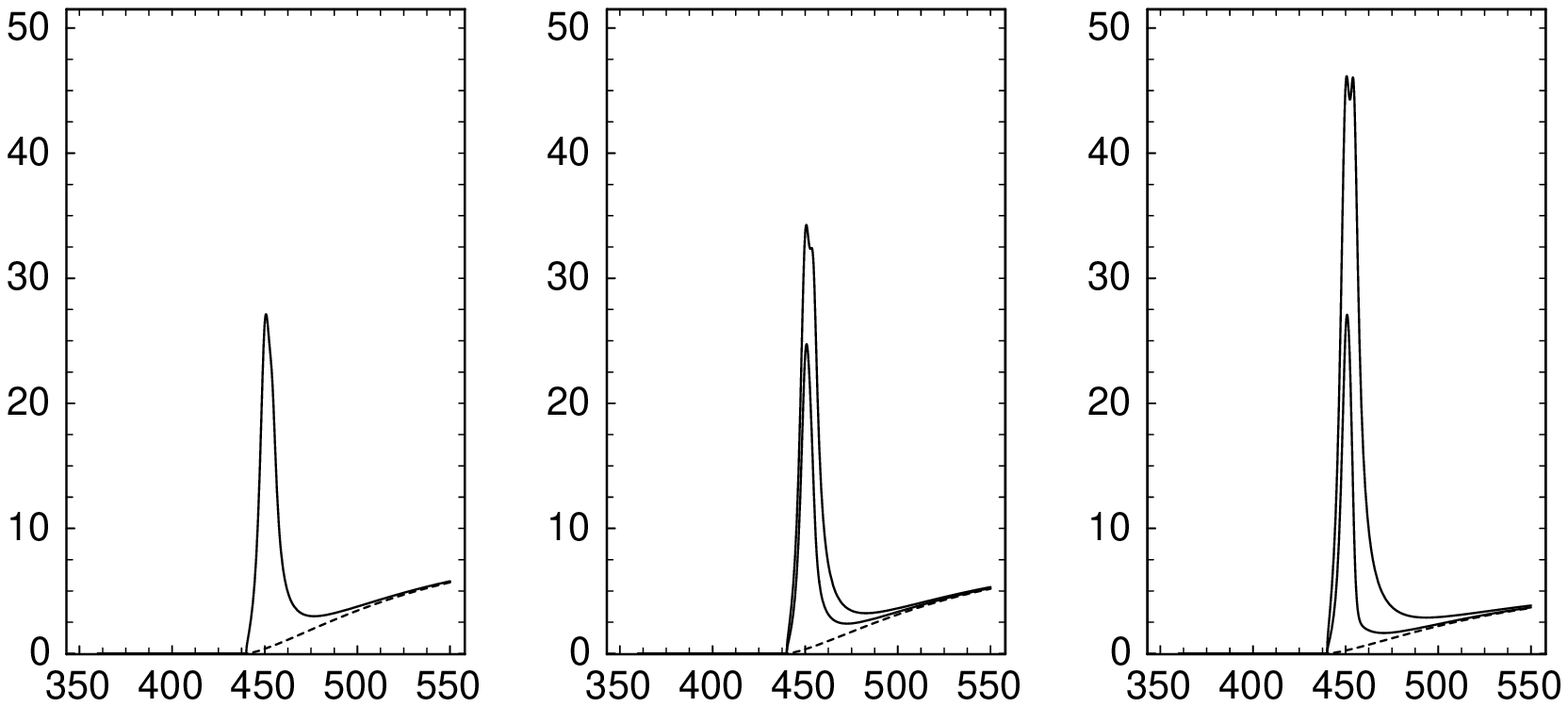,height=7.5cm}}}
\put(1,41){\makebox(0,0)[c]{\rotatebox{90}{$\sigma(\st_i\stbar_j)$~[f\/b]}}}
\put(32.5,0){\makebox(0,0)[cb]{$\sqrt{s}$~[GeV]}}
\put(81,0){\makebox(0,0)[cb]{$\sqrt{s}$~[GeV]}}
\put(129.5,0){\makebox(0,0)[cb]{$\sqrt{s}$~[GeV]}}
\put(17,64){\mbox{\footnotesize$\mbf{P=0}$}}
\put(67,64){\mbox{\footnotesize$\mbf{P=-0.3}$}}
\put(115,65){\mbox{\footnotesize$\mbf{P=-0.6}$}}
\put(23,48){\mbox{$\st_1\stbar_2=\st_2\stbar_1$}}
\put(84,49){\mbox{$\st_2\stbar_1$}}
\put(85,35){\mbox{$\st_1\stbar_2$}}
\put(86,34){\vector(-1,-1){5}}
\put(133,58){\mbox{$\st_2\stbar_1$}}
\put(133,35){\mbox{$\st_1\stbar_2$}}
\put(134,34){\vector(-1,-1){4}}
\end{picture}
\caption{Cross sections of $\mu^+\mu^-\to\st_1\bar{\st}_2$ and 
$\mu^+\mu^-\to\st_2\bar{\st}_1$ 
for unpolarized, 30\% and 60\% left--polarized beams as a function 
of $\sqrt{s}$, for $\mst{1}=180$ GeV, $\mst{2}=260$ GeV, $\cst=-0.55$, 
$\msb{1}=175$ GeV, $\msb{2}=195$ GeV, $\csb=0.9$, 
$M_{\ti L}=170$ GeV, $M_{\ti E}=150$ GeV, $A_\tau=300$ GeV,
$M=140$ GeV, $\mu=300$ GeV, $\tan\b=3$, and $m_A=450$ GeV;
${\mbf P}\equiv{\cal P}_{\!-}^{}={\cal P}_{\!+}^{}$. }
\label{fig:mustfig2}
\end{figure}
\begin{figure}[h!]
\center
\begin{picture}(70,70)
\put(4,4){\mbox{\psfig{figure=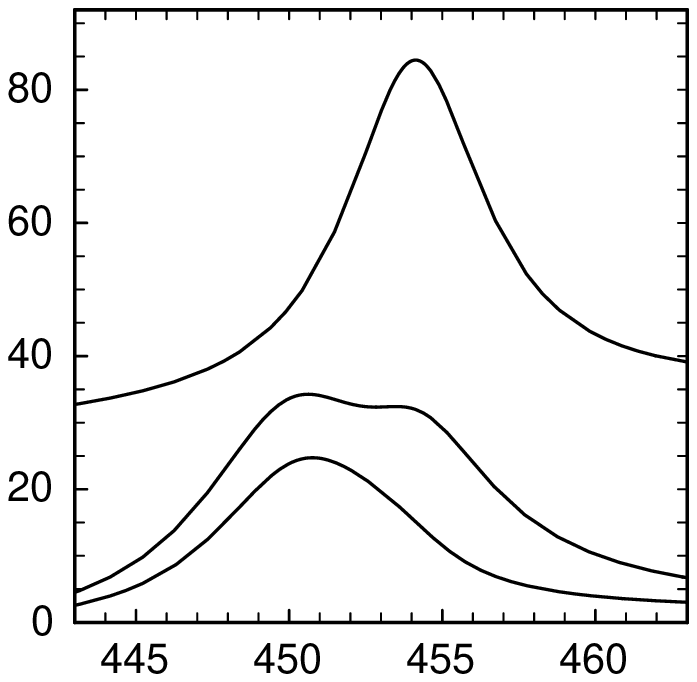,height=6.6cm}}}
\put(0,40){\makebox(0,0)[c]{\rotatebox{90}{$\sigma(\st_i\stbar_j)$~[f\/b]}}}
\put(40,0){\makebox(0,0)[cb]{$\sqrt{s}$~GeV}}
\put(50,58){\mbox{$\st_1\stbar_1$}}
\put(49,30){\mbox{$\st_2\stbar_1$}}
\put(33,19){\mbox{$\st_1\stbar_2$}}
\end{picture}
\caption{Cross sections of $\mu^+\mu^-\to\st_i\bar{\st}_j$
for 30\% left--polarized $\mu^\pm$ beams 
(${\cal P_{\!-}^{}}={\cal P_{\!+}^{}}=-0.3$) 
as a function of $\sqrt{s}$, 
for $\mst{1}=180$ GeV, $\mst{2}=260$ GeV, $\cst=-0.55$, 
$\msb{1}=175$ GeV, $\msb{2}=195$ GeV, $\csb=0.9$, 
$M_{\ti L}=170$ GeV, $M_{\ti E}=150$ GeV, $A_\tau=300$ GeV,
$M=140$ GeV, $\mu=300$ GeV, $\tan\b=3$, and $m_A=450$ GeV.}
\label{fig:mustfig3}
\end{figure}

In \fig{mustfig1} we show the $\sqrt{s}$ dependence of the total $\st_1\stbar_1$ 
production cross section for $\mst{1}=180\gev$, $\cst=-0.55$, and $m_A=450$ GeV. 
The three plots are for unpolarized 
($\mbf{P}\!\equiv{\cal P}_{\!-}^{}={\cal P}_{\!+}^{}=0$), 
30\% left--polarized ($\mbf{P}\!=-0.3$), and 60\% left--polarized ($\mbf{P}\!=-0.6$)
$\mu^+$ and $\mu^-$ beams, respectively. 
In addition to the cross sections at a $\mu^+\mu^-$ collider (full lines),  
the dashed lines show the cross sections in $e^+e^-$. 
For calculating the properties of the Higgs bosons we have taken 
$\mst{2}=260$ GeV, $\msb{1}=175$ GeV, $\msb{2}=195$ GeV, $\csb=0.9$, 
$M_{\ti L}=170$ GeV, $M_{\ti E}=150$ GeV, $A_\tau=300$ GeV,
$M=140$ GeV, $\mu=300$ GeV, and $\tan\b=3$. 
With this set of parameters we get $m_{H^0}=454\gev$, $\sin\a=-0.35$,
$\G_{\!H^0}^{}=5.4\gev$, and $\G_{\!A^0}^{}=7.3\gev$. 
As can be seen, a clear peak occurs at $\sqrt{s}=m_{H^0}$. 
The total cross section at the resonance varies only little 
with the polarization. However, the relative importance of the Higgs exchange 
strongly increases with increasing polarization. \\
Analogously, \fig{mustfig2} shows $\sqrt{s}$ dependence of 
$\s(\mu^+\mu^-\to\st_1\stbar_2)$ and $\s(\mu^+\mu^-\to\st_2\stbar_1)$ for 
the parameters of \fig{mustfig1}. 
In this case also $A^0$ exchange contributes. 
While $\s(\st_1\stbar_2)=\s(\st_2\stbar_1)$ for unpolarized beams, 
there is a difference between these cross sections for $\mbf{P}\!\not=0$. 
This difference strongly increases with increasing polarization.  
Notice, that while for $\st_1\stbar_2$ production the $A^0$ and $H^0$ 
resonances fully overlap this is not the case for $\st_2\stbar_1$ production.\\
This can be seen more clearly in \fig{mustfig3} which zooms on the 
resonances of Figs.~\ref{fig:mustfig1} and \ref{fig:mustfig2}: 
$\s(\st_1\stbar_1)$ peaks at $\sqrt{s}=m_{H^0}$, 
$\s(\st_1\stbar_2)$ has its maximum at $\sqrt{s}\simeq m_A$, 
and $\s(\st_2\stbar_1)$ shows two humps at $\sqrt{s}\simeq m_{A,H^0}$. 
With the excellent energy resolution of a muon collider it may thus be 
possible to determine the properties of heavy neutral SUSY Higgs bosons. 
In general, it is expected that this will be difficult at hadron and  
$e^+e^-$ colliders. 

\begin{figure}[h!]
\center
\begin{picture}(70,70)
\put(4,4){\mbox{\psfig{figure=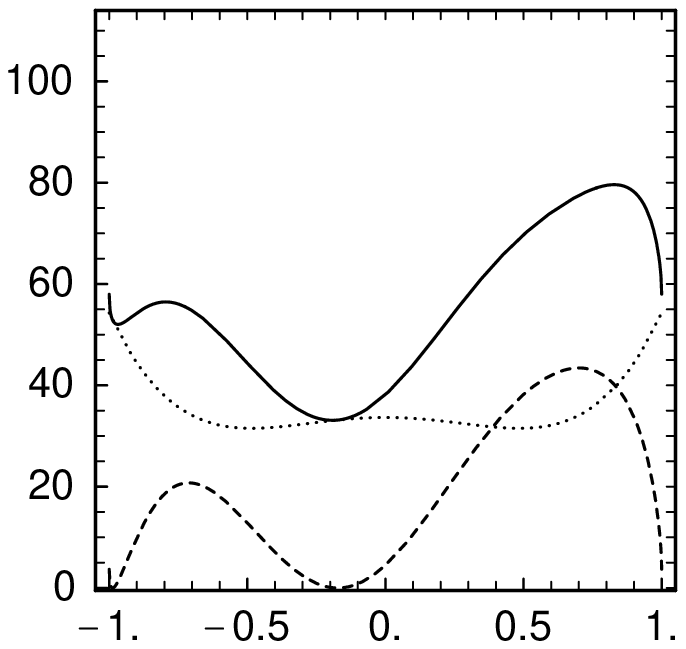,height=6.6cm}}}
\put(0,40){\makebox(0,0)[c]{\rotatebox{90}{$\sigma(\st_1\stbar_1)$~[f\/b]}}}
\put(41,0){\makebox(0,0)[cb]{$\cst$}}
\put(62,62){\mbox{\bf a}}
\end{picture}
\hspace*{10mm}
\begin{picture}(70,70)
\put(4,4){\mbox{\psfig{figure=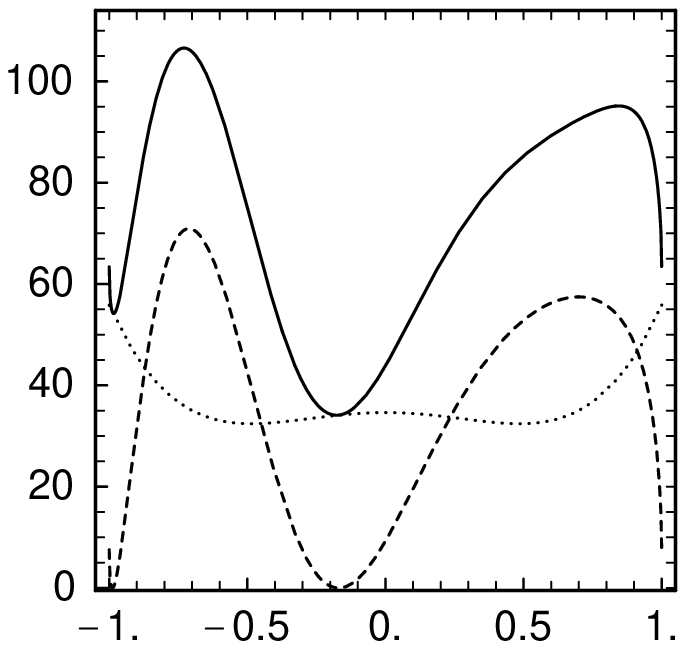,height=6.6cm}}}
\put(0,40){\makebox(0,0)[c]{\rotatebox{90}{$\sigma(\st_1\stbar_1)$~[f\/b]}}}
\put(41,0){\makebox(0,0)[cb]{$\cst$}}
\put(62,62){\mbox{\bf b}}
\end{picture}
\caption{Cross section of $\mu^+\mu^-\to\st_1\bar{\st}_1$ 
for 30\% left--polarized beams as a function 
of $\cst$, for $\mst{1}=180$ GeV, $\mst{2}=260$ GeV, 
$M_{\ti D}=1.12\,M_{\ti Q}$, $A_t=A_b$,  
$M_{\ti L}=170$ GeV, $M_{\ti E}=150$ GeV, $A_\tau=300$ GeV,
$M=140$ GeV, $\mu=300$ GeV, $\tan\b=3$, and $m_A=450$ GeV;
(a) $\sqrt{s}=450$~GeV, (b) $\sqrt{s}=454$~GeV. 
The dashed lines show the Higgs boson contributions, 
the dotted lines are the $(\gamma,\,Z)$ contributions, 
and the full lines are the total cross sections.}
\label{fig:mustfig4}
\end{figure}

\begin{figure}[h!]
\center
\begin{picture}(70,70)
\put(4,4){\mbox{\psfig{figure=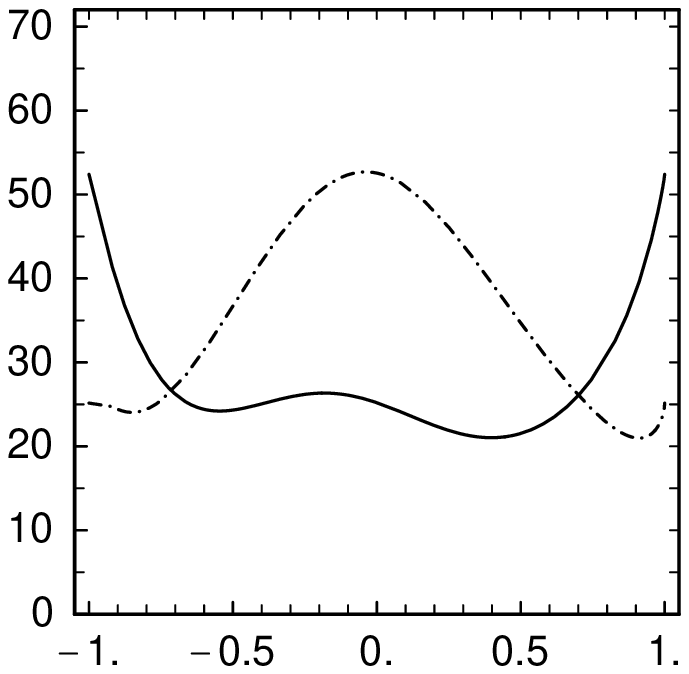,height=6.6cm}}}
\put(0,39){\makebox(0,0)[c]{\rotatebox{90}{$\sigma(\st_i\stbar_j)$~[f\/b]}}}
\put(40,0){\makebox(0,0)[cb]{$\cst$}}
\put(62,62){\mbox{\bf a}}
\put(15,50){\mbox{$\st_1\stbar_2$}}
\put(47,51){\mbox{$\st_2\stbar_1$}}
\end{picture}
\hspace*{10mm}
\begin{picture}(70,70)
\put(4,4){\mbox{\psfig{figure=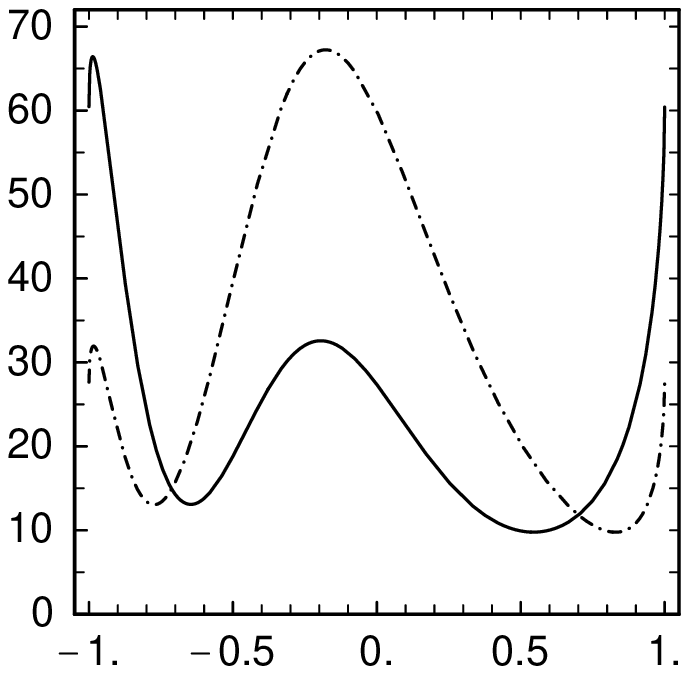,height=6.6cm}}}
\put(0,39){\makebox(0,0)[c]{\rotatebox{90}{$\sigma(\st_i\stbar_j)$~[f\/b]}}}
\put(40,0){\makebox(0,0)[cb]{$\cst$}}
\put(62,62){\mbox{\bf b}}
\put(32,39){\mbox{$\st_1\stbar_2$}}
\put(45,56){\mbox{$\st_2\stbar_1$}}
\end{picture}
\caption{Cross sections of $\mu^+\mu^-\to\st_1\bar{\st}_2$ 
and $\mu^+\mu^-\to\st_2\bar{\st}_1$ 
for 30\% polarized beams as a function of $\cst$, 
for $\mst{1}=180$ GeV, $\mst{2}=260$ GeV, 
$M_{\ti D}=1.12\,M_{\ti Q}$, $A_t=A_b$,  
$M_{\ti L}=170$ GeV, $M_{\ti E}=150$ GeV, $A_\tau=300$ GeV,
$M=140$ GeV, $\mu=300$ GeV, $\tan\b=3$, and $m_A=450$ GeV;
in (a) $\sqrt{s}=450$~GeV and in (b) $\sqrt{s}=454$~GeV.}
\label{fig:mustfig5}
\end{figure}

\noi
The $\cst$ dependence of $\s(\mu^+\mu^-\to\st_1\stbar_1)$ is shown 
in \fig{mustfig4} for 30\% left--polarized $\mu^\pm$ beams and 
$\mst{1}=180$ GeV, $\mst{2}=260$ GeV, 
$M_{\ti D}=1.12\,M_{\ti Q}$, $A_t=A_b$,  
$M_{\ti L}=170$ GeV, $M_{\ti E}=150$ GeV, $A_\tau=300$ GeV,
$M=140$ GeV, $\mu=300$ GeV, $\tan\b=3$, and $m_A=450$ GeV. 
In \fig{mustfig4}\,a $\sqrt{s}=m_A=450$~GeV and 
in \fig{mustfig4}\,b $\sqrt{s}=m_{H^0}=454$~GeV. 
The dashed lines show the Higgs boson contributions, 
the dotted lines the $(\gamma,\,Z)$ contributions, 
and the full lines the total cross sections.
While the vector boson contributions depend only on $\cos^2\tst$ 
the Higgs boson contributions induce a distinct dependence also on 
the sign of the stop mixing angle. 
Therefore, the $\cst$ dependence of the $\st_1\stbar_1$ production 
cross section at a $\mu^+\mu^-$ collider with $\sqrt{s}\sim m_A$ 
is very different from that at an $e^+e^-$ collider. \\
The same is true for $\st_1\st_2$ production. 
This can be seen in \fig{mustfig5} where we plot 
$\s(\mu^+\mu^-\to\st_1\stbar_2)$ and $\s(\mu^+\mu^-\to\st_2\stbar_1)$ 
as a function of $\cst$ for the parameters of \fig{mustfig4}. 
Here note, that the leading term of the $\st_1\st_2H^0$ coupling 
is $h_t(\mu\cos\a-A_t\sin\a)\cos2\tst$ 
while the $\st_1\st_2A^0$ coupling is $h_t(\mu\sin\b-A_t\cos\b)$.  
Since $\mu$ is constant in \fig{mustfig5}, $A_t$ varies with $\cst$.

\begin{figure}[h!]
\center
\begin{picture}(150,70)
\put(4,1){\mbox{\psfig{figure=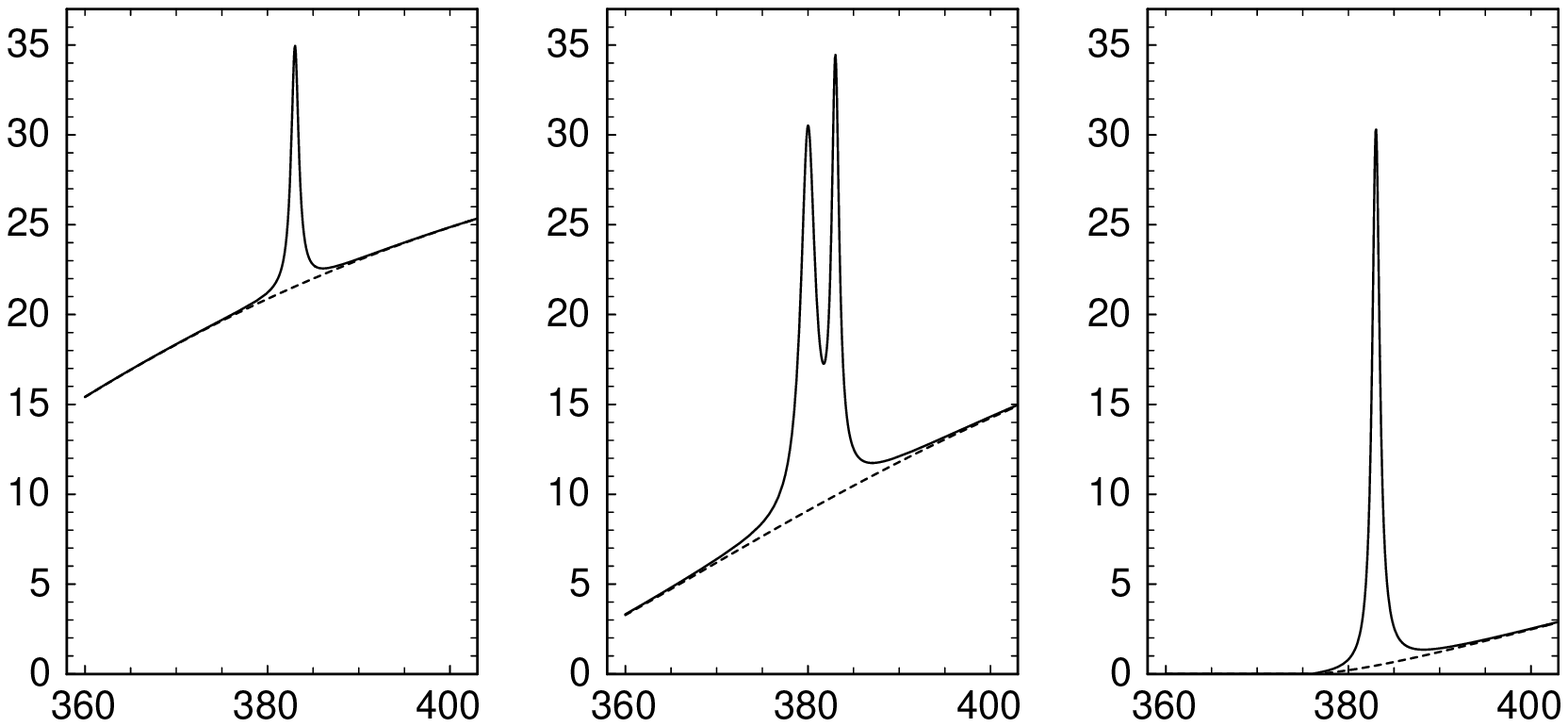,height=7.5cm}}}
\put(0,41){\makebox(0,0)[c]{\rotatebox{90}{$\sigma(\sb_i\sb_j)$~[f\/b]}}}
\put(32.5,0){\makebox(0,0)[cb]{$\sqrt{s}$~[GeV]}}
\put(81,0){\makebox(0,0)[cb]{$\sqrt{s}$~[GeV]}}
\put(130,0){\makebox(0,0)[cb]{$\sqrt{s}$~[GeV]}}
\put(18,65){\mbox{\bf a}}
\put(66,65){\mbox{\bf b}}
\put(115,65){\mbox{\bf c}}
\put(40,16){\mbox{$\sb_1\sbbar_1$}}
\put(89,16){\mbox{$\sb_1\sb_2$}}
\put(116,16){\mbox{$\sb_2\sbbar_2$}}
\end{picture}
\caption{Cross sections of (a) $\mu^+\mu^-\to\sb_1\bar{\sb}_1$, 
(b) $\mu^+\mu^-\to\sb_1\sb_2$ , and (c) $\mu^+\mu^-\to\sb_2\bar{\sb}_2$, 
for 30\% left--polarized beams (${\cal P}_{\!-}^{}={\cal P}_{\!+}^{}=-0.3$) 
as a function of $\sqrt{s}$, for 
$\msb{1}=157$ GeV, $\msb{2}=188$ GeV, $\csb=0.78$, 
$\mst{1}=197$ GeV, $\mst{2}=256$ GeV, $\cst=-0.66$, 
$M_{\ti L}=160$ GeV, $M_{\ti E}=155$ GeV, $A_\tau=100$ GeV,
$M=140$ GeV, $\mu=300$ GeV, $m_A=380$ GeV, and $\tan\b=4$.}
\label{fig:musbfig1}
\end{figure}

\begin{figure}[h!]
\center
\begin{picture}(70,70)
\put(4,3){\mbox{\psfig{figure=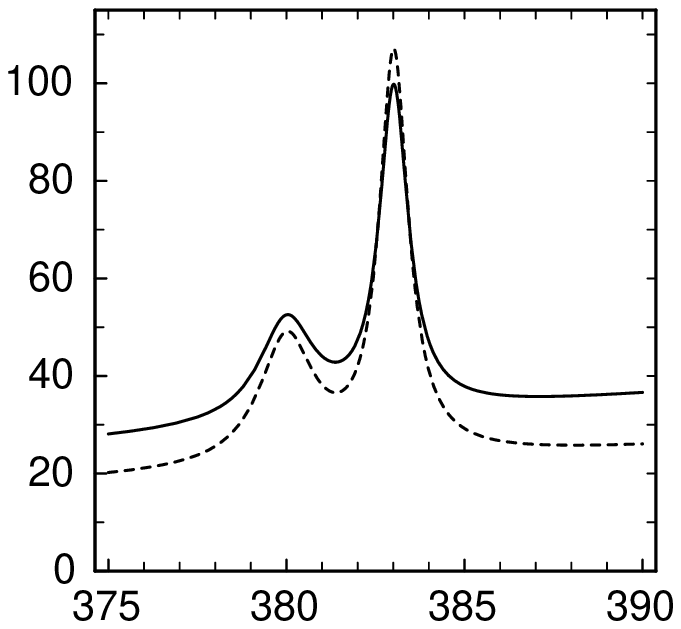,height=6.85cm}}}
\put(0,39){\makebox(0,0)[c]{\rotatebox{90}{$\sigma(\sb\sbbar)$~[f\/b]}}}
\put(42,0){\makebox(0,0)[cb]{$\sqrt{s}$~[GeV]}}
\put(62,62){\mbox{\bf a}}
\put(51,34){\mbox{\footnotesize$\mbf{P=-0.3}$}}
\put(51,21){\mbox{\footnotesize$\mbf{P=-0.6}$}}
\end{picture}
\hspace*{10mm}
\begin{picture}(70,70)
\put(4,4){\mbox{\psfig{figure=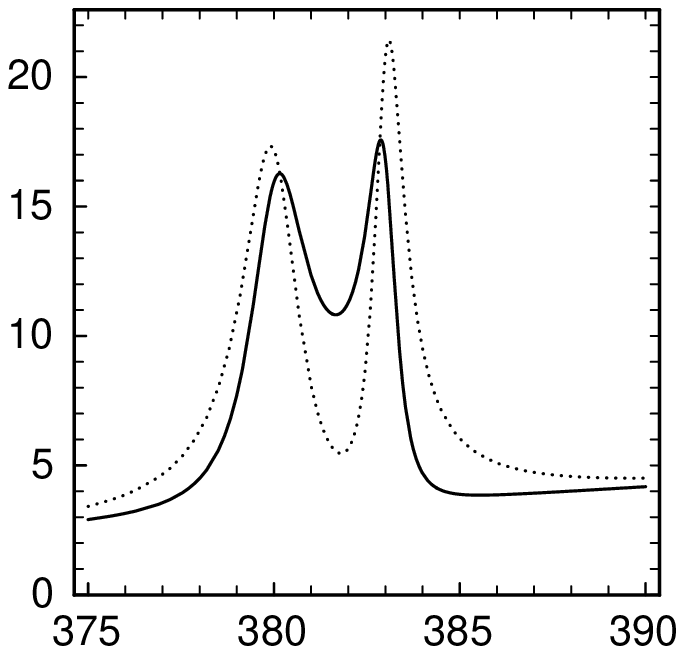,height=6.6cm}}}
\put(0,39){\makebox(0,0)[c]{\rotatebox{90}{$\sigma(\sb_i\sbbar_j)$~[f\/b]}}}
\put(39,0){\makebox(0,0)[cb]{$\sqrt{s}$~[GeV]}}
\put(61,62){\mbox{\bf b}}
\put(52,40){\mbox{$\sb_1\sbbar_2$}}
\put(46,57){\mbox{$\sb_2\sbbar_1$}}
\put(53,39){\vector(-1,-1){9}}
\put(47,57){\vector(-1,-1){3}}
\put(48,16){\mbox{\footnotesize$\mbf{P=-0.6}$}}
\end{picture}
\caption{(a) Total sbottom production cross section and 
(b) cross sections of $\mu^+\mu^-\to\sb_1\bar{\sb}_2$ and 
$\mu^+\mu^-\to\sb_2\bar{\sb}_1$ 
as a function of $\sqrt{s}$, for 
$\msb{1}=157$ GeV, $\msb{2}=188$ GeV, $\csb=0.78$, 
$\mst{1}=197$ GeV, $\mst{2}=256$ GeV, $\cst=-0.66$, 
$M_{\ti L}=160$ GeV, $M_{\ti E}=155$ GeV, $A_\tau=100$ GeV,
$M=140$ GeV, $\mu=300$ GeV, $m_A=380$ GeV, and $\tan\b=4$.}
\label{fig:musbfig2}
\end{figure}

An example for sbottom production is shown in \fig{musbfig1}  
where we plot the $\sqrt{s}$ dependence of the cross sections of 
$\sb_1\sbbar_1$, $\sb_1\sbbar_2+\sb_2\sbbar_1$, and $\sb_2\sbbar_2$ 
production for 30\% left--polarized $\mu^\pm$ beams.  
The input parameters are: 
$\msb{1}=157$ GeV, $\msb{2}=188$ GeV, $\csb=0.78$, 
$\mst{1}=197$ GeV, $\mst{2}=256$ GeV, $\cst=-0.66$, 
$M_{\ti L}=160$ GeV, $M_{\ti E}=155$ GeV, $A_\tau=100$ GeV,
$M=140$ GeV, $\mu=300$ GeV, $m_A=380$ GeV, and $\tan\b=4$.
In this case we get $m_{H^0}=383\gev$, $\sin\a=-0.28$,
$\G_{\!H^0}^{}=1\gev$, and $\G_{\!A^0}^{}=1.9\gev$. 
Note, that for $\sb_1\sb_2$ production (\fig{musbfig1}\,b)
two distinct peaks occur at the $A^0$ and $H^0$ resonances!
However, for the parameters of \fig{musbfig1} 
it might be difficult to distinguish $\sb_1$ and $\sb_2$ 
because their decay properties are very similar. 
We therefore show in \fig{musbfig2}\,a the total sbottom 
production cross section, $\sum_{i,j} \s(\mu^+\mu^-\to\sb_i\sbbar_j)$, 
for the parameters of \fig{musbfig1}. 
In addition to $\mbf{P}=-0.3$ (full line) we also show the case 
$\mbf{P}=-0.6$ (dashed line). As can be seen, the two peaks 
are still separated. 
\Fig{musbfig2}\,b compares the $\sb_1\sbbar_2$ and $\sb_2\sbbar_1$ 
production cross sections for $\mbf{P}=-0.6$ and the other parameters 
as in \fig{musbfig2}\,a. For $\mbf{P}=-0.3$ the difference between 
$\s(\sb_1\sbbar_2)$ and $\s(\sb_2\sbbar_1)$ is less pronounced.


\chapter{SUSY--QCD Corrections to Squark Decays}

\section {Introduction}

We now turn to the stop and sbottom 2--body decays.  
As the squarks of the 1st and 2nd generation, stops and sbottoms  
can decay into quarks plus charginos, neutralinos, or gluinos. 
In addition to these ``coventional'' decays into fermions, 
stops and sbottoms may decay into bosons, \ie into a 
lighter squark plus a gauge or Higgs boson. 
These decays are possible in the parameter space where large 
Yukawa couplings and $\sq_L^{}$--$\sq_R^{}$ mixing generate 
a large mass splitting between $\sq_1^{}$ and $\sq_2^{}$ 
and/or $\sq_i^{}$ and $\sq_j'$. 
Moreover, Yukawa couplings and $\sq_L^{}$--$\sq_R^{}$ mixing 
have a strong influence on the branching ratios of the 
various decays modes. 
Thus, stops and sbottoms can have quite complex decay patterns.

\noi
In summary, there are the decays ($i,\,j=1,\,2$; $k=1\ldots4$):
\begin{alignat}{2}
  \st_i\, & \to\, t\,\nt_k\,,\;  b\,\chp_j , & \qquad 
  \sb_i\, & \to\, b\,\nt_k\,,\;  t\,\chm_j ,
  \label{eq:ncmodes}
\\
  \st_i\, & \to\, t\,\sg\,, & \qquad 
  \sb_i\, & \to\, b\,\sg\,, 
  \label{eq:gluinomode}
\\
  \st_i\, & \to\, \sb_j + (W^+,\,H^+), & \qquad 
  \sb_i\, & \to\, \st_j + (W^-,\,H^-),  
  \label{eq:bosmodes1}
\\
  \st_2 & \to\, \st_1 + (Z^0,\, h^0,\, H^0,\, A^0) , & \qquad 
  \sb_2 & \to\, \sb_1 + (Z^0,\, h^0,\, H^0,\, A^0) .
  \label{eq:bosmodes2}
\end{alignat}

\noi
All these decays were first discussed at tree level in \cite{bmp}. 
They were studied in more detail in 
\cite{lep2paper,desy123D,desy123E,nlcpaper,bosdec}.  
For the lighter mass eigenstates $\st_1$ and $\sb_1$ the decays into 
fermions of Eqs.~\eq{ncmodes} and \eq{gluinomode} 
are in general the most important ones \footnote{If, however, 
  all 2--body decay modes are kinematically forbidden, 
  loop--induced decays such as $\st_1\to c\nt_1$ and 3--body decays 
  become important. The higher order decays of $\st_1$ have been studied 
  in \cite{hikasa-kobayashi,porod-woermann,werner-diss,werner-stop3}.}. 
If the strong decays into gluinos are kinematically allowed 
they play an important r\^ole. 
For the heavier $\st_2$ and $\sb_2$ it turned out the bosonic decays of 
Eqs.~\eq{bosmodes1} and \eq{bosmodes2} can be dominant in a wide range 
of the parameter space; see \cite{bosdec} for a detailed analysis.  
Moreover, also the decays $\sb_1\to\st_1 W^-$ and $\sb_1\to\st_1 H^-$ 
can be important if the $\st\mbox{--}\sb$ mass splitting 
is large enough \cite{desythtalk}. 

The 1--loop SUSY--QCD corrections to the decays of 
Eqs.~\eq{ncmodes} to \eq{bosmodes2}
have all been calculated within the last two years (1996--98). 
For the decays into charginos and neutralinos this was done 
in \cite{qcdnc-paper,qcdnc-djouadi,qcd-zerwas2},
for the decays into $W^\pm$ and $Z$ bosons in \cite{qcdwz-paper}, 
for the decays into Higgs bosons in \cite{qcdhx-djouadi,qcdhx-paper}, 
and for the decays into gluinos in \cite{qcd-zerwas1,qcd-zerwas2}.
(Here note that the decays into photon or gluon, which are absent at 
tree level, are not induced by these corrections either.)

\noi
In this chapter we study the $\Oas$ SUSY--QCD corrections 
to the decays of Eqs.~\eq{ncmodes}, \eq{bosmodes1}, and \eq{bosmodes2}. 
In Sect.~\ref{sect:qcd-general} we first discuss some general aspects of 
these calculations. 
Notations and conventions are clarified in Sect.~\ref{sect:convention};    
the formulae for the self--energies of squarks and squarks are given 
in Sect.~\ref{sect:selfenergies}. 
In Sect.~\ref{sect:subtleties} we discuss the subtleties that have to 
be taken into account for calculating the $\Oas$ SUSY--QCD corrections 
in the on--shell scheme. 
The specific squark decays are then treated in Sects.~\ref{sect:qcdnc} 
to \ref{sect:qcdhx}. For each process, we give the complete formulae 
for the $\Oas$ SUSY--QCD corrected decay widths and perform a 
detailed numerical analysis. 
Finally, the SUSY--QCD corrected branching ratios are discussed 
in Sect.~\ref{sect:branch}.

\section {General Aspects of SUSY--QCD Corrections}\label{sect:qcd-general}

For the corrections at $\Oas$ we have to consider 1--loop diagrams 
with gluons, gluinos, and squarks in the loops.  
The integration momenta in the loops run from zero to infinity,  
and some of the diagrams diverge. 
We therefore have to apply appropriate renormalization procedures to 
cosistently isolate and remove the infinities from the measurable quantities. 
The general approach is as follows:

\subsection{Renormalization}
We start with the bare Lagrangian, which consists of bare parameters 
(bare masses and couplings) and bare fields, and write it in terms 
of the renormalized quantities. The bare mass $m_0$ and  
bare coupling $c_0$, for instance, are replaced by the renormalized mass $m$ and 
coupling $c$ plus the associated counterterms:
\begin{equation}
  m_0 = m + \d m, \qquad c_0 = c + \d c.
\end{equation}
The bare fields are given by the renormalized ones multiplied by 
a wave--function renormalization factor \eg, 
\begin{equation}
  \phi_0^{} = Z^{\frac{1}{2}}\phi
\end{equation}
(where $Z^{\frac{1}{2}}$ is in general a matrix which mixes fields 
with the same quantum numbers into each other). 
With these replacements we can seperate the bare Lagrangian 
into the renormalized one and a part which contains 
the counterterms:
\begin{equation}
  \Lag_0 = \Lag + \d\Lag .
\end{equation}
Here we note that $\Lag$ and $\d\Lag$ have the same structure.
$\d\Lag$ gives rise to new (effective) vertices in each order 
of pertubation theory. 
Finally we apply appropriate renormalization conditions which 
determine the counterterms and the physical meaning of the
renormalized parameters. 

For our calculations we use the {\em on--shell renomalization} scheme. 
Within this scheme the renormalized mass is defined as the pole mass,
\ie the experimentally measured mass 
\footnote{This has to be distinguished from the minimal substraction 
method where the counterterms are `purely infinte', \ie they 
solely contain the divergent parts. In contrast to the on--shell 
scheme these counterterms are scale dependent.}. 
The on--shell renormalization conditions are: 
(i) the renormalized mass is the real part of the pole of the
propagator and 
(ii) the real part of the residue of the pole is unity.

\noi
Technically speaking, at first order we have to add 1--loop 
vertex and wave--function corrections to the tree--level diagram 
as well as the counterterms for the couplings. For the latter we have
to renormalize all parameters that enter the couplings.
The resulting 1--loop corrected decay width is then utraviolet (UV) finite.
In order to cancel also the infrared (IR) divergence we add the
emisson of real (hard and soft) gluons. 

\noi
The ${\cal O}(\a_{s})$ SUSY-QCD corrected squark decay width
can thus be decomposed in the following way:
\begin{equation}
  \G = \G^{0} + \d\G^{(v)} + \d\G^{(w)} + \d\G^{(c)} + \d\G_{real}.
  \label{eq:gencorr}
\end{equation}
Here $\G^0$ is the decay width at tree level and $\d\G^{(v,w,c)}$ are
the virtual corrections. 
The superscript $v$ denotes vertex corrections, 
$w$ wave function corrections, and 
$c$ the shift from bare to on--shell couplings. 
$\d\G_{real}$ is the correction due to real gluon emission. 

\subsection{Regularization: DREG vs. DRED}

The divergencies which arise in calculating the loop integrals must be
regulated. In the SM the standard technique is {\em dimensioal 
regularization} (DREG) which respects the gauge symmetry of the
Lagrangian (and thus also preserves the Ward identities). 
In this scheme, spacetime is continued to $D=4-\epsilon$ dimensions. 
For $D<4$ the divergencies then appear as simple poles in $\epsilon$. 
However, in DREG also the dimensionality of the fields is continued to
$D$ dimensions. As a consequence, the index of a gauge field $V_\mu$
runs from 0 to $D-1$, and also the Dirac algebra is $D$--dimensional
\eg, $\g_\mu\g^\mu=D$. 

This causes problems in supersymmetry:  
SUSY requires that the numbers of bosonic and fermionic degrees of
freedom be equal in each supermultiplet. 
Continuing a four--dimensional multiplet to $D=4-\epsilon$ dimensions
spoils this equality, introducing a mismatch of gauge boson and
gaugino degrees of freedom. It also spoils the SUSY--Ward identities.

\noi
A solution known as {\em dimensional reduction} (DRED) was proposed  
by Siegel, Capper, Jones, and van Nieuwenhuizen \cite{dred1,dred2}. 
In DRED spacetime is continued to $D=4-\epsilon$ dimensions while the
fields are not affected. 
Thus the index of the gauge fields runs from 0 to 3, and the Dirac
algebra is performed in four dimensions. 
This method nicely preserves both, gauge symmetry and SUSY 
--- at least up to 2--loop order \cite{dred3}.

\subsection{Notations, Conventions, etc.} \label{sect:convention}

In this thesis we work in the on--shell renormalization scheme
using dimensional reduction.
For the explicit calculation of the loop integrals we use 
Passarino--Veltman functions \cite{pave}  
in the notation and convention of A. Denner \cite{denner}.

\noi
The conceptual and calculational details of renormalization  
in supersymmetry have been thoroughly discussed 
in Refs.~\cite{pierce,helmut-diss}. 
We therefore here just present the final results of our calculations.
Analytical expressions of Passarino--Veltman integrals, 
generic diagrams, as well as details on how to calculate individual graphs, 
are given in \cite{helmut-diss}.

\noi 
For the numerical analysis 
we use on--shell squark masses $\msq{1,2}$ and mixing angles $\tsq$ 
($0\le\tsq<\pi$) as input parameters. 
Moreover, we take $m_{t} = 175$ GeV, $m_{b} = 5$ GeV, 
$m_{Z} = 91.2$ GeV, $\sin^{2}\t_{W} = 0.23$, 
$\a(m_{Z}) = 1/129$, and
$\a_{s}(m_{Z}) = 0.12$. 
For the running of $\a_{s}$ we use $\a_{s}(Q^{2}) = 
12\pi/[(33 - 2\,n_{f})\,\ln (Q^{2}/\Lambda^{2}_{n_{\!f}})]$ 
with $n_{\!f}$ the number of flavors. 
We take $\a_{s} = \a_{s}(\msq{i})$ for the $\sq^{}_{i}$ decay 
except for 
(i) the renormalization of $M_{\tilde Q}$ for which we take 
$\a_{s} = \a_{s}(M_{\tilde Q}(\st))$, see Sect.~\ref{sect:MQproblem}, 
and (ii) the calculation of the gluino mass, for which we take  
$\msg = \a_s(\msg)/\a_2 M$ with five steps of iteration. 
For the radiative corrections to the $h^0$ and $H^0$ masses and their 
mixing angle $\a$ ($-\frac{\pi}{2}\leq\a<\frac{\pi}{2}$ by convention) 
we use the formulae of \cite{radcorrh0}; 
for those to $m_{H^+}$ we follow \cite{radcorrhc} 
\footnote{Notice that \cite{radcorrh0,radcorrhc} have the opposite 
   sign convention for the parameter $\mu$.}. 

\noi
In order to respect the experimental mass bounds from LEP2 
\cite{LEPC50} and Tevatron \cite{tevlimits} we impose 
$m_{\st_1,\sb_1} > 85\gev$, 
$m_{h^0} > 90\gev$, 
$\mch{1} > 95\gev$, and 
$\msg > 300\gev$. 
Moreover, we require $\d\rho\,(\st\mbox{--}\sb) < 0.0012$ \cite{drhonum} 
from electroweak precision measurements 
using the one--loop formulae of \cite{Drees-Hagiwara} and 
$A_t^2 < 3\,(M_{\ti Q}^2(\st) + M_{\ti U}^2 + m_{H_2}^2)$, 
$A_b^2 < 3\,(M_{\ti Q}^2(\sb) + M_{\ti D}^2 + m_{H_1}^2)$ with 
$m_{H_2}^2=(m_{A}^2+m_{Z}^2)\cos^2\b-\frac{1}{2}\,m_Z^2$ and 
$m_{H_1}^2=(m_{A}^2+m_{Z}^2)\sin^2\b-\frac{1}{2}\,m_Z^2$ 
\cite{Deren-Savoy} from tree--level vacuum stability.

\section{Self Energies to $\Oas$} \label{sect:selfenergies}

\subsection{Squark Self Energy}

\begin{figure}[ht!]
\center
\begin{picture}(160,25)
\put(7,6){\mbox{\psfig{file=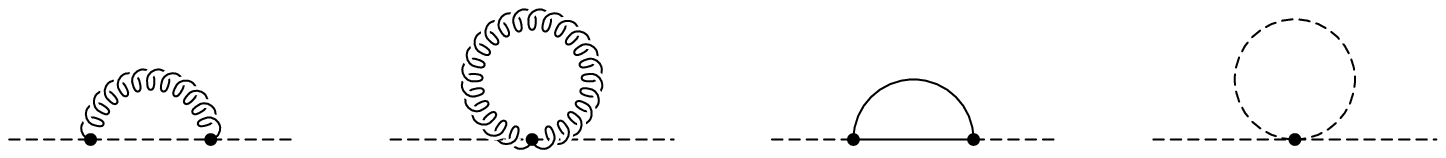}}}
\put(7,2){\mbox{$\sq_i$}}
\put(20,2){\mbox{$\sq_i$}}
\put(33,2){\mbox{$\sq_i$}}
\put(46,2){\mbox{$\sq_i$}}
\put(72,2){\mbox{$\sq_i$}}
\put(85,2){\mbox{$\sq_i$}}
\put(111,2){\mbox{$\sq_j$}}
\put(123,2){\mbox{$\sq_i$}}
\put(149.5,2){\mbox{$\sq_j$}}
\put(21,17){\mbox{$g$}}
\put(67,20){\mbox{$g$}}
\put(98,16){\mbox{$\sg$}}
\put(98,2){\mbox{$q$}}
\put(143,20){\mbox{$\sq_{1,2}$}}
\end{picture} 
\caption{Squark self--energy diagrams to $\Oas$.}
\label{fig:fd-sqself}
\end{figure}

The squark self energy to $\Oas$ gets contributions from gluon, 
gluino, and squark loops. The relevant Feynman digrams are shown 
in \fig{fd-sqself}.

\noi
The gluon--squark loop leads to
\begin{equation}
  \Sigma_{ii}^{(g)}(\msq{i}^2) =
  -\frac{2}{3}\frac{\a_s}{\pi}\,\msq{i}^2\, 
  \{ 2B_0(\msq{i}^2,0,\msq{i}^2) + B_1(\msq{i}^2,0,\msq{i}^2) \} .
\end{equation}
The gluon loop due to the $\sq\sq gg$ interaction gives no
contribution: It is proportional to $A_0(\l^2)$, where $\l$ is the IR
gluon mass, and vanishes for $\l\to 0$.
The contribution due to the gluino--quark loop is
\begin{equation}
  \begin{split}
  \Sigma_{ii}^{(\sg)}(\msq{i}^2) = 
  & -\frac{4}{3}\frac{\a_s}{\pi} \:\{ 
         A_0(m_q^2) + \msq{i}^2 B_1(\msq{i}^2,\msg^2,m_q^2) \\
  & \hspace{20mm} +\; [ \,
    \msg^2 + (-)^i\,\msg m_q \sin 2\tsq\,]\,B_0(\msq{i}^2,\msg^2,m_q^2) 
    \} , 
  \end{split}
\end{equation}
\begin{equation}
  \Sigma_{12}^{(\sg)}(\msq{i}^2) = \Sigma_{21}^{(\tilde g)}(\msq{i}^2) =
  \frac{4}{3}\frac{\alpha_{s}}{\pi} \msg m_q \cos 2 \tsq 
  B_0(\msq{i}^2, \msg^2,m_q^2) ,
\end{equation}
and the squark bubble leads to $(i \not= i')$
\begin{equation}
  \Sigma_{ii}^{(\sq)}(\msq{i}^2) = \frac{\a_s}{3\pi} \:
   \{\cos^2 2\tsq\,A_0(\msq{i}^2) + \sin^2 2\tsq\,A_0(\msq{i'}^2)\} ,
\end{equation}
\begin{equation}
\Sigma_{12}\hsq (\msq{i}^2) = 
  \frac{\alpha_s}{6\pi}\sin 4\tsq
  \{ A_{0}(m_{{\sq}_2}^2) - A_{0}(m_{{\sq}_1}^2) \} 
  = \Sigma_{21}\hsq (\msq{i}^2).
\end{equation}
Note, that $\Sigma_{ii'}\hsq (\msq{1}^2) = \Sigma_{ii'}\hsq (\msq{2}^2)$.

\noi
$\d\msq{i}^2$, 
the shift from the bare to the on--shell squark mass, is given by: 
\begin{equation}
  \d\msq{i}^2 = \Re \left[\,
  \Sigma_{ii}^{(g)} (\msq{i}^2) + \Sigma_{ii}^{(\sg)} (\msq{i}^2) + 
  \Sigma_{ii}^{(\sq)} (\msq{i}^2)
  \,\right] .
\end{equation}

\noi
The squark wave--function renormalization constants 
$\ti Z_{ni}(\sq_i^{})$  are:
\begin{equation}
  \delta \ti Z_{ii}^{(g,\sg)} = 
    -\Re \left\{\dot\Sigma_{ii}^{(g,\sg)}(\msq{i}^{2}) \right\}\,, 
  \qquad
  \delta \ti Z_{i'i}^{(\sg,\sq)} =
    \frac{\Re \left\{\Sigma_{i'i}^{(\sg,\sq)}(\msq{i}^{2})\right\} 
          }{\msq{i'}^2 - \msq{i}^2} \,, 
  \quad i \neq i'
\label{eq:Zsq} 
\end{equation}
with $\dot\Sigma_{ii} (m^2) = 
\partial\Sigma_{ii} (p^2)/\partial p^2 |_{p^2=m^2}$:
\begin{equation}
  \dot\Sigma_{ii}^{(g)} (\msq{i}^2) = - \frac{2\alpha_s}{3\pi} 
  \left[
     B_{0}(\msq{i}^{2}, 0, \msq{n}^{2}) 
     + 2\msq{i}^{2} \dot B_{0} (\msq{i}^{2}, \l^{2}, \msq{i}^{2})
  \right] ,
\end{equation}
\begin{align}  
  \dot\Sigma_{ii}^{(\sg)} (\msq{i}^2) = & \frac{2\a_{s}}{3\pi} 
  \left[ B_0(\msq{i}^2, \msg^2,m_q^2) + (\msq{i}^2-m_q^2-\msg^2) 
         \dot B_0 (\msq{i}^2, \msg^2, m_q^2) \right.\nn\\
  & \left.\hspace*{1cm} 
    -2 m_q\msg (-1)^i\sin 2\tsq \dot B_0 (\msq{i}^2, \msg^2,m_q^2) 
  \right] ,
\end{align}
The four--squark intraction does not contribute to $\d\ti Z_{ii}$ 
because $\dot\Sigma_{ii}^{(\sq)}=0$.
On the other hand, the off--diagonal squark wave--function
renormalization constant $\d\ti Z_{ii'}$ gets no contribution
from gluon exchange diagrams because they do not mix $\sq_1^{}$ 
and $\sq_2^{}$.
Bare and renormalized squark fields ($\sq_i^0$ and $\sq_i^{}$) 
are related by
\begin{align}
   \sq_i^0 &= (1+\onehf\d\ti Z_{ii})\,\sq_i^{} 
              + \d\ti Z_{ii'}\,\sq_{i'}^{} \, ,\\[1mm]
   \sq_i^{0*} &= (1+\onehf\d\ti Z_{ii})\,\sq_i^* 
              + \d\ti Z_{i'i}\,\sq_{i'}^* \, .
\end{align}

\subsection{Quark Self Energy}

\begin{figure}[h!]
\center
\begin{picture}(92,23)
\put(6,6.5){\mbox{\psfig{file=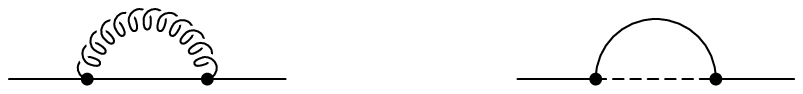}}}
\put(2,6.5){\mbox{$q$}}
\put(36,6.5){\mbox{$q$}}
\put(55,6.5){\mbox{$q$}}
\put(88,6.5){\mbox{$q$}}
\put(19,3){\mbox{$q$}}
\put(19,17){\mbox{$g$}}
\put(70.5,2){\mbox{$\sq_i$}}
\put(71,16){\mbox{$\sg$}}
\end{picture} 
\caption{Squark self energy diagrams to $\Oas$.}
\label{fig:fd-qself}
\end{figure}

The self energy of a quark gets contributions from the 
gluon and gluino loops shown in \fig{fd-qself}.

\noi
$\d m_{q}$, the shift from the bare to the pole mass of the quark 
$q$, thus has two contributions: 
The gluon exchange contribution is
\begin{equation}
  \d m_q^{(g)} = - \frac{2}{3}\frac{\a_s}{\pi}\; 
    m_q \left[\, B_0(m_q^2, 0, m_q^2) - B_1(m_q^2, 0, m_q^2) - r/2\, 
        \right] ,
\label{eq:dmqg}
\end{equation}
and the gluino contribution is
\begin{eqnarray}
  \d m_q^{(\ti g)} &=& - \frac{\a_s}{3\pi}\, \left\{\, 
  m_q\,\big[\, B_1(m_q^2, \msg^2, \msq{1}^2) 
         + B_1(m_q^2, \msg^2, \msq{2}^2) \,\big] \right. \nn \\
  & & \hspace{13mm} \left.
  +\,\msg \sin 2\tsq\,\big[\,B_0(m_q^2, \msg^2, \msq{1}^2) 
                - B_0(m_q^2, \msg^2, \msq{2}^2) \,\big]\, \right\}\, .
\end{eqnarray}
For the quark wave--function renormalization constants
due to gluon exchange one gets:
\begin{equation}
  \d Z^{L\,(g)}_{q} = \d Z^{R\,(g)}_{q} =
  - \frac{2}{3}\frac{\alpha_{s}}{\pi} 
    \left[\, B_{0} + B_{1} 
           - 2m_{q'}^{2} (\dot B_{0} - \dot B_{1}) - r/2 \,\right] 
\label{eq:qwave}
\end{equation}            
with $B_{m} = B_{m}(m_{q}^{2}, \l^{2}, m_{q}^{2})$,
$\dot B_{m} = \dot B_{m}(m_{q}^{2}, \l^{2}, m_{q}^{2})$;
those due to gluino exchange are: 
\begin{equation} 
  \d Z^{L\,(\sg)}_{q} = \frac{2}{3}\frac{\alpha_{s}}{\pi}\,
  \Big\{ \cos^{2}\t_{\ti q}B_{1}^{1} + \sin^{2}\t_{\ti q}B_{1}^{2}  
         + m_{q}^{2}\big[ \dot B_{1}^{1} + \dot B_{1}^{2} 
            + \frac{\msg}{m_{q}} \sin 2\t_{\ti q'}
              (\dot B_{0}^{1} - \dot B_{0}^{2}) \big] \Big\},  
   \hspace{-4cm}
\end{equation}
\begin{equation}           
  \d Z^{R\,(\sg)}_{q} = \frac{2}{3}\frac{\alpha_{s}}{\pi}\,
  \Big\{ \sin^{2}\t_{\ti q}B_{1}^{1} + \cos^{2}\t_{\ti q}B_{1}^{2} 
         + m_{q'}^{2}\big[ \dot B_{1}^{1} + \dot B_{1}^{2} 
            + \frac{\msg}{m_{q}} \sin 2\t_{\ti q}
              (\dot B_{0}^{1} - \dot B_{0}^{2}) \big] \Big\}, 
   \hspace{-4cm} 
\end{equation}  
where $B_m^i = B_m^{}(m_q^2,\,\msg^{2},\,\msq{i}^2)$ and 
$\dot B_m^i = \dot B_m^{}(m_q^2,\,\msg^{2},\,\msq{i}^2)$.  
The relation between the bare quark field $q^0$ and the renormalized
one $q$ is
\begin{gather}
  q^0 = (1 + \onehf\d Z_q^L P_L^{} + \onehf\d Z_q^R P_R^{})\,q\,,\\[1mm]
 \bar q^0 = \bar q\,(1 + \onehf\d Z_q^L P_R^{} + \onehf\d Z_q^R P_L^{})\,. 
\end{gather}

\noi
The parameter $r$ in \eq{dmqg} and \eq{qwave} exhibits the dependence 
on the regularization scheme: $r=0$ in DRED while $r=1$ in DREG
(remember that only DRED preserves SUSY!). 
Here we note that in our calculations $r$ does not cancel in the 
f\/inal results. Therefore, there are also numerical differences between the
two regularization schemes.

\section{Subtleties}\label{sect:subtleties}

All in all we have to consider ${\cal O}(10)$ loop diagrams per
process. This may not sound much; however, there are some subtleties 
which have to be taken into account:
 
\subsection{Renormalization of the Squark Mixing Angle}
\label{sect:tsq}

The stop and sbottom couplings depend on the respective mixing
angles which therefore must be renormalized. 
The first proper renormalization prescription for the squark mixing 
angle was given in \cite{ebm}. 
There the counterterm $\d\tsq$ was fixed in the process 
$e^+e^-\to \sq_i^{}\sqbar_j$ such that it cancels the 
off--diagonal part of the squark wave--function corrections 
(diagrams h and j of \fig{fd-eesq} with $k\not= i$). 
The idea is the following:
The variation of the squark--squark--$Z$ coupling $c_{ij}$ gives
\begin{equation}
  \d c_{ij}^{} = \left( \begin{matrix} 
     2\,c_{12}^{} & c_{22}^{}-c_{11} \\
     c_{22}^{}-c_{11} & -2\,c_{12}^{} \end{matrix}\right) \d\tsq
\label{eq:dcij}
\end{equation}
The requirement that the SUSY--QCD corrections to 
$e^+e^-\to\sq_i^{}\sqbar_j$ be UV finite leads to:
\begin{gather}
  \d\,c_{11} = 2\,c_{12}\,\d\tsq = -2\,\d\ti Z_{21}\,c_{12}, 
      \label{eq:dc11}\\
  \d\,c_{12} = (c_{22}-c_{11})\,\d\tsq 
             =  -\d\ti Z_{21}\,c_{22} - \d\ti Z_{12}\,c_{11}, 
      \label{eq:dc12}\\ 
  \d\,c_{22} = -2\,c_{12}\,\d\tsq = -2\,\d\ti Z_{12}\,c_{12},
      \label{eq:dc22}
\end{gather}
where the $\d\ti Z_{ij}$ are the squark wave--funktion renormalization
constants, see Eq.~\eq{Zsq}.
Using Eq.~\eq{dc12}, \ie the process $e^+e^-\to\sq_1^{}\sq_2^{}$, 
to fix the squark mixing angle we get   
(here and in the folowing $\Sigma_{12}\equiv\Re(\Sigma_{12})$):
\begin{equation}
  \d\tsq = 
  \frac{\Sigma_{12}(\msq{1}^2)\,c_{22}^{}-\Sigma_{12}(\msq{2}^2)\,c_{11}^{}
      }{(c_{22}^{}-c_{11}^{}) (\msq{1}^2-\msq{2}^2)}\,.
\end{equation} 
$\d\tsq$ thus gets contributions from gluon and gluino exchanges, 
$\d\tsq = \d\tsq\hsq + \d\tsq\hsg$. Explicetly: 
\begin{gather}  
  \d\tsq\hsq = \frac{\a_s}{6\pi} \,
    \frac{\sin 4\tsq}{\msq{1}^2 - \msq{2}^2}
    \left[ A_{0}(\msq{2}^2) - A_{0}(\msq{1}^2) \,\right] , \\[1mm]
  \d\tsq\hsg = \frac{4}{3}\frac{\a_s}{\pi}\,
    \frac{m_{\sg} m_q}{I_{3L}^q (m_{\sq_1}^2 - m_{\sq_2}^2)}
    \left[ B_{0}(m_{\sq_2}^2,m_{\sg}^2,m_q^2)\,c_{11} -
           B_{0}(m_{\sq_1}^2,m_{\sg}^2,m_q^2)\,c_{22}\,\right] .
\end{gather}
This scheme was also applied for SUSY--QCD corrections to 
squark decays in \cite{qcdnc-paper,qcdwz-paper,qcdhx-paper} 
as well as to Higgs decays into squarks in \cite{higgsdec}.
We will use this scheme in what follows.

There are also other possibilities of defining the on--shell squark 
mixing angle. 
In \cite{qcdnc-djouadi}, for instance, $\d\tsq$ was fixed such that 
the renormalized self energy of the squarks remains diagonal on the 
$\sq_2^{}$ mass shell (\ie using Eq.~\eq{dc22}). This leads to the condition
\begin{equation}
  \d\tsq\,(\cite{qcdnc-djouadi}) = 
  \frac{\Sigma_{12}(\msq{2}^2)}{\msq{1}^2-\msq{2}^2} \,. 
\end{equation} 
An analogous condition on the $\sq_1^{}$ mass shell was applied   
in \cite{qcdhx-djouadi}. 
In \cite{qcd-zerwas2} a scale--dependend definition was used: 
\begin{equation}
  \d\tsq(Q^2)\,(\cite{qcd-zerwas2}) = 
  \frac{\Sigma_{12}(Q^2)}{\msq{1}^2-\msq{2}^2} \,. 
\end{equation} 
A process independent ``democratic'' approach is to take 
the arithetic mean of Eqs.~\eq{dc11} and \eq{dc22}, 
as done in \cite{sola}. $\d\tsq$ is then given by 
\begin{equation}
  \d\tsq\,(\cite{sola})= 
  \frac{\Sigma_{12}(\msq{1}^2) + \Sigma_{12}(\msq{2}^2)
       }{2(\msq{1}^2 - \msq{2}^2)} \,.
\end{equation}
This is, by the way, the only definition that works for processes 
with charginos in the loops (Yukawa correctins \etc).

\begin{figure}[h]
\bce\begin{picture}(100,65)
\put(-2,5){\mbox{\epsfig{file=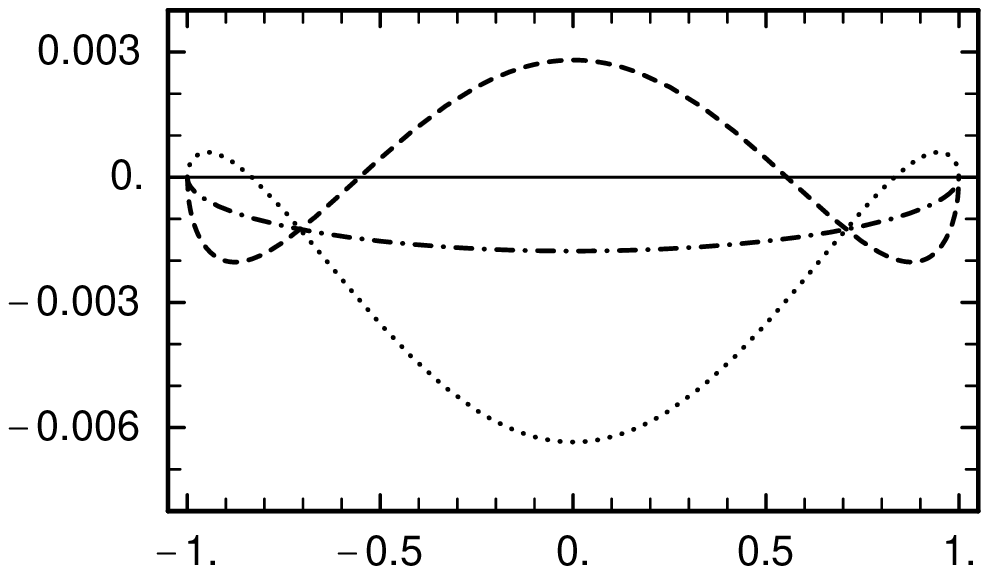,height=60mm}}}
\put(51,0){\mbox{$\cos\t_{\st}$}}
\put(-4,19){\makebox(0,0)[br]{{\rotatebox{90}{$\d\cth_{\st}[i]-\d\cth_{\st}$\cite{ebm} }}}}
\put(71,55){\mbox{\cite{qcdhx-djouadi}}}
\put(74,25){\mbox{\cite{qcdnc-djouadi}}}
\put(53,34){\mbox{\cite{sola}}}
\end{picture}\ece
\caption{Differences in $\d\cth_{\st}$ between 
   \cite{qcdnc-djouadi,qcdhx-djouadi,sola} 
   and our scheme \cite{ebm} as a function of $\cos\t_{\st}$, 
   for $\mst{1}=250\gev$, $\mst{2}=600\gev$, and $\msg=600\gev$.}
\label{fig:deltatheta}
\end{figure}

\noi
The differences between the various schemes are ultraviolet finite.
In Fig.~\ref {fig:deltatheta} we compare $\delta\t_{\st}$ of 
\cite{qcdnc-djouadi,qcdhx-djouadi,sola} to that of our scheme \cite{ebm}. 
Considering squark decays,  
 $\d\tsq(Q^2)\,(\cite{qcd-zerwas2})\equiv\d\tsq\,(\cite{qcdnc-djouadi})$ or 
 $\d\tsq(Q^2)\,(\cite{qcd-zerwas2})\equiv\d\tsq\,(\cite{qcdhx-djouadi})$.  
The numerical differences between the various 
schemes are very small, typically well below $1\%$. 

\subsection{Renormalization of \boldmath $M_{\ti Q}$}
\label{sect:MQproblem}

At tree--level and in the 
$\overline{{\rm DR}}$ \footnote{$\overline{{\rm DR}}$ = Minimal 
substraction with DRED.} renormalization scheme $SU(2)_L$ symmetry 
requires that the parameter $M_{\ti Q}$ in 
the $\st$ and $\sb$ mass matrices have the same value. 
This is, however, not the case at loop--level in the on--shell scheme 
due to different shifts $\d M_{\ti Q}^2$ in the $\st$ and in the 
$\sb$ sectors \cite{higgsdec,yamada}. 
Therefore, for a combined treatement of stops and sbottoms, 
an appropriate renormalization procedure for $M_{\ti Q}^2$ is
necessary. 

\noi
Taking on--shell squark masses and mixing angles as input parameters, 
we define the on--shell $M_{\ti Q}^2$ as
\begin{equation}
  M_{\ti Q}^2(\sq) = \msq{1}^2 \cos^2\t_{\sq} + 
  \msq{2}^2 \sin^2\t_{\sq} 
  - m_Z^2 \cos 2\b\, (I_{3L}^{q} - e_{q} \sW) - m_q^2\, ,  
\end{equation}
see Eqs.~\eq{MQst} and \eq{MQsb}. 
$M_{\ti Q}^2(\st)$ and $M_{\ti Q}^2(\st)$ are then related by 
\begin{equation}
  M_{\ti Q}^2(\sb) = 
  M_{\ti Q}^2(\st) + \d M_{\ti Q}^2(\st) - \d M_{\ti Q}^2(\sb) 
\label{eq:mQshift}
\end{equation}
with
\begin{equation}
  \d M_{\ti Q}^2(\sq) = 
  \d\msq{1}^2 \cos^2\tsq + \d\msq{2}^2 \sin^2\tsq - 
  (\msq{1}^2-\msq{2}^2) \sin 2\tsq\,\d\tsq - 2 m_q \d m_q .
\label{eq:dmQshift}
\end{equation}
The underlying $SU(2)_L$ symmetry is reflected by the fact that the 
shift $\d M_{\ti Q}^2(\st) - \d M_{\ti Q}^2(\sb)$ is finite.

\subsection{Renormalization of \boldmath $m_q A_q$}

The squark couplings to Higgs bosons involve the parameter $A_q$.
Therefore, also this parameter has to be renormalized at $\Oas$.
We proceed like for the definition of the on--shell $M_{\ti Q}^2$
and write the expression $m_q A_q$ in terms of on--shell squark masses
$\msq{i}$ and mixing angles $\tsq$, see Eqs. \eq{aq} and 
\eq{aqinv}. 
We thus get \cite{higgsdec}
\begin{equation} 
  \d (m_q A_q) = 
  \onehf\,(\d \msq{1}^2 - \d \msq{2}^2) \sin 2\tsq 
  + (\msq{1}^2 - \msq{2}^2) \cos 2\tsq\,\d\tsq
  + \mu\, \{\cot\b,\tan \b\}\,\d m_q  \hspace{-4cm}
\label{eq:dmqaq}
\end{equation}
where $\cot\b\;(\tan\b)$ has to be taken for $\sq = \st\;(\sb)$. 

\noi
The squark masses and couplings also depend on the parameters $\mu$ and
$\tan\b$. However, the on--shell $\mu$ and $\tan\b$ are defined via  
electroweak processes. 
Hence $\mu$ and $\tan\b$ are not renormalized at $\Oas$.

\clearpage 
\section {Decays into Charginos and Neutralinos} \label{sect:qcdnc}
 
\begin{figure}[p] 
\center 
\begin{picture}(113,172) 
\put(0,15){\mbox{\psfig{file=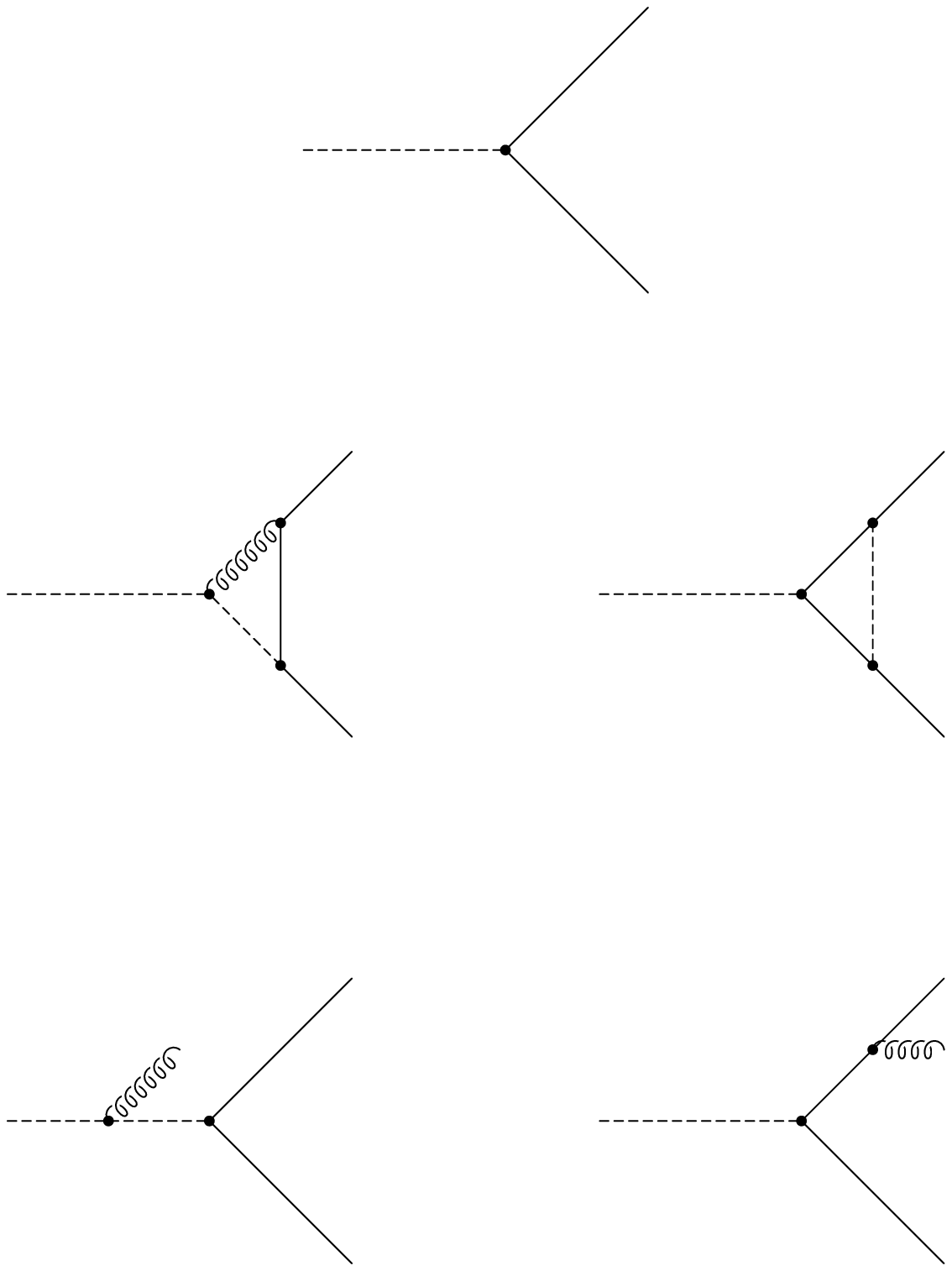}}} 
\put(59,128){\makebox(0,0)[t]{\bf (a)}} 
\put(34.5,149.5){\makebox(0,0)[r]{$\sq_i$}} 
\put(80,167.5){\makebox(0,0)[l]{$q'\,(q)$}} 
\put(80,133){\makebox(0,0)[l]{$\chpm_{j}\,(\nt_{k})$}} 
\put(47,148){\vector(1,0){4.3}} 
\put(49,146.5){\makebox(0,0)[t]{$k_1$}} 
\put(65.5,158){\vector(1,1){3.6}} 
\put(65,161){\makebox(0,0)[b]{$k_2$}} 
\put(66,141){\vector(1,-1){3.6}}
\put(65,138.5){\makebox(0,0)[t]{$k_3$}}  
\put(22,74){\makebox(0,0)[t]{\bf (b)}} 
\put(-1,96.5){\makebox(0,0)[r]{$\sq_i$}} 
\put(44,114){\makebox(0,0)[l]{$q'\,(q)$}} 
\put(44,79){\makebox(0,0)[l]{$\chpm_{j}\,(\nt_{k})$}} 
\put(27,104){\makebox(0,0)[br]{$g$}} 
\put(27,90){\makebox(0,0)[tr]{$\sq_i$}} 
\put(36,96){\makebox(0,0)[l]{$q'\,(q)$}} 
\put(94,74){\makebox(0,0)[t]{\bf (c)}} 
\put(70,96.5){\makebox(0,0)[r]{$\sq_i$}} 
\put(115,114){\makebox(0,0)[l]{$q'\,(q)$}} 
\put(115,79){\makebox(0,0)[l]{$\chpm_{j}\,(\nt_{k})$}} 
\put(100,103){\makebox(0,0)[br]{$\sg$}} 
\put(99,90){\makebox(0,0)[tr]{$q$}} 
\put(107,96){\makebox(0,0)[l]{$\sq_n'\,(\sq_n)$}}  
\put(22,11){\makebox(0,0)[t]{\bf (d)}} 
\put(-1,33){\makebox(0,0)[r]{$\sq_i$}} 
\put(44,50){\makebox(0,0)[l]{$q'\,(q)$}} 
\put(44,15){\makebox(0,0)[l]{$\chpm_{j}\,(\nt_{k})$}} 
\put(23,42){\makebox(0,0)[l]{$g$}} 
\put(94,11){\makebox(0,0)[t]{\bf (e)}} 
\put(70,33){\makebox(0,0)[r]{$\sq_i$}} 
\put(115,50){\makebox(0,0)[l]{$q'\,(q)$}} 
\put(115,15){\makebox(0,0)[l]{$\chpm_{j}\,(\nt_{k})$}} 
\put(115,41){\makebox(0,0)[l]{$g$}} 
\end{picture} 
\caption{Feynman diagrams for the ${\cal O}(\a_{s})$ SUSY-QCD  
  corrections to squark decays into charginos and neutralinos: 
  (a) tree level, (b) and (c) vertex corrections, and  
  (d) and (e) real gluon emission. For wave--function corrections 
  see Figs.~\ref{fig:fd-sqself} and \ref{fig:fd-qself}.} 
\label{fig:FD-qcdnc} 
\end{figure}

The tree--level amplitude for the decay $\sq_i^{}\to q'\chpm_j$ 
is (see Eq.~\eq{qsqch} and \fig{FD-qcdnc}\,a): 
\begin{equation} 
  {\cal M}^0(\sq_i^{}\to q'\chpm_j) =  
  ig\,\bar u(k_2)\left[\kij^{\sq}\PL + \lij^{\sq}\PR \right] v(k_3). 
\end{equation} 
 
\noi
The decay width at tree level is thus given by 
\begin{equation}
  \G^0 (\sq_i^{}\to q'\chpm_j) = 
  \frac{g^2 \kappa(\msq{i}^2,m_{q'}^2,\mch{j}^2)}{16\pi\msq{i}^3}\:
  \left( [(\lij^{\sq})^2 + (\kij^{\sq})^2]\,X 
         - 4 \lij^{\sq}\kij^{\sq} m_{q'}\mch{j} \right)
\end{equation}
with $X=\msq{i}^2-m_{q'}^2-\mch{j}^2$. 
The $\Oas$ loop corrected decay amplitude is obtained by the shifts  
\begin{gather} 
  \lij^{\st} \:\to \lij^{\st} + \d\lij^{\st\,(v)}  
               + \d\lij^{\st\,(w)} + \d\lij^{\st\,(c)}, \\[1mm] 
  \kij^{\st} \to \kij^{\st} + \d\kij^{\st\,(v)}  
               + \d\kij^{\st\,(w)} + \d\kij^{\st\,(c)}, 
\end{gather} 
where the superscript $v$ denotes vertex corrections, 
$w$ wave function corrections, and 
$c$ the shift from bare to on--shell couplings. 
The virtual corrections to the decay width, $\d\G^{(a)}$ ($a = v,\,w,\,c$), 
can thus be written as 
\begin{eqnarray} 
\d\G^{(a)} (\sq_i^{}\to q^\prime\chpm_j) &=& 
  \frac{g^2\,\kappa(\msq{i}^{2}, m_{q^{\prime}}^{2}, \mch{j}^{2})} 
       {16\pi\msq{i}^{3}} \\ 
 & &{\scriptsize\times} 
  \left[ (2\,\lij^{\sq}\,\d\lij^{\sq\,(a)}  
        + 2\,\kij^{\sq}\,\d\kij^{\sq\,(a)})\,X           
    - 4\,m_{q'}\mch{j} (\lij^{\sq}\,\d\kij^{\sq\,(a)}  
                      + \kij^{\sq}\,\d\lij^{\sq\,(a)})  
  \right] . \nn 
\end{eqnarray} 

\noi
Analogously, one gets for squark decays into neutralinos 
\begin{eqnarray} 
  \G^0 (\sq_i^{}\to q\nt_k) &=& 
  \frac{g^2 \kappa(\msq{i}^2,m_q^2,\mnt{k}^2)}{16\pi\msq{i}^3}\:
  \left( [(\aik^{\sq})^2 + (\bik^{\sq})^2]\,\hat X 
         - 4 \aik^{\sq}\bik^{\sq} m_q\mnt{k} \right) \,, 
  \\[3mm]
  \d\G^{(a)} (\sq_i^{}\to q\nt_k) &=& 
  \frac{g^2\,\kappa(\msq{i}^2, m_q^2, \mnt{k}^2)} 
       {16\pi\msq{i}^{3}} \\ 
  & &{\scriptsize\times} \left[  
    (2\,\aik^{\sq}\,\d\aik^{\sq\,(a)}  
      + 2\,\bik^{\sq}\,\d\bik^{\sq\,(a)})\,\hat X     
    - 4\,m_q^{}\mnt{k} (\aik^{\sq}\,\d\bik^{\sq\,(a)}  
                      + \bik^{\sq}\,\d\aik^{\sq\,(a)}) \right] \nn 
\end{eqnarray} 
where $\hat X=\msq{i}^2-m_q^2-\mnt{k}^2$.  
In the following, we give our results for squark decays into 
charginos. The analogous expressions for decays into neutralinos  
are obtained by the replacements 
$\chpm_{j} \to \nt_{k}$, $q' \to q$, $X\to \hat X$,  
$\ell_{ij}^{\sq} \to a_{ik}^{\sq}$, $k_{ij}^{\sq} \to b_{ik}^{\sq}$,  
$\d\ell_{ij}^{\sq\,(a)} \to \d a_{ik}^{\sq\,(a)}$, and  
$\d k_{ij}^{\sq\,(a)} \to \d b_{ik}^{\sq\,(a)}$. 
The shifts $\d\lij^{\sq\,(a)}$, $\d k_{ij}^{\sq\,(a)}$,  \etc  
get contributions from gluon exchange,  
gluino exchange, and the four--squark interaction.  
As we will see, in the renormalization scheme used   
the contribution due to the four--squark interaction cancels.

\subsection{Vertex Corrections} 
 
The gluonic vertex correction (\fig{FD-qcdnc}\,b) yields 
\begin{align} 
  \d\ell_{ij}^{\sq\,(v,g)}\,=\; 
  &\frac{\alpha_{s}}{3\pi}\, \Big\{ 
   \big[(4m_{q'}^{2} + 2X) (C_{0}+C_{1}+C_{2}) + 
        (2\mch{j}^{2} + X) C_{1} + B_{0} \big]\,\ell_{ij}^{\sq} 
   \nn  \\ 
  &\hspace{13mm} + \big[ 2m_{q'}\mch{j}C_{2} \big]\,k_{ij}^{\sq} 
   \Big\},  
  \\[2mm] 
  \d k_{ij}^{\sq\,(v,g)}\,=\; 
  &\frac{\alpha_{s}}{3\pi}\, \Big\{ 
   \big[(4m_{q'}^{2} + 2X) (C_{0}+C_{1}+C_{2}) + 
        (2\mch{j}^{2} + X) C_{1} + B_{0} \big]\,k_{ij}^{\sq}  
   \nn \\ 
  &\hspace{13mm} + \big[ 2m_{q'}\mch{j}C_{2} \big]\,\ell_{ij}^{\sq} 
   \Big\}.    
\end{align} 
$B_{0}$, $C_{0}$, $C_{1}$, and $C_{2}$ are the standard  
two-- and three--point functions \cite{pave}.  
In this case, $B_{0} = B_{0}(\mch{j}^{2}, \msq{i}^{2}, m_{q'}^{2})$  
and $C_{m} = C_{m}(\msq{i}^{2}, \mch{j}^{2}, m_{q'}^{2}; 
\l^{2}, \msq{i}^{2}, m_{q'}^{2})$, 
where we follow the conventions of \cite{denner}. 
As usually, we introduce a gluon mass $\l$ for the regularization of  
the infrared divergence.  
The contribution to due to the graph of  
\fig{FD-qcdnc}\,c with a gluino and a squark $\ti q'_{n}$ $(n = 1, 2)$  
in the loop is: 
\begin{align} 
  \d{\ell_{ij}^{\sq}}^{(v,\sg)}\,&=\: 
  \frac{2}{3}\frac{\alpha_{s}}{\pi} \Big\{ 
  \mch{j}\big[ (m_{q'}\a_{LR}+m_{q}\a_{RL}-\msg\a_{LL})\,\ell_{nj}^{\sq'}  
            + \mch{j}\a_{RL} k_{nj}^{\sq'} \big] C_{1}   
  \nn \\ 
  &+m_{q'}\big[ (m_{q'}\a_{RL}-m_{q}\a_{LR}+\msg\a_{RR}) k_{nj}^{\sq'}  
            - \mch{j}\a_{LR}\,\ell_{nj}^{\sq'} \big] (C_{1}+C_{2})  
   \\  
  &+\msg\big[ (m_{q'}\a_{RR}-m_{q}\a_{LL}+\msg\a_{RL}) k_{nj}^{\sq'}  
            - \mch{j}\a_{LL}\,\ell_{nj}^{\sq'} \big] C_{0}  
    + (X C_{1} + B_{0})\,\a_{RL} k_{nj}^{\sq'}  
    \Big\},  \nn 
  \\[2mm] 
  \d{k_{ij}^{\sq}}^{(v,\sg)}\,&=\; 
  \frac{2}{3}\frac{\alpha_{s}}{\pi} \Big\{ 
  \mch{j}\big[ (m_{q'}\a_{RL}+m_{q}\a_{LR}-\msg\a_{RR}) k_{nj}^{\sq'}  
            + \mch{j}\a_{LR}\,\ell_{nj}^{\sq'} \big] C_{1}  
  \nn \\ 
  &+ m_{q'}\big[ (m_{q'}\a_{LR}-m_{q}\a_{RL}+\msg\a_{LL})\,\ell_{nj}^{\sq'}  
            - \mch{j}\a_{RL} k_{nj}^{\sq'} \big] (C_{1}+C_{2})   
  \hspace{-5cm}\\              
  &+ \msg\big[ (m_{q'}\a_{LL}-m_{q}\a_{RR}+\msg\a_{LR})\,\ell_{nj}^{\sq'}  
            - \mch{j}\a_{RR} k_{nj}^{\sq'} \big] C_{0}  
     + (X C_{1} + B_{0})\,\a_{LR}\,\ell_{nj}^{\sq'} 
   \Big\} \nn       
\end{align} 
with  
\begin{equation} \begin{array}{ll} 
  \a_{LL} = \left( \a_{LL}\right)_{in} =  
    {\cal R}^{\sq}_{i1}\,{\cal R}^{\sq '}_{n1}, & \quad 
  \a_{LR} = \left( \a_{LR}\right)_{in} =  
    {\cal R}^{\sq}_{i1}\,{\cal R}^{\sq '}_{n2}, \\    
  \a_{RL} = \left( \a_{RL}\right)_{in} =  
    {\cal R}^{\sq}_{i2}\,{\cal R}^{\sq '}_{n1}, & \quad  
  \a_{RR} = \left( \a_{RR}\right)_{in} =  
    {\cal R}^{\sq}_{i2}\,{\cal R}^{\sq '}_{n2}.    
\end{array} \end{equation} 
 
\noi Here,  
$B_{0} = B_{0}(\mch{j}^{2}, m_{\sq'_{n}}^{2}, m_{q}^{2})$, and 
$C_{m} = C_{m}(\msq{i}^{2}, \mch{j}^{2}, m_{q'}^{2}; 
\msg^{2}, m_{q}^{2}, m_{\sq'_{n}}^{2})$. \\

\subsection{Wave--Function Correction} 
 
The wave--function correction is given by $(i\not= i')$ 
\begin{align} 
  \d\lij^{\sq\,(w)} &= \onehf\, \big[  
    \d Z^{L^{\dagger}}_{q'} + \d\ti Z_{ii}(\sq_i^{}) \big]\,\lij^{\sq}  
    + \d\ti Z_{i'i}(\sq_i^{})\,\ell_{i'j}^{\sq},  
  \\[1mm] 
  \d\kij^{\sq\,(w)} &= \onehf\, \big[  
    \d Z^{R^{\dagger}}_{q'} + \d\ti Z_{ii}(\sq_i^{}) \big]\,\kij^{\sq}  
    + \d\ti Z_{i'i}(\sq_i^{})\,k_{i'j}^{\sq}.                        
\end{align} 
$Z_{q'}^{L,R}$ are the quark wave--function renormalization constants  
due to gluon and gluino exchange (\fig{fd-sqself}). 
The squark wave--function renormalization constants $\ti Z_{ni}(\sq_i^{})$  
stem from gluon, gluino, and squark loops (\fig{fd-qself}).  
See Sect.~\ref{sect:selfenergies} for the explizit expressions.

\subsection{Renormalization of the Bare Couplings} 
 
In order to make the shift from the bare to the on--shell couplings  
it is necessary to renormalize the quark mass as well as  
the squark mixing angle: 
\begin{alignat}{2}  
  \d\lij^{\sq\,(c)} & =  
    {\cal O}_{jn}^{q}\,\d\Rsq_{in} 
    + {\cal R}^{\sq}_{i2}\,\d{\cal O}^{q}_{j2}, 
  & \qquad 
  \d\kij^{\sq\,(c)} & =  
    {\cal O}_{j2}^{q'}\,\d\Rsq_{i1}  
    + {\cal R}^{\sq}_{i1}\,\d{\cal O}^{q'}_{j2}, \\[1mm] 
  \d a_{ik}^{\sq\,(c)} & =  
    {\cal A}_{kn}^{q}\,\d\Rsq_{in} + 
    {\cal R}^{\sq}_{i2}\,\d h_{Rk}^{q},  
  & \qquad 
  \d b_{ik}^{\sq\,(c)} & =  
    {\cal B}_{kn}^{q}\,\d\Rsq_{in} +  
    {\cal R}^{\sq}_{i1}\,\d h_{Lk}^{q}, 
\end{alignat} 
with 
\begin{gather} 
  \d{\cal R}^{\sq} = 
  \left(\baa{lr} -\sth_{\sq} &  \cth_{\sq} \\  
                 -\cth_{\sq} & -\sth_{\sq} \eaa\right) \d\tsq , 
  \\[2mm] 
  \d{\cal O}^{t}_{j2} = \frac{V_{j2}}{\sqrt{2}\,m_{W}\sin\b}\:\d m_{t}, 
  \qquad 
  \d{\cal O}^{b}_{j2} = \frac{U_{j2}}{\sqrt{2}\,m_{W}\cos\b}\:\d m_{b},   
\end{gather} 
and analogously for $\d h_{Lk}^{q}$ and $\d h_{Rk}^{q}$  
according to Eqs.~\eq{fLk} -- \eq{hLk}.  
 
\noi 
For the renormalization of the squark mixing angle we use  
the scheme of \cite{ebm} as explained in Sect.~\ref{sect:tsq}. 
With this choice of $\d\tsq$ the squark contribution 
to the correction is zero: $\d\G^{(w,\sq)} + \d\G^{(c,\sq)} = 0$. 
Moreover, the off--diagonal contribution $(i \neq j)$ of  
the gluino--quark loop in \fig{fd-qself} vanishes in this scheme.  
 
\subsection{Real Gluon Emission} 
 
The total virtual correction $\d\G_{\!virt}=\G^{(v)}+\G^{(w)}+\G^{(w)}$  
is UV finite. In order to cancel 
also the infrared divergence due to $\l\to 0$ we include the 
emission of real (hard and soft) gluons, see \fig{FD-qcdnc}\,d,\,e. 
$\d\G_{\!real}(\sq_i^{}\to q'\ch_j)\equiv\G(\sq_i^{}\to g q' \ch_j)$ is: 
\begin{align} 
  \d\G_{\!real}(\sq_i^{}\to q'\ch_j) = 
  &-\frac{g^{2}\a_{s}}{6\pi^{2}\msq{i}}\, \Big\{ 
   \big[ (\kij^{\sq})^{2} + (\lij^{\sq})^{2} \big] (I_{1}^{0}+I) \\ 
  &\hspace*{20mm} 
   + 2 Z \big[ \msq{i}^{2} I_{00} + m_{q'}^{2} I_{11} +  
   (\msq{i}^{2}+m_{q'}^{2}-\mch{j}^{2}) I_{01} + I_{0} + I_{1} \big]  
   \Big\} \nn 
\end{align} 
where $Z = \big[ (\kij^{\sq})^{2} + (\lij^{\sq})^{2} \big] X -  
4\,\kij^{\sq} \lij^{\sq} m_{q'} \mch{j}$.  
The phase space integrals $I$, $I_{n}$, $I_{nm}$, and $I_{n}^{m}$  
have $(\msq{i}, m_{q'}, \mch{j})$ as arguments and are  
given in \cite{denner}. 
An analogous expression holds for  
$\d\G_{\!real}(\sq_i^{}\to q\nt_k)\equiv\G(\sq_i^{}\to g q \nt_k)$.  
The complete $\Oas$ correction $\d\G=\d\G_{\!virt}+\d\G_{\!real}$  
is UV and IR f\/inite.  
 
\subsection {Numerical Results} 
 
Let us now turn to the numerical analysis.  
Masses and couplings of charginos and neutralinos depent on the  
parameters $M$, $\mu$, and $\tan\b$ ($M'=\frac{5}{3}\tan^2\tW M$).  
For the stop sector we use $\mst{1}$, $\mst{2}$, $\cst$,  
$\mu$, and $\tan\b$ as input values. The sbottom masses and  
mixing angle are fixed by the assumptions $M_{\ti D}=1.12\,M_{\ti Q}(\st)$  
and $A_b = A_t$.   
The other parameters are taken as explained in \ref{sect:convention}. 
 
\begin{figure}[p] 
\center 
\begin{picture}(100,67) 
\put(1,6){\mbox{\epsfig{figure=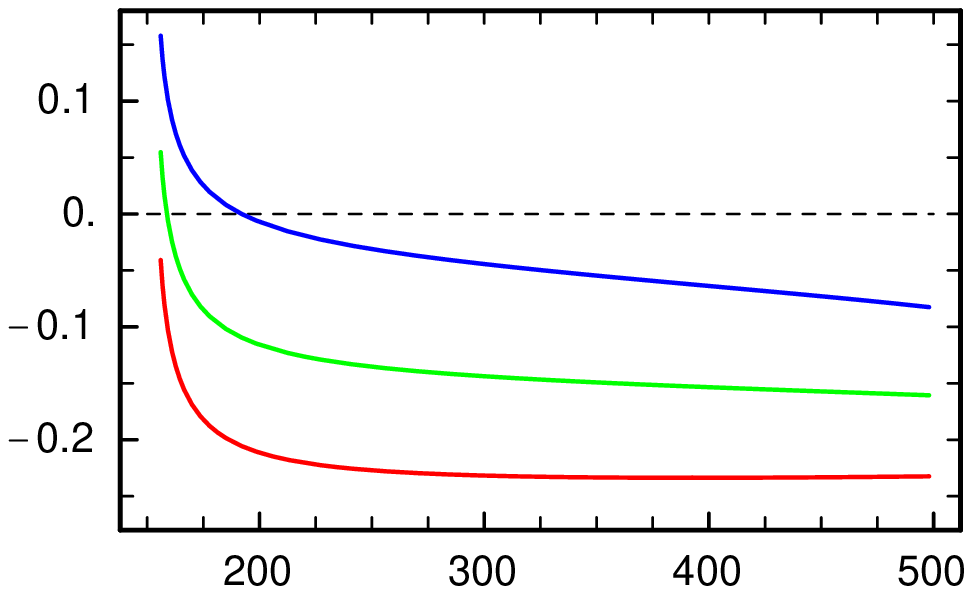,height=6cm}}} 
\put(57,2){\makebox(0,0)[c]{{$\mst{1}$~[GeV]}}} 
\put(-1,24){\makebox(0,0)[br]{{\rotatebox{90}{$\d\G/\G^0\,(\st_1\to b\,\chp_1)$}}}} 
\put(72,40){ 
  \makebox(0,0)[bl]{{\footnotesize (a) $M \ll |\mu|$}}} 
\put(71,33){ 
  \makebox(0,0)[tr]{{\footnotesize (c) $M \simeq |\mu|$}}} 
\put(50,22){ 
  \makebox(0,0)[br]{{\footnotesize (b) $|\mu| \ll M$}}} 
\end{picture} 
\caption{SUSY--QCD corrections for the decay $\st_1\to b\,\chp_1$  
  relative to the tree--level width as a function of $\mst{1}$,  
  for $\mch{1} = 150$ GeV, $\mst{2} = 500$ GeV, $\cst = 0.65$, 
  $M_{\ti D} = 1.12\,M_{\ti Q}$, $A_t=A_b$, and $\tan\b = 3$.   
  Three scenarios are studied:  
  (a) $M =  163$ GeV, $\mu = 500$ GeV,  
  (b) $M =  500$ GeV, $\mu = 163$ GeV, and  
  (c) $M = \mu = 219$ GeV.} 
\label{fig:ncfig1} 
\end{figure}

\begin{figure}[p] 
\center 
\begin{picture}(100,67) 
\put(4,5){\mbox{\epsfig{figure=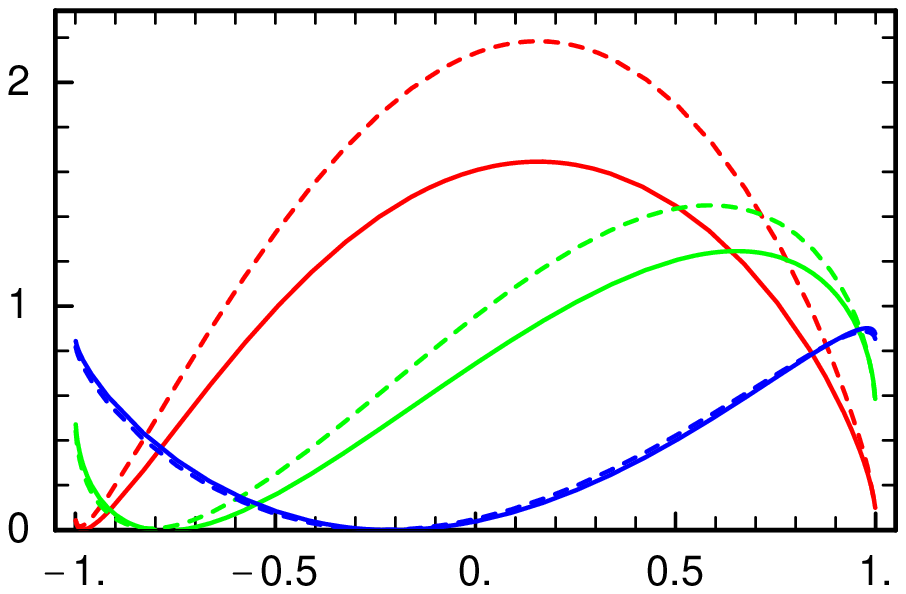,height=6cm}}} 
\put(61.3,0){\makebox(0,0)[br]{{$\cth_{\st}$}}} 
\put(5,20){\makebox(0,0)[br]{{\rotatebox{90}{$\Gamma\,(\st_1\to b\,\chp_1)$~[GeV]}}}} 
\put(61.5,24){ 
  \makebox(0,0)[bl]{{\scriptsize (a) $M \ll |\mu|$}}} 
\put(61,35){ 
  \makebox(0,0)[tr]{{\scriptsize (c) $M \simeq |\mu|$}}} 
\put(68,52){ 
  \makebox(0,0)[br]{{\scriptsize (b) $|\mu| \ll M$}}} 
\end{picture} 
\caption{Tree--level (dashed lines) and SUSY--QCD corrected (full lines)  
  widths of the decay $\st_1\to b\,\chp_1$ in GeV as a function  
  of $\cst$, for $\mch{1} = 150$ GeV,  
  $\mst{1} = 250$ GeV, $\mst{2} = 500$ GeV,  
  $M_{\ti D} = 1.12\,M_{\ti Q}$, $A_t=A_b$, and $\tan\b = 3$.   
  Three scenarios are studied:  
  (a) $M =  163$ GeV, $\mu = 500$ GeV,  
  (b) $M =  500$ GeV, $\mu = 163$ GeV, and  
  (c) $M = \mu = 219$ GeV.} 
\label{fig:ncfig2} 
\end{figure}

\Fig{ncfig1} shows the SUSY--QCD corrections $\d\G\equiv\G-\G^0$   
for the decay $\st_1\to b\,\chp_1$ relative to the tree--level width  
as a function of $\mst{1}$ for $\mch{1} = 150$ GeV, $\mst{2} = 500$ GeV,  
$\cst = 0.65$, and $\tan\b = 3$.   
In order to study the dependence on the nature of the chargino  
(gaugino-- or higgsino--like), we have chosen three sets of $M$ and  
$\mu$ values:  
$M\ll |\mu|$ ($M =  163$ GeV, $\mu = 500$ GeV),  
$M\gg |\mu|$ ($M =  500$ GeV, $\mu = 163$ GeV), and  
$M\simeq |\mu|$ ($M = \mu = 219$ GeV).  
Near the threshold, the corrections strongly depend on $\mst{1}$;  
for larger mass differences this dependence is much weaker.  
We consider $\mst{1}\gsim 200$~GeV:  
For a gaugino--like $\chp_1$ the correction is up to $-10\%$   
while for $M\simeq |\mu|$ the correction typically amounts to $-10\%$ to  
$-15\%$. The biggest effect is found for a higgsino--like chargino due  
to the large top Yukawa coupling; in this case we have  
$\d\G/\G^0 \sim -20\%$ to $-25\%$.  
The dependence on $\cst$ is shown in \fig{ncfig2}  
where we plot the $\Oas$ corrected decay widths (full lines)  
together with the tree--level widths (dashed lines)  
for $\mst{1} = 250$ GeV and the other parameters as above.  
Again, one can see that the SUSY--QCD corrections can considerably change  
the decay width of $\st_1$. This is in particular true for the case   
$|\mu|\gg M$ for which the correction is up to $-40\%$.  
However, also for $M\sim |\mu|$ and $M\gg |\mu|$ one can have large  
corrections, especially if the tree--level width is very small.  
In our examples, the SUSY--QCD corrections to the $\st_1\to b\chp_1$  
decay width are mostly negative.  
However, they can also enhance the tree--level  
width, see \eg~ Figs. 2 and 3 of Ref.~\cite{qcdnc-paper}.  
 
\begin{figure}[p] 
\center 
\begin{picture}(100,67) 
\put(1,6){\mbox{\epsfig{figure=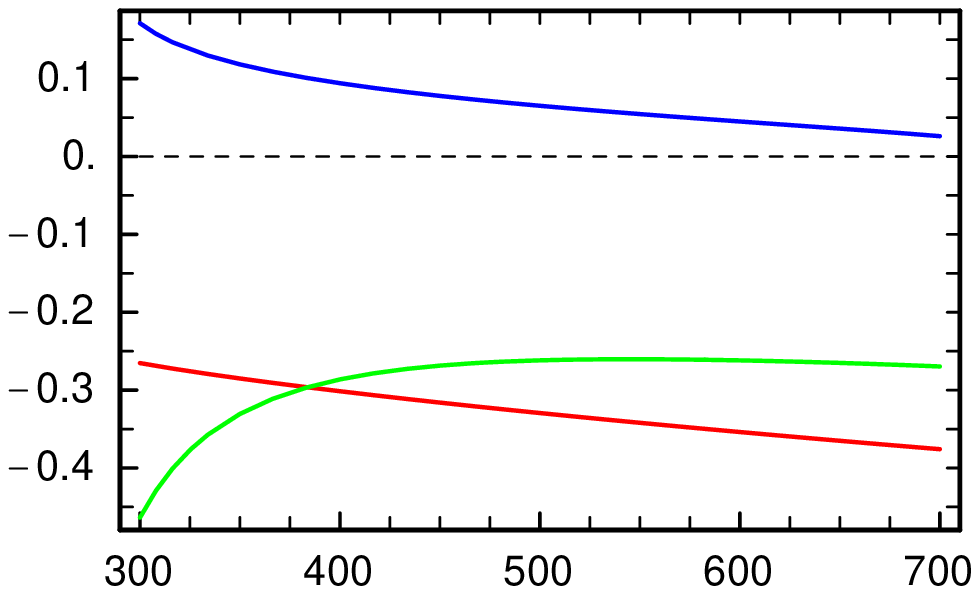,height=6cm}}} 
\put(57,2){\makebox(0,0)[c]{{$\mst{2}$~[GeV]}}} 
\put(-1,24){\makebox(0,0)[br]{{\rotatebox{90}{$\d\G/\G^0\,(\st_2\to t\,\nt_1)$}}}} 
\put(72,58){ 
  \makebox(0,0)[bl]{{\footnotesize (a) $M \ll |\mu|$}}} 
\put(72,32){ 
  \makebox(0,0)[bl]{{\footnotesize (c) $M \simeq |\mu|$}}} 
\put(64,19){ 
  \makebox(0,0)[br]{{\footnotesize (b) $|\mu| \ll M$}}} 
\end{picture} 
\caption{SUSY--QCD corrections for the decay $\st_2\to t\,\nt_1$  
  relative to the tree--level width as a function of $\mst{2}$,  
  for $\mnt{1} = 100$ GeV, $\mst{1} = 250$ GeV, $\cst = 0.65$, 
  $M_{\ti D} = 1.12\,M_{\ti Q}$, $A_t=A_b$, and $\tan\b = 3$.   
  Three scenarios are studied:  
  (a) $M =  208$ GeV, $\mu = 500$ GeV,  
  (b) $M =  500$ GeV, $\mu = 123$ GeV, and  
  (c) $M = \mu = 230$ GeV.} 
\label{fig:ncfig3} 
\end{figure}

\begin{figure}[p] 
\center 
\begin{picture}(100,67) 
\put(4,5){\mbox{\epsfig{figure=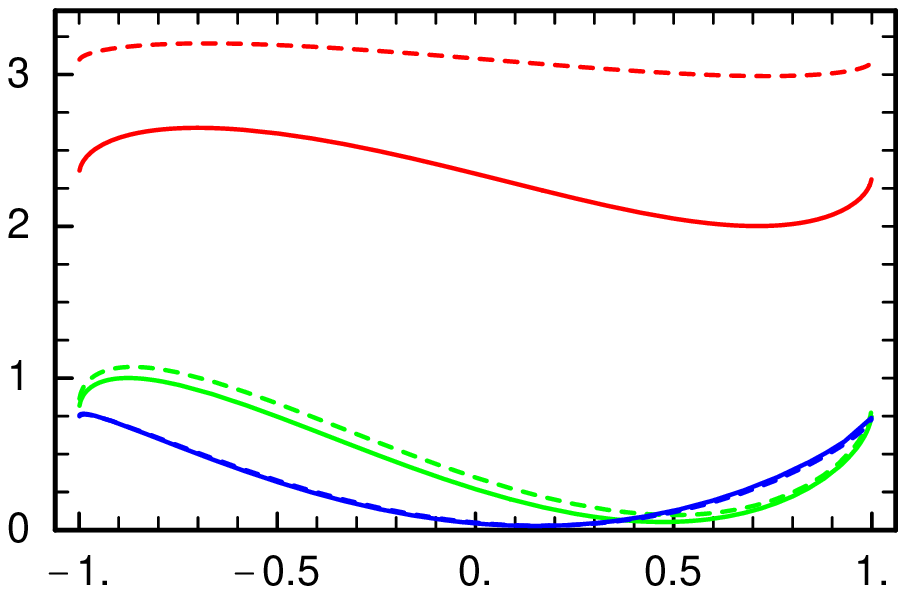,height=6cm}}} 
\put(61.3,0){\makebox(0,0)[br]{{$\cth_{\st}$}}} 
\put(5,20){\makebox(0,0)[br]{{\rotatebox{90}{$\Gamma\,(\st_2\to t\,\nt_1)$~[GeV]}}}} 
\put(68,20.5){ 
  \makebox(0,0)[bl]{{\footnotesize (a) $M \ll |\mu|$}}} 
\put(47,31){ 
  \makebox(0,0)[tr]{{\footnotesize (c) $M \simeq |\mu|$}}} 
\put(70,53){ 
  \makebox(0,0)[br]{{\footnotesize (b) $|\mu| \ll M$}}} 
\end{picture} 
\caption{Tree--level (dashed lines) and SUSY--QCD corrected (full lines)  
  widths of the decay $\st_2\to t\,\nt_1$ in GeV as a function  
  of $\cst$, for $\mnt{1} = 100$ GeV,  
  $\mst{1} = 250$ GeV, $\mst{2} = 500$ GeV,  
  $M_{\ti D} = 1.12\,M_{\ti Q}$, $A_t=A_b$, and $\tan\b = 3$.   
  Three scenarios are studied:  
  (a) $M =  208$ GeV, $\mu = 500$ GeV,  
  (b) $M =  500$ GeV, $\mu = 123$ GeV, and  
  (c) $M = \mu = 230$ GeV.} 
\label{fig:ncfig4} 
\end{figure}

As an example for stop decays into neutralinos we discuss the decay  
$\st_2\to t\nt_1$. Here we fix $\mnt{1}=100$ GeV. Again, we take  
$\tan\b = 3$ and study three different scenarios:  
$M\ll |\mu|$ ($M =  208$ GeV, $\mu = 500$ GeV),  
$M\gg |\mu|$ ($M =  500$ GeV, $\mu = 123$ GeV), and  
$M\simeq |\mu|$ ($M = \mu = 230$ GeV).  
\Fig{ncfig3} shows the SUSY--QCD corrections relative to the tree--level  
decay width as a function of $\mst{2}$, for $\mst{1} = 250$ GeV and  
$\cst = 0.65$.  
In the shown range of $\mst{2}$, $\d\G/\G^0$ varies from  
$18\%$ to $3\%$ in case of a gaugino--like $\nt_1$.  
In case of a higgsino--like neutralino $\d\G/\G^0 = -26\%$ to  
$-37\%$, and in the mixed scenario $\d\G/\G^0 = -46\%$ to  
$-26\%$.  
The dependence on $\cst$ is shown in \fig{ncfig4}.   
Analogously to \fig{ncfig2} we here plot the $\Oas$ corrected decay  
widths (full lines) together with the tree--level widths (dashed lines)   
for $\mst{2} = 500$ GeV and the other parameters as in  
\fig{ncfig3}. Again, a striking effect can be seen for $|\mu|\ll M$. 
The neutralino, chargino, and sbottom masses for the various 
($M,\,\mu$) values are listed in Tab.~\ref{tab:ncszens}. 

\renewcommand{\arraystretch}{1.4}
\begin{table}[h!] \center 
\begin{tabular}{|c|c||c|c|c|c||c|c||c|c|c|} 
\hline
  $M$ & $\mu$ & $\mnt{1}$ & $\mnt{2}$ & $\mnt{3}$ & $\mnt{4}$ 
              & $\mch{1}$ & $\mch{2}$ & $\msb{1}$ & $\msb{2}$ \\
\hline\hline
  163 & 500 & \hphantom{1}78& 151& 503& 518& 150& 516& 384& 427 \\  
  500 & 163 &            136& 166& 262& 517& 150& 516& 391& 424 \\  
  219 & 219 & \hphantom{1}94& 161& 223& 296& 150& 293& 388& 425 \\
\hline  
  208 & 500 & 100& 194& 502& 520& 193& 518& 385& 427 \\  
  500 & 123 & 100& 126& 259& 515& 112& 515& 391& 424 \\  
  230 & 230 & 100& 171& 234& 307& 161& 304& 388& 425 \\  
\hline \hline
\end{tabular}
\caption{Neutralino, chargino, and sbottom masses for the scenarios 
  discussed above ($\mst{1} = 250$ GeV, $\mst{2} = 500$ GeV, $\cst = 0.65$, 
  $M_{\ti D} = 1.12\,M_{\ti Q}$, $A_t=A_b$, and $\tan\b = 3$). 
  All values in [GeV].} 
\label{tab:ncszens} 
\end{table} 
\renewcommand{\arraystretch}{1}

For completeness, we also briefly discuss sbottom decays into   
charginos/neutralinos for the parameters of Tab.~\ref{tab:ncszens}.  
In these scenarios, $\msb{1}\sim 390$ GeV, $\msb{2}\sim 425$ GeV,   
and $\csb\simeq 1$. 
Table~\ref{tab:sbtonc} lists the SUSY--QCD corrected widths 
together with the relative corrections for all $\sb_i\to t\chm_j$  
and $\sb_i\to b\nt_k$ decays that are kinematically allowed.   
As can be seen, also for sbottom decays the SUSY--QCD corrections  
can be of either sign and amount to a few ten percent.  
Note, that in Tab.~\ref{tab:sbtonc} $\tan\b =3$.  
For larger values of $\tan\b$ the corrections to the sbottom decay  
widths become even more important since the bottom Yukawa  
coupling increases.

\renewcommand{\arraystretch}{1.4}   
\begin{table}[h!] \center 
\begin{tabular}{|c|c||c|r||c|r|} 
\hline  
 \multicolumn{2}{|c||}{~} & \multicolumn{2}{|c||}{$\sb_1$}  
                          & \multicolumn{2}{|c|}{$\sb_2$}  \\ 
\hline                        
  $(M,\,\mu)$~[GeV] & channel & $\G$~[GeV] & $\d\G/\G^0$  
                              & $\G$~[GeV] & $\d\G/\G^0$ \\
\hline\hline
  (163,\,500) & $b\nt_1$ & 0.11\hphantom{0}& $ 0.02$            
                         & 0.2\hphantom{00}& $-0.02\hphantom{0}$ \\
              & $b\nt_2$ & 1.01\hphantom{0}& $ 0.02$            
                         & 0.09\hphantom{0}& $ 0.45\hphantom{0}$ \\
              & $t\chm_1$& 0.83\hphantom{0}& $ 0.18\hphantom{0}$ 
                         & 0.09\hphantom{0}& $ 0.56\hphantom{0}$ \\
\hline
  (500,\,163) & $b\nt_1$ & 0.09\hphantom{0}& $-0.09\hphantom{0}$ 
                         & 0.01\hphantom{0}& $-0.54\hphantom{0}$\\
              & $b\nt_2$ & (0.006)         & $(-0.69)\!\!\hphantom{0}$ 
                         & 0.009           & $-0.66\hphantom{0}$\\
              & $b\nt_3$ & (0.003)         & $(0.57)\!\!\hphantom{0}$
                         & 0.09\hphantom{0}& $ 0.02\hphantom{0}$\\
              & $t\chm_1$& 2.03\hphantom{0}& $-0.11\hphantom{0}$ 
                         & 0.11\hphantom{0}& $-0.16\hphantom{0}$\\
\hline
  (219,\,219) & $b\nt_1$ & 0.22\hphantom{0}& $ 0.01\hphantom{0}$ 
                         & 0.16\hphantom{0}& $ 0.01\hphantom{0}$\\
              & $b\nt_2$ & 0.45\hphantom{0}& $ 0.05\hphantom{0}$ 
                         & 0.04\hphantom{0}& $ 0.05\hphantom{0}$\\
              & $b\nt_3$ & (0.007)         & $(-0.55)\!\!\hphantom{0}$ 
                         & 0.008           & $-0.62\hphantom{0}$\\
              & $b\nt_4$ & 0.14\hphantom{0}& $ 0.21\hphantom{0}$ 
                         & 0.02\hphantom{0}& $-0.13\hphantom{0}$\\
              & $t\chm_1$& 0.76\hphantom{0}& $-0.01\hphantom{0}$ 
                         & 0.04\hphantom{0}& $-0.18\hphantom{0}$\\
\hline\hline
  (208,\,500) & $b\nt_1$ & 0.1\hphantom{00}& 0.05\hphantom{0} 
                         & 0.19\hphantom{0}& -0.004\\
              & $b\nt_2$ & 0.8\hphantom{00}& 0.06\hphantom{0} 
                         & 0.07\hphantom{0}& 0.58\hphantom{0}\\
              & $t\chm_1$& 0.31\hphantom{0}& 0.68\hphantom{0} 
                         & 0.06\hphantom{0}& 0.76\hphantom{0}\\
\hline
  (500,\,123) & $b\nt_1$ & 0.09\hphantom{0}& $-0.14\hphantom{0}$ 
                         & 0.009           & $-0.69\hphantom{0}$\\
              & $b\nt_2$ & (0.009)         & $(-0.66)\!\!\hphantom{0}$
                         & 0.01\hphantom{0}& $-0.66\hphantom{0}$\\
              & $b\nt_3$ & (0.006)         & $(0.42)\!\!\hphantom{0}$ 
                         & 0.09\hphantom{0}& $ 0.03\hphantom{0}$\\
              & $t\chm_1$& 2.9\hphantom{00}& $-0.14\hphantom{0}$ 
                         & 0.11\hphantom{0}& $-0.16\hphantom{0}$\\
\hline
  (230,\,230) & $b\nt_1$ & 0.2\hphantom{00}& $ 0.02\hphantom{0}$ 
                         & 0.16\hphantom{0}& $ 0.01\hphantom{0}$ \\
              & $b\nt_2$ & 0.44\hphantom{0}& $ 0.06\hphantom{0}$ 
                         & 0.03\hphantom{0}& $ 0.08\hphantom{0}$ \\
              & $b\nt_3$ & (0.006)        &  $(-0.57)\!\!\hphantom{0}$ 
                         & 0.008           & $-0.63\hphantom{0}$\\
              & $b\nt_4$ & 0.12\hphantom{0}& $ 0.25\hphantom{0}$ 
                         & 0.02\hphantom{0}& $-0.12\hphantom{0}$\\
              & $t\chm_1$& 0.6\hphantom{00}& $ 0.02\hphantom{0}$ 
                         & 0.04\hphantom{0}& $-0.2\hphantom{00}$\\
\hline\hline
\end{tabular} 
\caption{Sbottom decay widths at ${\Oas}$ and  
  the relative corrections for the scenarios discussed above 
  ($\mst{1} = 250$ GeV, $\mst{2} = 500$ GeV, $\cst = 0.65$, 
  $M_{\ti D} = 1.12\,M_{\ti Q}$, $A_t=A_b$, and $\tan\b = 3$). 
  The values in brackets correspond to branching ratios below 1\%.} 
\label{tab:sbtonc} 
\end{table} 
\renewcommand{\arraystretch}{1}

\clearpage 
\section {Decays into {\em W} and {\em Z} Bosons} \label{sect:qcdwz}
 
\begin{figure}[p] 
\center 
\begin{picture}(145,210) 
\put(0,8){\mbox{\psfig{file=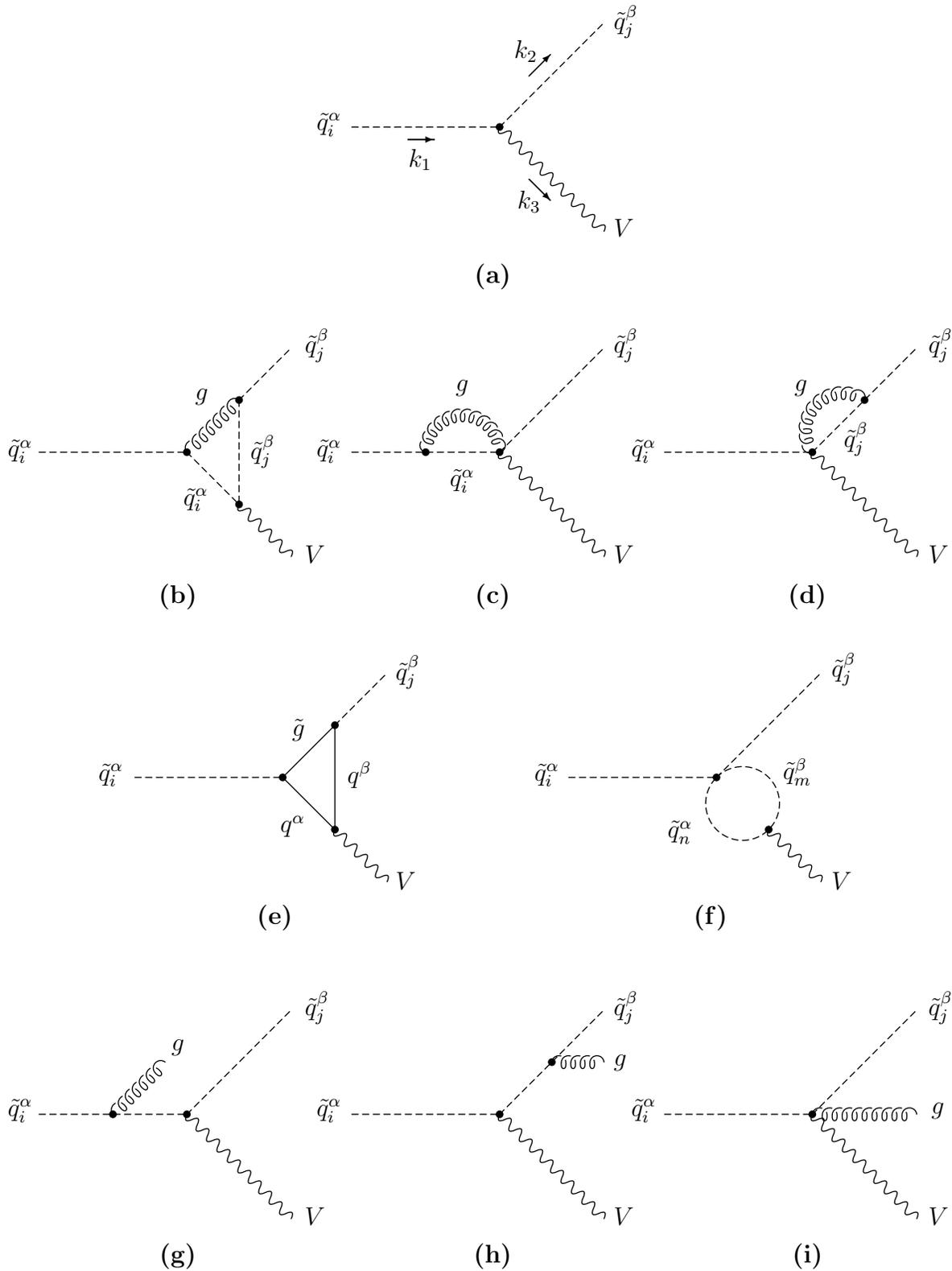}}}
\put(75,167){\makebox(0,0)[t]{\bf{(a)}}} 
\put(23,114){\makebox(0,0)[t]{\bf{(b)}}} 
\put(75,114){\makebox(0,0)[t]{\bf{(c)}}} 
\put(127,114){\makebox(0,0)[t]{\bf{(d)}}} 
\put(39,61){\makebox(0,0)[t]{\bf{(e)}}} 
\put(111,61){\makebox(0,0)[t]{\bf{(f)}}} 
\put(23,5){\makebox(0,0)[t]{\bf{(g)}}} 
\put(75,5){\makebox(0,0)[t]{\bf{(h)}}} 
\put(127,5){\makebox(0,0)[t]{\bf{(i)}}} 
\put(50,190){\makebox(0,0)[r]{$\sq_i^\a$}} 
\put(95,207){\makebox(0,0)[l]{$\sq_j^\b$}} 
\put(95,173){\makebox(0,0)[l]{$V$}} 
\put(61,187.5){\vector(1,0){4.3}} 
\put(63,186){\makebox(0,0)[t]{$k_1$}} 
\put(81,198){\vector(1,1){3.6}} 
\put(80.5,201){\makebox(0,0)[b]{$k_2$}} 
\put(81,181){\vector(1,-1){3.6}} 
\put(81,179){\makebox(0,0)[t]{$k_3$}} 
\put(-1,136){\makebox(0,0)[r]{$\sq_i^\a$}} 
\put(44,153.5){\makebox(0,0)[l]{$\sq_j^\b$}} 
\put(44,119){\makebox(0,0)[l]{$V$}} 
\put(28,145){\makebox(0,0)[r]{$g$}} 
\put(28,128){\makebox(0,0)[r]{$\sq_i^\a$}} 
\put(35,136){\makebox(0,0)[l]{$\sq_j^\b$}} 
\put(50,136){\makebox(0,0)[r]{$\sq_i^\a$}} 
\put(95,153.5){\makebox(0,0)[l]{$\sq_j^\b$}} 
\put(95,119){\makebox(0,0)[l]{$V$}} 
\put(70,133){\makebox(0,0)[tc]{$\sq_i^\a$}} 
\put(70,146){\makebox(0,0)[bc]{$g$}} 
\put(102,136){\makebox(0,0)[r]{$\sq_i^\a$}} 
\put(147,153.5){\makebox(0,0)[l]{$\sq_j^\b$}} 
\put(147,119){\makebox(0,0)[l]{$V$}} 
\put(133,138){\makebox(0,0)[l]{$\sq_j^\b$}} 
\put(127,146){\makebox(0,0)[br]{$g$}} 
\put(14,82.5){\makebox(0,0)[r]{$\sq_i^\a$}} 
\put(59,99.5){\makebox(0,0)[l]{$\sq_j^\b$}} 
\put(59,65){\makebox(0,0)[l]{$V$}} 
\put(44,90){\makebox(0,0)[r]{$\sg$}} 
\put(44,74){\makebox(0,0)[r]{$q^\a$}} 
\put(51,82.5){\makebox(0,0)[l]{$q^\b$}} 
\put(86,82.5){\makebox(0,0)[r]{$\sq_i^\a$}} 
\put(131,100){\makebox(0,0)[l]{$\sq_j^\b$}} 
\put(131,65){\makebox(0,0)[l]{$V$}} 
\put(108,73){\makebox(0,0)[r]{$\sq_n^\a$}} 
\put(123,83){\makebox(0,0)[l]{$\sq_m^\b$}} 
\put(-1,27){\makebox(0,0)[r]{$\sq_i^\a$}} 
\put(44,44){\makebox(0,0)[l]{$\sq_j^\b$}} 
\put(44,9.5){\makebox(0,0)[l]{$V$}} 
\put(23,37){\makebox(0,0)[b]{$g$}} 
\put(50,27){\makebox(0,0)[r]{$\sq_i^\a$}} 
\put(95,44){\makebox(0,0)[l]{$\sq_j^\b$}} 
\put(95,9.5){\makebox(0,0)[l]{$V$}} 
\put(95,35){\makebox(0,0)[l]{$g$}} 
\put(102,27){\makebox(0,0)[r]{$\sq_i^\a$}} 
\put(147,44){\makebox(0,0)[l]{$\sq_j^\b$}} 
\put(147,9.5){\makebox(0,0)[l]{$V$}} 
\put(147.5,27){\makebox(0,0)[l]{$g$}} 
\end{picture} 
\caption{Feynman diagrams for the ${\cal O}(\a_{s})$ SUSY-QCD  
  corrections to squark decays into vector bosons:
  (a) tree level, (b)--(f) vertex corrections, 
  (g)--(i) real gluon emission. For wave--function corrections 
  see Figs.~\ref{fig:fd-sqself} and \ref{fig:fd-qself}.} 
\label{fig:FD-qcdwz} 
\end{figure}

At tree level the amplitude of a squark decay into  
a $W^\pm$ or $Z$ boson has the general form 
\begin{equation} 
  {\cal M}^0(\sq^\a_i \to \sq^\b_j V) =  
  -ig\,c_{ijV}\, (k_1 + k_2)^\mu\, \epsilon_\mu^* (k_3), 
\label{eq:MtreeWZ}   
\end{equation} 
with $k_1$, $k_2$, and $k_3$ the four--momenta of $\sq^\a_i$,  
$\sq^\b_j$, and the vector boson $V$ ($V = W^\pm,Z^0$), respectively  
(\fig{FD-qcdwz}\,a). $\a$ and $\b$ are flavor indices.  
In the following we define $m_i = m_{\sq^\a_i}$, $m_j = m_{\sq^\b_j}$,  
$\R_{ik} = \R_{ik}^{\sq^\a}$, $\R_{jk} = \R_{jk}^{\sq^\b}$ for simplicity.  
Moreover, we shall use primes to explicitly distinguish between different  
flavors. 
With this notation the $\sq^\a_i\sq^\b_jV$ couplings  
$c_{ijV}$ are, see Eqs.~\eq{cij} and \eq{csqW}: 
\begin{equation} 
  c_{ijZ} = 
   \smaf{1}{\cos\tW}\,  
    \left( \begin{array}{cc} 
      I_{3L}^{q}\,\cos^2\tsq - e_{q} \sin^2\tW  
        & -\onehf\, I_{3L}^{q}\,\sin 2\tsq \\ 
    -\onehf\, I_{3L}^{q}\,\sin 2\tsq  
        & I_{3L}^{q}\,\sin^2\tsq - e_{q} \sin^2\tW 
    \end{array} \right)_{ij} ,  
\end{equation} 
\begin{equation} 
  c_{ijW} =  
    \smaf{1}{\rzw}\R_{i1}^{}\R_{j1}' =  
    \smaf{1}{\rzw}\,  
    \left( \begin{array}{rr} 
       \cos\tsq \cos\t_{\sq'} & -\cos\tsq \sin\t_{\sq'} \\ 
      -\sin\tsq \cos\t_{\sq'} &  \sin\tsq \sin\t_{\sq'} 
    \end{array} \right)_{ij} . 
\end{equation}  
 
\noi
The tree--level decay width can thus be written as:  
\begin{equation}
  \G^0(\sq_i^{} \to \sq_j^{(')} V) =  
  \frac{g^2\,(c_{ijV})^2\kappa^3(m_i^2,m_j^2,\mV^2)
       }{16\pi\,\mV^2\,m_i^3}\,.
\end{equation}

 
\noi 
The ${\cal O}(\a_{s})$ loop corrected decay amplitude is obtained by  
the shift   
\begin{equation} 
  c_{ijV}\to c_{ijV} + \d c_{ijV}^{(v)} + \d c_{ijV}^{(w)}  
                     + \d c_{ijV}^{(c)}   
\end{equation} 
in \eq{MtreeWZ}.  
The virtual correction to the decay width thus has the form  
\begin{equation} 
  \d\G^{(a)}(\sq_i^{} \to \sq_j^{(')} V) =  
  \frac{g^2\,\kappa^3(m_i^2,m_j^2,\mV^2)}{8\pi\,\mV^2\,m_i^3}\; 
  c_{ijV}\:\Re\left\{\d c_{ijV}^{(a)}\right\}   
  \hspace{6mm}(a = v,w,c).  
\label{eq:deltaG} 
\end{equation}

 
\noi          
The {\bf vertex correction} stems from the five diagrams shown in  
Figs.~\ref{fig:FD-qcdwz}\,b--f.  
The gluon exchange graphs of Figs.~\ref{fig:FD-qcdwz}\,b--d yield  
\begin{align} 
  \d c_{ijV}^{(v,g)} = 
  & -\frac{\a_s}{3\pi}\: c_{ijV} \left[ \, 
    B_0(m_i^2,\l^2,m_i^2) + B_0(m_j^2,\l^2,m_j^2) \right. \nn\\ 
  & \hspace{40mm}\left. 
    -\,2\,(m_i^2 + m_j^2 - \mV^2)\,(C_0 + C_1 + C_2)\, \right]  
    \hspace{-40mm} 
\end{align} 
with $C_{m} = C_{m}(m_i^2, \mV^2, m_j^2; \l^{2}, m_i^2, m_j^2)$.  
A gluon mass $\l$ is introduced to regularize the infrared 
divergence.  
The gluino--exchange contribution, \fig{FD-qcdwz}\,e, gives  
\begin{align} 
  \d c_{21Z}^{(v,\sg)} = 
  &-\,\frac{\a_s}{3\pi\cos\t_W}\: \Big\{ 
   2\,\msg\,m_q\,(I_{3L}^q - 2e_q\sin^2\t_W)\, 
                       (C_0+C_1+C_2)\,\cos 2\tsq \\ 
  &\hspace{8mm} 
   + I_{3L}^q\,\big[\,2\,\msg^2\,C_0 + m_{\sq_2}^2\,C_1 + m_{\sq_1}^2\,C_2  
   +(\msg^2 - m_q^2)\,(C_1+C_2) + B_0\,\big]\,\sin 2\tsq         
   \Big\} , \nn 
\end{align} 
with  
$C_{m} = C_{m}(m_{\sq_2}^2, m_{\!Z}^2, m_{\sq_1}^2; \msg^{2}, m_q^2, m_q^2)$  
and $B_0 = B_0(m_{\!Z}^2, m_q^2, m_q^2)$,  
for the decay $\sqz \to \sqe\, Z^0$, and  
\begin{align} 
  \d c_{ijW}^{(v,\sg)} = 
  & -\,\frac{\sqrt{2}}{3}\, \frac{\a_s}{\pi}\: \Big\{ 
    \msg\,(C_0 + C_1 + C_2)\, 
          (m_q\,\R_{i2}^{}\,\R'_{j1} + m_{q'}\,\R_{i1}^{}\,\R'_{j2}) \nn\\ 
  & \hspace{25mm} 
  -\,\big[\,m_i^2\,C_1 + m_j^2\,C_2 + \msg^2\,(2C_0+C_1+C_2) + B_0\,\big]\, 
     \R_{i1}^{}\,\R'_{j1} \nn\\ 
  & \hspace{25mm} -\, m_q\,m_{q'}\,(C_1 + C_2)\,\R_{i2}^{}\,\R'_{j2}\,  
  \Big\} 
\end{align} 
with $C_{m} = C_{m}(m_i^2, m_{\!W}^2, m_j^2; \msg^{2}, m_q^2, m_{q'}^2)$ 
and $B_0 = B_0(m_{\!W}^2, m_q^2, m_{q'}^2)$,  
for the decay $\sqi\to\sq'_j\, W^\pm$.  
The squark loop of \fig{FD-qcdwz}\,f does not contribute because  
it is proportional to the four--momentum of the vector boson. 
The total vertex correction is thus given by:  
\begin{equation} 
  \d c_{ijV}^{(v)} =  
  \d c_{ijV}^{(v,g)} + \d c_{ijV}^{(v,\sg)}. 
\end{equation}

\noi 
The {\bf wave--function correction} is given by 
$(i\neq i', j\neq j')$ 
\begin{equation}  
  \d c_{ijV}^{(w)} =  
  \onehf \left[  
    \d\ti Z_{ii}(\sq_i^\a) + \d\ti Z_{jj}(\sq_j^\b) \right] c_{ijV} 
  + \d\ti Z_{i'i}(\sq_i^\a)\,c_{i'jV} + \d\ti Z_{j'j}(\sq_j^\b)\,c_{ij'V}   
\end{equation} 
with $\ti Z_{nm}(\sq_n)$ the squark wave--function  
renormalization constants (see Sect.~\ref{sect:selfenergies}).
 
\noi 
The {\bf counterterms for the couplings} are:  
\begin{equation} 
  \d c_{21Z}^{(c)} = - \smaf{1}{\cos\t_W}I_{3L}^{q} \cos 2\tsq\, \d\tsq , 
\label{eq:dcZ} 
\end{equation} 
and 
\begin{align} 
  \d c_{ijW}^{(c)} = 
  & \smaf{1}{\rzw}\, \left[ 
    \left( \begin{array}{rr} 
      -\sin\tsq \cos\t_{\sq'} & \sin\tsq \sin\t_{\sq'} \\ 
      -\cos\tsq \cos\t_{\sq'} & \cos\tsq \sin\t_{\sq'} 
    \end{array} \right) \d\tsq \right.\nn \\[2mm] 
  & \hspace{30mm} + \left. \left( \begin{array}{rr} 
      -\cos\tsq \sin\t_{\sq'} & -\cos\tsq \cos\t_{\sq'} \\ 
       \sin\tsq \sin\t_{\sq'} &  \sin\tsq \cos\t_{\sq'} 
    \end{array} \right) \d\t_{\sq'} \right]_{ij}  
\end{align} 
with $\d\tsq$ as defined in Sect.~\ref{sect:tsq}. 
For the decay $\sq_2^{}\to\sq_1^{}\,Z^0$, the counterterm \eq{dcZ}  
completely cancels the off--diagonal wave function corrections  
($i\not=j$ in \fig{fd-sqself}).  
In case of $\sqi \to \sq_j^{\prime}\,W^\pm$ the contribution of the squark  
bubble in \fig{fd-sqself} is cancelled.  
Thus, in both cases the total squark loop  
contribution to the correction is zero, $\d\G^{(\sq)}\equiv 0$. 
 
\noi 
Finally, we add the {\bf emission of real gluons}  
(Figs.~\ref{fig:FD-qcdwz}\,g--i) 
in order to cancel the infrared divergence: 
\begin{align} 
  \d\G_{\!real}  
  &\equiv \G(\sq_i^\a \to g\sq_j^\b V) \nn\\[1mm] 
  &= \frac{g^2\,c_{ijV}^2\,\a_s}{3\pi^2 m_i}\:\Big\{  
     2 I - \frac{\kappa^2}{m_V^2} 
     \left[\, I_0 + I_1 + m_i^2\,I_{00} + m_j^2\,I_{11}  
               + (m_i^2 + m_j^2 - m_V^2)\,I_{01} \right]  
  \Big\}  \hspace{-5cm} 
\end{align} 
with $\kappa = \kappa(m_i^2,\,m_j^2,\,m_V^2)$;  
the phase space integrals $I$, $I_{n}$, and $I_{nm}$  
have $(m_i, m_j, m_V)$ as arguments.  
 
 
Let us now turn to the {\bf numerical results}.  
As the squark couplings to vector bosons depend only on  
the squark mixing angles, we just need the on--shell squark masses $\msq{1,2}$  
and mixing angles $\tsq$, and the gluino mass as input parameters.  
 
\begin{figure}[h!]\center 
\begin{picture}(97,65) 
\put(0,5){\mbox{\epsfig{figure=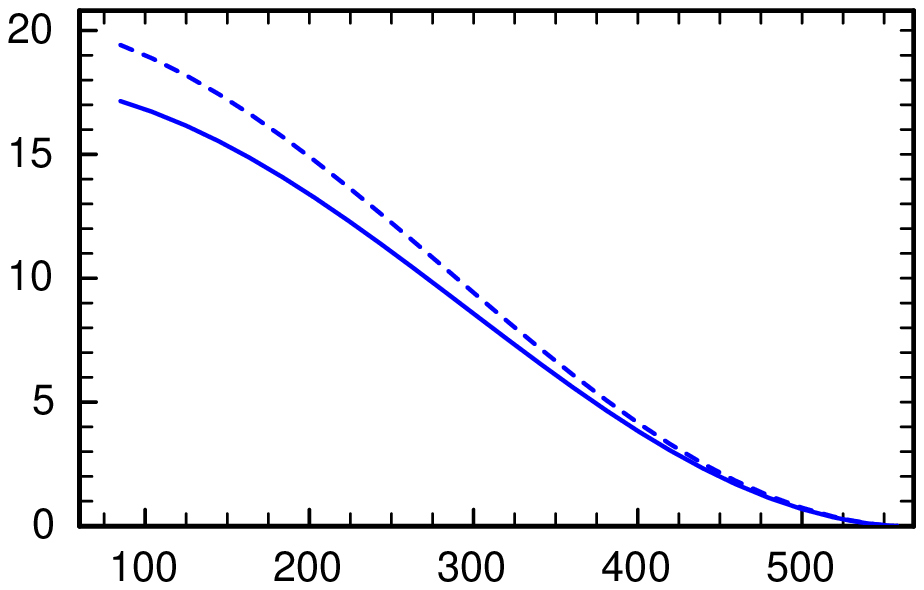,height=6cm}}} 
\put(0,21){\makebox(0,0)[br]{\rotatebox{90}{$\Gamma\,(\st_2\to\st_1 Z)$~[GeV]}}} 
\put(54,0){\makebox(0,0)[bc]{$\mst{1}$~[GeV]}} 
\end{picture} 
\caption{Tree--level (dashed line) and SUSY--QCD corrected (full line)  
  widths of the decay $\st_2\to \st_1 Z$ as a function  
  of $\mst{1}$, for $\mst{2} = 650\gev$, $\cst = -0.6$, and  
  $\msg=500\gev$.} 
\label{fig:qcdwz1} 
\end{figure}

\begin{figure}[h!]\center 
\begin{picture}(97,65) 
\put(0,5){\mbox{\epsfig{figure=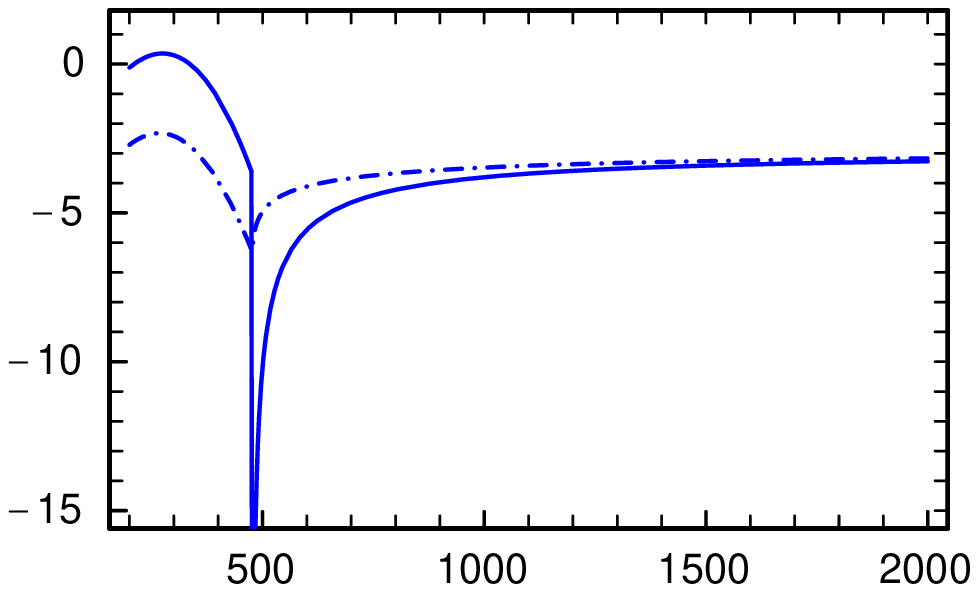,height=6cm}}} 
\put(-1,19){\makebox(0,0)[br]{\rotatebox{90}{$\d\G/\G^0\,(\st_2\to\st_1 Z)$~[\%]}}} 
\put(54,0){\makebox(0,0)[bc]{$\msg$~[GeV]}} 
\put(32,51){\mbox{\small $\cst=0.6$}} 
\put(35,40){\mbox{\small $\cst=-0.6$}} 
\end{picture} 
\caption{Tree--level (dashed line) and SUSY--QCD corrected (full line)  
  widths of the decay $\st_2\to \st_1 Z$ as a function  
  of $\msg$, for $\mst{1} = 200\gev$, $\mst{2} = 650\gev$, and 
  $\cst = \pm 0.6$.} 
\label{fig:qcdwz2} 
\end{figure}

\begin{figure}[h!]\center 
\begin{picture}(97,65) 
\put(0,5){\mbox{\epsfig{figure=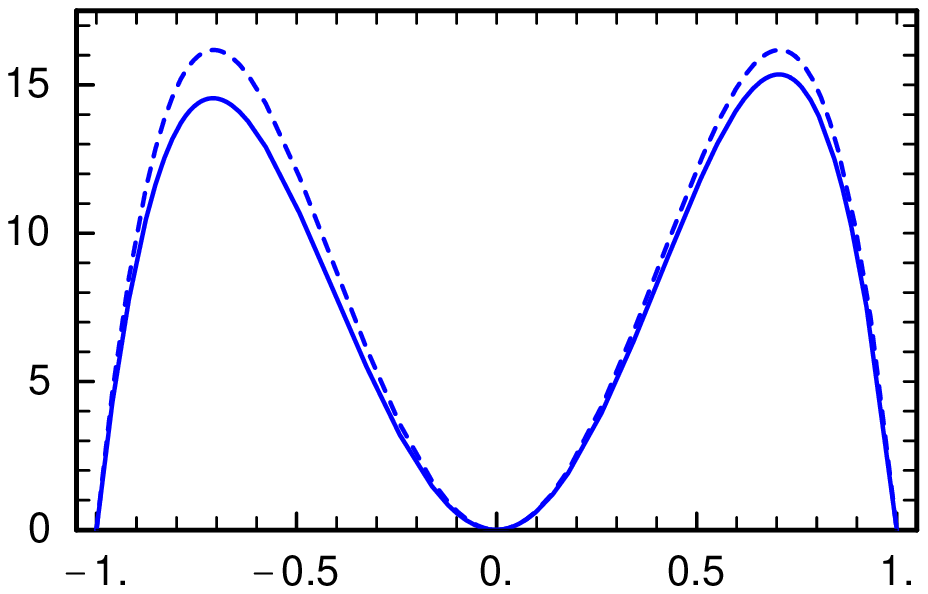,height=6cm}}} 
\put(0,21){\makebox(0,0)[br]{\rotatebox{90}{$\Gamma\,(\st_2\to\st_1 Z)$~[GeV]}}} 
\put(54,0){\makebox(0,0)[bc]{$\cst$}} 
\end{picture} 
\caption{Tree--level (dashed line) and SUSY--QCD corrected (full line)  
  widths of the decay $\st_2\to \st_1 Z$ as a function of $\cst$,  
  for $\mst{1}=200\gev$, $\mst{2} = 650\gev$, and $\msg=500\gev$.} 
\label{fig:qcdwz3a} 
\end{figure}

\begin{figure}[h!]\center 
\begin{picture}(97,65) 
\put(0,4.5){\mbox{\epsfig{figure=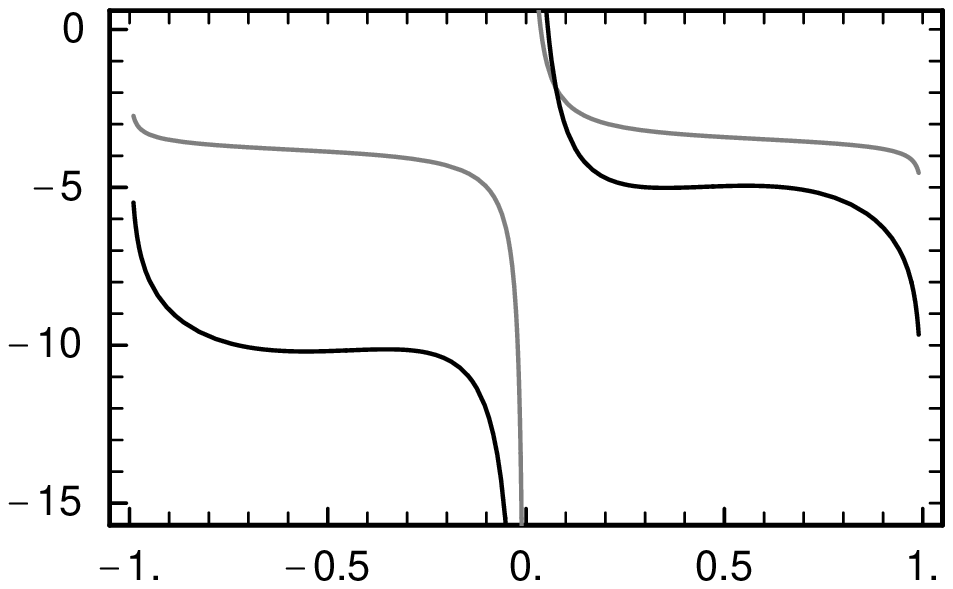,height=6cm}}} 
\put(0,17){\makebox(0,0)[br]{\rotatebox{90}{$\d\G/\G^0\,(\st_2\to\st_1 Z)$~[GeV]}}} 
\put(54,0){\makebox(0,0)[bc]{$\cst$}} 
\end{picture} 
\caption{SUSY--QCD correction relative to the tree--level   
  width of the decay $\st_2\to \st_1 Z$ as a function of $\cst$, 
  for $\mst{1}=200$ GeV and $\mst{2} = 650$ GeV.   
  The black (gray) lines are for $\msg=500$ (1000) GeV.} 
\label{fig:qcdwz3b} 
\end{figure}

 
\noi 
We first discuss the decay $\st_2\to\st_1 Z$.  
\Fig{qcdwz1} shows the tree--level and the ${\cal O}(\a_{s})$  
SUSY--QCD corrected widths of this decay as a function  
of the lighter stop mass $\mst{1}$, for $\mst{2}=650\gev$, $\cst=-0.6$,  
and $\msg=500\gev$.  
SUSY--QCD corrections reduce the tree--level width by $-11.7\%$ to $-6.8\%$  
in the range of $\mst{1} = 85$ to 558~GeV.   
It is interesting to note that the gluonic correction decreases  
quickly with increasing $\mst{1}$ while the correction due to gluino  
exchange varies only little with $\mst{1}$.  
In our example, $\d\G^{(g)}/\G^0=-4.5\%$ and $\d\G^{(\sg)}/\G^0=-7.2\%$  
at $\mst{1} = 85\gev$;   
at $\mst{1} = 550\gev$ the gluonic correction is negligible whereas  
$\d\G^{(\sg)}/\G^0$ is still $-6.8\%$. \\ 
Taking a closer look on the gluino mass dependence we find that   
the gluino decouples slowly. This is visualized in \fig{qcdwz2}   
where we plot the SUSY--QCD correction $\d\G\equiv \G-\G^0$ of  
the decay $\st_2\to\st_1 Z$ relative to its tree--level width  
for $\mst{1} = 200\gev$, $\mst{2} = 650\gev$, and $\cst = \pm 0.6$.  
As can be seen, for large gluino masses $\d\G/\G^0$ approaches $\sim-3\%$.  
The negative spike at $\msg=475$~GeV is due to the $\st_2\to t\sg$  
threshold. Notice, that this threshold is less pronounced for $\cst>0$. \\ 
The dependence on the stop mixing angle is shown  
in Figs.~\ref{fig:qcdwz3a} and \ref{fig:qcdwz3b}. 
\Fig{qcdwz3a} shows the tree--level width together with  
the ${\cal O}(\a_{s})$ corrected width of $\st_2\to\st_1 Z^0$  
as a function of $\cst$, for $\mst{1}=200\gev$, $\mst{2}=650\gev$,  
and $\msg=500\gev$.  
With the $\st_1\st_2 Z$ coupling proportional to  
$\sin 2\t_{\st}$ the decay width has maxima at  
$\cst=\pm \frac{1}{\sqrt{2}}$ (maximal mixing) and vanishes  
in case of no mixing ($\cst=0,\,\pm1$).  
SUSY--QCD corrections reduce the tree--level width by  
about $-5\%$ to $-10\%$.  
The relative correction $\d\G/\G^0$ for the parameters of \fig{qcdwz3a}  
can be seen explicetly in \fig{qcdwz3b} (black lines).  
In addition, we here also show the case $\msg=1$ TeV (gray lines).  
As the gluonic correction has the same $\t_{\st}$ dependence as  
the tree--level width (in our example: $\d\G^{(g)}/\G^0 = -3\%$)  
the $\t_{\st}$ dependence in \fig{qcdwz3b} comes only from the correction  
due to gluino exchange.  
For $\msg=500\gev$ and negative $\cst$ the correction  
is of the order of $-10\%$ while for positive $\cst$ it is roughly $-5\%$. 
For $\msg=1\tev$ the correction is about $-3\%$ to $-4\%$ with much less  
dependence on the stop mixing angle.  
Approaching $\cst=0$ or $\pm 1$, $\d\G/\G^0$ diverges  
because the tree--level coupling $c_{21Z}$ vanishes while $\d c_{21Z}\ne 0$.  
In this case the decay width becomes of ${\cal O}(\a_s^2)$.  
Note, however, that the appearance of this divergence, as well as the  
condition $\cst=\{0,\,\pm 1\}$, is renormalization scheme dependent.  
 
 
\noi 
The decay $\sb_2\to\sb_1 Z$ can be important for large $\tan\b$  
due to large mass splitting and mixing of $\sb_{L,R}$.  
The SUSY--QCD corrections to this decay are similar to those to  
$\st_2\to\st_1 Z$.  
However, the corrections due to gluino exchange are in general smaller  
and thus the dependece of $\d\G/\G^0$  
on the sbottom mixing angle is very weak. 
 
\clearpage

\noi 
We next turn to the squark decays into $W^\pm$ bosons.  
Here we discuss two special cases: \\ 
(i) $\sb_1$ and $\sb_2$ decaying into a relatively light $\st_1$ plus  
$W^-$ for small sbottom mixing (small $\tan\b$ scenario).  
In this case the mass difference of $\sb_1$ and $\sb_2$  
is expected to be rather small and thus the decays  
$\sb_2\to\sb_1\, (Z, h^0\!, H^0\!, A^0)$  
should be kinematically suppressed or even forbidden. \\ 
(ii) A heavy $\st_2$ decaying into a relatively light $\sb_1$ plus  
$W^+$ for large sbottom mixing (large $\tan\b$ scenario). \\ 
Note, that here we have to take into account that  
(for a given value of $\tan\b$) one of the parameters  
$\mst{1}$, $\mst{2}$, $\t_{\st}$, $\msb{1}$, $\msb{2}$, and $\t_{\sb}$  
is fixed by the others due to $SU(2)_L^{}$ gauge symmetry,  
see Sect.~\ref{sect:MQproblem}.  

\begin{figure}[p]\center 
\begin{picture}(97,65) 
\put(0,5){\mbox{\epsfig{figure=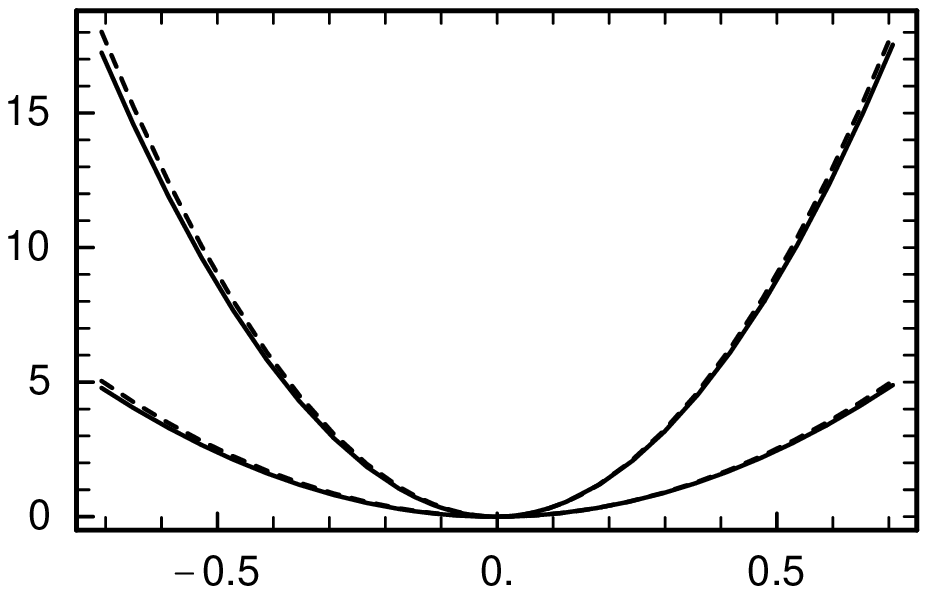,height=6cm}}} 
\put(-1,17){\makebox(0,0)[br]{\rotatebox{90}{$\G\,(\sb_i\to\st_1 W^-)$~[\%GeV]}}} 
\put(54,0){\makebox(0,0)[bc]{$\cst$}} 
\put(25,45){\makebox(0,0)[bl]{{\boldmath{$\sb_1$}}}} 
\put(20,25){\makebox(0,0)[bl]{{\boldmath{$\sb_2$}}}} 
\end{picture} 
\caption{Tree--level (dashed lines) and SUSY--QCD corrected (full lines)  
  widths of the decays $\sb_{1,2}\to \st_1 W^-$ as a function  
  of $\cst$, for $\msb{1}=500\gev$, $\msb{2} = 520\gev$, $\csb = -0.9$,  
  $\mst{1}=200\gev$, $\msg=520\gev$, and $\tan\b=2$. $\mst{2}$ is  
  a function of the other parameters and varies with $\cst$.} 
\label{fig:qcdwz4} 
\end{figure}

\noi
The decay widths of $\,\sb_{1,2}\to\st_1 W^-$ are shown in \fig{qcdwz4} 
as a function of the stop mixing angle, for $\msb{1}=500\gev$,  
$\msb{2}=520\gev$, $\csb=-0.9$ (at tree--level),  
$\mst{1}=200\gev$, $\msg=520\gev$, and $\tan\b=3$.  
The value of $\mst{2}$ is determined by the other parameters.  
Hence $\mst{2}$ varies from 533 GeV to 733 GeV depending on the stop  
mixing angle.  
Despite the larger phase space for the $\sb_{2}$ decay, the width of  
$\sb_{2}\to\st_1 W^-$ is smaller than that of $\sb_{1}\to\st_1 W^-$  
because the $W$ couples only to the ``left'' components of the squarks  
($\sb_{1}\sim\sb_{L}^{}$ and $\sb_{2}\sim\sb_{R}^{}$ for $\csb=-0.9$). 
For  $|\cst|\gsim 0.1$, the SUSY--QCD corrections  
change the tree--level widths by about $-11\%$ to $+4\%$.  
For $|\cst|\to 0$ $\d\G/\G^0$ again diverges because the tree--level  
coupling vanishes.  
The corrections slowly decrease with increasing gluino mass  
\eg, $\d\G/\G^0 \simeq -5\%$ to $-1\%$ for $\msg=1\tev$  
and the other parameters as in \fig{qcdwz4}.  
The dependence of $\d\G/\G^0$ on the sbottom mixing angle  
is, in general, much weaker than that on the stop mixing angle  
(apart from a singularity in case of $\sb_i=\sb_R$).  
The overall dependence on the gluino mass is in general similar to that  
of the $\st_2^{}\to\st_1^{}Z$ decay. 
However, the threshold effect at $\msg = \msb{i}-m_b$ is less pronounced. 

\begin{figure}[p]\center 
\begin{picture}(97,65) 
\put(0,5){\mbox{\epsfig{figure=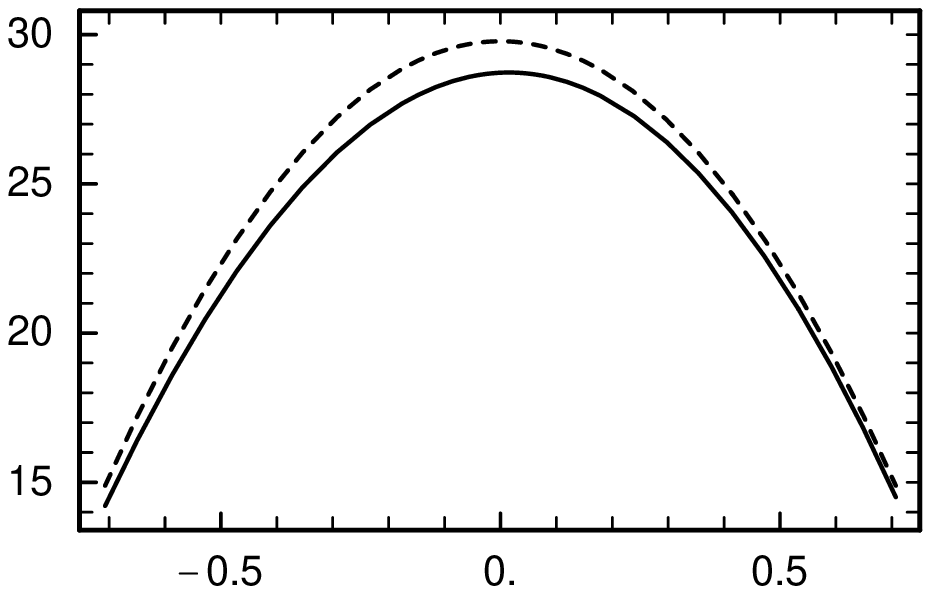,height=6cm}}} 
\put(0,19){\makebox(0,0)[br]{\rotatebox{90}{$\G\,(\st_2\to\sb_1 W^+)$~[GeV]}}} 
\put(54,0){\makebox(0,0)[bc]{$\cst$}} 
\end{picture}   
\caption{Tree--level (dashed line) and SUSY-QCD corrected (full line)  
  widths of the decay $\st_2\to \sb_1 W^+$ as a function  
  of $\cst$, for $\mst{1}=300\gev$, $\mst{2} = 650\gev$,  
  $\msb{1}=380\gev$, $\csb = -0.8$, $\msg=500\gev$, and $\tan\b=40$.  
  $\msb{2}$ is a function of the other parameters and varies  
  with $\cst$.} 
\label{fig:qcdwz6} 
\end{figure}

\noi
An example for large $\tan\b$ is shown in \fig{qcdwz6}.   
Here we plot the tree--level and the SUSY--QCD corrected widths  
of $\st_2\to\sb_1 W^+$ as a function of $\cst$ for $\mst{1}=300\gev$,  
$\mst{2}=650\gev$, $\msb{1}=380\gev$, $\csb=-0.8$, $\msg=500\gev$,  
and $\tan\b=40$.  
$\msb{2}$ is calculated from the other parameters and thus  
varies from 615 GeV to 918 GeV.   
As expected, the decay width is maximal for $\st_2=\st_L$ and vanishes  
for $\st_2=\st_R$. In the example chosen, the SUSY--QCD corrections are  
$-2.4\%$ to $-4.7\%$. 
For $\msg=1\tev$ (and the other parameters as above),  
they are about $-1\%$ to $-1.5\%$.  
Again, there is almost no dependence on $\csb$ apart from a  
singularity at $\csb=0$, \ie $\sb_1=\sb_R$.   
 
 
\noi 

\clearpage 
\section {Decays into Higgs Bosons}   \label{sect:qcdhx}

\begin{figure}[p]   
\bce 
\begin{picture}(145,155) 
\put(0,10){\mbox{\psfig{file=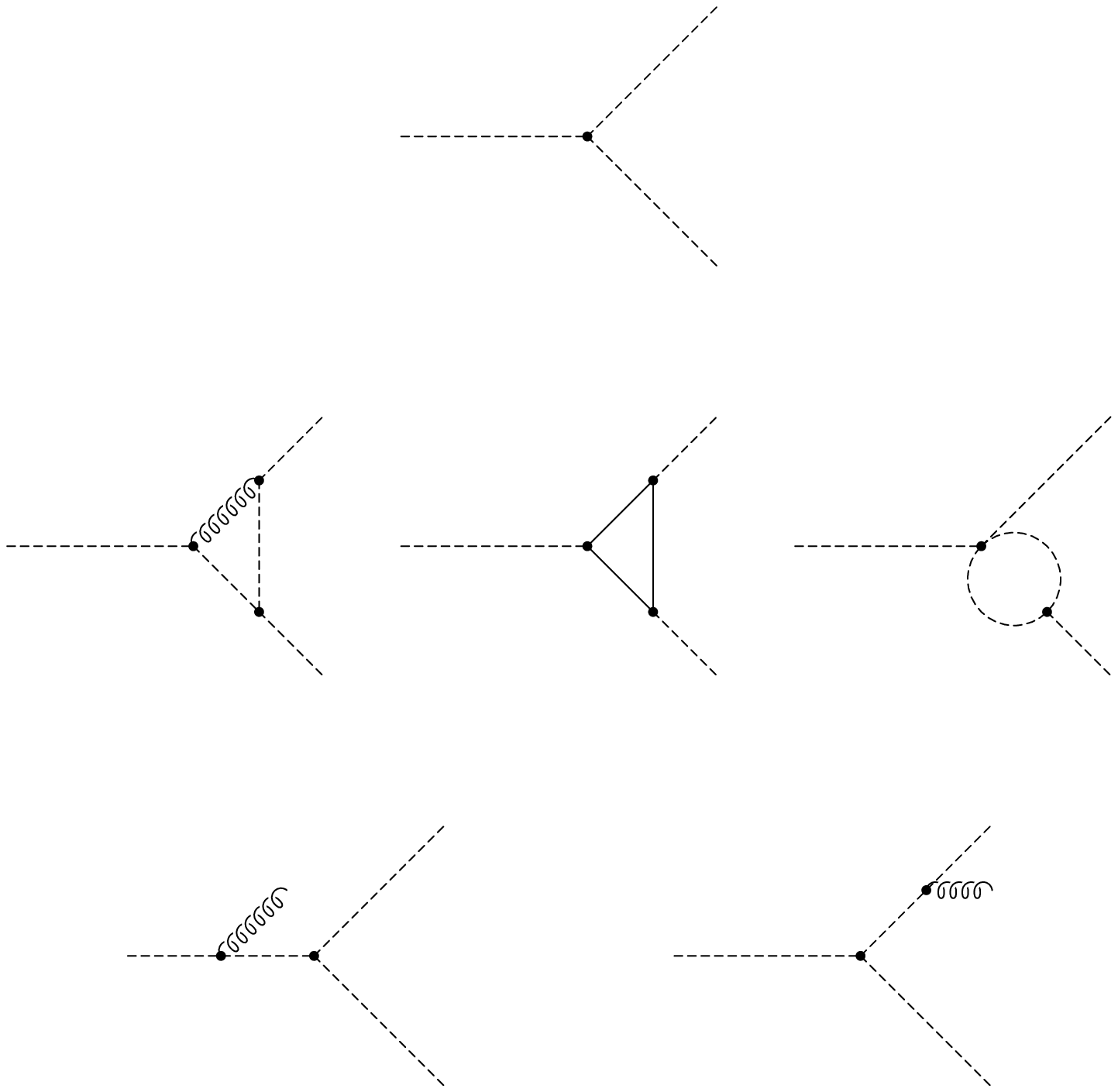}}} 
\put(75,111){\makebox(0,0)[tc]{\bf{(a)}}} 
\put(23,60){\makebox(0,0)[ct]{\bf{(b)}}}
\put(75,60){\makebox(0,0)[ct]{\bf{(c)}}}
\put(127,60){\makebox(0,0)[ct]{\bf{(d)}}} 
\put(39,6){\makebox(0,0)[ct]{\bf{(e)}}}
\put(111,6){\makebox(0,0)[ct]{\bf{(f)}}} 
%
%
\put(50,134){\makebox(0,0)[r]{$\sq_i^\a$}} 
\put(95,152){\makebox(0,0)[l]{$\sq_j^\b$}} 
\put(95,117){\makebox(0,0)[l]{$\Hk$}} 
%
%
\put(-1,81.5){\makebox(0,0)[r]{$\sq_i^\a$}} 
\put(44,99){\makebox(0,0)[l]{$\sq_j^\b$}} 
\put(44,63){\makebox(0,0)[l]{$\Hk$}} 
\put(28,89){\makebox(0,0)[r]{$g$}} 
\put(28,74){\makebox(0,0)[r]{$\sq_i^\a$}} 
\put(35,81.5){\makebox(0,0)[l]{$\sq_j^\b$}} 
\put(50,81.5){\makebox(0,0)[r]{$\sq_i^\a$}} 
\put(95,99){\makebox(0,0)[l]{$\sq_j^\b$}} 
\put(95,63){\makebox(0,0)[l]{$\Hk$}} 
\put(80,88){\makebox(0,0)[r]{$\sg$}} 
\put(80,74){\makebox(0,0)[r]{$q^\a$}} 
\put(87,81.5){\makebox(0,0)[l]{$q^\b$}} 
\put(102,81.5){\makebox(0,0)[r]{$\sq_i^\a$}} 
\put(147,99){\makebox(0,0)[l]{$\sq_j^\b$}} 
\put(147,63){\makebox(0,0)[l]{$\Hk$}} 
\put(125,74){\makebox(0,0)[tr]{$\sq_n^\a$}} 
\put(139,80){\makebox(0,0)[bl]{$\sq_m^\b$}}  
%
%
\put(14,28){\makebox(0,0)[r]{$\sq_i^\a$}} 
\put(60,45){\makebox(0,0)[l]{$\sq_j^\b$}} 
\put(60,9.5){\makebox(0,0)[l]{$\Hk$}} 
\put(39,37){\makebox(0,0)[bl]{$g$}} 
\put(86,28){\makebox(0,0)[r]{$\sq_i^\a$}} 
\put(132,45){\makebox(0,0)[l]{$\sq_j^\b$}} 
\put(132,9.5){\makebox(0,0)[l]{$\Hk$}} 
\put(132,35.5){\makebox(0,0)[bl]{$g$}}  
\end{picture} 
\ece 
\caption{Feynman diagrams for the ${\cal O}(\a_{s})$ SUSY-QCD  
  corrections to squark decays into Higgs bosons:
  (a) tree level, (b)--(f) vertex corrections, 
  (g)--(i) real gluon emission. For wave--function corrections 
  see Figs.~\ref{fig:fd-sqself} and \ref{fig:fd-qself}.} 
\label{fig:FD-qcdhx} 
\end{figure}

At tree level, amplitude and width of a squark decay into a Higgs boson, 
\fig{FD-qcdhx}\,a, are given by 
\begin{equation} 
  {\cal M}^0(\sq^\a_i \to \sq^\b_j \Hk) = i (G_k^{\,\a})_{ij}^{}
\label{eq:MtreeHx}   
\end{equation} 
and
\begin{equation} 
  \G^0(\sq^\a_i \to \sq^\b_j \Hk) = 
  \frac{|(G_k^{\,\a})_{ij}^{}|^2\;\kappa(m_i^2,m_j^2,m_{H_{k}}^2)
       }{16\pi m_i^3}  
\label{eq:GamtreeHx}   
\end{equation} 
where $m_i \equiv m_{\sq^\a_i}$ and $m_j \equiv m_{\sq^\b_j}$.  
$\a$ and $\b$ are flavour indices and  
$\Hk=\{h^0,H^0,A^0,H^\pm\}$, $k=1...4$. 
For $k=1,2,3$ we have of course $\a=\b$ and $i=2$, $j=1$.  
For $k=4$ we have $(\sq_{i}^{\alpha}, \sq_{j}^{\beta})$ = 
$(\st_{i},\sb_{j})$ or $(\sb_{i},\st_{j})$. 
The $H_{\!k}^\dagger\sq_j^{\b\dagger}\sq_i^\a$ couplings
$(G_k^\a)_{ij}$ are given in Sect.~\ref{sect:feyn-ssH}.
In the following, we will omit flavor indices when possible  
(flavor $=\a$ if not given otherwise).  

\noi
With the shift  
\begin{equation} 
  (G_k)_{ij}^{} \to  
  (G_k)_{ij}^{} + \d (G_k)_{ij}^{(v)} + \d (G_k)_{ij}^{(w)} + \d (G_k)_{ij}^{(c)} 
\label{eq:Gijkcorr} 
\end{equation} 
the ${\cal O}(\a_{s})$ corrected decay amplitude can  be expressed as  
\begin{equation} 
  \d\G^{(a)} (\sq_i^\a\to\sq_j^\b\Hk)= 
  \frac{ \kappa (m_i^2,m_j^2,m_{H_{k}}^2)}{8\pi\,m_i^3 }\;\, 
  \Re\left\{ (G_k)_{ij}^{\,*}\:\d (G_k)_{ij}^{(a)}\right\}  
  \hspace{6mm} 
  (a = v,\,w,\,c)  .
\end{equation}   
  
 
\noi 
The {\bf vertex correction} due to the gluon--squark--squark loop in  
\fig{FD-qcdhx}\,b is 
\begin{equation} 
  \d (G_k)_{ij}^{(v,g)} = \frac{\a_s}{3\pi}\: G_{ijk} \left[\, 
    B_0(m_i^2,0,m_i^2) + B_0(m_j^2,0,m_j^2)  
    - B_0(m_{H_{k}}^2,m_i^2,m_j^2) + 2\,X\,C_0 
  \,\right] 
\end{equation} 
with $X = m_i^2 + m_j^2 - m_{H_{k}}^2$ and  
$C_0 = C_0(m_i^2, m_{H_{k}}^2, m_j^2; \l^{2}, m_i^2, m_j^2)$.  
Again, we introduce a gluon mass $\l$ to regularize the  
infrared divergence.  
 
\noi 
The graph with the gluino--quark--quark loop in \fig{FD-qcdhx}\,c
leads to 
\begin{eqnarray} 
  \d (G_\ell)_{21}^{(v,\sg)} &=& 
  -\frac{2}{3}\frac{\a_s}{\pi}\;\msg\,\cos 2\tsq\:s_\ell^{\a}\,  
  \Big[\,B_0(\msq{2}^2,\msg^2,m_q^2) + B_0(\msq{1}^2,\msg^2,m_q^2) \nn\\ 
  & & \hspace{81mm} +\, (4\,m_q^2 - m_{\!H_\ell}^2)\,C_0 \,\Big]  
\label{eq:loopsgA} 
\end{eqnarray} 
for the decays into $h^0$ and $H^0$ ($\ell = 1,2$), 

\clearpage

\begin{equation}
\begin{split} 
  \d (G_3)_{21}^{(v,\sg)} =&  
  -\frac{2}{3}\frac{\a_s}{\pi}\;s_3^{\a} \,\left\{ 
    \,m_q\sin 2\tsq  
    \left[\, B_0(\msq{2}^2,\msg^2,m_q^2) - B_0(\msq{1}^2,\msg^2,m_q^2)   
             + (\msq{2}^2 - \msq{1}^2)\,C_0 
  \,\right] \right. \\ 
   & \hspace{23mm}\left. +\,\msg \left[\,  
    B_0(\msq{2}^2,\msg^2,m_q^2) + B_0(\msq{1}^2,\msg^2,m_q^2)  
    - m_{\!A}^2\,C_0 \,\right] \,\right\}, \hspace{-3cm}
\label{eq:loopsgB} 
\end{split} 
\end{equation}

for the decay into $A^0$, and  

\begin{eqnarray} 
\d (G_4)_{ij}^{(v,\sg)} &=& \frac{2}{3}\frac{\a_s}{\pi}\,\left\{\,  
  \left[\, (m_{q^\a} A_{11} + m_{q^\b} A_{22})\,s_4^\a 
          +(m_{q^\a} A_{22} + m_{q^\b} A_{11})\,s_4^\b 
  \,\right] B_0(m_{\!H^+}^2,m_t^2,m_b^2) \right. \nn\\ 
  & & \hspace{11mm}  
  +\,\left[\, (m_{q^\a} A_{11} - \msg A_{21})\,s_4^\a 
             +(m_{q^\a} A_{22} - \msg A_{12})\,s_4^\b 
  \,\right] B_0(m_i^2,\msg^2,m_{q^\a}^2) \nn\\  
  & & \hspace{11mm}  
  +\,\left[\, (m_{q^\b} A_{22} - \msg A_{21})\,s_4^\a 
             +(m_{q^\b} A_{11} - \msg A_{12})\,s_4^\b 
  \,\right] B_0(m_j^2,\msg^2,m_{q^\b}^2) \nn\\  
  & & \hspace{11mm}  
  +\,\left[\,m_{q^\b}\,(m_{q^\a}^2 - m_i^2 + \msg^2)\: 
                       (A_{22}\,s_4^\a + A_{11}\,s_4^\b) \right. \nn\\ 
  & & \hspace{22mm}  
          +\,m_{q^\a}\,(m_{q^\b}^2 - m_j^2 + \msg^2)\: 
                      (A_{11}\,s_4^\a + A_{22}\,s_4^\b)  \nn\\ 
  & & \hspace{22mm}    
          +\,\msg\,(m_{\!H^+}^2 - m_{q^\a}^2 - m_{q^\b}^2)\: 
                  (A_{21}\,s_4^\a + A_{12}\,s_4^\b) \nn\\ 
  & & \hspace{22mm} \left. \left.   
          -\,2\msg\,m_{q^\a}\,m_{q^\b}\:(A_{12}\,s_4^\a + A_{21}\,s_4^\b) 
  \right] C_0 \,\right\}  
\label{eq:loopsgC}   
\end{eqnarray} 
with $A_{nm} = \R_{in}^\a \R_{jm}^\b$ for the decay into a charged  
Higgs boson. \\ 
In Eqs.~(\ref{eq:loopsgA}) to (\ref{eq:loopsgC})  
$C_{0} = C_{0}(m_i^2, m_{H_{k}}^2, m_j^2; \msg^{2}, m_{q^\a}^2, m_{q^\b}^2)$.      
The factors $s_k^{q}$ are the Higgs couplings to quarks, see Eq.~\eq{qqHcop}.   
 
\noi 
The vertex correction due to the four--squark interaction  
in \fig{FD-qcdhx}\,d is 
\begin{equation} 
  \d (G_k)_{ij}^{(v,\sq)} = -\frac{\a_s}{3\pi} 
  \sum_{n,m=1,2}{\cal S}_{in}^{\a}\,{\cal S}_{jm}^{\b}\,G_{nmk}^{ }\, 
  B_0(m_{H_{k}}^2,m_{\sq_m^\b}^2,m_{\sq_n^\a}^2) 
\end{equation} 
with  
\begin{equation} 
  {\cal S}^\a_{in} =  \left( \begin{array}{rr}   
      \cos 2\tsq  & -\sin 2\tsq  \\ 
      -\sin 2\tsq & -\cos 2\tsq 
  \end{array} \right)_{in}^\a 
\end{equation} 
 
\noi 
The {\bf wave--function correction} is given by 
\begin{equation}  
  \d (G_k)_{ij}^{(w)} =  
  \onehf \left[  
    \d\ti Z_{ii}^{\alpha} + \d\ti Z_{jj}^{\beta} \right] (G_k^\a)_{ij} 
    + \d\ti Z_{i'i}^{\alpha}\,(G_k^\a)_{i'j}  
    + \d\ti Z_{j'j}^{\beta}\,(G_k^\a)_{ij'}\, ,   
  \hspace{8mm} {\footnotesize 
  \begin{array}{ll} i \not= i' \\ j \not= j' \end{array} }  
\end{equation} 
where the $\ti Z_{nm}^{\alpha}$ are the squark wave--function renormalization  
constants for $\sq^{\alpha}$.  
 
\noi 
We next fix the {\bf shift from the bare to the on--shell couplings} 
$\d (G_k)_{ij}^{(c)}$ in Eq.~\eq{Gijkcorr}.  
From Eqs.~\eq{Gksq} and \eq{Rsq} we get for the squark decays  
into $h^0$ or $H^0$ ($\ell=1,2$) 
\begin{eqnarray} 
  \d (G_\ell)_{21}^{\,\sq (c)} &=& \left[\, 
    \R^{\sq}\, \d \hat G_\ell^{\,\sq}\, (\R^{\sq})^{\rm T}  
    +\d\R^{\sq}\, \hat G_\ell^{\,\sq}\, (\R^{\sq})^{\rm T}  
    +\R^{\sq}\,   \hat G_\ell^{\,\sq}\, \d (\R^{\sq})^{\rm T}  
    \,\right]_{21} \nn \\[1mm] 
  &=&  
  \cos 2\tsq \,\left[\,\d \hat G_\ell^{\,\sq}\,\right]_{21}^{}  
  - \left[ (G_\ell^{\,\sq})_{11}^{} - (G_\ell^{\,\sq})_{22}^{} \right]\, \d\tsq \, 
\end{eqnarray} 
with $\d \hat G_\ell^{\,\sq}$ obtained by varying  
Eqs.~(\ref{eq:GLR1}) and (\ref{eq:GLR2}) \eg,  
\begin{eqnarray} 
  \d \hat G_2^{\,\st} &=& -\smaf{g}{2 m_W^{} {\rm s}_\b}  
     \left( \begin{array}{cc} 
        4 m_t\,{\rm s}_\a\,\d m_t  
             & \d (m_t A_t)\,{\rm s}_\a - \mu\,{\rm c}_\a\,\d m_t  \\ 
        \d (m_t A_t)\,{\rm s}_\a  - \mu\,{\rm c}_\a\,\d m_t  
             & 4 m_t\,{\rm s}_\a\,\d m_t 
     \end{array} \right)\, ,  
  \label{eq:dGLR2st} 
  \\[3mm] 
  \d \hat G_2^{\,\sb} &=& -\smaf{g}{2 m_W^{} {\rm c}_\b}  
     \left( \begin{array}{cc} 
        4 m_b {\rm c}_\a\d m_b   
             & \d (m_b A_b) {\rm c}_\a - \mu {\rm s}_\a \d m_b \\ 
        \d (m_b A_b) {\rm c}_\a - \mu {\rm s}_\a \d m_b   
             & 4 m_b {\rm c}_\a \d m_b 
     \end{array} \right)\,  
  \label{eq:dGLR2sb} 
\end{eqnarray} 
with ${\rm s}_\b = \sin\b$, ${\rm c}_\b = \cos\b$, 
${\rm s}_\a = \sin\a$, ${\rm c}_\a = \cos\a$,  
and $\a$ the Higgs mixing angle.  
$\d \hat G_{1}^{\,\sq}$ is obtained from 
Eqs.~\eq{dGLR2st} and \eq{dGLR2sb} by 
\begin{equation} 
  \d \hat G_{1}^{\,\sq} = \left( 
  \d \hat G_{2}^{\,\sq} \;\mbox{with}\; \a \to \a + \pi/2 \right)\, .  
\end{equation} 
For the couplings to the $A^0$ boson we have explicitly  
\begin{equation}  
  \d (G_3^{\,\sq})_{21}^{\,(c)} = \smaf{ig}{2 m_W} \,\left[\, 
    \d(m_q A_q)\, \{\cot\b, \tan\b\} + \mu\, \d m_q \,\right], 
\end{equation} 
where $\cot\b$ ($\tan\b$) is for $\sq = \st$ ($\sb$).  
For the decay $\st_i \to \sb_j\, H^+$ ($k = 4$) we get 
\begin{equation}  
  \d (G_4^{\,\st})_{ij}^{\,(c)} =  
  \left[\,\R^{\st}\, \d \hat G_4\, (\R^{\sb})^{\rm T} \,\right]_{ij} 
  - (-1)^i\, (G_4^{\,\st})_{i'j}^{}\, \d\t_{\st}  
  - (-1)^j\, (G_4^{\,\st})_{ij'}^{}\, \d\t_{\sb} \, , 
  \hspace{8mm} {\footnotesize 
  \begin{array}{ll} i \not= i' \\ j \not= j' \end{array} }  
  \hspace{-20mm}
\end{equation} 
with  
\begin{equation} 
  \d \hat G_4 = \smaf{g}{\sqrt{2}\,m_W^{}}  
  \left( \begin{array}{cc} 
     2 m_b \d m_b\tan\b + 2 m_t \d m_t \cot\b   
     & \d (m_b A_b)\tan\b + \mu\, \d m_b \\ 
     \d (m_t A_t)\cot\b + \mu\,\d m_t    
     & 2(\d m_t m_b + m_t \d m_b)/\sin 2\b 
  \end{array}\right) , 
\end{equation} 
and analogously the expression 
for $\sb_i \to \st_j\, H^-$ according to \eq{G4sq}. 
 
\noi 
For the renormalization of $\,m_q A_q$ and the squark mixing  
angels $\tsq$ we use the prescriptions described in  
Sect.~\ref{sect:subtleties}. 
 
\noi 
In order to cancel the infrared divergence we again  
include the {\bf emission of real gluons} (\fig{FD-qcdhx}\,e):  
\begin{eqnarray} 
  \d\G_{real} &\equiv& \G(\sq_i^\a \to \sq_j^\b\,\Hk\, g) \nn\\[2mm] 
  &=& -\:\frac{\a_s\,|(G_k^{\,\a})_{ij}^{}|^2}{3\pi^2 m_i}\: \left[\,  
    I_0 + I_1 + m_i^2\,I_{00} + m_j^2\,I_{11} + X\,I_{01} \right] \, . 
\end{eqnarray} 
Again, $X = m_i^2 + m_j^2 - m_{H_{k}}^2$. 
The phase space integrals $I_{n}$, and $I_{nm}$  
have $(m_i, m_j, m_{H_{k}})$ as arguments. 
 
 
Let us now turn to the {\bf numerical analysis}.   
As input parameters we use $\mst{1}$, $\mst{2}$, $\cos\t_{\st}$,   
$\tan\beta$, $\mu$, $m_A$, and $\msg$.  
From these we calculate the values of the  
soft SUSY--breaking parameters $M_{\ti Q}(\st)$, $M_{\ti Q}(\sb)$,  
$M_{\ti U, \ti D}$, and $A_{t,b}$, taking $M_{\ti D}=1.12\,M_{\ti Q}(\st)$ 
and $A_b=A_t$ for simplicity.    
  
\noi
As a reference point we take  
$\mst{1}=250$ GeV, $\mst{2}=600$ GeV,  
$\cos\t_{\st}=0.26$ ($\t_{\st}\simeq 75^\circ$),  
$\tan\b=3$, $\mu=550$ GeV, $m_A=150$ GeV, and $\msg=600$ GeV.  
This leads to $\msb{1}=563$ GeV \footnote{Notice that  
at tree level one has $\msb{1}=560$ GeV because  
$M_{\ti Q}=558$ GeV for both the $\st$ and $\sb$ mass matrices.   
At $\Oas$, however, one gets $M_{\ti Q}(\st)=558$ GeV and  
$M_{\ti Q}(\sb)=561$ GeV.},  
$\msb{2}=627$ GeV, $\cos\t_{\sb}=0.99$, $A_{t,b}=-243$ GeV,  
$m_{h^0}=100$ GeV, $m_{H^0}=162$ GeV,  
$\sin\a = -0.58$, and $m_{H^+}=164$ GeV.  
Thus $\st_2$ can decay into $\st_1 + (h^0,H^0,A^0)$, and  
$\sb_{1,2}$ can decay into $\st_1 H^-$. 
 
\begin{figure}[h!]
\bce\begin{picture}(95,65)
\put(0,5){\mbox{\epsfig{file=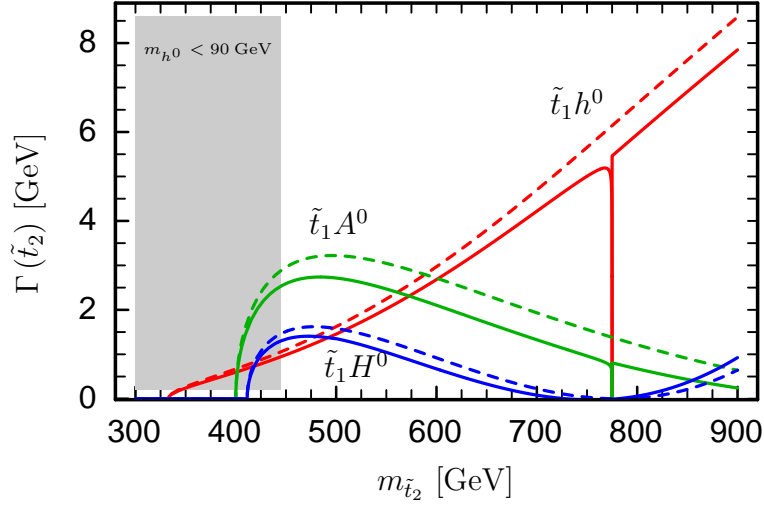,height=60mm}}}
\put(44,0){\mbox{$\mst{2}~[{\rm GeV}]$}}
\put(0,26){\makebox(0,0)[br]{{\rotatebox{90}{$\G\,(\st_2)~[{\rm GeV}]$}}}}
\put(67,50){\mbox{$\st_1 h^0$}}
\put(37,15){\mbox{$\st_1 H^0$}}
\put(35,34){\mbox{$\st_1 A^0$}}
\put(13,58){\mbox{\tiny $m_{h^0}<90\gev$}}
\end{picture}\ece
\caption{Tree--level (dashed lines) and $\Oas$ SUSY--QCD corrected 
(full lines) decay widths of $\st_2 \to \st_1 + (h^0\!,\: H^0\!,\: A^0)$ 
as a function of $\mst{2}$, for $\mst{1}=250\gev$, $\cos\t_{\st}=0.26$, 
$\mu=550\gev$, $\tan\beta=3$, $m_A=150\gev$, and $\msg=600\gev$. 
The grey area is excluded by the bound $m_{h^0}<90\gev$.}
\label{fig:sthxmst2}
\end{figure}

\begin{figure}[h!]
\bce\begin{picture}(95,65)
\put(0,5){\mbox{\epsfig{file=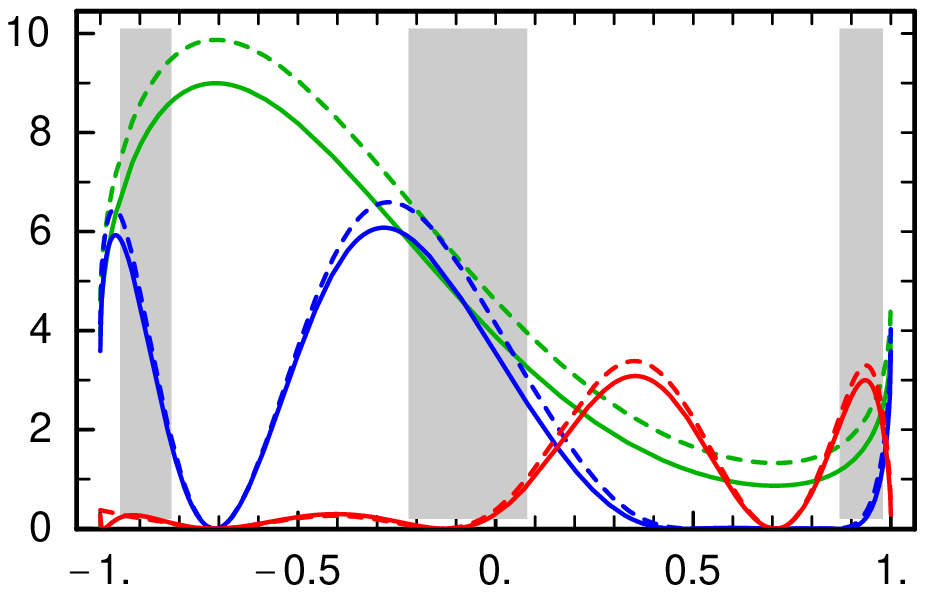,height=60mm}}}
\put(49,0){\mbox{$\cos\t_{\st}$}}
\put(0,26){\makebox(0,0)[br]{{\rotatebox{90}{$\G\,(\st_2)~[{\rm GeV}]$}}}}
\put(70,30){\mbox{$\st_1 h^0$}}
\put(35,25){\mbox{$\st_1 H^0$}}
\put(36,58){\mbox{$\st_1 A^0$}}
\put(16,60){\mbox{\tiny\bf (a)}}
\put(52.5,60){\mbox{\tiny\bf (b)}}
\put(89.3,60){\mbox{\tiny\bf (c)}}
\end{picture}\ece
\caption{Tree--level (dashed lines) and $\Oas$ SUSY--QCD corrected 
(full lines) decay widths of $\st_2 \to \st_1 + (h^0\!,\: H^0\!,\: A^0)$ 
as a function of $\cos\t_{\st}$, for $\mst{1}=250\gev$, $\mst{2}=600\gev$, 
$\mu=550\gev$, $\tan\beta=3$, $m_A=150\gev$, and $\msg=600\gev$.
The grey areas are excluded by the constraints given in 
Sect.~\ref{sect:convention}: $\d\rho(\st \mbox{--} \sb) > 0.0012$ in (a),
$m_{h^0}<90\gev$ in (b), and unstable vacuum in (c).}
\label{fig:sthxcosth}
\end{figure}

\begin{figure}[h!]
\bce\begin{picture}(95,65)
\put(0,5){\mbox{\epsfig{file=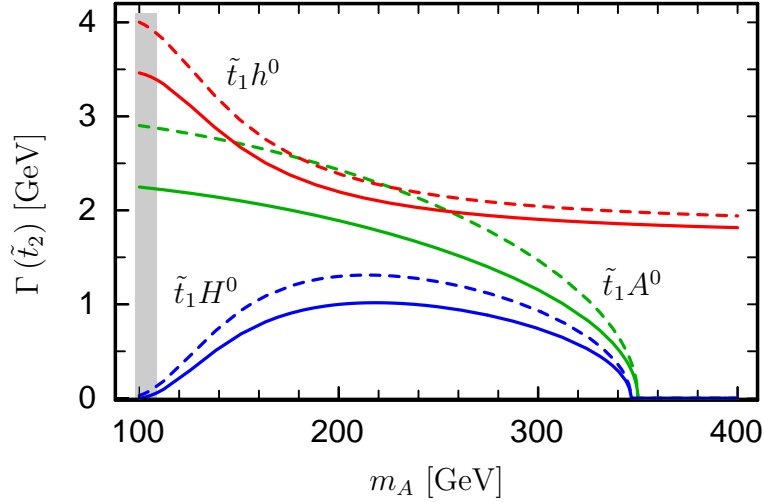,height=60mm}}}
\put(43,0){\mbox{$m_A~[{\rm GeV}]$}}
\put(0,26){\makebox(0,0)[br]{{\rotatebox{90}{$\G\,(\st_2)~[{\rm GeV}]$}}}}
\put(24,54){\mbox{$\st_1 h^0$}}
\put(17,25){\mbox{$\st_1 H^0$}}
\put(74,26){\mbox{$\st_1 A^0$}}
\end{picture}\ece
\caption{Tree--level (dashed lines) and $\Oas$ SUSY--QCD corrected 
(full lines) decay widths of $\st_2 \to \st_1 + (h^0\!,\: H^0\!,\: A^0)$ 
as a function of $m_A$, for $\mst{1}=250\gev$, $\mst{2}=600\gev$, 
$\cos\t_{\st}=0.26$, $\mu=550\gev$, $\tan\beta=3$, and $\msg=600\gev$. 
The grey area is excluded by the bound $m_{h^0}>90\gev$.}
\label{fig:sthxmA}
\end{figure}

\begin{figure}[h!] 
\bce\begin{picture}(95,65) 
\put(0,5){\mbox{\epsfig{file=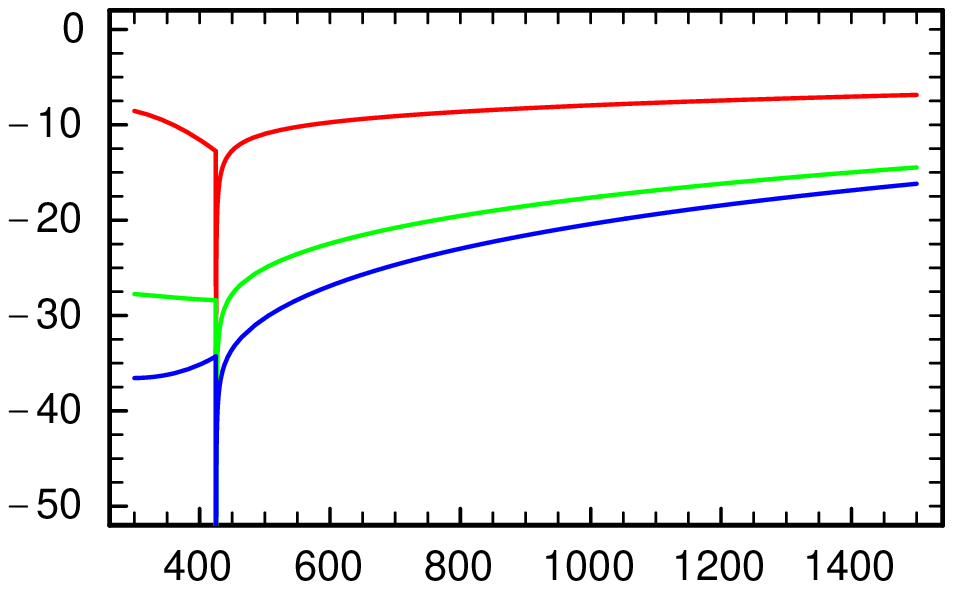,height=60mm}}} 
\put(47,0){\mbox{$\msg~[{\rm GeV}]$}} 
\put(0,24){\makebox(0,0)[br]{{\rotatebox{90}{$\d\G/\G^0\,(\st_2)~[\%]$}}}} 
\put(27,54){\mbox{$\st_1 h^0$}} 
\put(52,36){\mbox{$\st_1 H^0$}} 
\put(39,45){\mbox{$\st_1 A^0$}} 
\end{picture}\ece 
\caption{$\Oas$ SUSY--QCD corrections (in \%) to the widths of   
$\st_2 \to \st_1 + (h^0\!,\: H^0\!,\: A^0)$  
as a function of $\msg$, for $\mst{1}=250\gev$, $\mst{2}=600\gev$,  
$\cos\t_{\st}=0.26$, $\mu=550\gev$, $\tan\beta=3$, and $m_A=150\gev$.} 
\label{fig:sthxmsg} 
\end{figure}

\noi 
We first discuss the parameter dependence of the widths of 
$\st_2$ decays into neutral Higgs bosons by varying 
one of the input parameters of the reference point.  
\Fig{sthxmst2} shows the tree--level and the SUSY--QCD  
corrected widths of the decays $\st_2\to\st_1+(h^0,H^0,A^0)$  
as a function of $\mst{2}$.  
The relative corrections $\d\G/\G^0 \equiv (\G-\G^0)/\G^0$ are about  
$-10\%$ for the decay into $h^0$ and $-9\%$ to $-62\%$ 
for the decay into $A^0$.  
The corrections for $\st_2\to\st_1 H^0$ are  
$-9\%$, $-45\%$, and $+45\%$ for  
$\mst{2} =420, 670$, and $900$~GeV, respectively.  
The spikes in the corrected decay widths for $\mst{2}=775$ GeV are due  
to the $\st_2\to t\sg$ threshold. 
The different shapes of the decay widths can be understood by the  
wide range of the parameters entering the Higgs couplings to stops.  
In the range $\mst{2}=300$ GeV to 900 GeV, we have  
$A_t=144$ GeV to $-889$ GeV and $\sin\a=-0.52$ to $-0.73$  
($m_{h^0}=81$ GeV to 114 GeV, and $m_{H^0}=163$ GeV to 170 GeV). \\ 
\Fig{sthxcosth} shows the $\cos\theta_{\st}$  
dependence of the tree level and the  
corrected widths of $\st_2\to\st_1+(h^0,H^0,A^0)$ decays.  
Again the shapes of the decay widths reflect 
their dependence on the underlying SUSY parameters in a  
characteristic way.  
In particular we have  
$A_t= 1033$, $183$, $-666$~GeV and  
$\sin\a= -0.748$, $-0.565$, $-0.726$ for  
$\cos\t_{\st}=-0.7$, $0$, $0.7$, respectively.  
Apart from the points where the tree--level  
decay amplitudes vanish the relative corrections  
range from $-40\%$ to $20\%$. \\ 
In \fig{sthxmA} we show the tree--level and the SUSY--QCD corrected decay  
widths as a function of $m_A$.  
For $m_A=100$, 200, 300 GeV we have  
$m_{h^0}=85$, 104, 105 GeV,  
$m_{H^0}=128$, 207, 304 GeV, and  
$\sin\a=-0.87$, $-0.45$, $-0.37$, respectively.  
The corrections to $\G^0(\st_2\to\st_1 h^0)$ range  
from $-15\%$ to $-7\%$ for $m_A=100$ GeV to 400 GeV.  
Those to $\G^0(\st_2\to\st_1 H^0)$ are $-50\%$ to $-22\%$  
for $m_A\gsim 114$ GeV, and those to $\G^0(\st_2\to\st_1 A^0)$ are  
about $-25\%$. \\ 
As for the dependence on the gluino mass, the gluino decouples very 
slowly, as can be seen in \fig{sthxmsg}:  
In the range $\msg=300$ GeV to 1500 GeV $\d\G/\G^0$ varies from  
($-9\%$, $-37\%$, $-28\%$) to ($-7\%$, $-16\%$, $-14\%$)  
for the decays  
$\st_2\to\st_1+(h^0,H^0,A^0)$, apart from the  
$\st_2\to t\sg$ threshold at $\msg=425$ GeV. \\ 
As for the dependence on $\tan\b$, we get  
$\G(\st_2\to\st_1 h^0) = 2.68$, $2.09$, $1.42$~GeV with  
$\d\G/\G^0 \simeq -10\%$, $-7\%$, $-5\%$ for $\tan\b= 3, 10$, $30$, 
respectively.  
Likewise, we get  
$\G = 0.67$, $1.61$, $2.45$~GeV with  
$\d\G/\G^0 \simeq -27\%$, $-19\%$, $-17\%$  
for the decay into $H^0$ and  
$\G = 2.1$, $2.64$, $2.92$~GeV with  
$\d\G/\G^0 \simeq -22\%$, $-19\%$, $-18\%$ 
for the decay into $A^0$, respectively. 
 
\begin{figure}[h!]
\bce\begin{picture}(95,65)
\put(0,5){\mbox{\epsfig{file=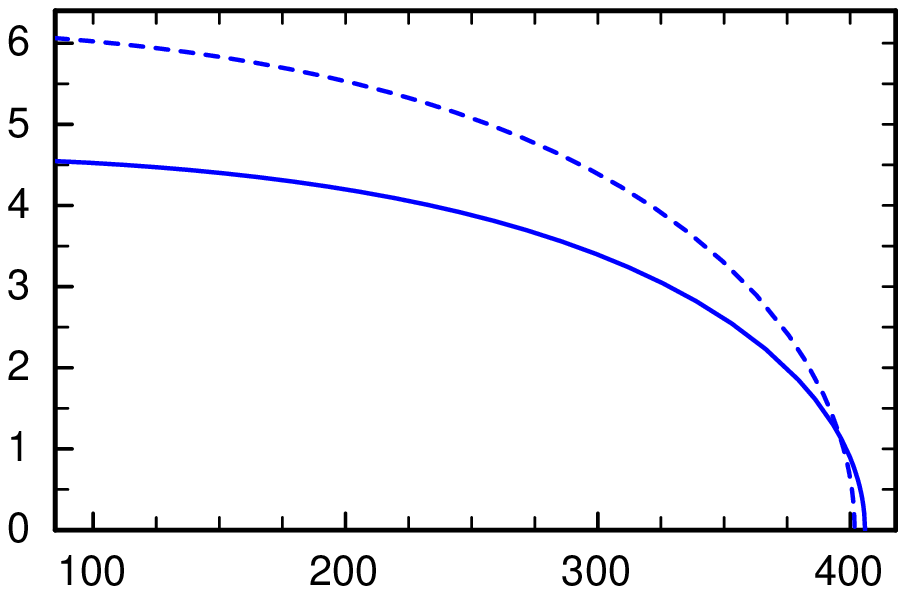,height=60mm}}}
\put(18,18){\mbox{\epsfig{file=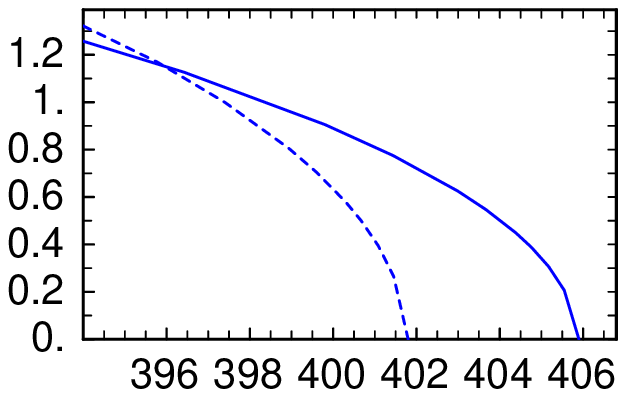,height=17mm}}}
\put(43.5,0){\mbox{$\mst{1}~[{\rm GeV}]$}}
\put(0,20){\makebox(0,0)[br]{{\rotatebox{90}{$\G\,(\sb_1\to\st_1H^-)~[{\rm GeV}]$}}}}
\end{picture}\ece
\caption{Tree--level (dashed line) and $\Oas$ SUSY--QCD corrected 
(full line) decay widths of $\sb_1 \to \st_1 H^-$ 
as a function of $\mst{1}$, for $\mst{2}=600\gev$, $\cos\t_{\st}=0.26$, 
$\mu=550\gev$, $\tan\beta=3$, $m_A=150\gev$, and $\msg=600\gev$. 
The insert zooms on the different thresholds at tree-- and one--loop level.}
\label{fig:sbhxmst1} 
\end{figure}

\begin{figure}[h!] 
\bce\begin{picture}(95,65) 
\put(0,5){\mbox{\epsfig{file=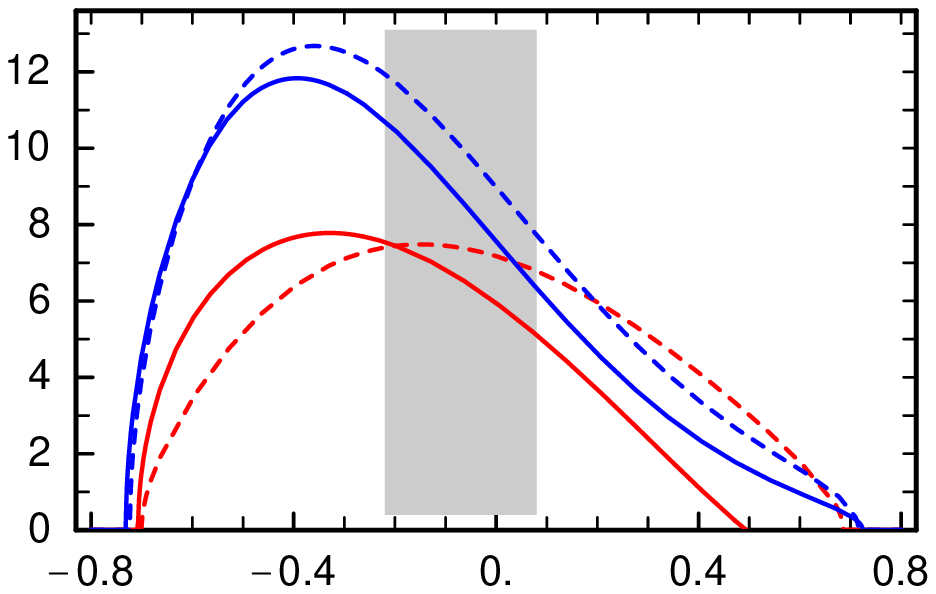,height=60mm}}} 
\put(48.3,0){\mbox{$\cos\t_{\st}$}} 
\put(0,20){\makebox(0,0)[br]{{\rotatebox{90}{$\G\,(\sb_1\to\st_1H^-)~[{\rm GeV}]$}}}} 
\put(26,50){\mbox{\footnotesize$\tan\b=3$}}
\put(24,24){\mbox{\footnotesize$\tan\b=10$}}
\end{picture}\ece 
\caption{Tree--level (dashed lines) and $\Oas$ SUSY--QCD corrected  
(full lines) decay widths of $\sb_1 \to \st_1 H^-$  
as a function of $\cos\t_{\st}$, for $\mst{1}=250\gev$, $\mst{2}=600\gev$,  
$\mu=550\gev$, $m_A=150\gev$, $\msg=600\gev$, and $\tan\beta=3,\,10$. 
The grey area is excluded for $\tan\b=3$ by the bound $m_{h^0}>90\gev$.} 
\label{fig:sbhxcosth} 
\end{figure}

\noi 
Let us now turn to the sbottom decays.   
We start again from the reference point given above.  
For the decay $\sb_1 \to \st_1 H^-$ we get 
$\G=3.88$ GeV with $\d\G/\G^0=-24\%$, and 
for the decay $\sb_2 \to \st_1 H^-$ we get $\G=0.08$ GeV with   
$\d\G/\G^0=+87\%$.
As in our examples the width of the latter decay 
is usually quite small (because $\sb_2\simeq\sb_R$ and
$\st_1 \sim \st_R$) 
we will discuss only the parameter dependence of the $\sb_1$ decay. \\ 
\Fig{sbhxmst1} shows the tree--level and the SUSY--QCD 
corrected widths of this decay as a function of $\mst{1}$. 
The SUSY--QCD corrections are about $-25\%$.
Notice that at tree--level we have
$\msb{1}=556\gev$ to 566 GeV for $\mst{1}=85$~GeV to 400~GeV,
whereas at $\Oas$ we have $\msb{1}=561\gev$ to 570 GeV. 
Therefore, the thresholds at tree--level and one--loop level are 
slightly different. \\ 
The dependence on the stop mixing angle is shown in \fig{sbhxcosth}  
for $\tan\b=3$ and 10, and the other parameters as given above. 
(For $|\cos\t_{\st}|\gsim 0.72$ the decay $\sb_1\to\st_1 H^-$ is  
kinematically not allowed.) 
In case of $\tan\b=3$, the SUSY--QCD corrections range  
from about $-40\%$ to $26\%$, 
with $\d\G/\G^0 > 0$ for $\cos\t_{\st} \lsim -0.6$. 
In case of $\tan\b=10$ $\d\G/\G^0$ is much larger. 
For $\cos\t_{\st} \gsim 0.5$ and $\tan\b=10$ 
we even get a negative corrected decay width. 
This is mainly due to a large contribution stemming from the term 
$\d(m_b A_b)\sim \mu\tan\b \,\d m_b$ of Eq.~\eq{dmqaq} 
and was already mentioned in \cite{higgsdec}. 
The same problem can occur for the decays $\sb_2\to\sb_1+(h^0,H^0,A^0)$ 
and $\st_2\to\sb_1 H^+$ which may be important for large $\tan\b$  
(due to the large bottom Yukawa coupling and the 
large $\sb_1$--$\sb_2$ mass splitting).\\  
We have also studied the dependence on $m_A$. 
In the case $\tan\b=3$ (10) we have found that 
$\d\G/\G^0(\sb_1\to\st_1 H^-) \sim -20\%$ $(-40\%)$ 
for $m_A=100$ GeV to 285 GeV and the other parameters as given above.\\
As for the dependence on the gluino mass, 
$\d\G/\G^0(\sb_1\to\st_1H^-)$ ranges from $-26\%$ to $-14\%$ 
($-47\%$ to $-39\%$) for $\msg=300$ GeV to 1500 GeV and $\tan\b=3$ (10).

\begin{figure}[h!] 
\bce\begin{picture}(100,65) 
\put(1,5){\mbox{\epsfig{file=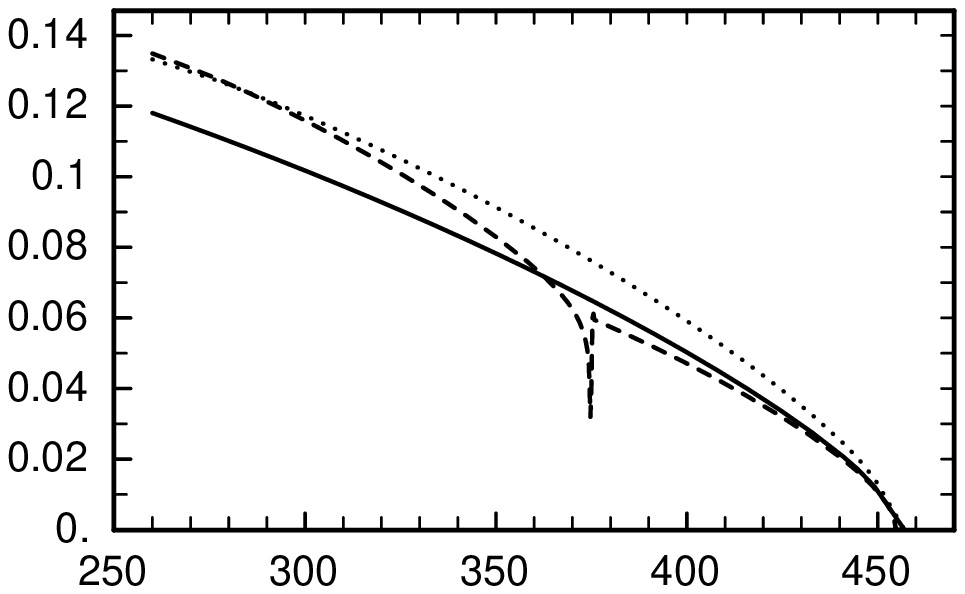,height=60mm}}} 
\put(48,0){\mbox{$\mst{2}~[{\rm GeV}]$}}
\put(0,21.5){\makebox(0,0)[br]{{\rotatebox{90}{$\G\,(\st_2\to\st_1 h^0)~[{\rm GeV}]$}}}} 
\end{picture}\ece 
\caption{Tree--level (full line) and $\Oas$ SUSY--QCD corrected  
(dashed and dotted lines) decay widths of $\st_2\to\st_1 h^0$  
as a function of $\mst{1}$ for the parameters used in Fig.~6\,a of 
Ref.~\cite{qcdhx-djouadi}. In our notation: $M_{\ti U}=500\gev$, 
$A_t=\mu=300\gev$, $\tan\b=1.6$, and $m_A=400\gev$.
The dashed line is for $\msg=500\gev$ 
and the dotted one for $\msg=1000\gev$. } 
\label{fig:sqh0comparison} 
\end{figure}

In Ref.~\cite{qcdhx-djouadi} a numerical analysis was made 
for the decays $\st_2\to \st_1 + (h^0\!,\,A^0)$.  
Whereas we fairly agree with their Fig.~6\,b for $\st_2\to\st_1 A^0$ 
(apart from the fact that the spike is due to the $\st_1\to t\sg$ threshold), 
we find a difference of about 10\% in both the tree--level and 
the corrected widths of $\st_2\to\st_1 h^0$ (their Fig.~6\,a). 
This may be due to a different treatment of the radiative corrections 
to the $h^0$ mass and mixing angle $\alpha$ \cite{dh-private}. 
For comparison, we show in \fig{sqh0comparison} our results on 
the $\st_2\to\st_1 h^0$ decay for the parameters used in 
Fig.~6\,a of Ref.~\cite{qcdhx-djouadi}. 
In our notation these parameters are: $M_{\ti U}=500\gev$, 
$A_t=\mu=300\gev$, $\tan\b=1.6$, and $m_A=400\gev$. Moreover, 
$\a_s\equiv 0.12$. The shown range of $\mst{1}$ is obtained 
by varying $M_{\ti Q}$ from 200 to 430 GeV. 
In this range, we get $\mst{2}=531$ to 534 GeV, $\cst\simeq -1$, 
$m_{h^0}=72$ to 80 GeV, and $\sin\a\simeq -0.56$. 
We have checked our results with two independent numerical programs.
As a sidemark, we here also note that in the case of $\sq_i\to\sq_j'H^\pm$   
Eq.~(43) of Ref.~\cite{qcdhx-djouadi} is only correct if one 
takes the transposed couplings. 
 
 
\clearpage
\section{Branching Ratios} \label{sect:branch}

\begin{figure}[b!]
\bce\begin{picture}(97,65)
\put(0,5){\mbox{\epsfig{file=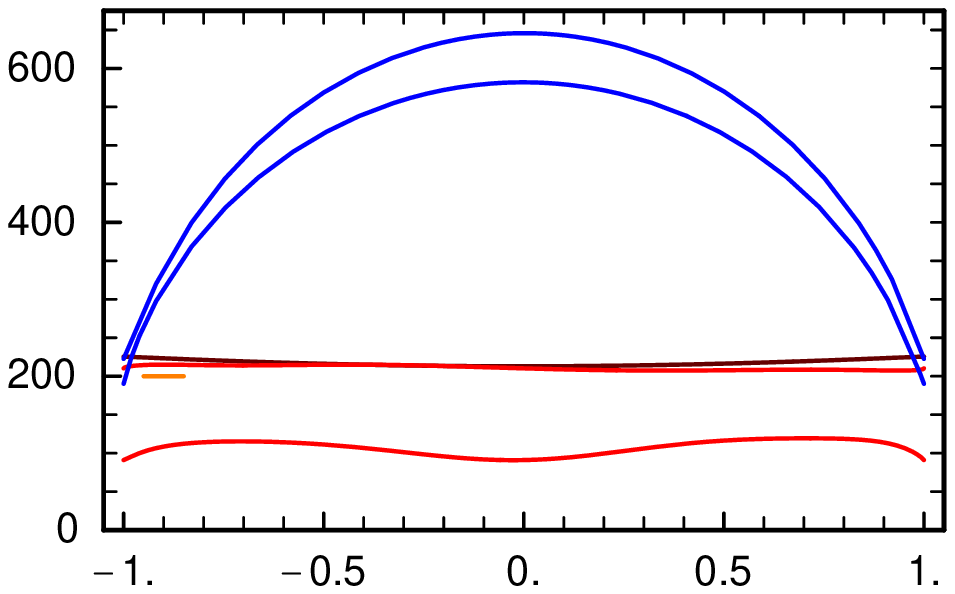,height=60mm}}}
\put(50,0){\mbox{$\cst$}}
\put(0,29){\makebox(0,0)[br]{{\rotatebox{90}{$m$~[GeV]}}}}
\put(26,54){\mbox{\footnotesize $\sb_2$}}
\put(33,46.5){\mbox{\footnotesize $\sb_1$}}
\put(85,30.6){\mbox{\footnotesize $H^\pm$}}
\put(81,24.3){\mbox{\footnotesize $H^0$}}
\put(30,22){\mbox{\footnotesize $h^0$}}
\put(21,25.3){\mbox{\scriptsize $A^0$}}
\end{picture}\ece
\caption{Sbottom and Higgs boson masses as a function of $\cst$ 
for $\mst{1}=250\gev$, $\mst{2}=600\gev$, $\mu=500\gev$, $\tan\b=3$, 
and $m_A=200\gev$; $M_{\ti D}=1.12 M_{\ti Q}(\st)$ and $A_t=A_b$.}
\label{fig:BR-scen}
\end{figure}

In this section we present the SUSY--QCD corrected branching ratios 
of stop and sbottom decays. 
We proceed as before, taking~
$\mst{1}$, $\mst{2}$, $\cst$, $\tan\b$, $\mu$, $M$, and $m_A$ 
as input parameters. 
The sbottom sector is fixed by 
$M_{\ti D}=1.12 M_{\ti Q}(\st)$ and $A_t=A_b$. 
\Fig{BR-scen} shows the sbottom and Higgs boson masses obtained in this way 
as a function of the stop mixing angle for $\mst{1}=250\gev$, $\mst{2}=600\gev$,   
$\mu=500\gev$, $\tan\b=3$, and $m_A=200\gev$. 

\begin{figure}[p]
\bce\begin{picture}(95,215)
\put(0,155){\mbox{\epsfig{file=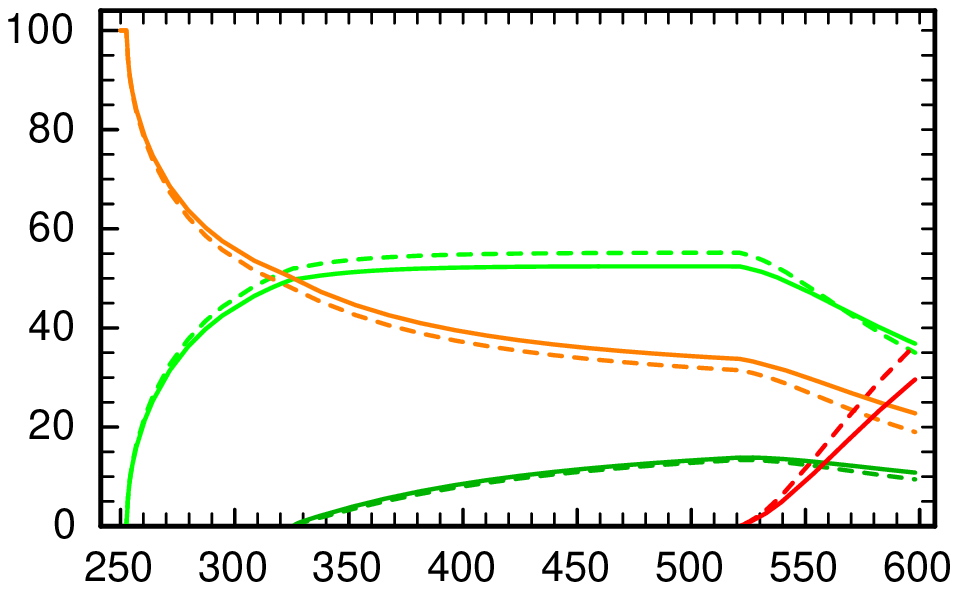,height=60mm}}}
\put(0,80){\mbox{\epsfig{file=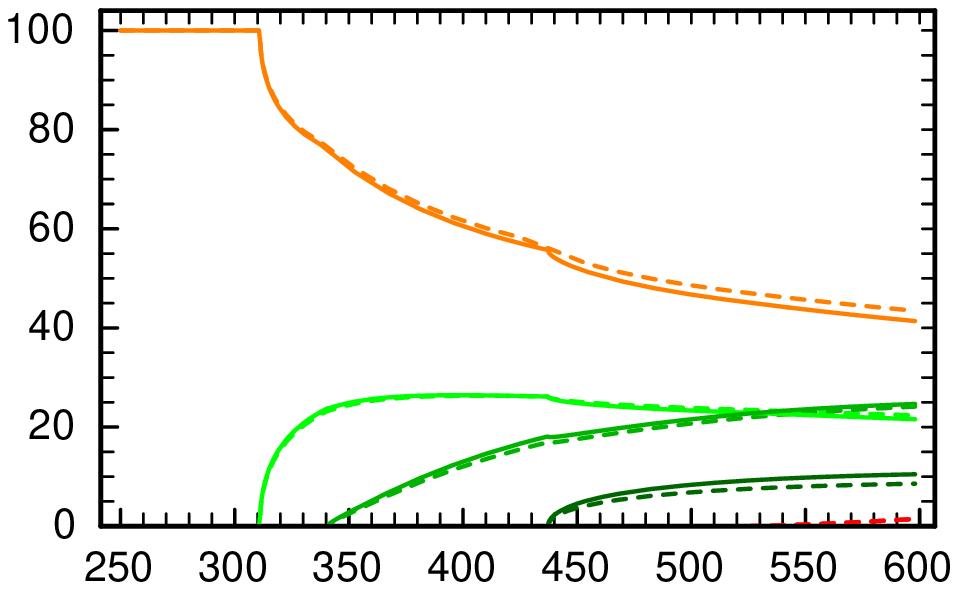,height=60mm}}}
\put(0,5){\mbox{\epsfig{file=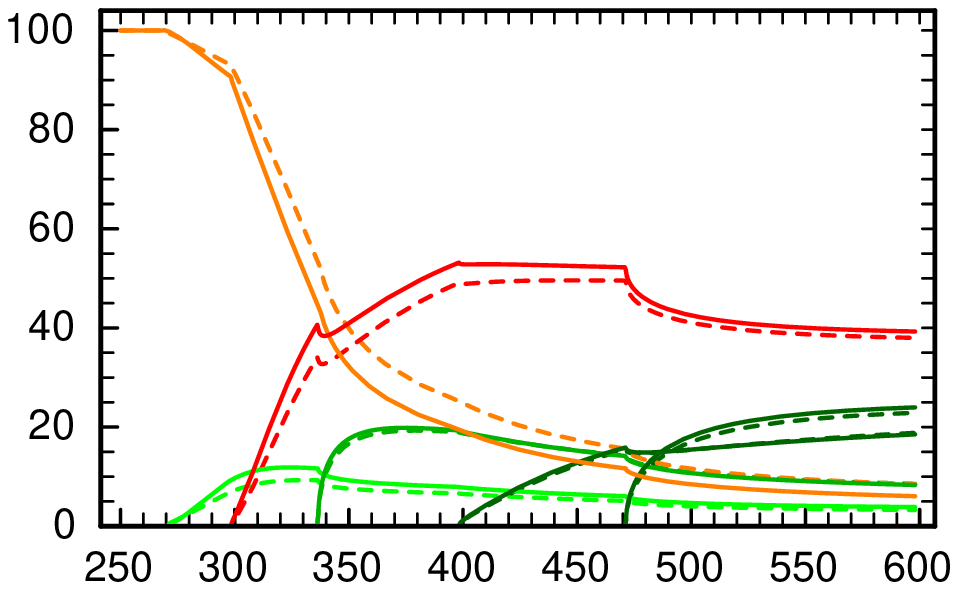,height=60mm}}}
\put(45,150){\mbox{$\mst{1}~[{\rm GeV}]$}}
\put(45,75){\mbox{$\mst{1}~[{\rm GeV}]$}}
\put(45,0){\mbox{$\mst{1}~[{\rm GeV}]$}}
\put(0,178){\makebox(0,0)[br]{{\rotatebox{90}{BR~$(\st_1)$~[\%]}}}}
\put(0,103){\makebox(0,0)[br]{{\rotatebox{90}{BR~$(\st_1)$~[\%]}}}}
\put(0,28){\makebox(0,0)[br]{{\rotatebox{90}{BR~$(\st_1)$~[\%]}}}}
\put(88,207){\mbox{\bf a}}
\put(88,131){\mbox{\bf b}}
\put(88,56.5){\mbox{\bf c}}
\put(18,203){\mbox{$b\chp_1$}}
\put(45,192){\mbox{$t\nt_1$}}
\put(54,170.5){\mbox{$t\nt_2$}}
\put(84,164.5){\mbox{\scriptsize $b\chp_2$}}
\put(33,129){\mbox{$b\chp_1$}}
\put(36,102){\mbox{$t\nt_1$}}
\put(43,95){\mbox{\footnotesize $t\nt_2$}}
\put(80,94){\mbox{\footnotesize $t\nt_3$}}
\put(29,54){\mbox{$b\chp_1$}}
\put(50,41){\mbox{$b\chp_2$}}
\put(19,18){\mbox{\footnotesize $t\nt_1$}}
\put(33,22){\mbox{\footnotesize $t\nt_2$}}
\put(63.5,21.5){\mbox{\footnotesize $t\nt_3$}}
\put(82,25.5){\mbox{\footnotesize $t\nt_4$}}
\end{picture}\ece
\caption{Tree level (dashed lines) and SUSY--QCD corrected (full lines) 
branching ratios of $\st_1$ decays as a function of $\mst{1}$ 
for $\mst{2}=600\gev$, $\cst=-0.5$, $\tan\b=3$, $m_A=500\gev$,    
(a) $M=163\gev$, $\mu=500\gev$, (b) $M=500\gev$, $\mu=163\gev$, and
(c) $M=\mu=219\gev$.}
\label{fig:BRst1-A}
\end{figure}

For discussing the {\bf branching ratios of \boldmath $\st_1$ decays} 
we take the parameter sets of Sect.~\ref{sect:qcdnc} with $\mch{1}=150\gev$. 
Figures~\ref{fig:BRst1-A}\,a--c show the tree--level and the 
SUSY--QCD corrected branching ratios of $\st_1$ as functions of $\mst{1}$ 
for $\mst{2}=600\gev$, $\cst=-0.5$, $\tan\b=3$, and $m_A=500\gev$.
Moreover, $M=163\gev$, $\mu=500\gev$ in \fig{BRst1-A}\,a; 
$M=500\gev$, $\mu=163\gev$ in \fig{BRst1-A}\,b; and 
$M=\mu=219\gev$ in \fig{BRst1-A}\,c. 
If only the decay into $b\chp_1$ is kinematically allowed it has 
a branching ratio of practically 100\%. 
Since the chargino decays further into $\nt_i q\bar q'$ or $\nt_i\ell^+\bar\nu$ 
the resulting signature is two acoplanar $b$--jets plus additional jets and/or 
charged leptons plus large missing (transverse) energy, $\Etmiss\ge 2\,\mnt{1}$. 
The decays $\st_1\to t\nt_k$ ($k=1,\ldots4$) and $\st_1\to b\chp_2$ 
can also have large branching ratios if they are kinematically allowed. 
Again, the decay cascades result in multi--jet/lepton signatures 
plus large missing energy. 
SUSY--QCD corrections change the tree--level branching ratios 
by ${\cal O}(\pm10\%)$ in Figs.~\ref{fig:BRst1-A}\,a,b. 
However, they can also have larger effects. 
This can be seen in \fig{BRst1-A}\,c where the corrections go up to $\pm 30\%$.

\begin{figure}[p]
\bce\begin{picture}(95,215)
\put(-1,155){\mbox{\epsfig{file=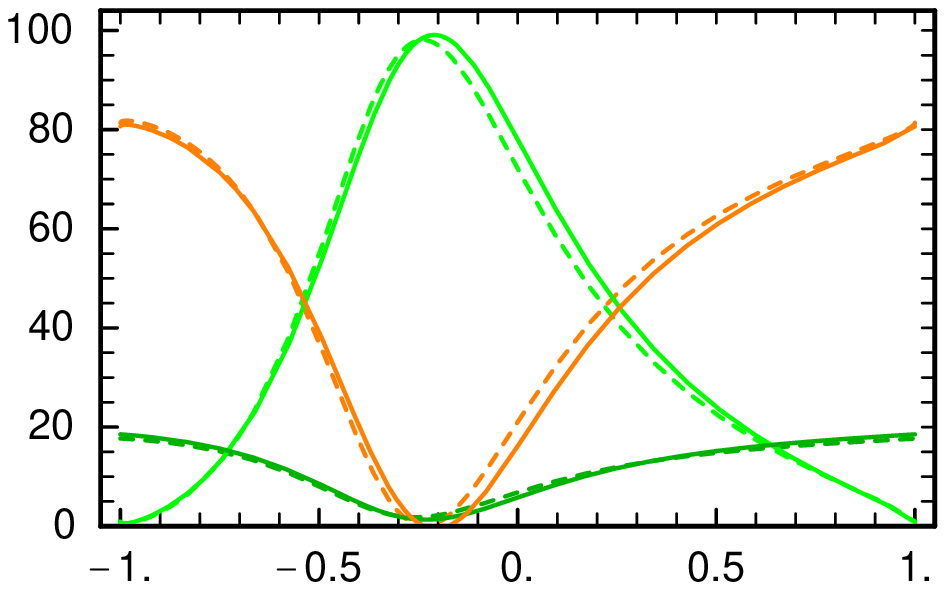,height=60mm}}}
\put(0,80){\mbox{\epsfig{file=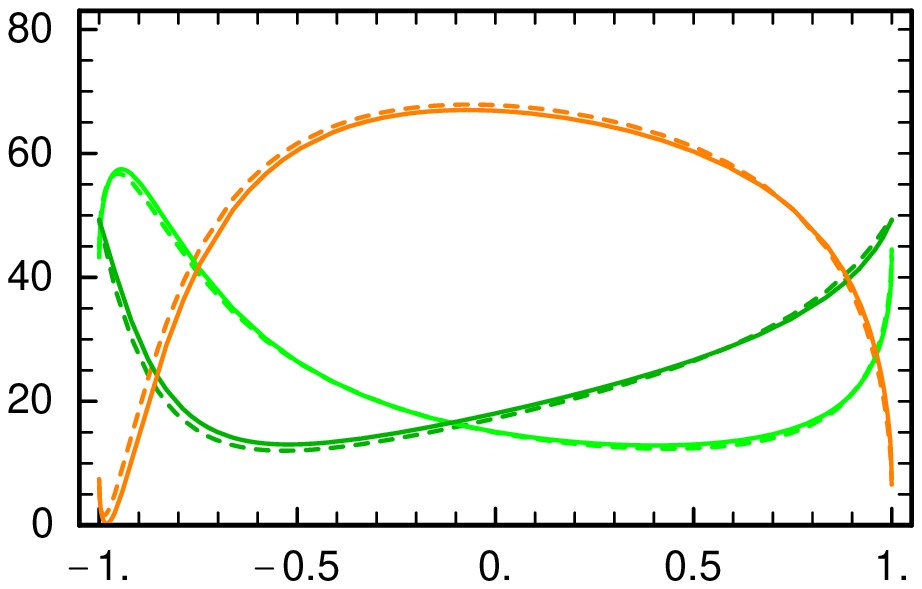,height=60mm}}}
\put(0,5){\mbox{\epsfig{file=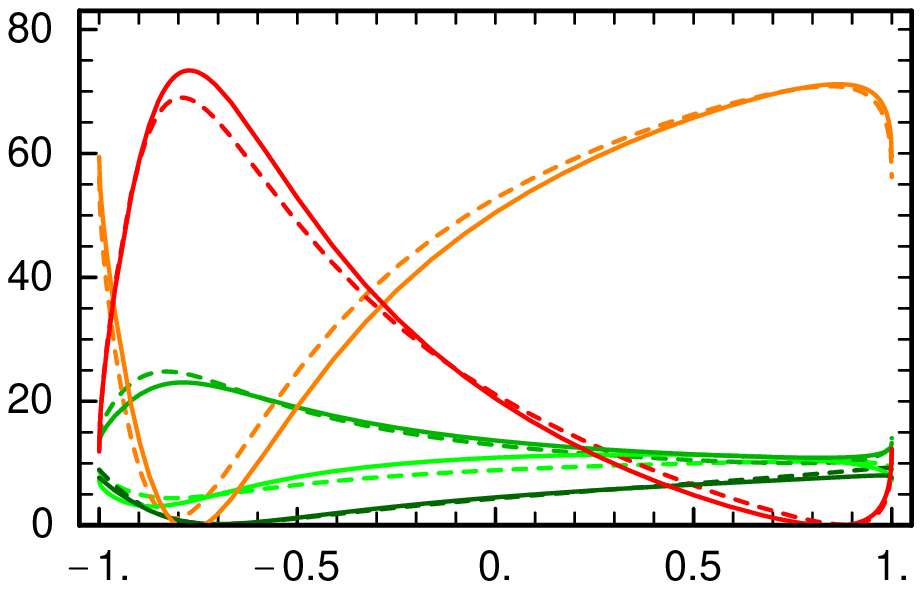,height=60mm}}}
\put(49,150){\mbox{$\cst$}}
\put(49,75){\mbox{$\cst$}}
\put(49,0){\mbox{$\cst$}}
\put(0,178){\makebox(0,0)[br]{{\rotatebox{90}{BR~$(\st_1)$~[\%]}}}}
\put(0,103){\makebox(0,0)[br]{{\rotatebox{90}{BR~$(\st_1)$~[\%]}}}}
\put(0,28){\makebox(0,0)[br]{{\rotatebox{90}{BR~$(\st_1)$~[\%]}}}}
\put(88,207){\mbox{\bf a}}
\put(88,131){\mbox{\bf b}}
\put(88,58){\mbox{\bf c}}
\put(77,200){\mbox{$b\chp_1$}}
\put(54,205){\mbox{$t\nt_1$}}
\put(86,173){\mbox{$t\nt_2$}}
\put(72,128){\mbox{$b\chp_1$}}
\put(32,108){\mbox{$t\nt_1$}}
\put(70,106.5){\mbox{$t\nt_2$}}
\put(63,54){\mbox{$b\chp_1$}}
\put(30,54){\mbox{$b\chp_2$}}
\put(20,29.5){\mbox{$t\nt_2$}}
\put(34,18.5){\mbox{\footnotesize $t\nt_1$}}
\put(84,26){\mbox{\footnotesize $t\nt_3$}}
\put(85,25){\vector(-1,-1){8}}
\end{picture}\ece
\caption{Tree level (dashed lines) and SUSY--QCD corrected (full lines) 
branching ratios of $\st_1$ decays as a function of $\cst$ 
for $\mst{1}=400\gev$, $\mst{2}=600\gev$, $\tan\b=3$, $m_A=500\gev$,  
(a) $M=163\gev$, $\mu=500\gev$, (b) $M=500\gev$, $\mu=163\gev$, and
(c) $M=\mu=219\gev$.}
\label{fig:BRst1-B}
\end{figure}

\noi
The dependence on the stop mixing angle is shown in Figs.~\ref{fig:BRst1-B}\,a--c. 
In order to have several decay modes open, we choose $\mst{1}=400\gev$ and 
the other parameters as in \fig{BRst1-A}. 
The branching ratios of $\st_1$ strongly depend on both the $\st_L$--$\st_R$ 
and the gaugino--higsino mixing. 
In \fig{BRst1-B}\,a, $\chp_1$ and $\nt_{1,2}$ are gaugino--like. 
For $-0.5\lsim \cst\lsim 0.2$ ($\st_1\sim \st_R$) the decay $\st_1\to t\nt_1$ 
dominates. Otherwise, $\st_1\to b\chp_1$ is the dominant decay mode. 
${\rm BR}(\st_1\to t\nt_2)$ has a similar $\cst$ dependence as 
${\rm BR}(\st_1\to b\chp_1)$ but is phase--space suppressed. \\
For higgsino--like $\chp_1$ and $\nt_{1,2}$, as in \fig{BRst1-B}\,b, 
the situation is quite different. 
In this case $\st_1$ decays mainly into $b\chp_1$ unless $\st_1\simeq\st_L$. 
Moreover, ${\rm BR}(\st_1\to t\nt_1) > (<)\:{\rm BR}(\st_1\to t\nt_2)$ for 
$\cst\lsim (\gsim) -0.1$. \\
In the mixed gaugino--higgsino scenario of \fig{BRst1-B}\,c, 
$\st_1$ decays mainly into $b\chp_{1,2}$ with 
${\rm BR}(\st_1\to b\chp_1) > 50\%$ for $\cst\gsim 0$. 
The decays into neutralinos altogether do not exceed 30\%. \\
In Figs.~\ref{fig:BRst1-B}\,a and \ref{fig:BRst1-B}\,b  
SUSY--QCD corrections change the tree--level branching ratios 
by ${\cal O}(\pm10\%)$. In \fig{BRst1-B}\,c they go up to $\pm 40\%$.

 
\begin{figure}[h!]
\bce\begin{picture}(96,140)
\put(0,80){\mbox{\epsfig{file=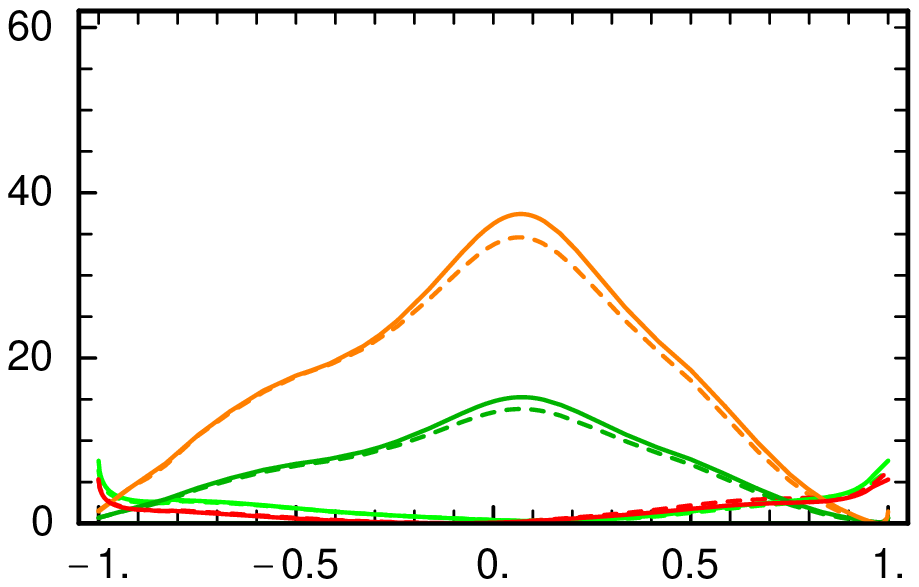,height=60mm}}}
\put(0,5){\mbox{\epsfig{file=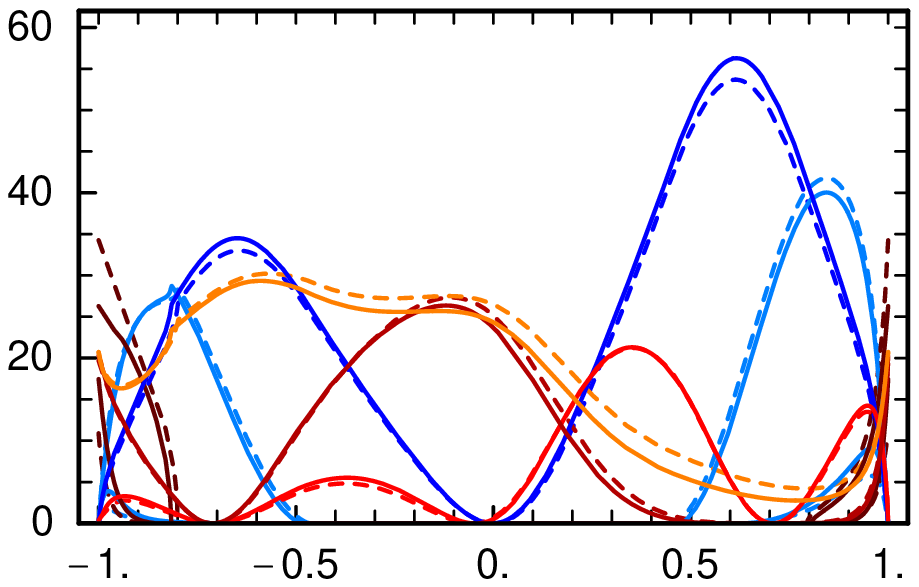,height=60mm}}}
\put(49,75){\mbox{$\cst$}}
\put(49,0){\mbox{$\cst$}}
\put(0,102.5){\makebox(0,0)[br]{{\rotatebox{90}{BR~$(\st_2)$~[\%]}}}}
\put(0,27.5){\makebox(0,0)[br]{{\rotatebox{90}{BR~$(\st_2)$~[\%]}}}}
\put(88,131){\mbox{\bf a}}
\put(88,56){\mbox{\bf b}}
\put(64,116){\mbox{\footnotesize$b\chp_1$}}
\put(49,101){\mbox{\footnotesize$t\nt_2$}}
\put(37,90){\mbox{\scriptsize$t\nt_2$}}
\put(65,90){\mbox{\scriptsize$b\chp_2$}}
\put(66,53){\mbox{\footnotesize$\st_1 Z$}}
\put(25,43){\mbox{\footnotesize$\st_1 Z$}}
\put(75,38){\mbox{\footnotesize$\sb_1 W^+$}}
\put(27,27){\mbox{\scriptsize$\sb_1 W^+$}}
\put(14,41){\mbox{\scriptsize$\sb_1 H^+$}}
\put(38,38){\mbox{\footnotesize$\st_1 A^0$}}
\put(47.5,25){\mbox{\footnotesize$\st_1 H^0$}}
\put(56,26){\vector(1,0){1}}
\put(37,18){\mbox{\scriptsize$\st_1 h^0$}}
\put(70,30.5){\mbox{\footnotesize$\st_1 h^0$}}
\put(83,22.5){\mbox{\scriptsize$\sb_2 W^+$}}
\put(85,21){\vector(1,-1){4}}
\end{picture}\ece
\caption{Tree level (dashed lines) and SUSY--QCD corrected (full lines) 
branching ratios of $\st_2$ decays as a function of $\cst$ 
for $\mst{1}=250\gev$, $\mst{2}=600\gev$,   
$M=163\gev$, $\mu=500\gev$, $\tan\b=3$, and $m_A=200\gev$;
(a) shows the decays into fermions and (b) the decays into bosons.}
\label{fig:BRst2-A}
\end{figure}

\begin{figure}[p]
\bce\begin{picture}(96,140)
\put(0,80){\mbox{\epsfig{file=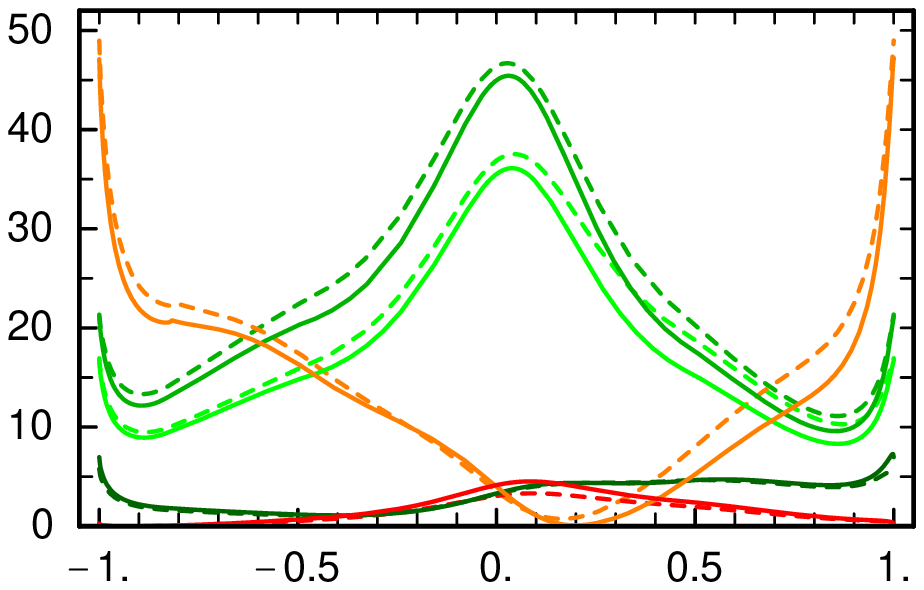,height=60mm}}}
\put(0,5){\mbox{\epsfig{file=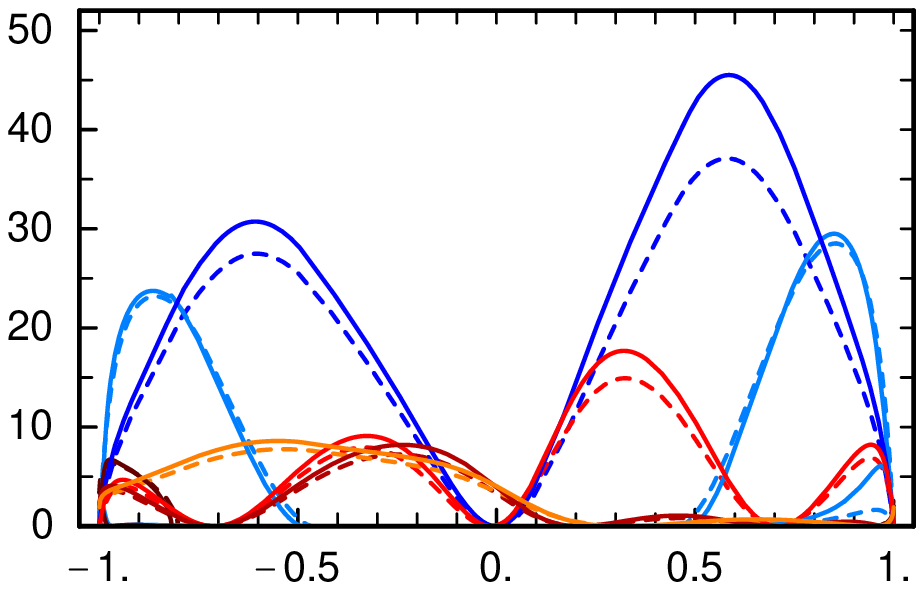,height=60mm}}}
\put(49,75){\mbox{$\cst$}}
\put(49,0){\mbox{$\cst$}}
\put(0,103){\makebox(0,0)[br]{{\rotatebox{90}{BR~$(\st_2)$~[\%]}}}}
\put(0,28){\makebox(0,0)[br]{{\rotatebox{90}{BR~$(\st_2)$~[\%]}}}}
\put(88,131){\mbox{\bf a}}
\put(88,56){\mbox{\bf b}}
\put(17,121){\mbox{\footnotesize$b\chp_1$}}
\put(86,121){\mbox{\footnotesize$b\chp_1$}}
\put(62,130){\mbox{\footnotesize$t\nt_2$}}
\put(50,112){\mbox{\footnotesize$t\nt_1$}}
\put(20,91){\mbox{\scriptsize$t\nt_3$}}
\put(89,89){\mbox{\tiny$b\chp_2$}}
\put(66,53){\mbox{\footnotesize$\st_1 Z$}}
\put(27,45){\mbox{\footnotesize$\st_1 Z$}}
\put(15,37){\mbox{\scriptsize$\sb_1 W^+$}}
\put(88,43){\mbox{\footnotesize$\sb_1 W$}}
\put(30,22){\mbox{\scriptsize$\st_1 A^0$}}
\put(38,22){\mbox{\scriptsize$\st_1 h^0$}}
\put(49,20){\mbox{\scriptsize$\st_1 H^0$}}
\put(70,30){\mbox{\footnotesize$\st_1 h^0$}}
\put(83,21){\mbox{\scriptsize$\sb_2 W^+$}}
\put(86,20){\vector(1,-1){4}}
\put(18,20){\mbox{\scriptsize$\sb_1 H^+$}}
\put(20,21){\vector(-1,-1){3}}
\end{picture}\ece
\caption{Tree level (dashed lines) and SUSY--QCD corrected (full lines) 
branching ratios of $\st_2$ decays as a function of $\cst$ 
for $\mst{1}=250\gev$, $\mst{2}=600\gev$,   
$M=500\gev$, $\mu=163\gev$, $\tan\b=3$, and $m_A=200\gev$;
(a) shows the decays into fermions and (b) the decays into bosons.}
\label{fig:BRst2-B}
\end{figure}

\begin{figure}[p]
\bce\begin{picture}(96,140)
\put(0,80){\mbox{\epsfig{file=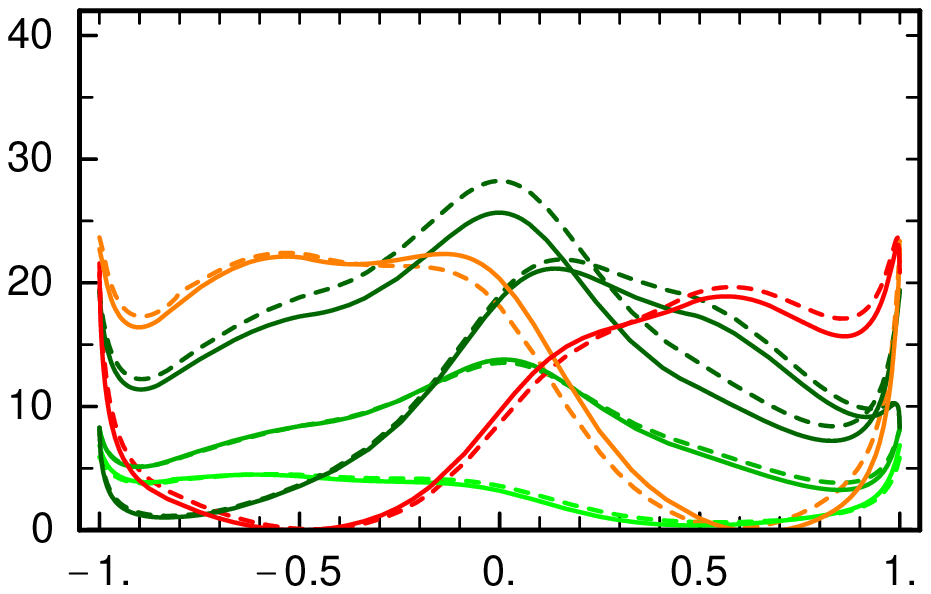,height=60mm}}}
\put(0,5){\mbox{\epsfig{file=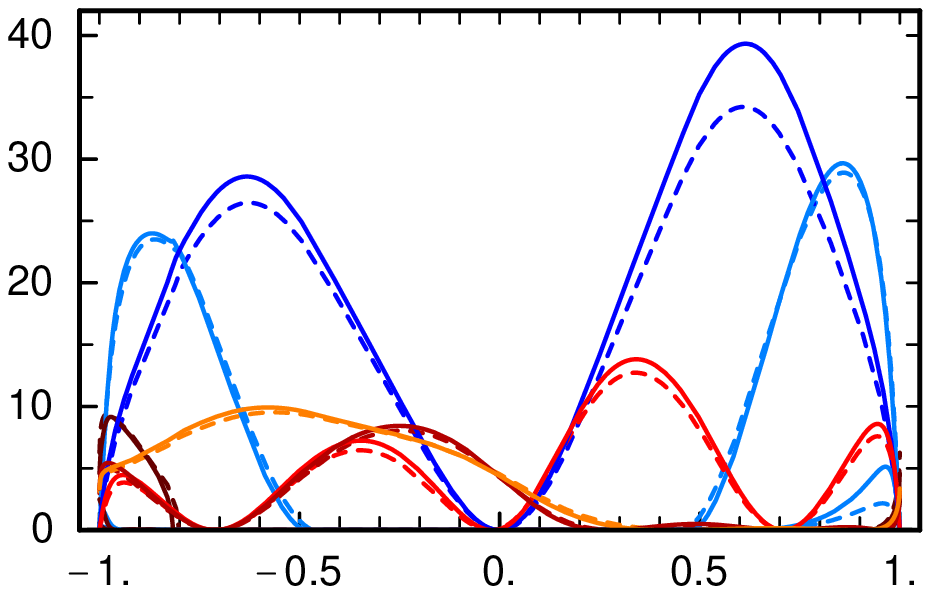,height=60mm}}}
\put(49,75){\mbox{$\cst$}}
\put(49,0){\mbox{$\cst$}}
\put(0,103){\makebox(0,0)[br]{{\rotatebox{90}{BR~$(\st_2)$~[\%]}}}}
\put(0,28){\makebox(0,0)[br]{{\rotatebox{90}{BR~$(\st_2)$~[\%]}}}}
\put(88,131){\mbox{\bf a}}
\put(88,56){\mbox{\bf b}}
\put(29,117){\mbox{\footnotesize$b\chp_1$}}
\put(81,113){\mbox{\footnotesize$b\chp_2$}}
\put(56,92){\mbox{\footnotesize$t\nt_1$}}
\put(33,100){\mbox{\footnotesize$t\nt_2$}}
\put(47,124){\mbox{\footnotesize$t\nt_3$}}
\put(66,115){\mbox{\footnotesize$t\nt_4$}}
\put(36,42){\mbox{\footnotesize$\st_1 Z$}}
\put(66,53){\mbox{\footnotesize$\st_1 Z$}}
\put(14,43){\mbox{\scriptsize$\sb_1 W^+$}}
\put(75,38){\mbox{\footnotesize$\sb_1 W^+$}}
\put(31,26){\mbox{\scriptsize$\st_1 A^0$}}
\put(39,16){\mbox{\scriptsize$\st_1 h^0$}}
\put(49,22){\mbox{\scriptsize$\st_1 H^0$}}
\put(70,30){\mbox{\footnotesize$\st_1 h^0$}}
\put(83,21){\mbox{\scriptsize$\sb_2 W^+$}}
\put(86,20){\vector(1,-1){4}}
\put(18,24){\mbox{\scriptsize$\sb_1 H^+$}}
\put(20,25){\vector(-1,-1){3}}
\end{picture}\ece
\caption{Tree level (dashed lines) and SUSY--QCD corrected (full lines) 
branching ratios of $\st_2$ decays as a function of $\cst$ 
for $\mst{1}=250\gev$, $\mst{2}=600\gev$,   
$M=\mu=219\gev$, $\tan\b=3$, and $m_A=200\gev$;
(a) shows the decays into fermions and (b) the decays into bosons.}
\label{fig:BRst2-C}
\end{figure}

Let us now turn to the {\bf branching ratios of \boldmath $\st_2$}. 
For the heavier stop mass eigenstate also decays into gauge and Higgs bosons 
may be important. We thus take $\mst{1}=250\gev$, $\mst{2}=600\gev$, 
$m_A=200\gev$, and $\tan\b=3$ in this discussion. 
Figures \ref{fig:BRst2-A}, \ref{fig:BRst2-B}, and \ref{fig:BRst2-C} 
show the $\cst$ dependence of the $\st_2$ branching ratios for 
$(M,\,\mu) = (163,\,500)\gev$, $(500,\,163)\gev$, and $(219,\,219)\gev$, 
respectively. \\
We first discuss the decays into charginos and neutralinos:  
In the case $M\ll |\mu|$, ${\rm BR}(\st_2\to b\chp_1)$ and 
${\rm BR}(\st_2\to t\nt_2)$ are maximal for $\st_2\sim\st_L$ 
$(\cst\sim 0)$. 
In the opposite case $|\mu|\ll M$, 
$\st_2\simeq\st_R$ mainly decays into $b\chp_1$ and 
$\st_2\simeq\st_L$ mainly decays into $t\nt_{1,2}$. 
For relatively small $M\sim|\mu|$ all decays into $b\chp_{1,2}$ and
$t\nt_{1...4}$ are kinematically allowed. The branching ratios show 
an intricate dependence on the stop mixing angle. In particular, 
${\rm BR}(\st_2\to t\nt_{2,3,4})$ is maximal for $\st_2\sim\st_L$ 
and ${\rm BR}(\st_2\to b\chp_2) > (<)\:{\rm BR}(\st_2\to b\chp_1)$  
for $\cst \gsim (\!\lsim)\:0.1$. \\
As for the bosonic decays, $\st_2\to\st_1 Z$ is important in case of 
large $\st_L$--$\st_R$ mixing. 
The decays into Higgs bosons are, quite generally, important for 
large values of $\mu$ and/or $A_t$. 
The stop--Higgs couplings involve combinations $\mu\pm A_t$. 
Since $A_t= (\mst{1}^2-\mst{2}^2)\sin 2\tst/(2m_t) + \mu\cot\b$ the 
branching ratios of $\st_2\to\st_1+(h^0,H^0,A^0)$ decays show a distinct 
dependence on the sign of $\cst$. Moreover, the $\st_1\st_2$ couplings to 
$h^0$ and $H^0$ vanish in case of maximal mixing.  
In our analysis, the sbottom masses also vary with $\cst$. 
For $|\cst|\gsim 0.5$ the decay $\st_2\to\sb_1 W^+$ is kinematically allowed. 
Its branching ratios go up to 40\%. 
For $|\cst|\gsim 0.8$ also the decays into $\sb_2 W^+$ and $\sb_i H^+$ can 
be relevant. 

\begin{figure}[p]
\bce\begin{picture}(95,65)
\put(0,5){\mbox{\epsfig{file=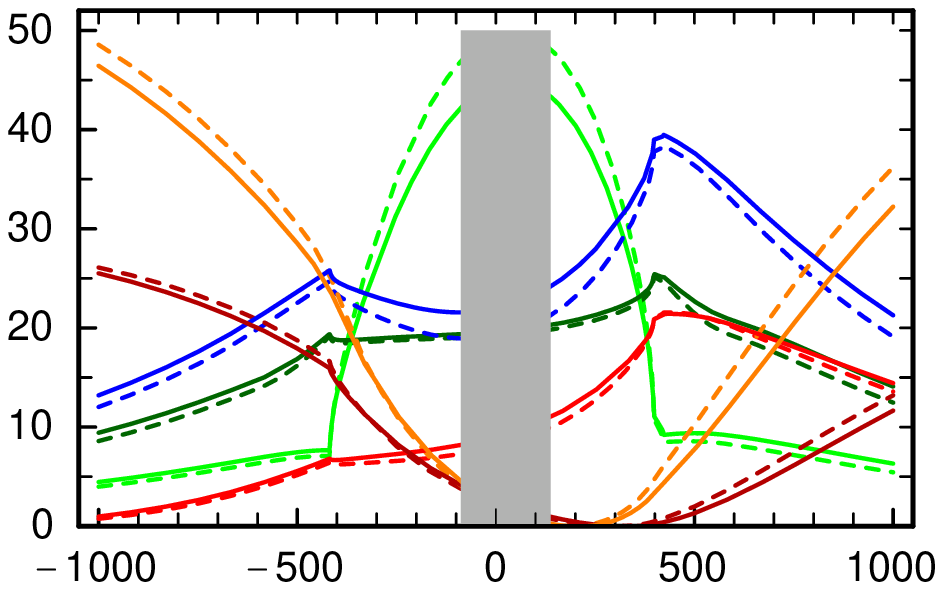,height=60mm}}}
\put(45,0){\mbox{$\mu$~[GeV]}}
\put(0,28){\makebox(0,0)[br]{{\rotatebox{90}{BR~$(\st_2)$~[\%]}}}}
\put(20,57){\mbox{\footnotesize$\st_1 A^0$}}
\put(15,39){\mbox{\footnotesize$\st_1 H^0$}}
\put(14,29){\mbox{\footnotesize$\st_1 Z$}}
\put(38.5,53){\mbox{\footnotesize$t\nt_k$}}
\put(28,23.5){\mbox{\scriptsize$b\chp_j$}}
\put(35,14.8){\mbox{\scriptsize$\st_1 h^0$}}
\put(73,51){\mbox{\footnotesize$\st_1 Z$}}
\put(85,48){\mbox{\footnotesize$\st_1 A^0$}}
\put(80,14){\mbox{\scriptsize$\st_1 H^0$}}
\put(59,27){\mbox{\scriptsize$\st_1 h^0$}}
\put(63,55){\mbox{\footnotesize$t\nt_k$}}
\put(70,38){\mbox{\footnotesize$b\chp_j$}}
\end{picture}\ece
\caption{Tree level (dashed lines) and SUSY--QCD corrected (full lines) 
branching ratios of $\st_2$ decays as a function of $\mu$ 
for $\mst{1}=250\gev$, $\mst{2}=600\gev$, $\cst=0.4$,  
$M=200\gev$, $\tan\b=3$, and $m_A=200\gev$. 
The grey area is excluded by the bound $\mch{1}>95\gev$.}
\label{fig:BRst2-D}
\end{figure}

\begin{figure}[p]
\bce\begin{picture}(95,65)
\put(0,5){\mbox{\epsfig{file=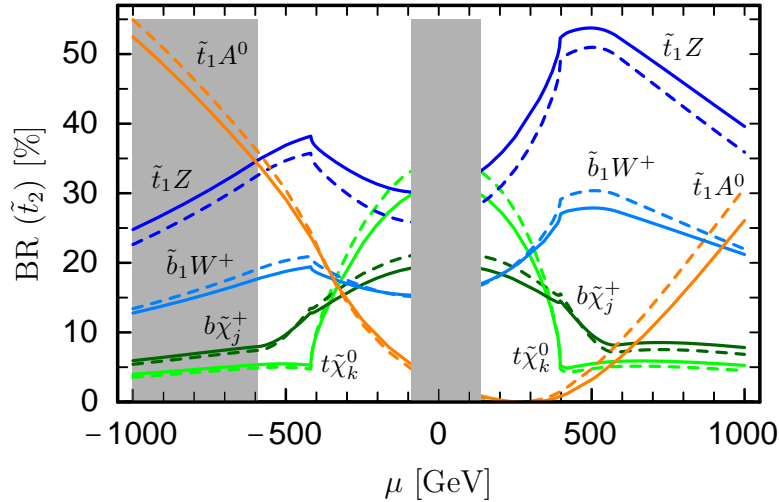,height=60mm}}}
\put(45,0){\mbox{$\mu$~[GeV]}}
\put(0,28){\makebox(0,0)[br]{{\rotatebox{90}{BR~$(\st_2)$~[\%]}}}}
\put(20,57){\mbox{\footnotesize$\st_1 A^0$}}
\put(14,40.5){\mbox{\footnotesize$\st_1 Z$}}
\put(16,29){\mbox{\footnotesize$\sb_1 W^+$}}
\put(21,21){\mbox{\footnotesize$b\chp_j$}}
\put(36.5,16){\mbox{\footnotesize$t\nt_k$}}
\put(82,58){\mbox{\footnotesize$\st_1 Z$}}
\put(72,42){\mbox{\footnotesize$\sb_1 W^+$}}
\put(86,39.5){\mbox{\footnotesize$\st_1 A^0$}}
\put(70.5,25){\mbox{\footnotesize$b\chp_j$}}
\put(62,17){\mbox{\footnotesize$t\nt_k$}}
\end{picture}\ece
\caption{Tree level (dashed lines) and SUSY--QCD corrected (full lines) 
branching ratios of $\st_2$ decays as a function of $\mu$ 
for $\cst=0.7$ and the other parameters as in \fig{BRst2-D}.  
The grey area around $\mu=0\gev$ is excluded by the bound $\mch{1}>95\gev$; 
in the gray area on the left the approximate condition for vacuum 
stability is violated.}
\label{fig:BRst2-E}
\end{figure}

\noi 
We have already noticed several times that the $\mu$ parameter plays a 
special r\^ole. Figures \ref{fig:BRst2-D} and \ref{fig:BRst2-E} show 
${\rm BR}(\st_2)$ as a function of $\mu$ for $\mst{1}=250\gev$, 
$\mst{2}=600\gev$, $M=200\gev$, $\tan\b=3$, and $m_A=200\gev$.  
In \fig{BRst2-D}, $\cst=0.4$. 
For small $|\mu|$ $(\!\lsim 400\gev)$, decays into neutralinos dominate. 
For increasing $|\mu|$ decays into Higgs bosons become important 
or even dominant. 
Notice the dependence of ${\rm BR}(\st_2\to\st_1+h^0,H^0)$ on 
the sign of $\mu$. 
For $300\gev < \mu < 800\gev$ the decay $\st_2\to\st_1 Z$ has the largest 
branching ratio. 
The decays $\st_2\to\sb_i+(W^+,H^+)$ are kinematically not allowed.\\
\Fig{BRst2-D} has to be compared with \Fig{BRst2-E} where $\cst=0.7$. 
Here the decays into $\st_1 h^0$ and $\st_1 H^0$ have branching ratios 
below 1\%; only the decay into $\st_1 A^0$ can still have a large 
branching ratio.   
Moreover, the decay into $\sb_1 W^+$ is possible in this case and has 
a branching ratio of about 10\% to 30\%. 
The decay into $\st_1 Z$ is even more important than in \fig{BRst2-D}.

\noi
For smaller $\mst{1}$--$\mst{2}$ mass splitting and larger $m_A$ the bosonic 
decays are of course less important or kinematically forbidden. 
However, if $m_A$ is large the $\st_2$ decays into $\st_1 Z$, $\st_1 h^0$, 
and $\sb_i W^+$ can still have large branching ratios. 

\noi
In all cases shown, SUSY--QCD corrections are important for precision measurements.


Typical signals of $\st_2$ decays into bosons 
and of those into fermions are shown in 
Table~1 of Ref.~\cite{st2sig}. 
In principle, the final states of both types of decays can be identical.  
For example, the final state of the decay chain 
\begin{gather}
    \st_2 \to \st_1 + (h^0,\,H^0,\,A^0~\mbox{or}~Z) 
    \to (t \nt_1) + (b \bar{b}) 
    \to (b q \bar{q}' \nt_1) + (b \bar{b}) \label{eq:chainA} \\
\intertext{has the same event topology as that of}
    \st_2 \to t + \nt_{2,3,4}
    \to t + ((h^0,\,H^0,\,A^0~\mbox{or}~Z) + \nt_1) 
    \to t + (b\bar{b} \nt_1) 
    \to (b q\bar{q}') + (b\bar{b} \nt_1). \hspace{-3cm}
    \label{eq:chainB} \\
\intertext{Likewise,} 
    \st_2 \to \sb_{1,2} + (H^+~\mbox{or}~W^+) 
    \to (b \nt_1) + (q \bar{q}') \label{eq:chainC} \\
\intertext{has the same event topology as}
    \st_2 \to b + \chp_{1,2}
    \to b + ((H^+~\mbox{or}~W^+) + \nt_1)
    \to b + (q \bar{q}' \nt_1). \label{eq:chainD}
\end{gather} 
However, the decay structures and kinematics of the two modes 
\eq{chainA} and \eq{chainB} (\eq{chainC} and \eq{chainD}) 
are quite different from each other, 
since the $\nt_1$ is emitted from $\st_1$ and $\nt_{2,3,4}$ 
($\sb_{1,2}$ and $\chp_{1,2}$), respectively. 
This could result in significantly different event distributions 
(\eg, missing energy--momentum distribution) of the $\st_2$ decays 
into gauge or Higgs bosons compared to the decays into fermions.
Hence the possible dominance of the former decay modes could have 
an important impact on the search for $\st_2$, 
and on the measurement of the MSSM parameters. 
The effects of the bosonic decays should therefore be included 
in the Monte Carlo studies of $\st_2$ and $\sb_2$ decays. 

 
\clearpage

\begin{figure}[h!]
\bce\begin{picture}(96,140)
\put(-1,80){\mbox{\epsfig{file=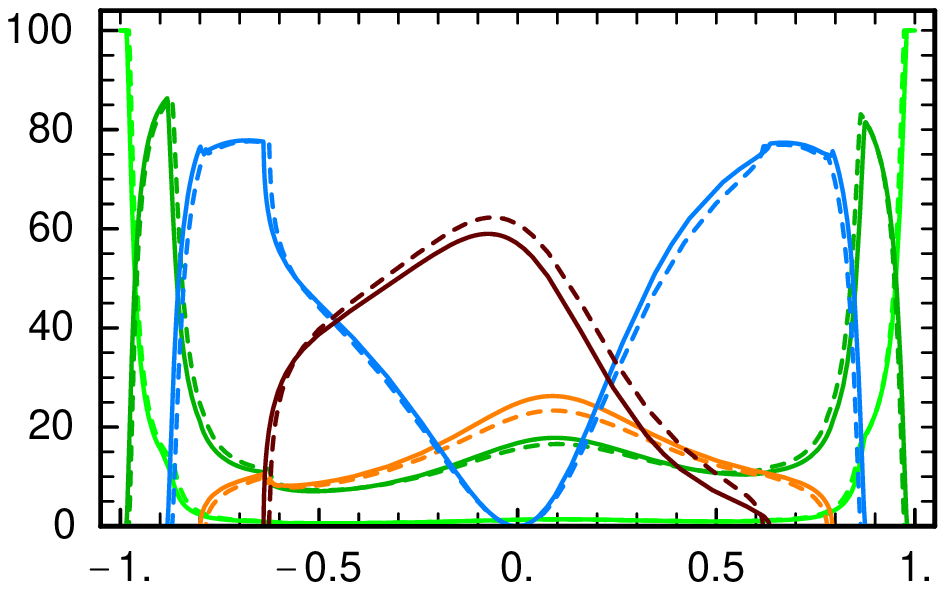,height=60mm}}}
\put(-1,5){\mbox{\epsfig{file=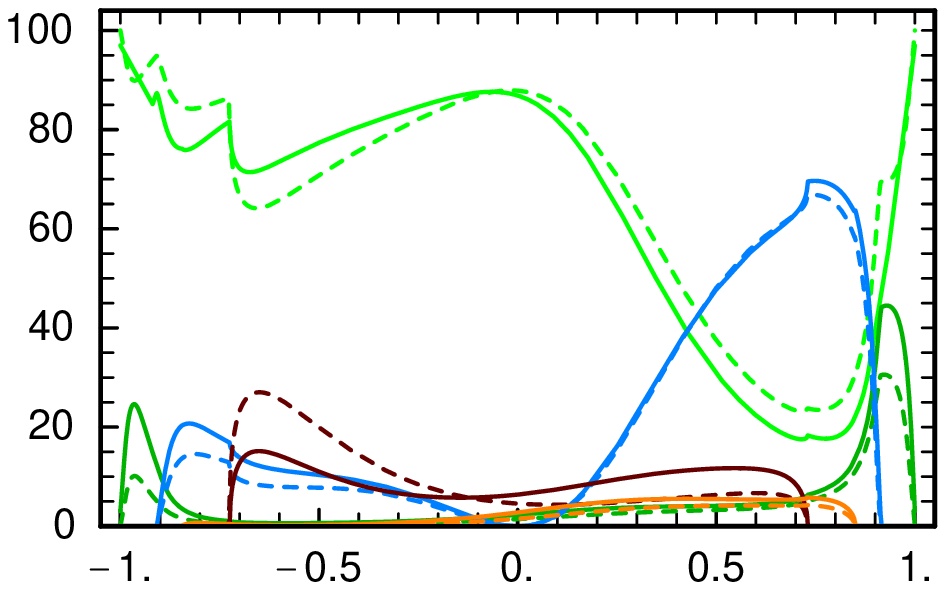,height=60mm}}}
\put(49,75){\mbox{$\cst$}}
\put(49,0){\mbox{$\cst$}}
\put(0,103){\makebox(0,0)[br]{{\rotatebox{90}{BR~$(\sb_1)$~[\%]}}}}
\put(0,28){\makebox(0,0)[br]{{\rotatebox{90}{BR~$(\sb_2)$~[\%]}}}}
\put(16,134){\mbox{\footnotesize$b\nt_1$}}
\put(22,129){\mbox{\footnotesize$b\nt_2$}}
\put(20,130){\vector(-1,0){0}}
\put(30,124){\mbox{\footnotesize$\st_1 W^-$}}
\put(42,119){\mbox{\footnotesize$\st_1 H^-$}}
\put(50,102){\mbox{\footnotesize$t\chm_1$}}
\put(68,124){\mbox{\footnotesize$\st_1 W^-$}}
\put(87,134){\mbox{\footnotesize$b\nt_1$}}
\put(82,129){\mbox{\footnotesize$b\nt_2$}}
\put(40,57){\mbox{\footnotesize$b\nt_1$}}
\put(14,27){\mbox{\scriptsize$b\nt_2$}}
\put(19.5,23.5){\mbox{\scriptsize$\st_1 W^-$}}
\put(33,24){\mbox{\footnotesize$\st_1 H^-$}}
\put(70,21){\mbox{\scriptsize$t\chm_1$}}
\put(71.5,19.5){\vector(1,-1){5}}
\put(76,47){\mbox{\footnotesize$\st_1 W^-$}}
\put(82,34){\mbox{\scriptsize$b\nt_2$}}
\put(84,32.5){\vector(3,-1){9}}
\end{picture}\ece
\caption{Tree level (dashed lines) and SUSY--QCD corrected (full lines) 
branching ratios of $\sb_1$ and $\sb_2$ decays as a function of $\cst$ 
for $\mst{1}=250\gev$, $\mst{2}=600\gev$,   
$M=230\gev$, $\mu=500\gev$, $\tan\b=3$, and $m_A=200\gev$.}
\label{fig:BRsb-A}
\end{figure}

Last but not least we discuss {\bf branching ratios of sbottom decays}. 
\Fig{BRsb-A} shows the tree level and the SUSY--QCD corrected 
branching ratios of $\sb_1$ and $\sb_2$ decays as a function of 
the stop mixing angle for $\mst{1}=250\gev$, $\mst{2}=600\gev$,   
$M=230\gev$ \footnote{We choose a somewhat higher value of $M$ such 
that sbottom decays into gluinos are kinematically not allowed.}, 
$\mu=500\gev$, $\tan\b=3$, and $m_A=200\gev$. 
The corresponding sbottom and Higgs boson masses are shown 
in \fig{BR-scen}. Moreover, we have 
$\mnt{k}=\{111,\,215,\,502,\,522\}\gev$,  
$\mch{j}=\{214,\,520\}\gev$, and $\msg=645\gev$. 
For $\sb_1$, which is mainly a $\sb_L$, bosonic decays are important  
where they are kinematically allowed. 
To be concrete, in \fig{BRsb-A} $\sb_1\to\st_1 W^-$ is the dominant decay 
mode in case of large stop mixing because the $W$ only couples to the 
``left'' components of the squarks. 
For $\st_1\sim\st_R$ the decay $\sb_1\to\st_1 H^-$ is the dominant one 
because (i) $\sb_1\st_1 W^\pm$ coupling vanishes and 
(ii) the coupling to $H^\pm$ favours $\sq_L^{}\sq'_R$ combinations. \\
The situation is quite different for $\sb_2$ which is mainly a $\sb_R$. 
Here the decay into $b\nt_1$ dominates in most of the $\cst$ range. 
Only for $0.4\lsim\cst\lsim 0.9$ it has a branching ratio below 50\%. 
In this range the decay into $\st_1 W^-$ dominates. 

\begin{figure}[h!]
\bce\begin{picture}(96,140)
\put(0.5,80){\mbox{\epsfig{file=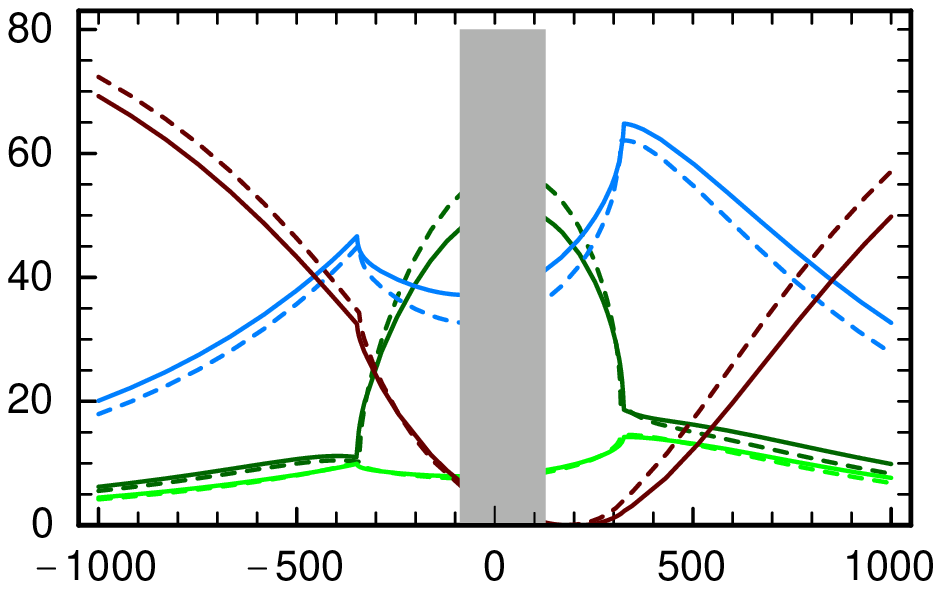,height=60mm}}}
\put(-0.5,5){\mbox{\epsfig{file=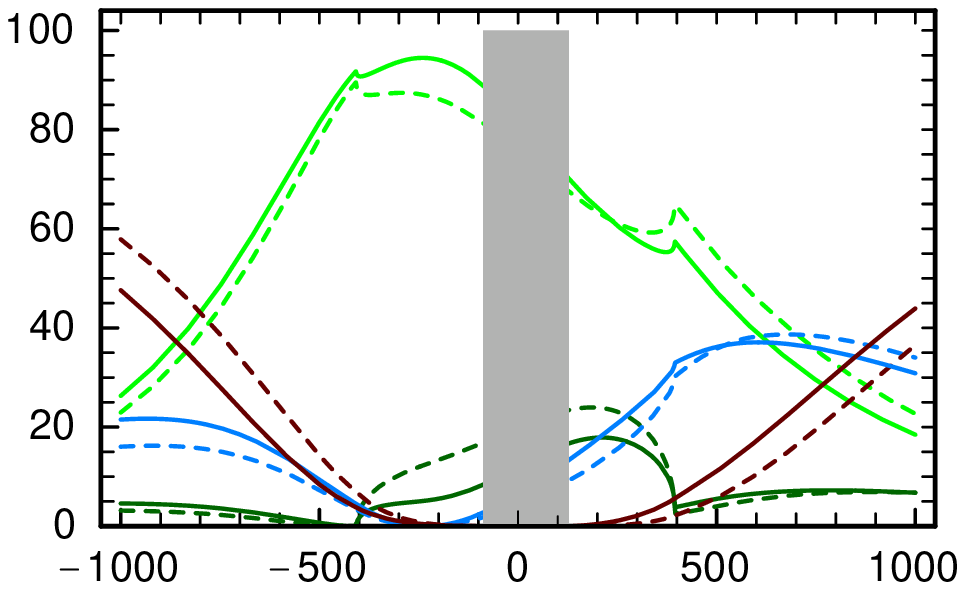,height=60mm}}}
\put(46,75){\mbox{$\mu$~[GeV]}}
\put(46,0){\mbox{$\mu$~[GeV]}}
\put(0,103){\makebox(0,0)[br]{{\rotatebox{90}{BR~$(\sb_1)$~[\%]}}}}
\put(0,28){\makebox(0,0)[br]{{\rotatebox{90}{BR~$(\sb_2)$~[\%]}}}}
\put(22,128){\mbox{\footnotesize$\st_1 H^-$}}
\put(35,118){\mbox{\footnotesize$\st_1 W^-$}}
\put(33,97){\mbox{\footnotesize$t\chm_j$}}
\put(39,89.5){\mbox{\scriptsize$b\nt_k$}}
\put(84,122){\mbox{\footnotesize$\st_1 H^-$}}
\put(73,125){\mbox{\footnotesize$\st_1 W^-$}}
\put(66,108){\mbox{\footnotesize$t\chm_j$}}
\put(59,96){\mbox{\scriptsize$b\nt_k$}}
\put(26,54){\mbox{\footnotesize$b\nt_k$}}
\put(17,16){\mbox{\scriptsize$t\chm_j$}}
\put(41.6,21){\mbox{\footnotesize$t\chm_j$}}
\put(27,28){\mbox{\footnotesize$\st_1 H^-$}}
\put(16.6,23.7){\mbox{\tiny$\st_1 W^-$}}
\put(62.5,45.5){\mbox{\footnotesize$b\nt_k$}}
\put(85,18){\mbox{\footnotesize$t\chm_j$}}
\put(86,34.3){\mbox{\footnotesize$\st_1 H^-$}}
\put(64,30){\mbox{\footnotesize$\st_1 W^-$}}
\end{picture}\ece
\caption{Tree level (dashed lines) and SUSY--QCD corrected (full lines) 
branching ratios of $\sb_1$ and $\sb_2$ decays as a function of $\mu$ 
for $\mst{1}=250\gev$, $\mst{2}=600\gev$, $\cst=0.4$,  
$M=200\gev$, $\tan\b=3$, and $m_A=200\gev$. 
The grey area is excluded by the bound $\mch{1}>95\gev$.}
\label{fig:BRsb-C}
\end{figure}

\noi
The $\mu$ dependence of ${\rm BR}(\sb_{1,2})$ is shown in 
\fig{BRsb-C} for $\mst{1}=250\gev$, $\mst{2}=600\gev$, 
$\cst=0.4$, $M=200\gev$, $\tan\b=3$, and $m_A=200\gev$. 
Again, bosonic decays are very important for $\sb_1$ 
while for $\sb_2$ decays into neutralinos are dominant 
in a wide range of $\mu$. 
${\rm BR}(\sb_{1,2}\to\st_1 H^-)$ increases with increasing 
$|\mu|$ because this parameter directly enters the 
$\st\sb H^\pm$ couplings.  

\noi
In both Figs.~\ref{fig:BRsb-A} and \ref{fig:BRsb-C}, SUSY--QCD corrections 
change the individual branching ratios by up to $\pm 50\%$.

As for the signatures of sbottom decays, analogous arguments apply as for 
the stop decays: If $\sb_{1,2}$ decays into $b\nt_1$ the signature 
is two acoplanar $b$--jets plus large missing energy. In all other cases 
one has multiple jets (and leptons) plus missing energy. 
Again, bosonic and fermionic decays can lead to the same final 
states. However, the various decay modes might be disentangled by 
their different event distributions due to the different 
decay structures and kinematics.


\clearpage
\section*{Open Questions}
\addcontentsline{toc}{chapter}{Open Questions}

In the recent years, extensive studies have been done for both   
stop and sbottom production and decays. 
However, some questions still remain open:

\begin{itemize}

\item The radiative corrections to squark prodution in $\mu^+\mu^-$ 
  have not yet been calculated. It can be expected that 
  --- similar to the $e^+e^-$ case --- SUSY--QCD corrections and 
  photon radiation are important. Indeed, a first analysis has shown 
  that photon radiation can reduce the total production cross section 
  significantly \cite{mumu-photonrad}. 

\item Since Yukawa couplings have a crucial influence on the 
  $\st$ and $\sb$ phenomenology, also Yukawa coupling corrections 
  to the production and decay processes should be studied.
  So far, this has only been done for $\sb_i\to t\chm_j$ \cite{sola} 
  and $e^+e^-\to \sq_i^{}\sqbar_j$ \cite{our-yuk-corr}. 
  In both cases it has turned out that these corrections are 
  important. 

\item For SUSY--QCD corrections to squark interactions with Higgs bosons 
  in the on--shell scheme, the parameter $A_q$ has to be renormalized.   
  As mentioned in Sect.~\ref{sect:qcdhx}, for large $\tan\b$ the term 
  $\d(m_bA_b)\sim \mu\tan\b\,\d m_b$ can become extremly large, leading to 
  negative decay widths at $\Oas$. 
  The inclusion of Yukawa corrections may help solve this problem.
  Another possible solution might be to take the bottom mass running. 
  However, this is difficult to accomplish in a consistent way in the 
  on--shell scheme. 

\end{itemize}



\clearpage
\addcontentsline{toc}{chapter}{Bibliography}

\begin{thebibliography}{99}


\bibitem{nilles}  
H.\,P.~Nilles \PRep{110} (1984) 1.

\bibitem{haberkane}  
H.\,E. Haber and G.\,L. Kane, \PRep{117} (1985) 75.

\bibitem{barbieri}
R.~Barbieri, {\em Riv. Nuov. Cim.} {\bf 11} (1988) 1.

\bibitem{drees}
M.~Drees, {\tt hep-ph/9611409}.

\bibitem{martin}
S.~Martin, {\tt hep-ph/9709356}, to be published in 
{\em Perspectives on Supersymmetry}, ed. G.\,L.~Kane, World Scientific.

\bibitem{bailin-love}
D.~Bailin and A.~Love, 
{\em Supersymmetric Gauge Field Theory and String Theory}, 
Institute of Physics Publishing, Bristol and Philadelphia, 1994.

\bibitem{CKM}
M.~Kobayashi and T.~Maskawa, {\em Prog. Theor. Phys} {\bf 49} (1973) 652;\\
N.~Cabbibo, \PRL{10} (1993) 531.

\bibitem{neutral-LSP}
J.~Ellis, J.\,S.~Hagelin, D.\,V.~Nanopoulos, K.~Olive 
and M.~Srednicki, \NPB{238} (1984) 453. 

\bibitem{hillwalkers}
See \eg, J.~Ellis, {\tt hep-ph/9812235},
for more details and references to the original literature.

\bibitem{Rparity}
H.~Dreiner, {\tt hep-ph/9707435}, to be published in 
  {\em Perspectives on Supersymmetry}, ed. G.\,L.~Kane, World Scientific; \\
R.~Barbier \etal, {\tt hep-ph/9810232}; \\
J.W.F.~Valle, {\tt hep-ph/9603307}, {\tt hep-ph/9802292}. 
 
\bibitem{softbreaking}
L.~Girardello and M.~Grisaru, \NPB{194} (1982) 65.

\bibitem{guni}
J.~Ellis, S.~Kelly and D.\,V.~Nanopoulos, \PLett{260} (1991) 131;\\
U.~Amaldi, W.~de~Boer and H.~Furstenau, \PLett{260} (1991) 447;\\
P.~Langacker and M.~Luo, \PRD{44} (1991) 817.

\bibitem{RGE}
V.~Barger, M.\,S.~Berger and P.~Ohmann, \PRD{49} (1994) 4908; \\
D.\,J.~Casta\~no, E.\,J.~Piard and P.~Ramond \PRD{49} (1994) 4882.

\bibitem{inoue}
K.~Inoue, A.~Kakuto, H.~Komatsu and S.~Takeshita, 
\emph{Prog. Theor. Phys.}~{\bf 67} (1982) 1889, 
\emph{ibid.}~{\bf 68} (1982) 927. 

\bibitem{ibanez-ross}
L.~Ib\'a\~nez and G.\,G.~Ross, \PLB{110} (1982) 215.

\bibitem{dred1}
W. Siegel, {\em Phys. Lett.} {\bf 84B} (1979) 193; \\
D.\,M. Capper, D.R.T. Jones and P. van Nieuwenhuizen, \NPB{167} (1980) 479.

\bibitem{dred2} 
I. Jack, D.R.T. Jones, S.\,P. Martin, M.\,T. Vaughn and Y. Yamada, 
\PRD{50} (1994) 5481.

\bibitem{dred3}
I. Jack and D.R.T. Jones, {\tt hep-ph/9707278}.

\bibitem{superHiggs}
D.\,Z.~Freedman, P.~van~Nieuwenhuizen and S.~Ferrara, \PRD{14} (1976) 912; 
S.~Deser and B.~Zumino, \PRL{38} (1977) 1433.

\bibitem{gravmed}
A.\,H.~Chamseddine, R.~Arnowitt and P.~Nath, \PRL{49} (1982) 970;\\
L.\,J.~Hall, J.~Lykken and S.~Weinberg, \PRD{27} (1983) 2359.

\bibitem{msugra}
P.~Nath, R.~Arnowitt and A.\,H.~Chamseddine, 
{\em Applied N=1 Supergravity}, World Scientific, Singapore, 1984;\\
W.~de~Boer, \emph{Prog. in Part. Nucl. Phys.} {\bf 33} (1994) 201.

\bibitem{gaugemed}
M.~Dine and A.~Nelson, \PRD{48} (1993) 1277, \PRD{51} (1995) 1362;\\
M.~Dine, A.~Nelson, Y.~Nir and Y.~Shirman, \PRD{53} (1996) 2658.

\bibitem{DPF95}
H.\,A.~Baer \etal, {\tt hep-ph/9503479}; 
M.~Drees and S.\,P.~Martin, {\tt hep-ph/9504324}; 
in: \emph{Report of the Working Group on Electroweak Symmetry Breaking 
and New Physics} of the 1995 DPF study of the future of particle physics 
in the USA, World Scientific, Singapore.

\bibitem{snowmass}
J.~Amundson \etal, {\tt hep-ph/9609374}, 
in: Proceedings of the \emph{1996 DPF/DPB Summer Study on High--Energy Physics}, 
Snowmass, Colorado, 1996, Eds. D.G.~Cassel, L.~Trindle Gennari, R.H. Siemann, p.~655; \\ 
A.~Bartl \etal, ibid., p.~693.

\bibitem{giudice}
G.\,F.~Giudice and R.~Rattazzi, {\tt hep-ph/9801271}, to be published in 
{\em Perspectives on Supersymmetry}, ed. G.\,L.~Kane, World Scientific. 

\bibitem{radcorrh0}
J.~Ellis, G.~Ridolfi and F.~Zwirner, \PLB{262} (1991) 477.  
   
\bibitem{radcorrhc}  
A.~Brignole, \PLB{277} (1992) 313. 

\bibitem{masiero}
F.~Gabbiani, E.~Gabriello, A.~Masiero and L.~Silvestrini, 
\NPB{477} (1996) 321;\\
J.~Ellis and D.\,V.~Nanopoulos, \PLB{110} (1982) 44.

\bibitem{ellis-rudaz}
J.~Ellis and S.~Rudaz, \PLB{128} (1993) 248.

\bibitem{superK}
Super--Kamiokande collab., Y.~Fukuda \etal, \PRL{81} (1998) 1562.
 
\bibitem{charginos} 
A. Bartl, H. Fraas, W. Majerotto and B. M\"o{\ss}lacher, \ZPC{55} (1992) 257.

\bibitem{garfield}
A. Bartl, E. Christova, T. Gajdosik and W. Majerotto,  
Nucl. Phys. B 507 (1997) 35.

\bibitem{neutralinos} 
A. Bartl, H. Fraas and W. Majerotto, \NPB{278} (1986) 1.


\bibitem{hadron}
M.~Carena, R.L.~Culbertson, S.~Eno, H.J.~Frisch and S.~Mrenna, {\tt hep-ex/9712022};\\
F.\,E.~Paige, BNL-HET-98/1, in \emph{TASI97: Supersymmetry, Supergravity and 
Supercolliders}, Boulder, CO, 1997. 

\bibitem{hikasa-kobayashi}
K.\,I. Hikasa and M.\,Kobayashi, \PRD{36} (1987) 724.

\bibitem{acd}
A.~Arhrib, M.~Capdequi--Peyranere and A.~Djouadi, \PRD{52} (1995) 1404.

\bibitem{ebm}
H. Eberl, A. Bartl and W. Majerotto, \NPB{472} (1996) 481.

\bibitem{lep2paper}
A. Bartl, H. Eberl, S. Kraml, W. Majerotto and W. Porod, \ZPC{73} (1997) 469.

\bibitem{nlcpaper}
A. Bartl, H. Eberl, S. Kraml, W. Majerotto and W. Porod, \ZPC{76} 
(1997) 549.

\bibitem{drees-hikasa}
M. Drees and K.\,I. Hikasa, \PLB{252} (1990) 127.

\bibitem{bhz}
W. Beenakker, R. H\"opker and P.\,M. Zerwas, \PLB{349} (1995) 463.

\bibitem{helmut-diss}
H. Eberl, {\em Strahlungskorrekturen im minimalen supersymmetrischen 
Standardmodell}, PhD thesis, TU Vienna (1998).

\bibitem{peskin}
see \eg, M.\,E. Peskin, in 
{\em Physics at the 100 GeV mass scale},
Proceedings of the 17th SLAC Summer Institute on Particle Physics,
ed. E. C. Brennan, SLAC, Stanford, 1989.

\bibitem{phoisr}
R. Ridolfi, Monte Carlo routine PHOISR, 
LEP2 workshop and private communication.  


\bibitem{lep2workshop}
Proceedings of the Workshop {\em Physics at LEP2},
CERN 96-01, Vol.~1, eds. G. Altarelli, T. Sj\"ostrand and F. Zwirner.

\bibitem{wolfgang}
W. Adam, private communication.

\bibitem{desch-vancouver}
K. Desch, Talk at the XXIX International Conference on 
High Energy Physics (ICHEP'98), Vancouver, Canada, 23--29 July 1998.

\bibitem{pp-vancouver}
P. Rebecchi, Talk at the XXIX International Conference on 
High Energy Physics (ICHEP'98), Vancouver, Canada, 23--29 July 1998.

\bibitem{LEPC50}
E.~Lancon (ALEPH), V.~Ruhlmann--Kleider (DELPHI), 
R.~Clare (L3) and D.~Plane (OPAL), 
talks at the 50th CERN LEPC meeting, 12 Nov. 1998; 
for minutes and transparencies, see 
{\tt http://www.cern.ch/Committees/LEPC/minutes/LEPC50.html}

\bibitem{cernpresslep2}
CERN press release, 19 June 1998.

\bibitem{porod-woermann} 
W. Porod and T. W\"ohrmann, \PRD{55} (1997) 2907. 

\bibitem{werner-diss}
W. Porod, {\em Phenomenology of stops, sbottoms, staus and tau-sneutrino}, 
PhD thesis, Univ. Vienna (1997), {\tt hep-ph/9804208} .

\bibitem{werner-stop3}
W. Porod, {\tt hep-ph/9812230}.


\bibitem{LCph1}
Proceedings of the \emph{1982 DPF Summer Study on Elementary Particle Physics 
  and Future Facilities}, Snowmass, Eds. R.~Donaldson, R.~Gustafson and 
  F.~Paige, Fermilab 1982; \\
Proceedings of the \emph{Workshop on Physics at Future Accelerators},  
  La Thuile, Ed. J.\,H.~Mulvey, CERN Report 87-07;\\
Report, \emph{Oportunities and Requirements for Experimentation at a 
  Very High Energy $e^+e^-$ Collider}, SLAC-329 (1988);\\
H.~Baer, A.~Bartl, D.~Karatas, W.~Majerotto and X.~Tata,
  \emph{Int. J. Mod. Phys} {\bf A4} (1989) 4111.

\bibitem{LCph2}
Proceedings, 
  \emph{Physics and Experiments with Linear Colliders}, 
  Saariselk\"a, Finland, Sep 9--14, 1991, Eds. R.~Orava, P.~Eerola and M.~Nordberg, 
  World Scientific, Singapore (1992);\\
Proceedings, 
  \emph{Physics and Experiments with Linear $e^+e^-$ Colliders}, 
  Waikola, Hawaii, Apr 26--30, 1993, Eds. F.A.~Harris, S.L.~Olsen, S.~Pakvasa 
  and X.~Tata, World Scientific, Singapore (1993);\\
Proceedings, 
  \emph{Physics and Experiments with Linear Colliders: LCWS95}, 
  2 Vol., Morioka--Appi, Japan, Sep 8--12, 1995, Eds. A.~Miyamoto, Y~Fujii, 
  T.~Matsui and S.~Iwate, World Scientific, Singapore (1996).

\bibitem{LCph3}
Proceedings, 
  \emph{$e^+e^-$ Collisions at 500 GeV: The Physics Potential}, 
  Munich--Annecy--Hamburg, Feb~4 -- Sep~3, 1991, Ed. P.\,M.~Zerwas, 
  DESY~92-123A, DESY~92-123B (1992);\\
Proceedings, 
  \emph{$e^+e^-$ Collisions at 500 GeV: The Physics Potential}, 
  Munich--Annecy--Hamburg, Nov~1992 -- Sep~1993, Ed. P.\,M.~Zerwas, 
  DESY~93-123C (1993);\\
Proceedings, 
  \emph{$e^+e^-$ Collisions at TeV Energies: The Physics Potential}, 
  Annecy--Gran~Sasso--Hamburg, Feb--Sep 1995, Ed. P.\,M.~Zerwas, 
  DESY~96-123D (1996);\\
Proceedings, 
  \emph{$e^+e^-$ Linear Colliders: Physics and Detector Studies}, 
  Frascati--London--Munich--Hamburg, Feb--Nov 1996, Ed. R.~Settles, 
  DESY~97-123E (1997);\\
E.~Accomando \etal, \emph{Physics with $e^+e^-$ Linear Colliders}, 
  DESY~97-100, \PRep{299} (1998) 1.

\bibitem{LCph4}
NLC Physics Working Group (S.~Kuhlman \etal), 
\emph{Physics and Technology of the Next Linear Collider: 
A Report Submitted to Snowmass'96}, {\tt hep-ex/9605011}. 

\bibitem{LCinfo}
For the homepages of the various $e^-e^+$ Linear Collider projects and activities, 
see\\ {\tt http://wwwhephy.oeaw.ac.at/p3w/theory/susy/infoboard.html};\\
for a summary of SUSY simulation studies, see \\
{\tt http://hep-www.colorado.edu/${\small\sim}$uriel/SUSY\_studies/}.

\bibitem{nlclumi}
R.~Brinkmann,
2nd ECFA/DESY Study on Physics and Detectors for a Linear 
Electron--Positron Collider, LAL Orsay, France, April 1998.


\bibitem{desy123D}
A.~Bartl, H.~Eberl, S.~Kraml, W.~Majerotto, W.~Porod and A.~Sopczak, 
Proc. of the workshop 
\emph{$e^+e^-$ Collisions at TeV Energies: The Physics Potential}, 
DESY 92-123D, ed. P.~Zwerwas, p. 385.

\bibitem{desy123E}
A.~Bartl, H.~Eberl, T.~Gajdosik, S.~Kraml, W.~Majerotto, 
W.~Porod and A.~Sopczak, 
Proc. of the workshop 
\emph{$e^+e^-$ Linear Colliders: Physics and Detector Studies}, 
DESY 92-123E, ed. P.~Zwerwas, p. 471.


\bibitem{feng-LC} 
J.\,L.~Feng, D.\,L.~Finell, \PRD{49} (1994) 2369.

\bibitem{baer-LC} 
H.~Baer, R.~Munroe, X.~Tata, \PRD{54} (1996) 6735.

\bibitem{daniel-LC}
M. N. Danielson \etal, Proc. of the 1996 Snowmass Workshop on
``New Directions for High Energy Physics''



\bibitem{mu-snowmass}
{\em $\mu^+\mu^-$ Collider, A Feasibility Study}, 
Proceedings of the 1996 DPF/DPB Summer Study on High Energy Physics,
Snowmass 1996.

\bibitem{mu-coll-web}
Muon Collider Collaboration, {\tt http://www.cap.bnl.gov/mumu/}.

\bibitem{mu-ecfa-web}
Prospective Study on Muon Colliders,
{\tt http://nicewww.cern.ch/$\sim$autin/MuonsAtCERN}

\bibitem{mu-ecfa-report}
B.~Autin \etal, Prospective Study of Muon Storage Rings in Europe,
in preparation. 

\bibitem{cernfuture}
J. Ellis, E. Keil and G. Rolandi, CERN-SL-98-004.

\bibitem{bking}
B. King, {\tt http://pubweb.bnl.gov/people/bking/}.

\bibitem{mu-cerncourier}
S. Geer, Cern Courier, Dec. 1997.

\bibitem{palmer}
R.\,B. Palmer, {\tt physics/9802005}.

\bibitem{mumupaper}
A. Bartl, H. Eberl, S. Kraml, W. Majerotto and W. Porod, 
\PRD{58}:115002 (1998).




\bibitem{bmp}
A. Bartl, W. Majerotto and W. Porod, 
\ZPC{64} (1994) 499; erratum ibid. {\bf C68} (1995) 518.

\bibitem{bosdec}
A. Bartl, H. Eberl, K. Hidaka, T. Kon, S. Kraml, W. Majerotto, 
W. Porod and Y. Yamada, \PLB{435} (1998) 118.

\bibitem{desythtalk}
S.~Kraml, talk at the DESY Theory Workshop ``Directions 
Beyond the Standard Model'', 30 Sep. -- 2 Oct. 1998, 
Hamburg, Germany.


\bibitem{qcdnc-paper}
S.~Kraml, H.~Eberl, A.~Bartl, W.~Majerotto and W.~Porod, 
\PLB{386} (1996) 175.

\bibitem{qcdnc-djouadi}
A.~Djouadi, W.~Hollik and C.~J\"unger, \PRD{55} (1997) 6975.

\bibitem{qcd-zerwas2}
W.~Beenakker, R.~H\"opker, T.~Plehn and P.\,M.~Zerwas, 
\ZPC{75} (1997) 349.

\bibitem{qcdwz-paper}
A.~Bartl, H.~Eberl, K.~Hidaka, S.~Kraml, W.~Majerotto, W.~Porod 
and Y. Yamada, \PLB{419} (1998) 243.

\bibitem{qcdhx-djouadi}
A.~Arhib, A.~Djouadi, W.~Hollik and C.~J\"unger, \PRD{57} (1998) 5860.

\bibitem{qcdhx-paper}
A.~Bartl, H.~Eberl, K.~Hidaka, S.~Kraml, W.~Majerotto, W.~Porod 
and Y. Yamada, {\tt hep-ph/9806299}.

\bibitem{qcd-zerwas1}
W.~Beenakker, R.~H\"opker and P.\,M.~Zerwas, 
\PLB{378} (1996) 159.

\bibitem{sola}
J.~Guasch, J.~Sola and W.~Hollik, \PLB{437} (1998) 88.


\bibitem{pave}
G. Passarino and M. Veltman, \NPB{160} (1979) 151.

\bibitem{denner}
A. Denner, {\em Fortschr. Phys.} 41 (1993) 307. 

\bibitem{pierce}
D.\,M. Pierce, {\em Renormalization of Supersymmetric Theories}, 
1997 TASI lectures, {\tt hep-ph/980549}.

\bibitem{tevlimits}
D$\emptyset$ Collab., S. Abachi \etal, Phys. Rev. Lett. 75 (1995) 618;
Phys. Rev. Lett 76 (1996) 2222; \\
CDF Collab., F. Abe \etal, Phys. Rev. D 56 (1997) 1357; \\
J.\,A.~Valls, talk at the XXIX International Conference on High 
Energy Physics (ICHEP98), Vancouver, Canada, 23--29 July 1998, 
FERMILAB-CONF-98-292-E. 

\bibitem{drhonum}
G. Altarelli, R. Barbieri and F. Caravaglios,
 {\em Int. J. Mod. Phys.} A13 (1998) 1031.

\bibitem{Drees-Hagiwara}
M.~Drees and K.~Hagiwara, \PRD{42} (1990) 1709.

\bibitem{Deren-Savoy}
J. P. Derendinger and C. A. Savoy, Nucl. Phys. B237 (1984) 307.


\bibitem{higgsdec}
A.~Bartl, H.~Eberl, K.~Hidaka, T.~Kon, W.~Majerotto,  
and Y. Yamada, \PLB{402} (1997) 303.

\bibitem{yamada}
Y.~Yamada, \PRD{54} (1996) 1150.

\bibitem{dh-private}
A. Djouadi and W. Hollik, private communication.

\bibitem{st2sig}
A. Bartl, K. Hidaka, Y. Kizukuri, T. Kon, and W. Majerotto, 
Phys. Lett. B 315 (1993) 360;\\
A. Bartl, H. Eberl, K. Hidaka, T. Kon, W. Majerotto, and Y.~Yamada, 
Phys. Lett. B 389 (1996) 538.


\bibitem{our-yuk-corr}
H.~Eberl, talk at the 2nd ECFA/DESY Study on Physics and Detectors 
for a Linear Electron--Positron Collider, 3rd meeting, 6-8 Nov. 1998, 
FNL Frascati, Italy;\\
H.~Eberl, S.~Kraml, W.~Majerotto, in preparation.

\bibitem{mumu-photonrad}
D.~Perret--Gallix, private communication. 


\end{thebibliography}
\end{document}